%% file: main.tex
\documentclass[
11pt, 
english, 
onehalfspacing, 
nohyperref, 
headsepline, 
]{MastersDoctoralThesis} 

\usepackage[utf8]{inputenc} 
\usepackage[T1]{fontenc} 
\usepackage{babel}
\usepackage{mathpazo} 

\usepackage[backend=biber,style=authoryear,natbib=true, giveninits=true,maxnames=70, maxcitenames=1,sorting=nyt]{biblatex} 

\usepackage{bm}
\usepackage{amsmath}

\usepackage{pdfpages}

\usepackage{microtype}

\usepackage{hyperref}
\hypersetup{%
  colorlinks = true,
  linkcolor  = black,
  citecolor=black,
  urlcolor=black,
}

\usepackage[autostyle=true]{csquotes} 

\usepackage{makecell}

\defbibenvironment{bibliography}
{\enumerate{}
{\setlength{\leftmargin}{\bibhang}%
\setlength{\itemindent}{-\leftmargin}%
\setlength{\itemsep}{\bibitemsep}%
\setlength{\parsep}{\bibparsep}}}
{\endenumerate}
{\item}

\thesistitle{Automated detection of pronunciation errors in non-native English speech employing deep learning} 
\supervisor{Prof. Bożena \textsc{Kostek}} 
\examiner{} 
\degree{Doctor of Philosophy} 
\author{Daniel \textsc{Korzekwa}} 
\addresses{} 

\subject{Biological Sciences} 
\keywords{} 
\university{\href{http://www.university.com}{Gdansk University of Technology}} 
\department{\href{http://department.university.com}{Faculty of Electronics, Telecommunications and Informatics}} 
\group{\href{http://researchgroup.university.com}{ }} 
\faculty{\href{http://faculty.university.com}{ }} 

\AtBeginDocument{
\hypersetup{pdftitle=\ttitle} 
\hypersetup{pdfauthor=\authorname} 
\hypersetup{pdfkeywords=\keywordnames} 
}

\let\cite\parencite

\begin{document}

\frontmatter 

\pagestyle{plain} 

\includepdf[pages=-]{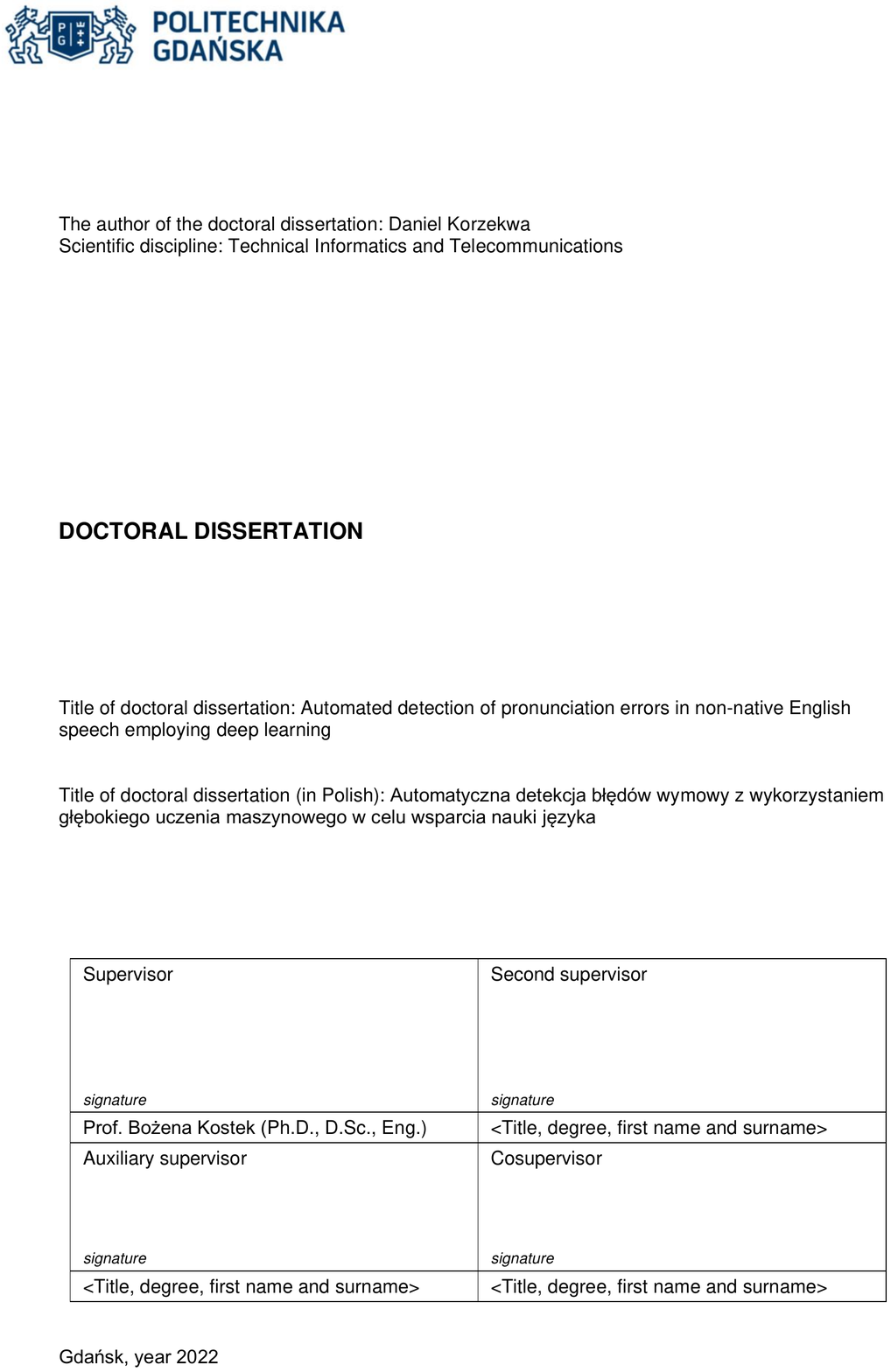}

\include{Chapters/extend_summary_in_polish}

\cleardoublepage

\begin{acknowledgements}
\addchaptertocentry{\acknowledgementname} 

I would like to thank many people who left their mark on this dissertation as well as on me as a person. Prof. Bożena Kostek, I could always count on your advice and you were always available to me no matter what time of day or day of the week. Roberto Barra-Chicote, I still remember our discussion in Cambridge on affective computing and empathetic AI, it was 2018. This discussion prompted me to do my Ph.D. research, and you, Roberto, have been with me all this time. Jaime Lorenzo-Trueba, thank you for our frequent discussions about the results of scientific experiments and research plans for the future. Thomas Drugman, thanks to you, my writing in English has improved significantly and the clarity of my publications is now much better. Szymon Zaporowski, you recorded the corpus of non-native speech, which gave me the data to evaluate my pronunciation error detection models. Shira Calamaro, you helped with the linguistic and phonetic parts of the research, and with understanding the nature of pronunciation errors made by non-native speakers. Grzegorz Beringer, your internship at Amazon on pronunciation error detection and our numerous discussions motivated me to take up this topic in my Ph.D. research. Alicja Serafinowicz, you gave me a unique perspective of an English teacher on computer-assisted pronunciation training, and created a list of words that are often mispronounced by your students. Jasha Droppo, you have advised me how important it is to keep the big picture of the research in mind. I still remember your remark about moving not only forward but in the right long-term direction. Gary Cook, Andrew Breen, Mateusz Łajszczak, Adam Nadolski, Jonah Rohnke, Viacheslav Klimkov, and Kayoko Yanagisava thank you for reviewing my publications and thesis, and providing many constructive comments. Thank you to Amazon company for allowing me to use the Amazon EC2 cloud to conduct research experiments and for 5 weeks to work on the final mile of my Ph.D. thesis. I would like to thank my beloved wife Luiza and my sons Kacper, Mateusz, and Tymoteusz. There are no words to express how grateful I am to them. Only they know how much support and dedication they gave me. Finally, I would like to thank my parents who persistently raised me in difficult times.
\end{acknowledgements}


\listoffigures 
\addchaptertocentry{List of Figures} 

\listoftables 
\addchaptertocentry{List of Tables} 


\begin{abbreviations}{ll} 
\addchaptertocentry{Abbreviations} 
& \textbf{Language Learning and Speech Processing}\\
& \\
\textbf{CALL} & \textbf{C}omputer-\textbf{A}ssisted \textbf{L}anguage \textbf{L}earning \\
\textbf{CAPT} & \textbf{C}omputer-\textbf{A}ssisted \textbf{P}ronunciation \textbf{T}raining \\
\textbf{EPI} & \textbf{E}nglish \textbf{P}roficiency \textbf{I}ndex \\
\textbf{ESL} & \textbf{E}nglish as a \textbf{S}econd \textbf{L}anguage \\
\textbf{GOP} & \textbf{G}oodness \textbf{o}f \textbf{P}ronunciation \\
\textbf{L1} & Native language \\
\textbf{L2} & Non-native language \\
\textbf{MCD} & \textbf{M}el \textbf{C}epstral \textbf{D}istortion \\
\textbf{MDN} & \textbf{M}ispronunciations \textbf{D}etection \textbf{N}etwork \\
\textbf{MDD} & \textbf{M}ispronunciation \textbf{D}etection and \textbf{D}iagnosis \\
\textbf{MOS} & \textbf{M}ean \textbf{O}pinion \textbf{S}core \\
\textbf{MUSHRA} & \textbf{MU}ltiple \textbf{S}timuli with \textbf{H}idden \textbf{R}eference and\textbf{ A}nchor \\
\textbf{PM} & \textbf{P}ronunciation \textbf{M}odel \\
\textbf{PR} & \textbf{P}honeme \textbf{R}ecognizer \\
\textbf{PRN} & \textbf{P}honeme \textbf{R}ecognition \textbf{N}etwork \\
\textbf{TTS} & \textbf{T}ext-\textbf{T}o-\textbf{S}peech \\
&\\
& \textbf{ Math, Stats, and Machine Learning} \\
&\\
\textbf{AUC} & \textbf{A}rea \textbf{U}nder the RO\textbf{C} Curve \\
\textbf{CDF} & \textbf{C}umulative \textbf{D}ensity \textbf{F}unction \\
\textbf{CPT} & \textbf{C}onditional \textbf{P}robability \textbf{T}able \\
\textbf{XOR} & \textbf{E}xclusive \textbf{OR} \\
\textbf{EM} & \textbf{E}xpectation \textbf{M}aximization \\
\textbf{FAR} & \textbf{F}alse \textbf{A}cceptance \textbf{R}ate \\
\textbf{FNR} & \textbf{F}alse \textbf{N}egative \textbf{R}ate \\
\textbf{FP} & \textbf{F}alse \textbf{P}ositives \\
\textbf{FPR} & \textbf{F}alse \textbf{P}ositive \textbf{R}ate \\
\textbf{FN} & \textbf{F}alse \textbf{N}egatives \\
\textbf{FRR} & \textbf{F}alse \textbf{R}ejection \textbf{R}ate  \\
\textbf{GMM} & \textbf{G}aussian \textbf{M}ixture \textbf{M}odel \\
\textbf{GP} & \textbf{G}aussian \textbf{P}rocess \\
\textbf{HMM} & \textbf{H}idden \textbf{M}arkov \textbf{M}odel \\
\textbf{iid} & \textbf{i}ndependently and \textbf{i}dentically \textbf{d}istributed \\
\textbf{KLD} & \textbf{K}ullback–\textbf{L}eibler \textbf{D}ivergence \\
\textbf{ML} & \textbf{M}achine \textbf{L}earning \\
\textbf{PDF} & \textbf{P}robability \textbf{D}ensity \textbf{F}unction \\
\textbf{PGM} & \textbf{P}robabilistic \textbf{G}raphical \textbf{M}odel \\
\textbf{PMF} & \textbf{P}robability \textbf{M}ass \textbf{F}unction \\
\textbf{RBF} & \textbf{R}adial \textbf{B}asis \textbf{F}unction \\
\textbf{TN} & \textbf{T}rue \textbf{N}egatives \\
\textbf{TP} & \textbf{T}rue \textbf{P}ositives \\
\textbf{TPR} & \textbf{T}rue \textbf{P}ositive \textbf{R}ate \\
&\\
& \textbf{Deep Learning}\\
&\\
\textbf{A-RNN} & \textbf{A}ttention-based \textbf{R}ecurrent \textbf{N}eural \textbf{N}etwork \\
\textbf{CNN} & \textbf{C}onvolutional  \textbf{N}eural \textbf{N}etwork \\
\textbf{CTC} & \textbf{C}onnectionist \textbf{T}emporal \textbf{C}lassification \\
\textbf{DGP} & \textbf{D}eep \textbf{G}aussian \textbf{P}rocesses \\
\textbf{DNN} & \textbf{D}eep \textbf{N}eural \textbf{N}etworks \\
\textbf{MLP} & \textbf{M}ulti-\textbf{L}ayer \textbf{P}erceptron \\
\textbf{NF} & \textbf{N}ormalizing \textbf{F}low \\
\textbf{RCNN} & \textbf{R}ecurrent \textbf{C}onvolutional \textbf{N}eural \textbf{N}etwork \\

\textbf{ReLU} & \textbf{R}ectified \textbf{L}inear \textbf{U}nit \\
\textbf{RNN} & \textbf{R}ecurrent \textbf{N}eural \textbf{N}etwork \\
\textbf{VAE} & \textbf{V}ariational \textbf{A}uto\textbf{E}ncoders \\
\textbf{VQ-VAE} & \textbf{V}ector-\textbf{Q}uantized \textbf{V}ariational-\textbf{A}uto-\textbf{E}ncoder \\
&\\
& \textbf{Other terms}\\
&\\
\textbf{GUT} & \textbf{G}dansk \textbf{U}niversity of \textbf{T}echnology \\
\textbf{ITU-R} & \textbf{I}nternational \textbf{T}elecommunication \textbf{U}nion – \textbf{R}adiocommunication\textbf{ S}ector \\
\textbf{UNESCO} & \textbf{U}nited \textbf{N}ations \textbf{E}ducational, \textbf{S}cientific and \textbf{C}ultural \textbf{O}rganization \\
\end{abbreviations}







\begin{symbols}{lll} 
\addchaptertocentry{Symbols} 
$x,y$ & local variables used across different chapter of the Ph.D. Thesis & \\
$\textbf{x},\textbf{y}$ & variables in bold indicate vectors or a list of variables & \\
$f: x \mapsto  y$ & a function mapping from $x$ to $y$ &\\
$y=f(x)$ & a function mapping from $x$ to $y$ & \\

\addlinespace
 \addlinespace

$\{1,2,3\}$ & a set of three numbers & \\
$(0,1)$ & an open interval from 0 to 1, excluding the values of 0 and 1 & \\
$[0,1]$ & a closed interval from 0 to 1, including the values of 0 and 1 & \\ 
$x\in \mathcal{R}^{D}$ & a variable $x$ is a member of a set of $D$-dimensional real numbers & \\
$x \in [0,1]$ & a variable $x$ is a member of a closed interval between 0 and 1 & \\
$\{1..N\} \backslash i$ & a set excluding the $i^{th}$  element \\

\addlinespace
 \addlinespace

$x^T$ & the transpose of the variable & \\
$x \odot y$  & element-wise matrix multiplication \\
($f_1 \circ f_2) (x)$ & $f_1(f_2(x))$ & \\
$|x-y|$ & Euclidean distance & \\
$|K|$ & determinant of matrix K & \\

\addlinespace
 \addlinespace

$p(x)$  & a probability distribution of the variable x & \\
$p(x,y)$  & a joint probability distribution of the variables $x$ and $y$ & \\
$p(x|y)$ & a conditional probability distribution ($x$ conditioned on $y$) & \\
$x \sim p(x|y)$ & a variable follows the probability distribution & \\
$p(y) \sim \int p(x,y)dx$ & a marginal distribution over the variable $y$ & \\
$p(x|y) \propto p(x)p(y|x)$ & the probability distribution p(x|y) is proportional to  & \\
$\mu$ & the mean value of a probability distribution & \\
$\sigma$ & standard deviation of a probability distribution & \\
$\tilde{x}, \tilde{\mu}, \tilde {\sigma_2}$ & a variable with tilde corresponds to the posterior value of that variable & \\
$\sigma^2$ & variance of a probability distribution & \\
$\mathcal{N}(\mu, \sigma^2)$ & Normal (Gaussian) probability distribution & \\
$\mathcal{L}(\theta)$ & the likelihood function parametrized by $\theta$& \\
$\mathcal{I}$ & identity matrix \\
$p-value$ & the probability value in the null-hypothesis statistical test  \\
$t-test$ & a statistical test checking for a significant difference in the two mean values \\ 

\addlinespace
 \addlinespace

$tp$ & the number of true positives,  $tp \in \mathbb{Z}$ \\
$tn$ & the number of true negatives, $tn \in \mathbb{Z}$ \\
$fp$ & the number of false positives, $fp \in \mathbb{Z}$ \\
$fn$ & the number of false negatives, $fn \in \mathbb{Z}$ \\

\addlinespace
 \addlinespace

$e$ & the probability of pronunciation error, $e \in (0,1)$ \\
 $t$ &  a threshold value \\
 $\theta$ & trainable parameters of a machine learning model \\
$\kappa$ & an activation function in neural networks \\
 
 \addlinespace
  \addlinespace


\end{symbols}



\tableofcontents 
\addchaptertocentry{Contents} 

\mainmatter 

\pagestyle{thesis} 


\include{Chapters/Introduction}
\include{Chapters/ResearchMethodology/ResearchMethodology}
\include{Chapters/PronunciationErrorDetection/PronunciationErrorDetection}
\include{Chapters/RelatedApplications}
\include{Chapters/Conclusions}


\appendix 


\include{Appendices/Appendix_author_statements}

\include{Appendices/AppendixC}

\include{Appendices/AppendixA}
\include{Appendices/AppendixB}

\printbibliography[heading=bibintoc,title=References]


\end{document}

%% file: Chapters/extend_summary_in_polish.tex
\chapter*{Rozszerzone streszczenie w j. polskim} 

\label{chapter:extended_abstract_in_polish} 

\newcommand{\tabhead}[1]{\textbf{#1}}


\section*{Cel pracy doktorskiej}

Język odgrywa kluczową rolę w edukacji, dając ludziom dostęp do dużej ilości informacji zawartych w książkach, notatkach i pamiętnikach spisywanych na przestrzeni wieków. Niestety edukacja nie jest dostępna jednakowo dla wszystkich ludzi. Według raportu UNESCO 40\% światowej populacji nie ma dostępu do edukacji w języku, który rozumieją \cite{unesco2016if}. Jeszcze trudniejszy wydaje się przypadek nauki języka obcego, bowiem, w tym przypadku działa zasada: „jeśli nie rozumiesz, jak możesz się uczyć?". Nauka języka wspomagana komputerowo (ang. Computer-Assisted Language Learning (CALL)) \cite{asrifan2020effects} jest jednym z możliwych rozwiązań, które mogą poprawić znajomość języka angielskiego w różnych regionach świata. CALL opiera się na narzędziach komputerowych, które są wykorzystywane przez uczniów do ćwiczenia języka, zwykle języka obcego (w mowie nierodzimej).

\begin{figure}[th]
\renewcommand{\figurename}{Rysunek}
\centering
\includegraphics[width=1\textwidth]{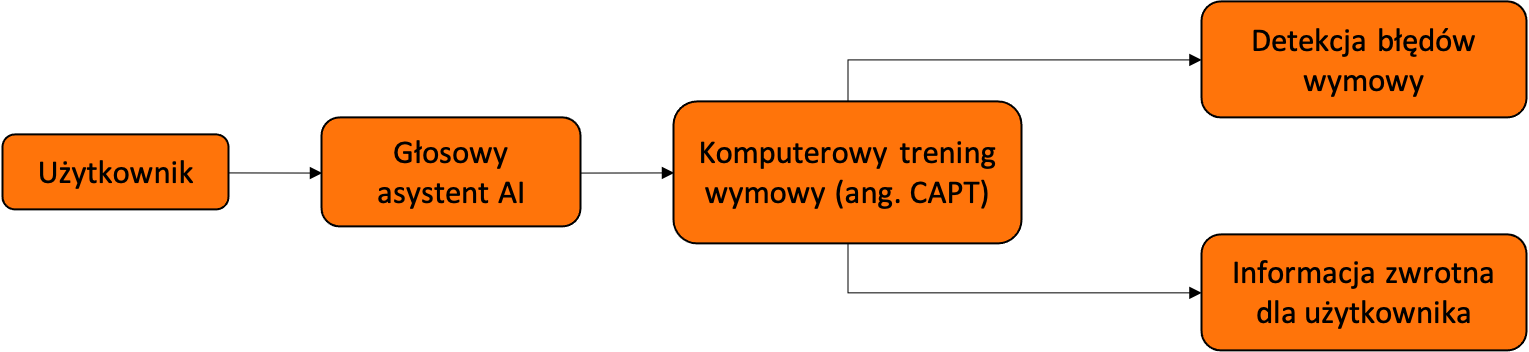}
\decoRule
\caption[]{Ogólny schemat komputerowego systemu do nauki wymowy (ang. CAPT).}
\label{fig:pl_alexa_capt}
\end{figure}

Niniejsza praca doktorska poświęcona jest zagadnieniom związanym z wykrywaniem  błędów w wymowie i treningiem wymowy przez osoby uczące się języka angielskiego (Computer-Assisted Pronunciation Training;  CAPT) \cite{fouz2015trends} - jest to element  systemu CALL. System CAPT składa się z dwóch części: modułu automatycznej oceny wymowy i modułu informacji zwrotnej, jak pokazano na rysunku \ref{fig:pl_alexa_capt}. Moduł automatycznej oceny wymowy jest odpowiedzialny za wykrywanie błędów wymowy, na przykład za wykrywanie niepoprawnie wymawianych fonemów lub słów. Moduł informacji zwrotnej informuje użytkownika o błędnie wymawianych słowach i podpowiada, jak je poprawnie wymówić.

W szczególności, niniejsza rozprawa koncentruje się na automatycznej ocenie wymowy. Pomimo badań poświęconych automatycznej ocenie wymowy prowadzonych intensywnie przez kilka ostatnich dekad, nadal istnieje duży potencjał w kontekście poprawy dokładności automatycznego wykrywania błędów wymowy. Istniejące metody wykrywają błędy wymowy ze stosunkowo niską dokładnością (precyzja rzędu 60\% przy wskaźniku czułości 40\%-80\%) \cite{leung2019cnn,korzekwa2021mispronunciation,zhang2021text}. Wskazywanie poprawnie wymawianych słów jako błędów wymowy przez narzędzie CAPT może zdemotywować osobę uczącą się języka i wpłynąć na jakość nauki. Z kolei, pomijanie błędów wymowy może spowolnić proces uczenia się.

\section*{Tezy i tło badawcze}

W odpowiedzi na cel badawczy, jakim jest poprawa dokładności wykrywania błędów wymowy w nierodzimej (L2) mowie angielskiej, sformułowano podstawową tezę badawczą:

\begin{center}
\textbf{Możliwe jest zwiększenie dokładności metod uczenia głębokiego do wykrywania błędów wymowy w nierodzimej mowie angielskiej poprzez zastosowanie syntetycznego generowania mowy i bezpośredniej detekcji błędów typu end-to-end, które zmniejszają zapotrzebowanie na nagrania i fonetyczną transkrypcję mowy.}
\end{center}

Oprócz podstawowej tezy badawczej, w celu zbadania możliwości uogólniania zaproponowanych metod wykrywania błędów wymowy w pokrewnym obszarze mowy dyzartrycznej, sformułowana została druga teza badawcza.

\begin{center}
\textbf{Metody uczenia głębokiego służące do wykrywania błędów wymowy w nierodzimej mowie angielskiej można przenieść na pokrewne zadania wykrywania i rekonstrukcji mowy dyzartrycznej.}
\end{center}

\subsection*{Wykrywanie błędów wymowy w nierodzimej mowie }

Błąd wymowy w mowie można zdefiniować jako przypadek, kiedy osoba wymawia słowo lub zdanie inaczej niż wymowa oczekiwana według kanonicznej transkrypcji fonetycznej \cite{witt2000phone}. Błędy w wymowie mogą odnosić się np. do błędnie wymawianych fonemów, np. błędne wymówienie  fonemu /eh/ jako /ey/ w zdaniu "I said" /ay s eh d/. Błąd akcentu leksykalnego \cite{ferrer2015classification} to inny rodzaj błędu wymowy, który pojawia się, gdy osoba podkreśla nieprawidłową sylabę w słowie, na przykład, niepoprawne podkreślenie pierwszej sylaby w słowie „remind” /r iy1 m ay0 n d/. Błędy wymowy mogą występować na różnych poziomach szczegółowości, na przykład, na poziomie fonemów \cite{leung2019cnn}, słów \cite{korzekwa21b_interspeech}, lub zdań \cite{Gong2022}.

Wydaje się, że wykrycie błędu wymowy na poziomie fonemów powinno być dobrym rozwiązaniem dla osoby uczącej się, ale system tego typu może okazać się zbyt skomplikowany. Rzadko kiedy osoby uczące się języka znają pojęcie fonemu. Ponadto, automatyczne rozpoznanie wymawianych fonemów nie jest proste \cite{zhang2021text}. Dlatego, nauczyciel języka nie zawsze przekazuje informację zwrotną na poziomie fonemów, zamiast tego, wskazuje źle wymówione słowo i pokazuje, jak je poprawnie wymówić. Podobnie, Asystent CAPT oparty na sztucznej inteligencji może przekazywać użytkownikowi  informację zwrotną za pomocą syntetycznego głosu. Dodatkowo, w ten sposób użytkownik może ćwiczyć umiejętność wymowy za pośrednictwem interfejsu głosowego.

W ramach pracy doktorskiej zaproponowano różne modele do wykrywania zarówno błędnie wymawianych fonemów \cite{BeringerAMLC2020, korzekwa2021mispronunciation, korzekwa21b_interspeech, korzekwa22_speechcomm}, jak i błędów akcentu leksykalnego \cite{korzekwa21_interspeech}, na poziomie fonemów i słów. Jednak kierunek w którym te modele ewoluują – w kierunku wykrywania błędów wymowy na poziomie słowa – jest motywowany przypadkiem użycia ćwiczenia umiejętności wymowy z wykorzystaniem interfejsu asystenta głosowego opartego na sztucznej inteligencji, jak pokazano na rysunku \ref{fig:pl_alexa_capt}.

\bigskip

Na podstawie literatury można zauważyć, że  istniejące metody wykrywania błędów wymowy nie sprawdzają się w różnym kontekście. Obserwacje te prowadzą do nowych modeli głębokiego uczenia się w celu poprawy dokładności wykrywania błędów wymowy i usprawnienia działania narzędzi CAPT:

\begin{enumerate}
\item Transkrypcja mowy obcej jest trudna i kosztowna

Końcowym wynikiem modelu wykrywania błędów wymowy jest prawdo-podobieństwo błędu wymowy na poziomie segmentu, takiego jak fonem lub słowo. Stworzenie modelu, który nie wymaga rozpoznania wypowiedzianych fonemów i bezpośrednio (ang. end-2-end) szacuje to prawdopodobieństwo, może sprawić, że transkrypcje fonetyczne mowy obcej staną się niepotrzebne \cite{zhang2021text, korzekwa21b_interspeech}.

\item Dokładne dopasowanie kanonicznych i rozpoznanych fonemów jest skomplikowane

Aby wykryć błędy wymowy, istniejące metody CAPT rozpoznają wymawiane fonemy, a następnie porównują je z oczekiwaną (kanoniczną) wymową osoby mówiącej w języku rodzimym \cite{witt2000phone,li2016mispronunciation,sudhakara2019improved,leung2019cnn}. Wykrywanie błędów wymowy bezpośrednio przez model (ang. end-to-end) może wyeliminować proces dopasowania fonemów jako potencjalne źródło błędów negatywnie wpływających na dokładność wykrywania błędów wymowy.

\item Nie wszystkie błędy wymowy są tak samo istotne dla osoby uczącej się języka

Niektóre błędy wymowy są bardziej istotne niż inne. Kategoryzacja błędów wymowy według poziomu istotności pozwala zgłaszać osobie uczącej się tylko poważniejsze błędy i zmniejsza ryzyko wykrycia poprawnie wymawianego tekstu jako błędu wymowy \cite{yan20_interspeech, korzekwa21b_interspeech}.   

\item Zdanie można wymówić poprawnie na wiele różnych sposobów

Osoby mówiące w języku rodzimym mogą wymawiać ten sam tekst poprawie na wiele sposobów. Model wykrywania błędów wymowy powinien brać to pod uwagę. Uwzględnienie zmienności w wymowie zmniejszy prawdopodo-bieństwo zgłaszania użytkownikowi fałszywych alarmów dotyczących jego wymowy \cite{qian2010capturing, korzekwa2021mispronunciation}.

\item Ćwiczenie akcentu leksykalnego jest ważną częścią CAPT

Istniejące metody CAPT koncentrują się na ćwiczeniu wymowy fonemów \cite{witt2000phone, leung2019cnn, korzekwa2021mispronunciation}. Niemniej jednak wykazano, że ćwiczenie akcentu leksykalnego poprawia zrozumiałość nierodzimej mowy w języku angielskim \cite{field2005intelligibility,lepage2014intelligibility}. Dobre modele uczenia głębokiego powinny być w stanie wykryć zarówno błędy w wymawianych fonemach, jak i błędy akcentu leksykalnego.

\item Dostępność mowy nierodzimej z błędami wymowy jest ograniczona

Modele uczenia głębokiego działają bardzo dobrze, gdy ilość danych trenin-gowych jest duża \cite{shah21_ssw}. Istnieją dowody w pokrewnej dziedzinie wizji komputerowej, że generowanie obrazów syntetycznych poprawia dokładność modeli klasyfikacyjnych \cite{wong2016understanding}. Dlatego, podobna technika może poprawić dokładność wykrywania błędów wymowy w nierodzimej mowie. Powielanie ilości danych (ang. data augmentation) \cite{badenhorst2017limitations} i generowanie danych (ang. data generation) \cite{lee2016language} to dwie techniki, które pomagają tworzyć syntetyczne błędy wymowy w celu uwzględnienia ograniczonej dostępności nierodzimej mowy z błędami wymowy. Ostatnie postępy w syntezie mowy \cite{fazel21_interspeech} i konwersji głosu \cite{shah21_ssw} otwierają możliwość generowania mowy syntetycznej, która będzie w stanie doskonale naśladować nierodzimą mowę i pozwoli na trenowanie modeli wykrywania błędów wymowy tylko na danych syntetycznych.

\item Wielozadaniowe uczenie maszynowe (ang. multi-tasking) jako podejście do walki z nadmiernym dopasowaniem (ang. overfitting) w metodach głębokiego uczenia się

W wielozadaniowym uczeniu maszynowym, oprócz podstawowego zadania wykrywania błędów wymowy w sygnale mowy, można dodać zadanie drugorzędne, takie jak rozpoznawanie wymawianych fonemów \cite{zhang2021text,korzekwa21b_interspeech}. Oba zadania będą ze sobą współdziałać, dzięki czemu model będzie mniej podatny na nadmierne dopasowanie.
\end{enumerate}

\subsection*{Wykrywanie i rekonstrukcja mowy dyzartrycznej}

Pożądanymi cechami metod uczenia maszynowego są możliwość łatwego uogólnienia oraz skalowalność w kontekście innych powiązanych problemów. Druga teza badawcza ma na celu zbadanie, czy metody głębokiego uczenia można stosować w zadaniach wykrywania i rekonstrukcji mowy dyzartrycznej.

Dyzartria jest motorycznym zaburzeniem mowy, które wynika z zaburzeń neurologicznych, takich jak porażenie mózgowe, udar mózgu/afazja, otępienie i torbiel mózgu \cite{cuny2017neuropsychological,banovic2018communication}. Z powodu uszkodzenia układu nerwowego, połączenia pomiędzy mózgiem a narządem mowy i ich mięśniami ulegają osłabieniu, co skutkuje zniekształceniem mowy \cite{asha2019}. W porównaniu z normalną mową, mowa dyzartryczna jest szorstka i zawiera zwiększoną ilość oddechów, zawiera błędy w wymowie, ma spłaszczoną intonację i zmniejszoną prędkość mówienia. 

Można postawić hipotezę, że modele uczenia głębokiego używane do automatycznego wykrywania błędów wymowy w nierodzimej mowie mogą zostać przeniesione do zadania wykrywania mowy dyzartrycznej, lub szerzej, upośledzonej mowy, takiej jak w chorobie Parkinsona (PD) \cite{korzekwa2019interpretable,romana2021automatically}. Zarówno w nierodzimej, jak i dyzartrycznej mowie, można zaobserwować podobne zniekształcenia mowy, takie jak błędna wymowa fonemów i nieprawidłowe wzorce prozodii. Dlatego wydaje się zasadne postawienie hipotezy badawczej, że podobne modele uczenia głębokiego mogą mieć zastosowanie w obu obszarach.

Osoby z dyzartrią mają trudności z porozumiewaniem się z innymi ludźmi, ponieważ ich mowa jest zniekształcona i mniej zrozumiała. Istnieją podobieństwa pomiędzy generowaniem mowy syntetycznej imitującej nierodzimą mowę w celu poprawy dokładności wykrywania błędów wymowy a rekonstrukcją mowy dyzar-trycznej w celu uczynienia mowy bardziej zrozumiałą. W scenariuszu detekcji błędów wymowy, system konwersji mowy na mowę (ang. speech-to-speech) służy do `niszczenia' poprawnie wymawianej mowy poprzez wprowadzanie błędów wymowy, albo poprzez  podmianę fonemów lub poprzez wprowadzenie niepoprawnego wzorca stresu leksykalnego. W scenariuszu mowy dyzartrycznej, mowa ta jest `naprawiana' tak aby była bardziej płynna, np., poprzez automatyczne usunięcie niepotrzebnych przerw między fonemami i sylabami, oraz aby wypowiedziane fonemy były bardziej  zrozumiałe. Można postawić hipotezę, że podobne techniki uczenia głębokiego powinny być skuteczne w obu scenariuszach.

\section*{Publikacje i wkład naukowy}

W ramach prowadzonych badań powstało sześć publikacji, w których autor rozprawy   jest głównym autorem \cite{korzekwa22_speechcomm, korzekwa21b_interspeech, korzekwa2021mispronunciation, korzekwa21_interspeech, korzekwa2019deep, korzekwa2019interpretable}. Publikacje te są bezpośrednio związane z tezami badawczymi przedstawionymi w rozdziale \ref{sec:research_theses} i stanowią główny wkład naukowy rozprawy doktorskiej.

Dodatkowo, z tematem rozprawy doktorskiej wiąże się dziewięć
publikacji, których współautorem jest Daniel Korzekwa. Pierwsze dwie publikacje poświęcone są tematyce automatycznej detekcji błędów wymowy w nierodzimej mowie \cite{Zhang2022_interspeech, BeringerAMLC2020}. Kolejne sześć publikacji dotyczy syntezy mowy i konwersji głosu, które kładą podwaliny pod generowanie syntetycznych błędów wymowy i rekonstrukcję mowy dyzartrycznej \cite{Bilinski2022_interspeech, merritt22_icassp_vc,jiao2021universal,gabrys2021improving,shah21_ssw,ezzerg2021enhancing}. Dziewiąta publikacja dotyczy kolekcji nienatywnego korpusu mowy, który został wykorzystany do oceny modeli wykrywania błędów wymowy \cite{Weber2020}.

\bigskip

\textbf{Wkład naukowy}

\bigskip

\begin{figure}[th]
\renewcommand{\figurename}{Rysunek}
\centering
\centerline{
\includegraphics[width=1.2\textwidth]{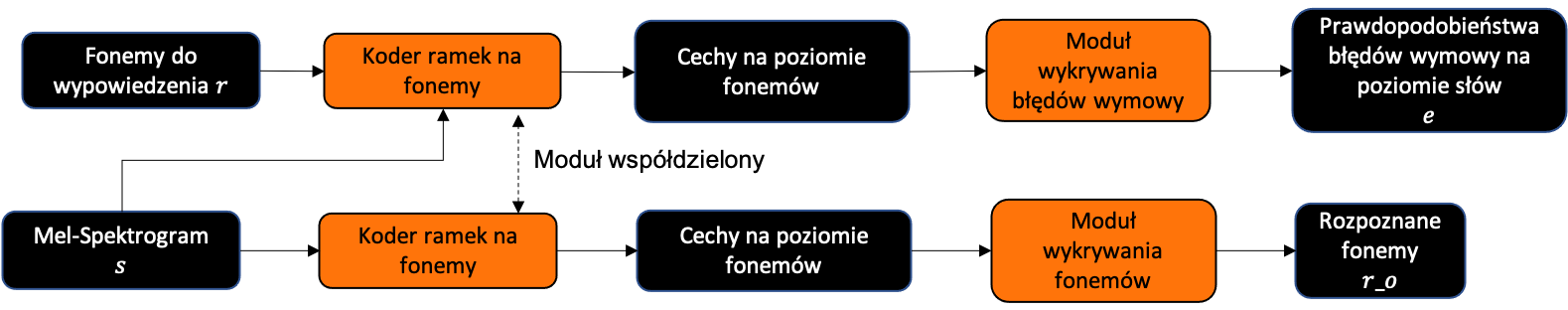}}
\decoRule
\caption[]{Architektura modelu WEAKLY-S do wykrywania błędów wymowy na poziomie wyrazów w konfiguracji wielozadaniowej. Zadanie 1 - wykrycie błędów w wymowie $e$. Zadanie 2 - rozpoznawanie fonemów $r_o$.}
\label{fig:pl_weakly_architecture}
\end{figure}

\begin{figure}[th]
\renewcommand{\figurename}{Rysunek}
\centering
\centerline{
\includegraphics[width=1.20\textwidth]{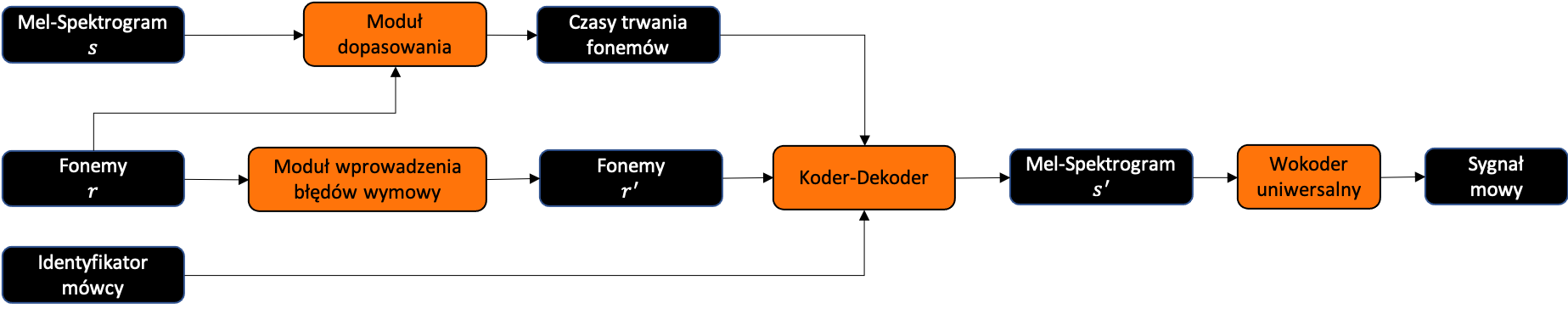}}
\decoRule
\caption[]{Architektura modelu zamiany mowy na mowę (ang. speech-to-speech) do generowania błędnie wymawianej mowy syntetycznej przy zachowaniu prozodii i barwy głosu mowy wejściowej. Czarne prostokąty reprezentują dane (tensory), a pomarańczowe prostokąty reprezentują bloki przetwarzania.}
\label{fig:pl_speech_to_speech_architecture}
\end{figure}

W ramach pracy doktorskiej zaproponowano oraz opracowano wiele nowatorskich metod uczenia głębokiego do wykrywania błędów wymowy w nierodzimej mowie angielskiej, podsumowanych poniżej.
\begin{enumerate}
\item
Wykonywanie transkrypcji fonetycznej nierodzimej mowy jest czasochłonne, a niekiedy transkrypcja ta jest niemożliwa ze względu na różnice pomiędzy językami mówionymi. Zaproponowano nową metodę do bezpośredniego (ang. end-2-end) wykrywania błędów wymowy, o nazwie WEAKLY-S (Weakly-supervised), pokazanej na rysunku \ref{fig:pl_weakly_architecture}. Ze względu na bezpośrednią detekcję błędów wymowy, metoda ta nie wymaga transkrypcji fonetycznej nierodzimej mowy \cite{korzekwa21b_interspeech}.

\item
Istniejące metody wykrywania błędów wymowy dopasowują kanoniczne i rozpoznane sekwencje fonemów w celu identyfikacji błędnie wymawianych segmentów mowy, takich jak fonemy i słowa. Wszelkie niedokładności wprowadzone w procesie dopasowania obniżyłyby dokładność wykrywania błędów wymowy. Zaproponowana metoda WEAKLY-S do bezpośredniego wykrywania błędów wymowy nie wymaga dopasowywania kanonicznych i rozpoznawanych sekwencji fonemów \cite{korzekwa21b_interspeech}. Metoda ta zwiększa dokładność wykrywania błędów wymowy w metryce AUC (Area under the Curve) nawet o 30\% w porównaniu do istniejących metod w literaturze.

\item
Istnieją dwa czynniki, które mogą wpływać na dokładność wykrywania błędów wymowy. Po pierwsze, to samo zdanie można wymówić na wiele poprawnych sposobów, co nie powinno powodować detekcji błędu wymowy. Po drugie, dokładne rozpoznanie fonemów wymawianych przez osobę uczącą się jest trudne i należy wziąć to pod uwagę. W tym celu zaproponowano nową metodę uwzględniającą: i) wiele poprawnych sposobów wymawiania tego samego zdania oraz ii) niepewność rozpoznawania fonemów \cite{korzekwa2021mispronunciation}. Zaproponowana metoda zwiększa precyzję wykrywania błędów w wymowie nawet o 18\% w porównaniu z istniejącym podejściem.

\item
Istniejące metody wykrywania błędów wymowy często opierają się na ekstrahowaniu cech mowy, takich jak f0, energia, dopasowanie fonemów do sygnału mowy, itd. W pracy zaproponowano nową metodę wykrywania błędów wymowy opartą na mechanizmie uwagi (ang. attention mechanism) do automatycznego wyodrębniania optymalnych cech mowy \cite{korzekwa21_interspeech}. Mechanizm uwagi odgrywa istotną rolę w proponowanych modelach głębokiego uczenia stosowanych do wykrywania niepoprawnie wymówionych fonemów i błędów akcentu leksykalnego.

\item
Mechanizm uwagi pomaga rozłożyć model głębokiego uczenia się na wiele zależnych składników w procesie zwanym faktoryzacją. Faktoryzacja prowadzi do lepszej interpretacji modelu głębokiego uczenia, na przykład, wizualizacji modelu do wykrywania błędów akcentu leksykalnego w celu lepszego zrozumienia, jak działa taki model i w jaki sposób podejmuje decyzje \cite{korzekwa21_interspeech}. Uczenie wielozadaniowe to rodzaj faktoryzacji modelu, który może sprawić, że model głębokiego uczenia będzie bardziej niezawodny i mniej podatny na nadmierne dopasowanie \cite{korzekwa21b_interspeech}. Zaproponowano wielozadaniowy model wykrywania błędów wymowy WEAKLY-S z dwoma zadaniami, a) rozpoznawanie fonemów i b) wykrywanie błędów wymowy, co zwiększa dokładność tego modelu. Faktoryzacja może również przybrać formę interpretowalnej warstwy ukrytej w modelu głębokiego uczenia (ang. latent space or hidden space), która może być wykorzystana do modyfikacji określonych cech sygnału. Na przykład, może uczynić mowę dyzartryczną bardziej płynną i zrozumiałą \cite{korzekwa2019interpretable}.

\item
Dostępność nierodzimej mowy jest ograniczona, a jej nagranie/zbieranie i wykonanie transkrypcji fonetycznej  jest czasochłonne. Odwołując się do teorii prawdopodobieństwa i reguły Bayesa, problem wykrywania błędów wymowy został przeformułowany jako zadanie generowania mowy, jak pokazano na rysunku \ref{fig:pl_speech_to_speech_architecture}. Intuicyjnie, w przypadku nieograniczonej ilości mowy syntetycznej, która mogłaby naśladować nierodzimą mowę, modele uczenia głębokiego do wykrywania błędów wymowy byłyby mniej podatne na nadmierne dopasowanie. Najlepsza zaproponowana metoda generowania nierodzimej mowy (ang. speech-to-speech) zwiększa dokładność wykrywania błędów wymowy w metryce AUC o 41\%, z 0.528 do 0.749, w porównaniu z istniejącym podejściem \cite{korzekwa22_speechcomm}.

\end{enumerate}

\section*{Zastosowanie}

\textbf{Modele wykrywania błędów wymowy}

\bigskip

Zaproponowane modele CAPT do wykrywania błędów wymowy w nierodzimej mowie angielskiej zastosowano do automatycznego wykrywania błędów wymowy w mowie syntetycznej w dwóch scenariuszach: 1) podczas wnioskowania (ang. during inference) i 2) podczas treningu modeli syntezy mowy. Celem modelu CAPT podczas wnioskowania jest automatyczna ocena jakości mowy generowanej przez modele syntezy mowy, to znaczy czy mowa jest zrozumiała i nie zawiera błędów wymowy. Po wytrenowaniu modelu syntezy mowy, duża liczba wypowiedzi jest syntetyzowana i automatycznie przetwarzana przez model wykrywania błędów wymowy. Automatyczne wykrywanie błędów wymowy umożliwia ocenę głosów syntetycznych na dużą skalę i znacznie zmniejsza liczbę testów odsłuchowych, które są przeprowadzane przez słuchaczy. Podczas trenowania, model wykrywania błędów wymowy jest używany do pomiaru czy wygenerowana mowa zawiera odpowiednie fonemy, dzięki czemu model syntezy mowy wygeneruje bardziej zrozumiałą mowę.

\bigskip

\textbf{ Synteza i konwersja mowy}
 
\bigskip

Systemy syntezy mowy i konwersji głosu składają się z dwóch części, modułu generowania kontekstu, który generuje spektrogram w skali melowej z wejściowego tekstu i/lub wejściowego sygnału mowy, oraz modułu wokodera, który wytwarza surowy sygnał mowy w dziedzinie czasu na podstawie spektrogramu w skali melowej. Oba komponenty zostały wdrożone na urządzeniach Alexa i obsługują miliony użytkowników Amazon na całym świecie. Ponadto, mowa syntetyczna generowana przez modele syntezy mowy i konwersji głosu została wykorzystana podczas trenowania modeli CAPT do wykrywania błędów wymowy, poprawiając ich dokładność.

\section*{Wnioski}
W ramach badań związanych z doktoratem opracowano nowatorskie metody głębokiego uczenia w celu automatycznego wykrywania błędów wymowy w nierodzimej (drugi język - L2) mowie angielskiej. Przeprowadzono rozległe eksperymenty, aby zmierzyć skuteczność proponowanych metod w CAPT. Do oceny zaproponowanych metod wykorzystano nierodzimą mowę angielską osób głównie posługujących się rodzimym językiem niemieckim, włoskim i polskim. Zarejestrowano dwa korpusy nierodzimej mowy angielskiej osób z rodzimym językiem słowiańskim i bałtyckim \cite{Weber2020}. Najlepsza zaproponowana metoda poprawia dokładność wykrywania błędów wymowy w metryce AUC o 41\%, z 0.528 do 0.749, w porównaniu z istniejącym podejściem \cite{korzekwa22_speechcomm}. Odpowiada to 80.45\% w metryce precyzji i 40.12\% w metryce czułości. Biorąc pod uwagę tylko poważne błędy wymowy, według subiektywnej oceny osób natywnie posługujących się językiem angielskim, AUC wzrasta z 0.749 do 0.834, co odpowiada 93.54\% precyzji i 40.15\% czułości. Dwie najważniejsze techniki zastosowane w tej metodzie to: 1) bezpośrednia detekcja błędów wymowy (ang. end-to-end) oraz 2) wykorzystanie techniki zamiany mowy na mowę (ang. speech-to-speech) do generowania syntetycznej mowy z błędami wymowy. Obie techniki zmniejszają zapotrzebowanie na nagrania i fonetyczną transkrypcję mowy, która jest potrzebna do trenowania modeli CAPT. Osiągnięcia te pozwoliły na udowodnienie pierwszej (głównej) tezy badawczej. 

Aby zbadać możliwości uogólniania, opracowane techniki uczenia głębokiego do wykrywania błędów wymowy zostały z powodzeniem zastosowane w pokrewnym obszarze wykrywania i rekonstrukcji mowy dyzartrycznej \cite {korzekwa2019interpretable}. Zaproponowano model autoenkodera (ang. autoencoder; uczenie nienadzorowane), aby przekodować cechy mowy dyzartrycznej na przestrzeń utajoną (ang. latent space). Kontrolując utajoną reprezentację, można poprawić płynność mowy, np., poprzez automatyczne usunięcie niepotrzebnych przerw pomiędzy fonemami i sylabami. Kontrola ta polega na automatycznym znalezieniu wektora przesunięcia w przestrzeni utajonej, który sprawi, że mowa stanie się bardziej płynna przy jednoczesnym zachowaniu innych parametrów mowy takich jak barwa głosu czy wypowiedziane fonemy. Utajoną reprezentację można wykorzystać do wykrywania mowy dyzartrycznej na poziomie słów z precyzją na poziomie 83.1\% i czułością na poziomie 91.1\%. Nowe techniki głębokiego uczenia zostały z powodzeniem zastosowane w temacie mowy dyzartrycznej, potwierdzając walidację drugiej tezy badawczej.

\section*{Plan na przyszłość}

W trakcie pracy doktorskiej wyłoniło się wiele interesujących kierunków badawczych. Najbardziej przyszłościowym pomysłem jest kontynuacja badania nad przeformuło-waniem problemu wykrywania błędów wymowy jako zadania generowania mowy \cite{korzekwa22_speechcomm}. Zaproponowana metoda zamiany mowy na mowę (ang. speech-to-speech, S2S) może generować syntetyczną błędną wymowę, ale nie jest w stanie w pełni naśladować mowę nierodzimą. Aby udoskonalić metodę S2S, należy stworzyć uniwersalny model, aby generować dowolny rodzaj mowy. Model ten powinien być w stanie przekształcić rodzimą mowę w nierodzimą mowę, odzwierciedlając tożsamość głosu, prozodię, styl mówienia i wymowę w nierodzimej mowie. Takie podejście może sprawić, że bazy danych mowy typu L2 (mowa nierodzima) będą zbędne, ponieważ model wykrywania błędów wymowy będzie trenowany tylko na danych syntetycznych. 

Innymi interesującymi kierunkami badawczymi jest zbadanie nienadzorowanych reprezentacji mowy, takich jak Wav2vec \cite {peng21e_interspeech}, oraz przeprowadzenie wielomodalnego (ang. multi-modal) wykrywania błędów wymowy poprzez wykorzystanie z audiowizualnych korpusów mowy \cite{czyzewski2017audio, oneata2022improving}. 

Przyszłe prace skoncentrują się również na opracowaniu kompletnego systemu CAPT opartego na sztucznej inteligencji w celu podniesienia znajomości języków obcych na świecie, nie tylko języka angielskiego. W tym celu powinien zostać utworzony agent konwersacyjny oparty na sztucznej inteligencji. Agent ten będzie składał się z dwóch elementów: modułu wykrywania błędów wymowy oraz modułu informacji zwrotnej. Moduł wykrywania błędów wymowy będzie oparty na wynikach badań zawartych w rozprawie doktorskiej, natomiast moduł informacji zwrotnej będzie wymagał dodatkowych badań. System CAPT będzie kontrolowany tylko za pomocą interfejsu głosowego, a uczeń będzie miał wrażenie zajęć prowadzonych przez nauczyciela języka obcego.

%% file: Chapters/Introduction.tex

\chapter{Introduction} 

\label{chapter:introduction} 


\section{Problem statement}
Language is a way of communication between people. Currently, about 7,139 languages are spoken in the world, English being the most dominant one with 1.348 billion speakers \cite{Ethnologue2021}. English has its written and spoken versions \cite{denham2012linguistics}. Written language is based on words and sentences made out of symbols called letters. Spoken language enables people to communicate verbally by producing a stream of sounds called phones that represent spoken words and sentences.

Language plays a key role in education, giving people access to a large amount of information contained in books, notes, and diaries written down through the ages. Thanks to spoken language, people can participate in interactive discussions with teachers and engage in lively brainstorming with other people. In the age of the Internet and online education, people can access books, articles, video lectures, and even get a university degree from almost anywhere in the world.

Unfortunately, education is not equally accessible to everybody. Regarding the UNESCO report, 40\% of the global population does not have access to education in a language they understand \cite{unesco2016if}.  ‘If you don’t understand, how can you learn?' the report says. This situation could be improved by popularizing education in the student's mother (native) language, but with more than seven thousand languages in the world, it may not be possible to increase the access to education significantly. 

Another approach to increasing access to education is to ensure that people learn at least one foreign language, such as English. Learning multiple languages has benefits beyond having access to better education. It has been reported that multilingualism can boost economic growth \cite{economicforum2018}, help find a better job \cite{EPI2020}, and protect against cognitive decline \cite{kroll2017benefits}. 

However, learning a foreign language seems easier than it is. The study by EF Education First \cite{EPI2020} shows a large disproportion in English proficiency in different countries and continents. The lowest English proficiency is in the Middle East region that falls into the 'very low' category of language proficiency. People falling into this category are not able to navigate an English-speaking country or understand a simple email from a colleague. The description of the English proficiency categories is shown in Table \ref{tab:epi_description}, while Table \ref{tab:epi_by_region} shows English proficiency by region. 

\begin{table}[h]
\caption{Description of EF English Proficiency Index (EPI) \cite{EPI2020}}
\label{tab:epi_description}
\centering
\begin{tabular}{l l}
\toprule
\tabhead{Proficiency Index (EPI) } & \tabhead{Sample tasks}\\
\midrule

\makecell[l]{\textbf{Very High} \\ Netherlands \\ Singapore \\ Sweden} &\makecell[l]{Use nuanced and appropriate language in social situations \\ Read advanced texts with ease \\ Negotiate a contract with a native English speaker}\\
\hline

\makecell[l]{\textbf{High} \\Hungary \\Kenya \\Philippines} & \makecell[l]{Make a presentation at work \\ Understand TV shows \\
Read a newspaper}\\
\hline

\makecell[l]{\textbf{Moderate} \\
China \\ Costa Rica \\Italy} &\makecell[l]{Participate in meetings in one’s area of expertise \\ Understand song lyrics \\ Write professional emails on
familiar subjects} \\
\hline
 
\makecell[l]{\textbf{Low} \\
Dominican Republic \\ Pakistan \\ Turkey} & \makecell[l]{Navigate an English-speaking country as a tourist \\ Engage in small talk with colleagues \\ Understand simple emails from colleagues}\\
\hline

\makecell[l]{\textbf{Very low}
\\ Cambodia \\Tajikistan \\United Arab Emirates} & \makecell[l]{Introduce oneself simply \\ Understand simple signs \\ Give basic directions to a foreign visitor}\\

\bottomrule\\
\end{tabular}
\end{table}

\begin{table}[h!]
\caption{English proficiency by region \cite{EPI2020}}
\label{tab:epi_by_region}
\centering
\begin{tabular}{l l}
\toprule
\tabhead{Region} & \tabhead{English Proficiency Index (EPI)}\\
\midrule
Europe & high\\
Asia & low\\
Africa & low\\
Latin America & low \\
Middle East & very low\\
\bottomrule\\
\end{tabular}
\end{table}

\section{Aim of the thesis}

Computer-Assisted Language Learning (CALL) \cite{asrifan2020effects} is a possible solution to improve English proficiency in different regions. CALL is based on self-service computer-based tools that are used by students to practice a language, usually a foreign (non-native) language. In CALL, students can practice multiple aspects of the language including grammar, vocabulary, writing, reading, and speaking. CALL can complement traditional language learning provided by teachers. It also has a chance to democratize second-language learning in places where traditional ways of learning languages are not possible due to the costs of learning or the lack of access to foreign language teachers.

\begin{figure}[th]
\centering
\includegraphics[width=1\textwidth]{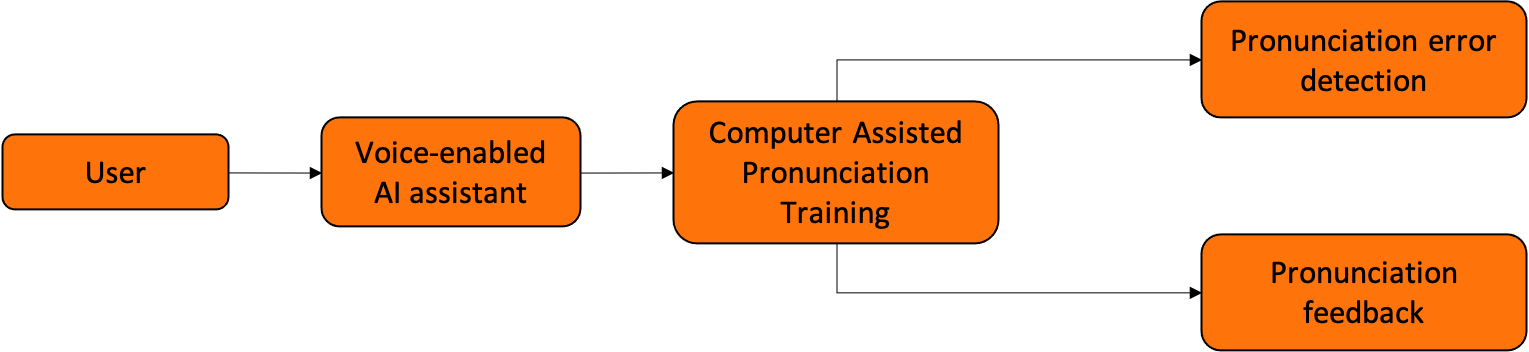}
\decoRule
\caption[Overview of the Computer-Assisted Pronunciation Training System]{Overview of the Computer-Assisted Pronunciation Training System.}
\label{fig:alexa_capt}
\end{figure}

This Ph.D. thesis has been completed within the  “Implementation doctorate” program, carried out by the Gdańsk University of Technology, and written in agreement with the Amazon company employing the Ph.D. candidate. It is devoted to CALL in the task of learning pronunciation skills by non-native speakers of English, also known as Computer-Assisted Pronunciation Training (CAPT) \cite{fouz2015trends}. In general, a CAPT system consists of two components: an automated pronunciation assessment component and a feedback component. The automated pronunciation assessment component is responsible for detecting pronunciation errors in the pronounced speech, for example, for detecting phonemes or words pronounced by the speaker incorrectly. The feedback component informs the speaker about mispronounced words and advises on how to pronounce them correctly. In the future, the CAPT system may be integrated into a voice-enabled AI assistant to let people practice pronunciation skills using a voice interface, as illustrated in Figure \ref{fig:alexa_capt}.

In particular, this dissertation focuses on the automated pronunciation assessment in CAPT. Despite decades of work in the scientific community devoted to automated pronunciation assessment, there is still a great potential to improve the accuracy to detect pronunciation errors in speech automatically. State-of-the-art methods detect pronunciation errors with a relatively low accuracy of 60\% precision at 40\%-80\% recall \cite{leung2019cnn,korzekwa2021mispronunciation,zhang2021text}. Highlighting correctly pronounced words as pronunciation errors by a CAPT tool can demotivate the language learner and affect the quality of learning. In contrast, missing pronunciation errors can slow down the learning process. The ultimate motivation behind the doctoral dissertation is twofold:

\begin{enumerate}

\item
\textbf{To raise foreign language proficiency in the global population by improving the accuracy of automated pronunciation assessment in CAPT.}

\item
\textbf{To apply the results obtained in the doctoral dissertation at Amazon company as the thesis was realized within the “Implementation doctorate” program, carried out by the Gdańsk University of Technology.}
\end{enumerate}

This thesis is organized as follows:
\begin{itemize}
\item Chapter \ref{chapter:introduction}: Introduction - The Ph.D. thesis begins with presenting the problem statement, motivation, and the aim of the thesis. Next, the subject of the dissertation is translated into research theses as well as the research background is presented. At the end of the chapter, a summary of the Ph.D. scientific contribution is included in the form of the most important co-authored publications presented at international conferences and in scientific journals.
\item Chapter \ref{chapter:research_methodology}: Research methodology - This chapter covers the fundamental topics related to the Ph.D. research, including speech generation process, the probability theory, probabilistic machine learning, deep learning, and evaluation metrics. This material lays the foundations for deep learning methods in automated detection of pronunciation errors, presented in the next chapter.
\item Chapter \ref{chapter:pron_error_detection}: Pronunciation error detection - This chapter constitutes the main scientific part of 1the doctoral dissertation. Original deep learning methods for detecting pronunciation and lexical stress errors in non-native English speech are presented. 
\item Chapter \ref{chapter:dysarthric_speech}: Generalization of deep learning methods for pronunciation error detection - This chapter explores the generalization capabilities of deep learning methods for detecting pronunciation errors in two related tasks: detection and reconstruction of dysarthric speech.
\item Chapter \ref{chapter:conclusions}: Summary and Conclusions - The final chapter summarizes the doctoral dissertation, presents the main conclusions, and draws a plan for the future.
\item References and Appendices.
\end{itemize}

\section{Research theses and background}
\label{sec:research_theses}

To address the research goal, which is to improve the accuracy of detecting pronunciation errors in non-native English speech, the primary research thesis is formulated. The primary aim of this Ph.D. work is to establish a new state-of-the-art deep learning method for the detection of pronunciation errors in non-native English, so the thesis is formulated as follows:

\begin{center}
\textbf{1. It is possible to improve the accuracy of deep learning methods for detecting pronunciation errors in non-native English by employing synthetic speech generation and end-to-end modeling techniques that reduce the need for phonetically transcribed mispronounced speech.}
\end{center}

In addition to the primary research thesis, the secondary research thesis is formulated to investigate the generalization capabilities of the invented methods of pronunciation error detection in the related area of dysarthric speech.

\begin{center}
\textbf{2. Deep learning methods for the detection of pronunciation errors in non-native speech are transferable to the related tasks of detection and reconstruction of dysarthric speech.}
\end{center}

\subsection{Pronunciation error detection in non-native speech}

\textbf{What are pronunciation errors and pronunciation error detection?}

\bigskip

A pronunciation error in speech occurs when a speaker pronounces a word or a sentence differently from the expected pronunciation provided by the canonical phonetic transcription \cite{witt2000phone}. Mispronunciations may refer to incorrectly pronounced phonemes, e.g., mispronouncing the phoneme /eh/ as /ey/ in the English sentence `I said' /ay s eh d/. 

Phonemes are abstract symbols that correspond to the mental representation of the pronunciation of a word. Phonemes are related to phones that  correspond to specific sounds made by a speaker. The way a word is pronounced is determined by its phonetic transcription. For example, the word ‘cat’ is transcribed as [k ae t] and the word ‘cell’ is transcribed as [s eh l]. Phoneme transcription is represented with slashes //, e.g., /s eh l/ as opposed to using brackets [] for phones. A more detailed description of a speech production process is presented in Section \ref{sec:speech_production}.

Lexical stress error \cite{ferrer2015classification} is another type of pronunciation error that occurs when a speaker stresses an incorrect syllable in a word, e.g., incorrectly stressing the first syllable in the word `remind' /r iy1 m ay0 n d/. Pronunciation errors can exist at different levels of granularity, for example, at the level of phonemes \cite{leung2019cnn}, words \cite{korzekwa21b_interspeech} and utterances \cite{Gong2022}.

Apparently, detecting a pronunciation error at the phoneme level provides a user with the most informative feedback, but it is more complicated. Not all language learners are familiar with the concept of a phoneme; secondly, sometimes, it may be very difficult to recognize the phoneme pronounced by a user \cite{zhang2021text}. Therefore, language teachers do not always provide users with the phoneme-level feedback. Instead, they simply point out a mispronounced word and use their voice to show how to pronounce it correctly. AI-based CAPT assistants can provide similar verbal feedback to a user using their synthetic voices. In this way, a user can practice pronunciation skills just from the comfort of the couch via the voice interface.

Within the Ph.D. thesis, various models were built to detect both mispronounced phonemes \cite{BeringerAMLC2020, korzekwa2021mispronunciation, korzekwa21b_interspeech, korzekwa22_speechcomm} and lexical stress errors \cite{korzekwa21_interspeech}, at the phoneme and word levels. However, the direction in which these models are evolving - towards detecting pronunciation errors at the word level - is motivated by the use case of practicing pronunciation skills based on AI-based voice assistant interface, as shown in Figure \ref{fig:alexa_capt}.

\bigskip

\textbf{How deep learning methods to detect pronunciation errors may be improved?}

\bigskip

Deep learning is often considered a universal machine that can automatically solve any problem if sufficient training data are available. However, deep learning models are generally data-hungry \cite{lake2015human,marcus2018deep}. They work well for speech processing tasks but require a large amount of training data to generalize to unseen data \cite{shah21_ssw}. In pronunciation error detection, existing deep learning methods detect pronunciation errors with a relatively low accuracy of 60\% precision at 40\%-80\% recall \cite{leung2019cnn,korzekwa2021mispronunciation,zhang2021text}. Many interesting statements can be made about existing methods of detecting pronunciation errors. These statements can lead to new designs of deep learning models to improve the accuracy of  pronunciation error detection models and ultimately improve the CAPT user experience.

These statements that constitute the background of this Ph.D. work are as follows:
\begin{enumerate}
\item Transcription of non-native speech is a difficult and costly process

The end result of the pronunciation error detection model is the probability of a pronunciation error at the segment level, such as a phoneme or a word. Creating an end-2-end model \cite{zhang2021text} that directly estimates this probability could make phonetic transcriptions of non-native speech redundant \cite{korzekwa21b_interspeech}.

\item Aligning canonical and recognized phonemes accurately is challenging

To detect pronunciation errors, existing methods recognize pronounced phonemes and then compare them with the expected (canonical) pronunciation of a native speaker \cite{witt2000phone,li2016mispronunciation,sudhakara2019improved,leung2019cnn}. Detecting pronunciation errors directly by an end-to-end model could eliminate the alignment as a potential source of errors affecting the accuracy of detecting pronunciation errors.

\item Not all pronunciation errors are the same

Some pronunciation errors are more severe than others. Categorizing pronunciation errors by severity level allows reporting only more severe errors to the user and reduces the risk of correctly pronounced text being detected as a pronunciation error \cite{yan20_interspeech, korzekwa21b_interspeech} 

\item A sentence can be pronounced correctly in multiple different ways

Native speakers can pronounce the same text in many correct ways. The pronunciation error detection model should take this observation into account and allow a language learner to pronounce the same text in different ways. Taking into account the variability of pronunciation will reduce the likelihood of reporting false pronunciation alarms to the user \cite{qian2010capturing, korzekwa2021mispronunciation}

\item Practicing lexical stress is an important part of CAPT

Existing CAPT methods concentrate on practicing the pronunciation of phonemes \cite{witt2000phone, leung2019cnn, korzekwa2021mispronunciation}. Nevertheless, it has been shown that practicing lexical stress improves the intelligibility of non-native English speech \cite{field2005intelligibility,lepage2014intelligibility}. Good deep learning models in CAPT should be capable of detecting both pronunciation and lexical stress errors.

\item The availability of non-native speech with pronunciation errors is limited

Deep learning models work very well when the amount of training data is large \cite{shah21_ssw}. There is evidence in the related field of computer vision that generating synthetic images improves the accuracy of classification models \cite{wong2016understanding}. Therefore, a similar technique may improve the accuracy of detecting pronunciation errors in non-native speech. Data augmentation \cite{badenhorst2017limitations,fu2022improving} and data generation \cite{lee2016language} are two techniques that can create synthetic pronunciation errors to account for the limited availability of non-native speech with pronunciation errors. Recent advances in speech synthesis \cite{fazel21_interspeech} and voice conversion \cite{shah21_ssw} open the door to the generation of synthetic speech, which eventually may be able to mimic non-native human speech perfectly and enable training pronunciation error detection models only on synthetic data.

\item Multi-task learning as an approach to tackling overfitting in deep learning methods

In multi-tasking, in addition to the primary task of detecting pronunciation errors in a speech signal, a secondary task can be added, such as recognizing pronounced phonemes \cite{zhang2021text,korzekwa21b_interspeech}. Both tasks will interact, making the model less prone to overfitting.

\end{enumerate}
To summarize the research thesis on pronunciation error detection, this Ph.D. research explores various deep learning methods related to probabilistic machine learning, multi-tasking, and data generation techniques. It has been hypothesized that by using these techniques, it should be possible to improve the accuracy of the state-of-the-art methods of detecting pronunciation errors. The proposed models for detecting pronunciation errors are evaluated on the non-native speech of multiple nationalities, including German, Italian, and Polish speakers, including a new corpus of non-native speech \cite{Weber2020} recorded at the Gdańsk University of Technology (GUT) to facilitate these evaluations. 

\subsection{Detection and reconstruction of dysarthric speech}
\label{subsec:related_research_thesis}

Good machine learning methods should be generic and scale to other related problems. The secondary research thesis aims to investigate whether deep learning methods can be transferred to the tasks of detecting and reconstructing dysarthric speech. 

\bigskip

\textbf{Detection of dysarthric speech}

\bigskip

Speech production begins in the brain, where the mental representation of a message is formed as a sequence of abstract symbols called phonemes. The brain then controls the speech organs to generate a spoken message. The lungs generate air that flows through the larynx, and the oral and nasal cavities, generating speech. Multiple muscles are involved in this process, such as lips, throat (pharynx), and jaw \cite{trujillo2006production}. 

Dysarthria is a motor speech disorder that results from neurological disorders such as cerebral palsy, brain stroke/aphasia, dementia, and brain cyst \cite{cuny2017neuropsychological,banovic2018communication}. Due to damage to the nervous system, the connections between the brain and the speech organs and their muscles are weakened, resulting in distorted speech \cite{asha2019}. Compared to normal speech, dysarthric speech is harsh and breathy, contains mispronunciations, has flattened intonation, and has a lower speech rate.

It can be hypothesized that deep learning models used to automatically detect pronunciation errors in non-native speech can be transferred to the dysarthric speech detection task, or more broadly, impaired speech, such as in Parkinson’s disease (PD) \cite{korzekwa2019interpretable,romana2021automatically}. In both non-native and dysarthric speech, similar distortions of speech, such as mispronunciations and incorrect prosody patterns, can be observed. Therefore, similar deep-learning models should apply in both areas.

\begin{figure}[th]
\centering
\includegraphics[width=1\textwidth]{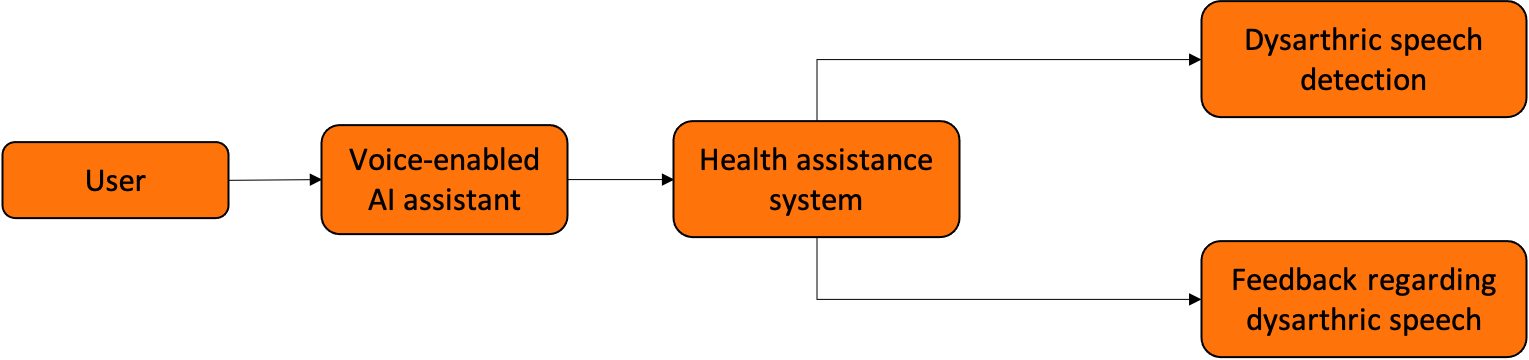}
\decoRule
\caption[Overview of the Dysarthric Speech Detection System]{Overview of the Dysarthric Speech Detection System.}
\label{fig:alexa_health_assistant}
\end{figure}

In Figure \ref{fig:alexa_capt}, it was shown that a voice-enabled AI assistant could be used to build a system for detecting pronunciation errors and providing feedback to a user. Such design can be adopted to create a health assistant system that can detect dysarthric speech and provide advice to a user to visit a consultant, as illustrated in Figure \ref{fig:alexa_health_assistant}. 

\bigskip

\textbf{Reconstruction of dysarthric speech}

\bigskip

People with dysarthria have difficulty communicating with other people because their speech is distorted and less intelligible. Speech therapy is one way to improve spoken communication skills; for example, when dysarthria results from a brain stroke causing an aphasia condition \cite{farrajota2012speech,koyuncu2016speech,brady2016speech}. In cases where speech therapy is not effective, it may still be possible to help people communicate by reconstructing their speech using a speech-to-speech approach \cite{korzekwa2019interpretable, huang2021preliminary}. The input to a speech-to-speech system is distorted speech spoken by a person with dysarthria, and the output is the reconstructed speech with improved intelligibility. Similar techniques can be applied to non-native speech. Radzikowski et al. \cite{radzikowski2016non} use Hidden Markov Models (HMM) to make corrections in non-native speech so that students and teachers can communicate more easily during lectures. 

There are similarities between the generation of synthetic speech errors in non-native speech for the detection of pronunciation errors and the reconstruction of dysarthric speech. In the synthetic speech scenario, the speech-to-speech system is used to 'destroy' correctly pronounced speech by introducing pronunciation errors. In the dysarthric speech scenario, speech is processed the other way round to improve the intelligibility of distorted speech. It can be hypothesized that a similar deep learning technique should be effective in both scenarios.

\section{Publications and scientific contribution}
\label{sec:publications}

In this Section, first, the articles co-authored by Daniel Korzekwa are listed, and then the main scientific contributions are presented in more detail in the following subsections.

Six articles were published or accepted for publication with Daniel Korzekwa as the primary author. The declaration of authorship is included in Appendix \ref{sec:authorship}. These publications are directly related to the research theses presented in Section \ref{sec:research_theses} and constitute the main scientific contribution of the doctoral dissertation: 
\begin{itemize}
\item Computer-assisted Pronunciation Training - Speech synthesis is almost all you need;  accepted for publication in Speech Communication Journal on June 17 ‘2022, in print \cite{korzekwa22_speechcomm}
\item Weakly-supervised word-level pronunciation error detection in non-native English speech, Interspeech, 2021 \cite{korzekwa21b_interspeech}
\item Mispronunciation Detection in Non-native (L2) English with Uncertainty Modeling, ICASSP, 2021 \cite{korzekwa2021mispronunciation}
\item Detection of Lexical Stress Errors in Non-native (L2) English with Data Augmentation and Attention, Interspeech, 2021 \cite{korzekwa21_interspeech}
\item Deep learning model for automated assessment of lexical stress of non-native English speakers, The Journal of the Acoustical Society of America, 2019 \cite{korzekwa2019deep}
\item Interpretable deep learning model for the detection and reconstruction of dysarthric speech, Interspeech, 2019 \cite{korzekwa2019interpretable}
\end{itemize}

Additionally, nine publications co-authored by Daniel Korzekwa are devoted to topics related to the doctoral dissertation. Two publications are devoted to pronunciation error detection in non-native English. Six publications concern speech synthesis and voice conversion, which lay the foundations for generating synthetic pronunciation errors and the reconstruction of dysarthric speech. The ninth publication concerns the collection of non-native speech corpus that was used to evaluate the pronunciation error detection models:
\begin{itemize}
\item L2-GEN: A Neural Phoneme Paraphrasing Approach to L2 Speech Synthesis for Mispronunciation Diagnosis, accepted to Interspeech, 2022 \cite{Zhang2022_interspeech}
\item Creating New Voices using Normalizing Flows, accepted to Interspeech, 2022 \cite{Bilinski2022_interspeech}
\item Text-free non-parallel many-to-many voice conversion using normalizing flows, ICASSP, 2022 \cite{merritt22_icassp_vc}
\item Universal neural vocoding with parallel wavenet, ICASSP, 2021 \cite{jiao2021universal}
\item Improving the expressiveness of neural vocoding with non-affine Normalizing Flows, Interspeech, 2021 \cite{gabrys2021improving}
\item Non-Autoregressive TTS with Explicit Duration Modelling for Low-Resource Highly Expressive Speech, ISCA Speech Synthesis Workshop – a satellite event at Interspeech, 2021 \cite{shah21_ssw}
\item Enhancing audio quality for expressive Neural Text-to-Speech, ISCA Speech Synthesis Workshop – a satellite event at Interspeech, 2021 \cite{ezzerg2021enhancing}
\item Constructing a dataset of speech recordings with Lombard effect, IEEE SPA 2020 \cite{Weber2020}
\item Extending Goodness of Pronunciation to generate mispronunciation hypotheses for pronunciation assessment in L2-English, AMLC, 2020 \cite{BeringerAMLC2020}
\end{itemize}

\subsection{Contributions from primary author publications}

Three publications are devoted to the automated detection of pronunciation errors (incorrectly pronounced phonemes) in non-native speech. 

\begin{center}
\textit{Korzekwa, D., J. Lorenzo-Trueba, T. Drugman, and B. Kostek (2022). “Computer-
assisted Pronunciation Training - Speech synthesis is almost all you need”. In: accepted for publication in Speech Communication Journal on June 17 ‘2022, in print.}
\end{center}

\textbf{Novelty:} The research community has long studied computer-assisted pronunciation training (CAPT) methods in non-native speech. Researchers focused on studying various model architectures, such as Bayesian networks and deep learning methods, as well as on the analysis of different representations of the speech signal. Despite significant progress in recent years, existing CAPT methods are not able to detect pronunciation errors with high accuracy (only 60\% precision at 40\%-80\% recall). One of the key problems is the low availability of mispronounced speech that is needed for the reliable training of pronunciation error detection models. If we had a generative model that could mimic non-native speech and produce any amount of training data, then the task of detecting pronunciation errors would be much easier. We present three innovative techniques based on phoneme-to-phoneme (P2P), text-to-speech (T2S), and speech-to-speech (S2S) conversion to generate correctly pronounced and mispronounced synthetic speech. We show that these techniques not only improve the accuracy of three machine learning models for detecting pronunciation errors but also help establish a new state-of-the-art in the field. Earlier studies have used simple speech generation techniques such as P2P conversion, but only as an additional mechanism to improve the accuracy of pronunciation error detection. We, on the other hand, consider speech generation to be the first-class method of detecting pronunciation errors. The effectiveness of these techniques is assessed in the tasks of detecting pronunciation and lexical stress errors. Non-native English speech corpora of German, Italian, and Polish speakers are used in the evaluations. The best proposed S2S technique improves the accuracy of detecting pronunciation errors in AUC metric by 41\% from 0.528 to 0.749 compared to the state-of-the-art approach.

\begin{center}
\textit{Korzekwa, D., J. Lorenzo-Trueba, T. Drugman, S. Calamaro, and B. Kostek (2021). “Weakly-Supervised Word-Level Pronunciation Error Detection in Non-Native English Speech”. In: Proc. Interspeech 2021, pp. 4408–4412. DOI: 10.21437/Interspeech.2021-38.}
\end{center}

\textbf{Novelty:} We propose a weakly-supervised model for word-level mispronunciation detection in non-native (L2) English speech. To train this model, phonetically transcribed L2 speech is not required and we only need to mark mispronounced words. The lack of phonetic transcriptions for L2 speech means that the model has to learn only from a weak signal of word-level mispronunciations. Because of that and due to the limited amount of mispronounced L2 speech, the model is more likely to overfit. To limit this risk, we train it in a multi-task setup. In the first task, we estimate the probabilities of word-level mispronunciation. For the second task, we use a phoneme recognizer trained on phonetically transcribed L1 speech that is easily accessible and can be automatically annotated. Compared to state-of-the-art approaches, we improved the accuracy of detecting word-level pronunciation errors in AUC metric by 30\% on the GUT Isle Corpus of L2 Polish speakers and by 21.5\% on the Isle Corpus of L2 German and Italian speakers.

\begin{center}
\textit{Korzekwa, D., J. Lorenzo-Trueba, S. Zaporowski, S. Calamaro, T. Drugman, and B. Kostek (2021). “Mispronunciation Detection in Non-Native (L2) English with Uncertainty Modeling”. In: ICASSP 2021-2021 IEEE International Conference on Acoustics, Speech and Signal Processing (ICASSP). IEEE, pp. 7738–7742. DOI: 10.1109/ICASSP39728.2021.9413953}
\end{center}

\textbf{Novelty:} A common approach to the automatic detection of mispronunciation in language learning is to recognize the phonemes produced by a student and compare them to the expected pronunciation of a native speaker. This approach makes two simplifying assumptions:  a) phonemes can be recognized from speech with high accuracy, b) there is a single correct way for a sentence to be pronounced. These assumptions do not always hold, which can result in a significant amount of false mispronunciation alarms. We propose a novel approach to overcome this problem based on two principles: a) taking into account uncertainty in the automatic phoneme recognition step, b) accounting for the fact that there may be multiple valid pronunciations. We evaluate the model on non-native (L2) English speech of German, Italian and Polish speakers, where it is shown to increase the precision of detecting mispronunciations by up to 18\% (relative) compared to the common approach.

\bigskip

Two publications relate to the detection of lexical stress errors.  Preliminary work was first presented in the Journal of the Acoustical Society of America in 2019. The final results were published at the Interspeech 2021 conference.

\begin{center}
\textit{Korzekwa, D. and B. Kostek (2019). “Deep learning model for automated assessment of lexical stress of non-native English speakers”. In: The Journal of the Acoustical Society of America 146.4, pp. 2956–2957. DOI: 10.1121/1.5137270}
\end{center}

\begin{center}
\textit{Korzekwa, D., R. Barra-Chicote, S. Zaporowski, G. Beringer, J. Lorenzo-Trueba, A. Serafinowicz, J. Droppo, T. Drugman, and B. Kostek (2021). “Detection of Lexical Stress Errors in Non-Native (L2) English with Data Augmentation and Attention”. In: Proc. Interspeech 2021, pp. 3915–3919. DOI: 10.21437/Interspeech.2021-86}
\end{center}

\textbf{Novelty:} We describe two novel complementary techniques that improve the detection of lexical stress errors in non-native (L2) English speech: attention-based feature extraction and data augmentation based on Neural Text-To-Speech (TTS). In a classical approach, audio features are usually extracted from fixed regions of speech, such as the syllable nucleus. We propose an attention-based deep learning model that automatically derives optimal syllable-level representation from frame-level and phoneme-level audio features. Training this model is challenging because of the limited amount of incorrect stress patterns. To solve this problem, we propose to augment the training set with incorrectly stressed words generated with Neural TTS. Combining both techniques achieves 94.8\% precision and 49.2\% recall for the detection of incorrectly stressed words in L2 English speech of Slavic and Baltic speakers.

\bigskip

 One publication deals with the detection and reconstruction of dysarthric speech - a topic of the secondary research thesis.

\begin{center}
\textit{Korzekwa, D., R. Barra-Chicote, B. Kostek, T. Drugman, and M. Lajszczak (2019). “Interpretable Deep Learning Model for the Detection and Reconstruction of Dysarthric Speech”. In: Proc. Interspeech 2019, pp. 3890–3894. DOI: 10.21437/ Interspeech.2019-1206}
\end{center}

\textbf{Novelty:} We present a novel deep learning model for the detection and reconstruction of dysarthric speech. We train the model with a multi-task learning technique to jointly solve dysarthria detection and speech reconstruction tasks. The model key feature is a low-dimensional latent space that is meant to encode the properties of dysarthric speech. It is commonly believed that neural networks are “black boxes” that solve problems but do not provide interpretable outputs. On the contrary, we show that this latent space successfully encodes interpretable characteristics of dysarthria, is effective at detecting dysarthria, and that manipulation of the latent space allows the model to reconstruct healthy speech from dysarthric speech. This work can help patients and speech pathologists to improve their understanding of the condition, lead to more accurate diagnoses, and aid in reconstructing healthy speech for afflicted patients.

\subsection{Contributions from additional co-authored publications}

\textbf{Publications related to pronunciation error detection:}

\begin{center}
\textit{Zhang, D., A. Ganesan, S. Campbell, and D. Korzekwa (2022). “L2-GEN: A Neural Phoneme Paraphrasing Approach to L2 Speech Synthesis for Mispronunciation Diagnosis”. In: accepted to Interspeech 2022.}
\end{center}

\textbf{Novelty:} In this paper, we study the problem of generating mispronounced speech mimicking non-native (L2) speakers learning English as a Second Language (ESL) for the mispronunciation detection and diagnosis (MDD) task. The paper is motivated by the widely observed yet not well addressed data sparsity issue in MDD research where both L2 speech audio and its fine-grained phonetic annotations are difficult to obtain, leading to unsatisfactory mispronunciation feedback accuracy. We propose L2-GEN, a new data augmentation framework to generate L2 phoneme sequences that capture realistic mispronunciation patterns by devising an unique machine translation-based sequence paraphrasing model. A novel diversified and preference-aware decoding algorithm is proposed to generalize L2-GEN to handle both unseen words and new learner population with very limited L2 training data. A contrastive augmentation technique is further designed to optimize MDD performance improvements with the generated synthetic L2 data. We evaluate
L2-GEN on public L2-ARCTIC and SpeechOcean762 datasets. The results have shown that L2-GEN leads to up to 3.9\%, and 5.0\% MDD F1-score improvements in in-domain and out-of-domain scenarios respectively.

\begin{center}
\textit{Beringer, G., D. Korzekwa, A. Sanchez, B.Wang, and J. Lorenzo-Trueba (2020). “Extending Goodness of Pronunciation to generate mispronunciationhypotheses for pronunciation assessment in L2-English”. In: Amazon Machine Learning Conference, Seattle.}
\end{center}

\textbf{Novelty:} We propose a method to extend Goodness of Pronunciation (GOP), a commonly used pronunciation scoring metric, to generate mispronunciation hypotheses, which are then used to find what the speaker has actually uttered. We show that this allows to alleviate GOP’s problem of being over-dependant on phone boundaries computed by force-alignment, leading to an improvement in mispronunciation detection and diagnosis. We also argue that introducing hypothesis prior could be used to improve the model in the context of pronunciation teaching, where high precision is required. We demonstrate that a method of increasing the prior of canonical hypothesis by a factor can enable us to have control over precision-recall trade-off. For our experiments, we use a dataset of isolated words, which contain recordings of 23 Polish-based speakers.

\bigskip

Six co-authored publications are devoted to the topic of speech synthesis and voice conversion. These methods are used in two areas of the Ph.D. thesis: generation of mispronounced non-native speech and reconstruction of dysarthric speech. In addition, speech synthesis technology is used in Alexa devices, serving millions of people worldwide.

A modern speech synthesis and voice conversion systems consist of two components: a context generator and a vocoder. The context generator creates a mel-spectrogram from the input text (text-to-speech mode) \cite{wang2017tacotron}. Alternatively, it can process the mel-spectrogram extracted from another speech signal (speech-to-speech mode) \cite{jia2019direct}. The mel-spectrogram created by the context generator is processed by a vocoder to generate the raw audio signal \cite{oord2018parallel, lorenzo2018towards}.

\bigskip

\textbf{Publications related to context generation:}

\begin{center}
\textit{Bilinski, P., T. Merritt, A. Ezzerg, K. Pokora, S. Cygert, K. Yanagisawa, R. Barra Chicote, and D. Korzekwa (2022). “Creating New Voices using Normalizing Flows”. In: accepted to Interspeech 2022.}
\end{center}

\textbf{Novelty:} Creating realistic and natural-sounding synthetic speech remains a big challenge for voice identities unseen during training. As there is growing interest in synthesizing voices of new speakers, here we investigate the ability of normalizing flows in text-to-speech (TTS) and voice conversion (VC) modes to extrapolate from speakers observed during training to create unseen speaker identities. Firstly, we create an approach for TTS and VC, and then we comprehensively evaluate our methods and baselines in terms of intelligibility, naturalness, speaker similarity, and ability to create new voices. We use both objective and subjective metrics to benchmark our techniques on 2 evaluation tasks: zero-shot and new voice speech synthesis. The goal of the former task is to measure the precision of the
conversion to an unseen voice. The goal of the latter is to measure the ability to create new voices. Extensive evaluations demonstrate that the proposed approach systematically allows to obtain state-of-the-art performance in zero-shot speech synthesis and creates various new voices, unobserved in the training
set. We consider this work to be the first attempt to synthesize new voices based on mel-spectrograms and normalizing flows, along with a comprehensive analysis and comparison of the TTS and VC modes.

\begin{center}
 \textit{Shah, R., K. Pokora, A. Ezzerg, V. Klimkov, G. Huybrechts, B. Putrycz, D. Korzekwa, and T. Merritt (2021). “Non Autoregressive TTS with Explicit Duration Modelling for Low-Resource Highly Expressive Speech”. In: Proc. 11th ISCA Speech Synthesis Workshop (SSW 11), pp. 96–101. DOI: 10.21437/SSW.2021-17}
\end{center}

\begin{center}
\textit{Ezzerg, A., A. Gabryś, B. Putrycz, D. Korzekwa, D. Saez Trigueros, D. McHardy, K. Pokora, J. Lachowicz, J. Lorenzo-Trueba, and V. Klimkov (2021). “Enhancing audio quality for expressive Neural Text-to-Speech”. In: Proc. 11th ISCA Speech Synthesis Workshop (SSW 11), pp. 78–83. DOI: 10.21437/SSW.2021-14}
\end{center}

\textbf{Novelty:} These two publications propose context generation models based on deep learning techniques, including VAE \cite{chorowski2019unsupervised,van2017neural}, attention mechanism \cite{vaswani2017attention}, sequence-to-sequence models \cite{sutskever2014sequence}, and controllable speech synthesis \cite{ren2019fastspeech}. The main novelties are: improving signal quality and stability of speech synthesis, and creating TTS voices of speakers with a limited amount of speech recordings. The proposed TTS models lay the foundations for the generation of mispronounced non-native speech (Section \ref{sec:speech_synthesis}) and improved intelligibility of dysarthric speech (Chapter \ref{chapter:dysarthric_speech}).

\begin{center}
\textit{Merritt, T., A. Ezzerg, P. Biliński, M. Proszewska, K. Pokora, R. Barra-Chicote, and D. Korzekwa (2022). “Text-free non parallel many-to-many voice conversion using normalising flows”. In: Acoustics, Speech and Signal Processing (ICASSP). DOI: 10.1109/ICASSP43922.2022.9746368}
\end{center}

\textbf{Novelty:} One of the trends deeply explored in the Ph.D. thesis concerns the use of speech conversion to generate synthetic pronunciation errors \cite{korzekwa21b_interspeech}. This task requires a speech-to-speech (S2S) technique that takes correctly pronounced native speech and converts it to mispronounced speech. This publication proposes a novel voice conversion technique that enables speech conversion without relying on phonetic transcriptions. Collecting phonetic transcriptions is very time-consuming and this method makes the process redundant, paving the way to much more efficient ways of generating mispronounced speech. Second, this method  can convert any input speaker to any output speaker, which is useful for generating a diverse range of speakers.

\bigskip

\textbf{Publications related to neural speech vocoding:}

\begin{center}
\textit{Jiao, Y., A. Gabryś, G. Tinchev, B. Putrycz, D. Korzekwa, and V. Klimkov (2021). “Universal neural vocoding with parallel wavenet”. In: ICASSP 2021-2021 IEEE International Conference on Acoustics, Speech and Signal Processing (ICASSP). IEEE, pp. 6044–6048. DOI: 10.1109/ICASSP39728.2021.9414444}
\end{center}

\begin{center}
\textit{Gabryś, A., Y. Jiao, V. Klimkov, D. Korzekwa, and R. Barra-Chicote (2021). “Improving the Expressiveness of Neural Vocoding with Non-Affine Normalizing Flows”. In: Proc. Interspeech 2021, pp. 1679–1683. DOI: 10.21437/Interspeech. 2021-1555}
\end{center}

\textbf{Novelty:} In the dissertation-based research, the speech-to-speech technique is used to generate pronunciation errors for hundreds of voices. The context generator creates a mispronounced speech spectrogram that is converted into a raw speech signal by the vocoder. Typically, a dedicated vocoder would have to  be trained for each unique voice, but that approach would not scale here. These two articles propose a universal neural vocoder that can transform any speaker's mel-spectrogram into a raw speech signal. The universal vocoder is a key component, enabling the generation of mispronounced speech for many speakers at scale.

\bigskip

In addition to four publications on speech synthesis, there is one publication on the non-native speech corpus collection. 

\begin{center}
\textit{Weber, D., S. Zaporowski, and D. Korzekwa (2020). “Constructing a Dataset of Speech Recordings with Lombard Effect”. In: 24th IEEE SPA. DOI: 10.23919/ SPA50552.2020.9241266}
\end{center}

\textbf{Novelty:} The speech corpus of non-native speech was collected and used to evaluate the proposed pronunciation error detection models.

\section{Applicability}
Note: Due to confidentiality reasons, only selected use-cases are provided. 

The results of the doctoral dissertation are widely applicable at Amazon in many use cases. The pronunciation error detection models are used to detect pronunciation errors in speech synthesis automatically. The speech synthesis and voice conversion models are used in Alexa devices to serve millions of Amazon customers around the world. In addition, speech synthesis and voice conversion are used as a data augmentation technique to improve the accuracy of the pronunciation error detection models.
 
\subsection{Pronunciation error detection}

Scientists who work on new speech synthesis models often explore many research hypotheses and train many machine learning models for speech synthesis. Being able to get quick feedback on the quality of the generated speech is crucial for rapid progress in research. Traditionally, the quality of speech generated by speech synthesis models is evaluated by humans. Humans listen to synthesized utterances and evaluate them in terms of naturalness and speech intelligibility. Typically, multiple voices are scored within a single evaluation to understand which of the voices perform best with respect to certain aspects of speech quality. This manual perceptual evaluation process is a bottleneck that is slowing down research into new models of speech synthesis.

\begin{figure}[th]
\centering
\includegraphics[width=1\textwidth]{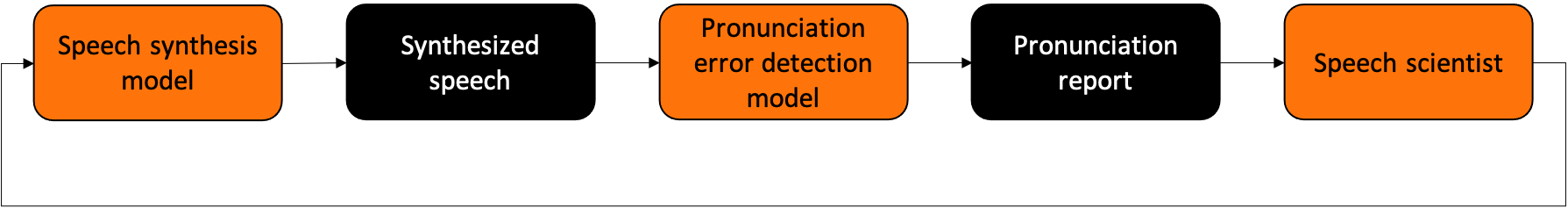}
\decoRule
\caption[The use of a pronunciation error detection model to evaluate speech synthesis]{The use of a pronunciation error detection model to evaluate speech synthesis.}
\label{fig:pron_error_detection_in_tts}
\end{figure}

The pronunciation error detection model \cite{korzekwa21b_interspeech} can complement manual perceptual evaluation, speeding up the research work on speech synthesis at Amazon in four different languages (detailed locations cannot be provided for confidentiality reasons). Many utterances can be synthesized and automatically checked for pronunciation errors, as illustrated in Figure \ref{fig:pron_error_detection_in_tts}.  Scientists  are researching new models of speech synthesis working in a closed-loop cycle. They conduct perceptual evaluations and use their results to design and create new speech synthesis models. 

\begin{figure}[th]
\centering
\includegraphics[width=0.8\textwidth]{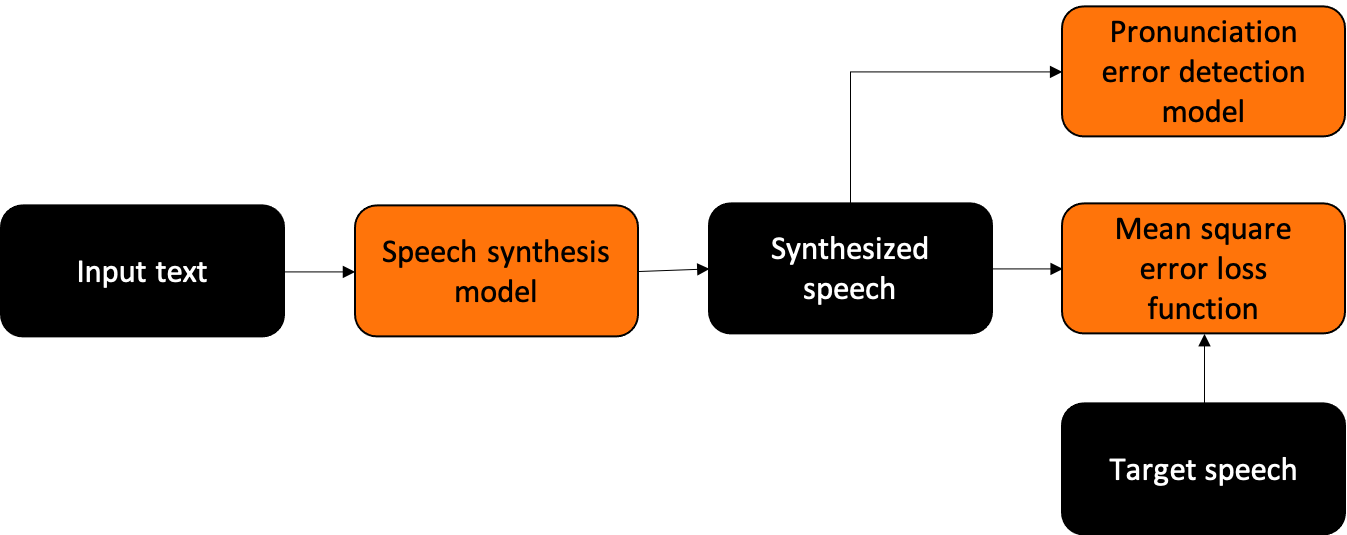}
\decoRule
\caption[The use of pronunciation error detection at the training time of a speech synthesis model]{The use of pronunciation error detection at the training time of a speech synthesis model.}
\label{fig:pron_error_detection_in_tts_training}
\end{figure}

 The pronunciation error detection model \cite{korzekwa2021mispronunciation} can be used at the training time of speech synthesis models. Traditionally, a speech synthesis model is trained in a supervised way by minimizing the mean square error between the synthesized and target speech signals, as shown in Figure \ref{fig:pron_error_detection_in_tts_training}. Adding another loss, which minimizes the probability of pronunciation errors, improves the stability of the synthesized speech.

\subsection{Speech synthesis and voice conversion}
The created speech synthesis and voice conversion models \cite{merritt22_icassp_vc, jiao2021universal,gabrys2021improving, shah21_ssw,ezzerg2021enhancing} serve two purposes. Firstly, many of them are used by Alexa devices to generate synthetic speech and communicate with people, but the second important application of these methods from the point of view of the Ph.D. thesis is the generation of synthetic mispronounced speech, which improves the accuracy of pronunciation error detection models. 

\begin{figure}[th]
\centering
\includegraphics[width=1\textwidth]{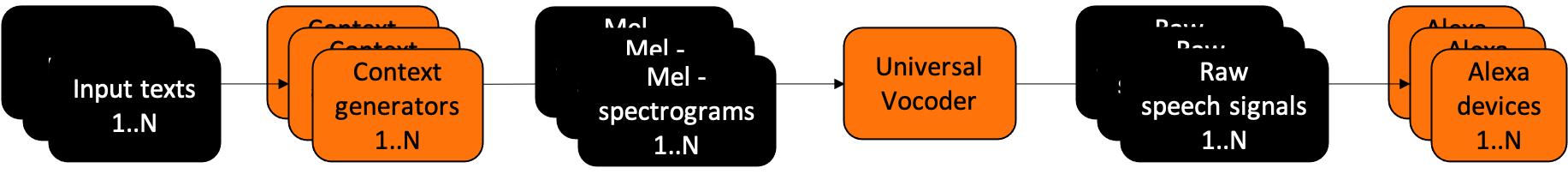}
\decoRule
\caption[A single universal vocoder serving multiple text-to-speech requests across multiple voices]{A single universal vocoder serving multiple text-to-speech requests across multiple voices.}
\label{fig:universal_vocoder_in_action}
\end{figure}

\begin{figure}[th]
\centering
\includegraphics[width=1\textwidth]{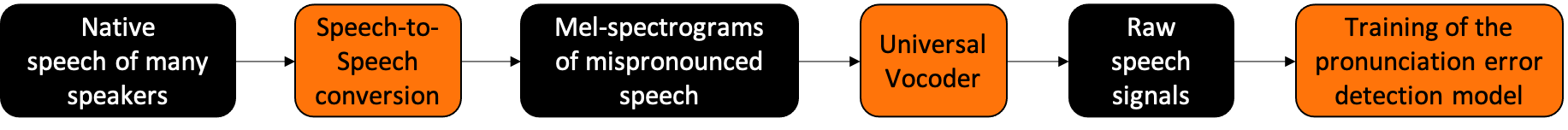}
\decoRule
\caption[A single universal vocoder serving the generation of mispronounced speech across multiple speakers]{A single universal vocoder serving the generation of mispronounced speech across multiple speakers.}
\label{fig:universal_vocoder_in_action_pron_training}
\end{figure}

Universal vocoder (UV) \cite{jiao2021universal,gabrys2021improving} is a model that converts a mel-spectrogram to a raw speech signal. A mel-spectrogram is generated based on the input text \cite{shah21_ssw,ezzerg2021enhancing}. The vocoder is universal because it supports all speakers and speaking styles, eliminating the need to train a dedicated vocoder for each speaker. The universal nature of the vocoder makes it much easier for the Alexa device to speak with multiple voices, as shown in Figure \ref{fig:universal_vocoder_in_action}. In addition, UV allows generating mispronounced speech for hundreds of speakers, which is used for training pronunciation error detection models \cite{korzekwa21b_interspeech}, as shown in Figure \ref{fig:universal_vocoder_in_action_pron_training}.


%% file: Chapters/ResearchMethodology/ResearchMethodology.tex

\chapter{Research methodology} 

\label{chapter:research_methodology} 

This chapter provides the basis for the topics of speech production, machine learning, and performance metrics to lay the groundwork for a detailed description of the research work and obtained dissertation results. Much of the material presented in this chapter is elaborated in the following chapters in relation to the doctoral dissertation.

\input{Chapters/ResearchMethodology/SpeechProduction}

\input{Chapters/ResearchMethodology/MachineLearning}

\input{Chapters/ResearchMethodology/PerformanceMetrics}

%% file: Chapters/ResearchMethodology/SpeechProduction.tex
\section{Speech production}
\label{sec:speech_production}
Spoken languages date back to 2000 B.C. The first spoken languages, Sumerian, Chinese, Mayan, used symbols to represent whole words \cite{Jurafsky:2009:SLP:1214993}. Modern spoken languages represent different parts of words with symbols. Japanese hiragana is a syllabic language in which one symbol corresponds to one syllable. In contrast, Roman languages, such as English, use an alphabet of letters to represent different words. In the Ph.D. thesis, the focus is on English, as this is important from the product applicability point of interest for Amazon’s Alexa. To recall, this thesis is realized within the  “Implementation doctorate” program, carried out by the Gdańsk University of Technology, and written in agreement with the Amazon company employing the Ph.D. candidate.

\begin{figure}[th]
\centering
\includegraphics[width=1\textwidth]{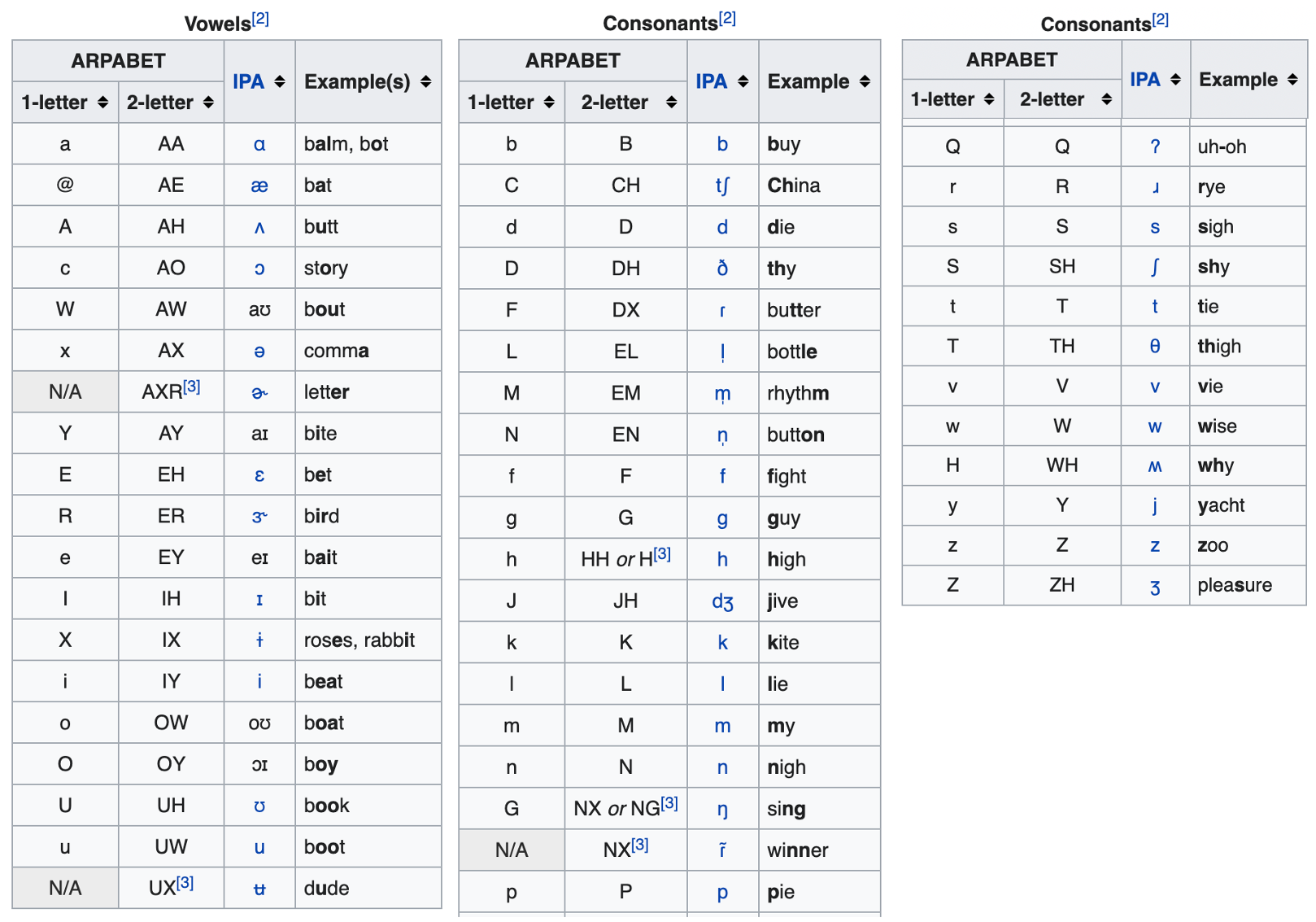}
\decoRule
\caption[Arpabet phonetic alphabet \cite{wikipedia_arpabet}]{Arpabet phonetic alphabet \cite{wikipedia_arpabet}.}
\label{fig:arpabet}
\end{figure}

The English alphabet consists of letters. Letters are the basic units of written words and then sentences. However, it is not enough to look at the letters to understand how to pronounce a word. The letter `c' may be pronounced differently in the words `cat' and `cell' and a similar observation applies to other letters. The phonetic alphabet is made of phones. Each phone corresponds to a specific sound made by a speaker. The way a word is pronounced is determined by its phonetic transcription. For example, the word `cat' is transcribed as [k ae t] and the word `cell' is transcribed as [s eh l]. Phonemes are abstract symbols that correspond to the mental representation of the pronunciation of a word. Phoneme transcription is represented with slashes //, e.g. /s eh l/ as opposed to using brackets [] for phones. Two popular phonetic alphabets are  International Phonetic Alphabet (IPA) and Arpabet \cite{Jurafsky:2009:SLP:1214993}, as shown in Figure \ref {fig:arpabet}. In this Ph.D. thesis, the Arpabet representation is used. Different ways a phoneme can be pronounced (i.e., phonetic variations of a phoneme that do not change spoken word meaning) are called allophones \cite{piotrowska2021evaluation}.

\begin{figure}[th]
\centering
\includegraphics[width=0.6\textwidth]{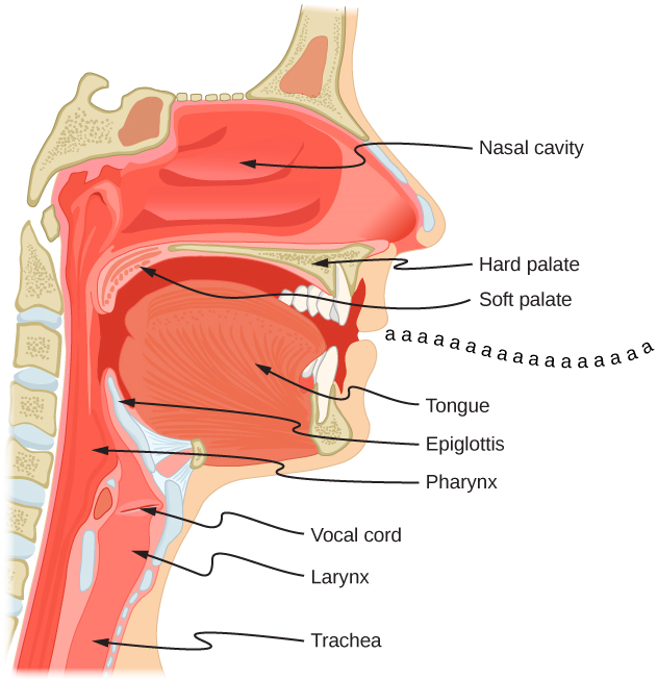}
\decoRule
\caption[Organs involved in the process of speech production \cite{alma991002741086804901}]{Organs involved in the process of speech production \cite{alma991002741086804901}.}
\label{fig:speech_prod_organs}
\end{figure}

Speech begins in the brain. The mental picture of the message is formed and represented by a phoneme sequence. The nervous system initiates the flow of air in the lungs. Air passes through the trachea, larynx, and then leaves the human body through the mouth and nose \cite{Jurafsky:2009:SLP:1214993}. All key parts of the human body involved in speech production are presented in Figure \ref{fig:speech_prod_organs}. The flow of air caries energy in the form of fluctuations in air molecules oscillating at specific frequencies, creating sound waves. Many sound waves that oscillate in parallel at certain frequencies over time and carry certain energy are called speech. Various unique speech sounds with different energy at different frequencies over time are called phones. Simply speaking, the human brain of the listener receives the incoming flow of air through the ears and decodes the message, creating its mental representation on the listener's side.

\subsection{Articulation}

There are two types of phones, voiced and unvoiced sounds. Voiced sounds are created by introducing vibrations into the vocal folds located in the larynx organ. In unvoiced sounds, the vocal folds do not vibrate. 

Phones are split into consonants and vowels. Consonants can be voiced, e.g. [b], [d], and unvoiced, e.g. [p], [t], while vowels, such as [a], [o], are generally voiced. In some languages, for example, Japanese, vowels can be unvoiced in a certain context. In English, all whispered vowels can be considered as unvoiced.

Consonants are formed by controlling different parts of the vocal tract. The vocal tract is the area above the trachea, which consists of the larynx and oral and nasal cavities. Depending on the location of the vocal tract that imposes the biggest restriction on airflow, consonants can be divided into labial, dental, alveolar, palatal, velar, and glottal categories. For example, [p] and [b] phones are called labial because they are generated by restricting airflow by putting lips together. Additionally, constants can be divided into stop, nasal, fricatives, sibilants, approximant, and tap categories, depending on the type of air restriction. For example, stop consonants, such as [b], [d], [p],  require that the airflow be completely blocked for a short time \ref{fig:speech_prod_organs}.

Vowels are formed similarly to consonants by changing the position of the articulators, mainly of the  tongue and lips. The tongue can be higher or lower in the nasal cavity, it can be moved forward to get closer to the lips or moved backward. Depending on the position of the tongue, vowels can be divided into front/back and high/medium/low categories. For example, in the word 'beet' [b iy t], the tongue is placed high and forward, whereas, in the word 'bat' [b ae t], the tongue is placed low and front. Changing the shape of lips is another important way to create vowels. For example, the phone [uw] requires the lips to be rounded off, as shown by the word 'tulip' [t uw l ix p]. Some vowels involve a change in the position of the articulators during the formation of a vowel, which corresponds to the production of two vowels immediately following each other. Such vowels are called diphthongs. For example, the diphthong [ow] in the word 'lotus' [l ow dx ax s].

Consonants and vowels make up syllables. Each syllable usually consists of one vowel and at least one consonant. The vowel part of a syllable is called the nucleus. Syllables form words and words form sentences. Words can consist of one, two, three, etc., syllables. For example, the word `napkin' [n a p - k ax n]  has two syllables.

Non-native English speakers can make pronunciation mistakes for a number of reasons. They can incorrectly map the written word into a phoneme representation. Even if the phonemes are pronounced correctly, meaning that the correct phones are produced, the word will be mispronounced due to a mismatch between the expected (canonical) phoneme representation and the corresponding phoneme representation made in the human brain. Correctly encoding a word in the human brain does not mean it is pronounced correctly. Different languages have different phone sets, therefore, people may not be able to pronounce all the phones in the non-native language correctly. An example of this is the phone [th] in English; this phone does not exist in the Polish phone set.  People may also skip phones while speaking or not be able to pronounce them because of various health problems such as dysarthria.

\subsection{Prosody}

Prosody is related to features of speech consisting of F0, energy, and duration. F0 is the fundamental frequency at which vocal folds vibrate. Energy is defined as the variance of a speech signal. Energy is usually expressed in decibels (dB), reflecting more human perception than raw energy values. The energy in dB is called loudness. Duration determines how long various sounds last, such as phones and silence between words and sentences \cite{Jurafsky:2009:SLP:1214993}.

Prosody emphasizes different parts of speech, which usually corresponds to raising F0, increasing loudness and extending the duration of speech sounds \cite{Jurafsky:2009:SLP:1214993}. Emphasizing syllables corresponds to lexical stress. English dictionary contains rules (lexical stress) that define which syllables in different words should be stressed. Sometimes, placing lexical stress on an incorrect syllable may change the meaning of the word, for example, the word 'produce' has two forms, the verb is stressed on the second syllable and the noun is stressed on the first syllable. In compound nouns, one word can be emphasized while the other is not. A compound noun is a noun that consists of two parts, two nouns or an adjective followed by a noun. For example, in the compound noun `bulldog' the first noun is stressed. 

Prosody is used to distinguish vocal patterns, e.g., to indicate whether a sentence is a question or not. Yes-no questions in English have raised intonation at the end, for example, `Can we meet tomorrow?'. On the contrary, in declarative sentences such as `We will meet tomorrow', the intonation falls down at the end of the sentence. Intonation can also be used to separate different words in enumerations. For example, the intonation slightly raises after each comma in the sentence 'One, two, three, start!'.

Non-native speakers can make prosodic mistakes in speech, for example, because they are not familiar with the rules defined in the language dictionary, such as which syllable to emphasize. Multiple studies have shown that correct prosody improves the intelligibility of speech \cite{field2005intelligibility,lepage2014intelligibility}, therefore, practicing the prosodic aspects of speech is an important part of CAPT.

%% file: Chapters/ResearchMethodology/MachineLearning.tex
\section{Machine learning techniques}


The doctoral thesis focuses on the application of deep learning techniques in automated pronunciation assessment. Deep learning is a branch of machine learning \cite{lecun2015deep}. In general, machine learning is the process in which a machine automatically learns how to perform a specific task. For example, learn to classify images into two categories of cats and dogs. In the context of the doctoral dissertation, it is about learning to detect pronunciation errors in speech. Deep learning is a multidisciplinary field rooted in multiple related fields, including machine learning, probability theory, statistics, and mathematics. To explain deep learning, there are other areas that need to be discussed first, notably the probability theory, machine learning, and probabilistic machine learning \cite{bishop2006pattern, murphy2012machine}. 

In the simplest scenario, both machine learning and its probabilistic variant aim to learn the function $y=f(x)$. The variable $x$ may represent an image, whereas the variable $y$ may represent a decision whether the image represents a dog or a cat. The function $f()$ represents the mapping between both variables.

In machine learning, the variables $x$ and $y$ are vectors $x\in \mathcal{R}^{D_{X}}$ and $y\in \mathcal{R}^{D_{Y}}$in multidimensional spaces $D_{X}$ and $D_{y}$, and the function f() can take any form. While in probabilistic machine learning, the variables  $x$ and $y$ are constrained to the form of certain probability distributions, denoted as $x\sim p(x)$, $y\sim p(y)$, and $y=f(x) \sim p(y|x)$. The main idea behind probabilistic modeling is to represent variables and dependencies with probability distributions, as opposed to using only scalar or vector variables. Intuitively, probabilistic models account for the uncertainty by looking at all possible values of the input and output variables, whereas non-probabilistic methods only consider input and output variables as point estimates. Representing variables as probability distributions helps to overcome the problem of overfitting in which a machine learning model does not generalize well to unseen data, e.g., the inability to correctly classify unseen images into the categories of dogs and cats.

Interestingly, there are many similarities between both non-probabilistic and probabilistic machine learning. For example, a machine learning technique called dropout introduces random noise in the training process and consequently makes the input and output variables more probabilistic. In recent years, there has been a trend of mixing the concepts of probabilistic and non-probabilistic machine learning, gradually blurring the lines between the two types of machine learning. Good examples of such models are the Variational Auto-Encoder (VAE) \cite{van2017neural} and Normalizing Flows (NF) \cite{kobyzev2020normalizing}. To understand existing modern machine learning architectures and design new ones, it is important to explore both non-probabilistic and probabilistic views on machine learning.

Deep learning differs from machine learning in the way the function $y=f(x)$ is defined. In deep learning, this function has multiple levels of nesting: $y=f(x)=f_1 \circ f_2 \circ ... \circ f_n$, whereas in the non-deep variant there is just one function mapping from $x$ to $y$. In the simplest possible scenario of two nested levels, the deep learning model is defined as $y=f_1(f_2(x))$. Deep learning is often equated with Deep Neural Networks (DNN) that have multiple hidden functions (neural network layers). However, there are other types of deep learning models, such as Deep Gaussian Processes (DGP) \cite{damianou2013deep}. Therefore, the term 'deep learning' should be considered more broadly. 

The following sections will introduce in detail various concepts of machine learning in that are used in the Ph.D. thesis, including the probability theory, probabilistic machine learning, deep learning, and the probabilistic perspective on deep learning.

\subsection{Probability theory}
\label{sub:probability_theory}
The probability theory provides a mathematical framework that enables modeling random events. A\textbf{ random event}, also known as a \textbf{random variable}, represents an event with an unknown outcome. Imagine you are selecting a ball from a container with two balls, one red and one blue. This is an example of a random variable $x$ with two possible outcomes $x \in \{red,blue\}$. If both balls are identical except for the color, the chances of blindly selecting red and blue balls will be the same. If there were three reds and one blue ball, the chances of choosing a red ball will be higher respectively. The chance that an event would lead to a certain outcome is also known as \textbf{likelihood} or \textbf{probability}.

The origins of the probability theory go back to the 16th century when Gerolamo Cardano studied games of chance such as roulette and dice, in which the outcome depends on random events \cite{ore2017cardano}. In the next century, Blaise Pascal made his first attempts to formulate the concept of expected value by studying a game of chance called `problem of points' \cite{todhunter2014history}. The expected value, also know as 'expectation', is an important part of the probability theory \cite{bishop2006pattern}. In the 18th and 19th centuries, Thomas Bayes and Pierre Laplace formulated the probability theory as we know it today \cite{bishop2006pattern}. 

The core of probabilistic machine learning is the probability theory, and in particular, its two concepts are very important: probability distribution and Bayes' theorem. Both concepts are described in this section, whereas a comprehensive look at the probability theory and probabilistic machine learning is presented in these three excellent books written in recent years. `Pattern Recognition and Machine Learning' by Christopher Bishop \cite{bishop2006pattern}, `Probabilistic Graphical Models: Principles and Techniques' by Daphne Koller \cite{koller2009probabilistic}, and `Machine Learning: a Probabilistic Perspective' by Kevin P. Murphy\cite{murphy2012machine}.

\subsubsection{Probability distribution}
Probability, also known as likelihood, or more colloquially a chance, is denoted as $p(x) \in [0,1]$. The probability of a random event (random variable) $x$ can be 0 - the event cannot take place, it can be higher than zero but less than 1 - the event may happen, or it can be exactly 1 - the event will always happen. A random variable can have multiple outcomes, for example, selecting a ball from three possible colors with the value $p(x=red)=0.2$ means that the probability of selecting the red ball is 20\%. A random variable can be discrete or continuous. Selecting a ball out of a finite set of possible colors $x \in \{red, blue, green\}$ is an example of a discrete random variable, while selecting a number from a set of real numbers $x \in \mathcal{R}$ corresponds to a continuous random variable.

The function that defines the probabilities for all possible outcomes of a random variable is called a \textbf{probability distribution}, denoted as $x \sim p(x)$. The probability distribution of a discrete random variable is called a \textbf{Probability Mass Function (PMF)}, whereas a \textbf{Probability Density Function (PDF)} defines the probability distribution of a continuous random variable. 

The PMF function can be presented in a tabular form (Table \ref{tab:ml_ball_selection_pt}). 

\begin{table}[ht]
\caption{The PMF function for the  $x$  variable presented in a tabular form (color of the ball selected from the container) .}
\label{tab:ml_ball_selection_pt}
\centering
\begin{tabular}{l l}
\toprule
\tabhead{x} & \tabhead{p(x)} \\
\midrule
red & 0.75 \\
blue & 0.25\\
\bottomrule\\
\end{tabular}
\end{table}

The PDF function is usually represented by a mathematical equation, as illustrated by a random variable following the Normal probability distribution (Eq. \ref{eqn:univariate_normal}).  The Normal distribution, also known as Gaussian, is one of the commonly used representations of random variables due to its simple form that makes mathematical computations possible in closed form \cite{bishop2006pattern}. In practice, other continuous probability distributions are also used, such as Beta and Gama distributions \cite{murphy2012machine,bishop2006pattern}.

\begin{equation}
p(x) \sim \mathcal{N}(\mu, \sigma)=\frac{1}{\sigma\sqrt{2\pi}}e^{-0.5(\frac{x-\mu}{\sigma})^2}
\label{eqn:univariate_normal}
\end{equation}
The PMF and PDF functions must satisfy two conditions. First, the probability value must be greater or equal to 0:

\begin{equation}
p(x)\geq 0 
\end{equation}Second, the sum of the probabilities for all possible outcomes of the event must be 1, represented as a sum function (Eq. \ref{eqn:prob_mass_func}) and an integral function (Eq. \ref{eqn:prob_density_func}) for discrete and continuous random variables, respectively.

\begin{equation}
\sum_x{p(x)}=1
\label{eqn:prob_mass_func}
\end{equation}
\begin{equation}
\int p(x)dx=1
\label{eqn:prob_density_func}
\end{equation}

\subsubsection{Conditional probability distribution}

In the real world, multiple random variables can interact with each other. The probability distribution of one random variable may depend on the outcome of another variable. This concept is illustrated in the coin game, where the goal is to guess whether a coin will land on heads or tails. The coin can be fair, resulting in equal probabilities for both possible outcomes. However, the coin may be biased with one outcome more likely than the other, e.g., the coin is made by a pirate who wants to win the game by cheating. This scenario can be modeled using two random variables. The variable $y \in \{heads,tails\}$ represents the two possible coin outcomes, and the variable $x \in \{true,false\}$, with the value of $true$ if the coin comes from a pirate, $false$ otherwise. 

The PMF functions for both $x$ and $y$ variables can be represented in tabular form, also known as a Conditional Probability Table (CPT) \cite{koller2009probabilistic}. Suppose that the probabilities of the $x$ variable are the same for both outcomes, which means that there are equal chances that the coin can come from a pirate or not. The CPT for the $x$ variable is shown in Table \ref{tab:ml_pirate_coin}. In addition, let's assume that the probabilities of the $y$ variable depend on the $x$ variable - if the coin comes from a pirate, it is more likely to land on heads than on tails (the CPT is shown in \ref{tab:ml_coin_outcome}). Both assumptions are known as \textbf{prior probabilities} or prior knowledge, giving information about the environment that is modeled with random variables. The probability distribution of the random variable $y$, denoted as $y \sim p(y|x)$, is called \textbf{conditional probability distribution} because it is conditioned on the variable $x$.

\begin{table}[ht]
\caption{The Conditional Probability Table (CPT) for the variable $x$ - whether the coin is biased (comes from a pirate) or not.}
\label{tab:ml_pirate_coin}
\centering
\begin{tabular}{l l}
\toprule
\tabhead{x} & \tabhead{p(x)}  \\
\midrule
false & 0.5 \\
true & 0.5 \\
\bottomrule\\
\end{tabular}
\end{table}

\begin{table}[ht]
\caption{The Conditional Probability Table (CPT) for the variable $y$ (the outcome of the coin) conditioned on the variable $x$ (the coin comes from a pirate or not).}
\label{tab:ml_coin_outcome}
\centering
\begin{tabular}{l l l}
\toprule
\tabhead{x} & \tabhead{y}  & \tabhead{p(y|x)}  \\
\midrule
false & heads & 0.5 \\
false & tails & 0.5\\
true & heads & 0.6\\
true & tails & 0.4 \\
\bottomrule\\
\end{tabular}
\end{table}

\subsubsection{Bayesian networks}

\begin{figure}[th]
\centering
\includegraphics[width=0.2\textwidth]{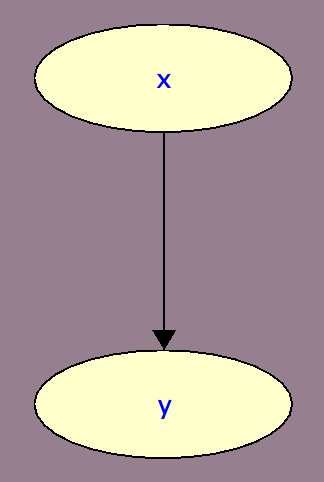}
\decoRule
\caption[Probabilistic Graphical Model (PGM) for the coin game, including two random variables; $x$ - whether the coin is biased (comes from a pirate), $y$ - the outcome of a single coin toss. The PGM image was created with the SamIam - a tool for modeling and reasoning in Bayesian Networks \cite{darwiche2009modeling}]{Probabilistic Graphical Model (PGM) for the coin game, including two random variables; $x$ - whether the coin is biased (comes from a pirate), $y$ - the outcome of a single coin toss. The PGM image was created with the SamIam - a tool for modeling and reasoning in Bayesian Networks \cite{darwiche2009modeling}}
\label{fig:ml_coin_game}
\end{figure}

Random variables and their dependencies can be represented graphically using the framework of Probabilistic Graphical Models (PGM) \cite{darwiche2009modeling,koller2009probabilistic}. The PGM graph for the coin game is depicted in Figure \ref{fig:ml_coin_game}. In PGM notation, circles represent random variables, whereas directed arrows represent dependencies between variables (conditional probability distributions). Each random variable can have many children and parent variables. PGM containing only directed arrows and no directed cycles is called Bayesian Network \cite{darwiche2009modeling}. The variant of PGM with unidirectional arrows is called Markov Network.

Random variables can be multiplied with each other, the concept is known as \textbf{product rule} \cite{bishop2006pattern}. The product of multiple random variables results in the \textbf{joined probability distribution} shown in Eq. \ref{eqn:product_rule}.
\begin{equation}
p(x,y)=p(x)p(y|x)=p(y)(x|y)
\label{eqn:product_rule}
\end{equation}
The random variable can be integrated out of the joined probability distribution, resulting in the \textbf{marginal probability distribution} over the remaining random variables. This process is known as the \textbf{sum rule} and is shown in Eq. \ref{eqn:sum_rule_x}.
 
\begin{equation}
p(x)=\int p(x,y)dy
\label{eqn:sum_rule_x}
\end{equation}
Both the sum and product rules can be combined to form the Bayes's theorem, also known as the Bayes rule, as shown in Eq. \ref{eqn:bayes_rule}.

\begin{equation}
p(x|y)=\frac{p(x)p(y|x)}{p(y)}
\label{eqn:bayes_rule}
\end{equation}
The sum rule, product rule, and Bayes rule provide a powerful framework for reasoning and making decisions under uncertainty. Reasoning, also known as \textbf{inference}, is the process of estimating the state of a random variable based on evidence provided by other dependent random variables. For example, the conditional probability $p(x|y)$ that the coin is biased given it has landed on heads can be estimated using the Bayes rule in Eq. \ref{eqn:bayes_rule}. This new state of the random variable given evidence is known as \textbf{posterior probability} or posterior probability distribution. The posterior probability contrasts with \textbf{prior probability} that represents the belief about the random variable before observing the outcomes of dependent variables. An unobserved variable is referred to as a \textbf{hidden variable} or a \textbf{latent variable}. The marginal probability of a latent variable can be estimated by integrating other latent variables using the sum rule in Eq. \ref{eqn:sum_rule_x}, the process also known as marginalization. For example, the probability that a coin will land heads is given by $p(y)=\int p(x,y)dx$.

\begin{figure}[th]
\centering
\includegraphics[width=0.8\textwidth]{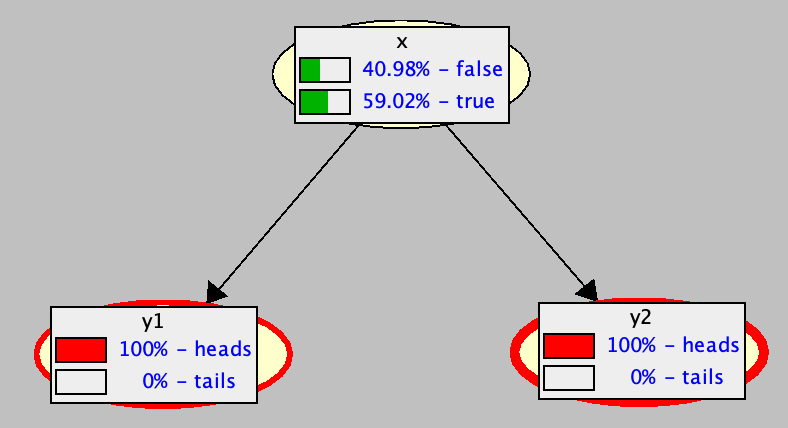}
\decoRule
\caption[Probabilistic Graphical Model (PGM) for the coin game, including three random variables. $x$ - whether the coin is biased (comes from a pirate), $y1,y2$ - the outcomes of two coin tosses. The PGM image was created with the SamIam - a tool for modeling and reasoning in Bayesian networks \cite{darwiche2009modeling}]{Probabilistic Graphical Model (PGM) for the coin game, including three random variables. $x$ - whether the coin is biased (comes from a pirate), $y1,y2$ - the outcomes of two coin tosses. The PGM image was created with the SamIam - a tool for modeling and reasoning in Bayesian networks \cite{darwiche2009modeling}}
\label{fig:ml_coin_game_two_tosses}
\end{figure}

In this coin game example, there is only one coin toss represented by a single random variable. To estimate the probability that the coin is biased, given it has landed on heads twice, another random variable is added to the PGM graph, as illustrated in Figure \ref{fig:ml_coin_game_two_tosses}. The Figure shows the posterior probability of the $x$ variable (whether the coin is biased) given it has landed heads twice.

The presented foundations of the probability theory form the basis of  both probabilistic and non-probabilistic machine learning. The concepts of graphical models and reasoning (inference) about latent variables enable the creation of different types of machine learning models for tasks such as prediction, detection, and classification.

\subsection{Probabilistic machine learning}

Probabilistic machine learning  is based on probability theory. The basic principle of modeling real problems with random variables and inferring the state of a hidden variable given some evidence can be applied to many practical problems. Different problems can be solved with different model architectures such as Hidden Markov Models, Gaussian Mixture Models, Kalman Filters, Gaussian Processes, and Variational Auto Encoders \cite{bishop2006pattern, goodfellow2016deep}. Each model can include different sets of latent variables and their dependencies (conditional probability distributions). Latent variables can have different probability distributions, discrete (Binomial, Multi-modal) and continuous (Gaussian, Beta, Gamma). Probabilistic models can be trained with the help of many algorithms such as Belief Propagation, Expectation Maximization, Expectation Propagation, and Variational Inference \cite{bishop2006pattern}. All of these considerations on how to apply in practice the basic principles of random variables and Bayes-rule define what probabilistic machine learning is about.

To see probabilistic machine learning in action, consider the problem of estimating temperature values from noisy observations. The training data consist of $N$ temperature measurements $y_i$, where $i=1..N$, collected at different $t_i$ time locations in the $(0,10)$ range. The task is to estimate the temperature values at unseen time locations within the range of the training data (interpolation task) and outside this range (extrapolation task). 

\begin{figure}[t!h]
\centering
\includegraphics[width=0.9\textwidth]{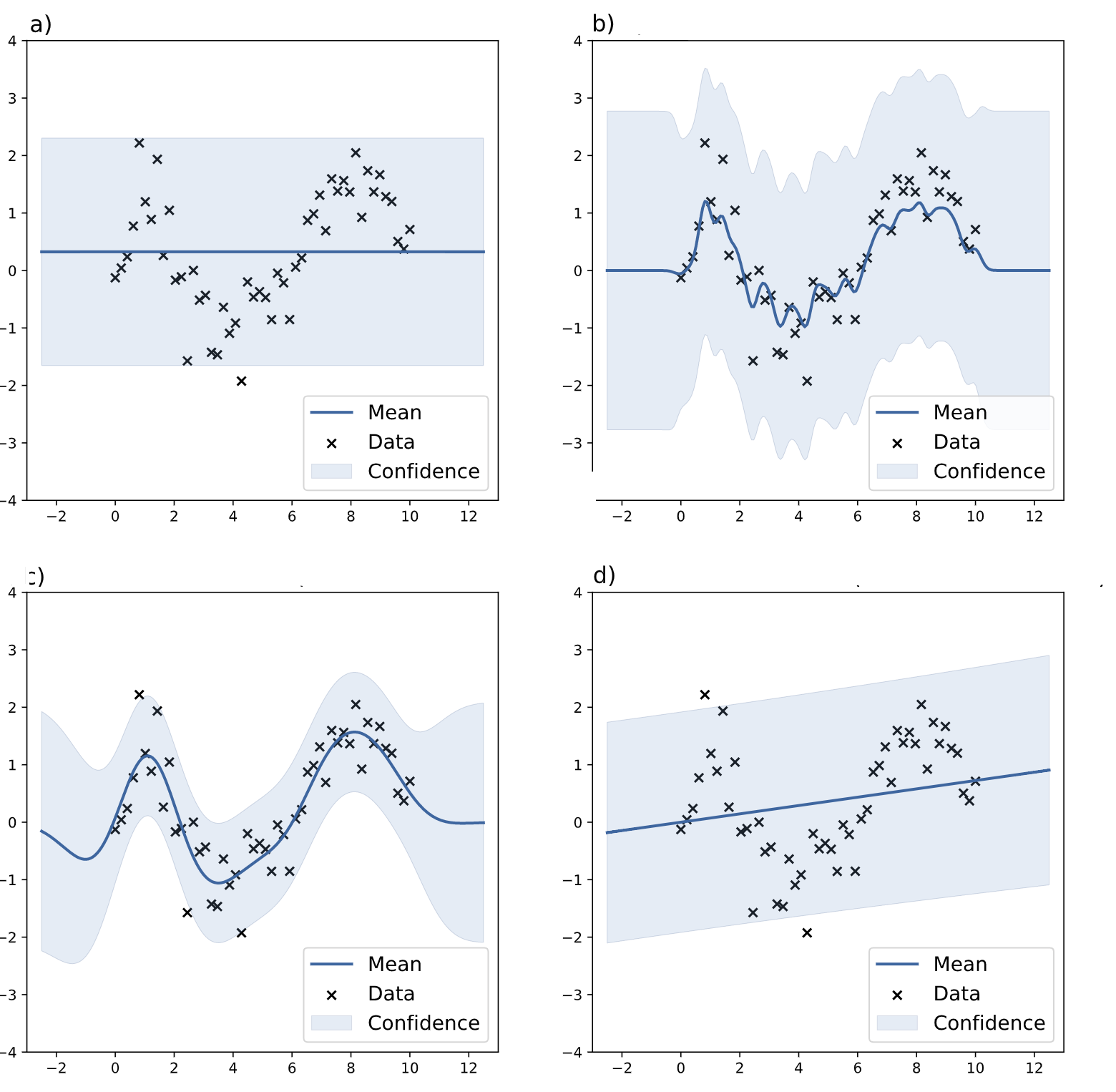}
\decoRule
\caption[Posterior plots for different probabilistic model architectures from Figure \ref{fig:probml_graphical_model_architectures}: a) Naive Bayes - a single latent variable $x$ estimated from multiple independent observations $\{y_1,y_2,...,y_N\}$, b) Hidden Markov Model - a latent variable $x_i$ conditioned on the local context of two neighboring variables $x_{i-1}$ and $x_{i+1}$, c) Gaussian Process with the RBF kernel - a model with an infinite number of latent variables $\{x_1,x_2,...,x_N\}$ conditioned on independent observations $\{y_1,y_2,...,y_N\}$, and d) Gaussian Process with the Linear kernel - a model with an infinite number of latent variables. Each plot contains observations from the training set, the predicted mean values, and the corresponding 95\% confidence interval]{Posterior plots for different probabilistic model architectures from Figure \ref{fig:probml_graphical_model_architectures}: a) Naive Bayes - a single latent variable $x$ estimated from multiple independent observations $\{y_1,y_2,...,y_N\}$, b) Hidden Markov Model - a latent variable $x_i$ conditioned on the local context of two neighboring variables $x_{i-1}$ and $x_{i+1}$, c) Gaussian Process with the RBF kernel - a model with an infinite number of latent variables $\{x_1,x_2,...,x_N\}$ conditioned on independent observations $\{y_1,y_2,...,y_N\}$, and d) Gaussian Process with the Linear kernel - a model with an infinite number of latent variables. Each plot contains observations from the training set, the predicted mean values, and the corresponding 95\% confidence interval.}
\label{fig:probml_posterior_plots_for_pgms}
\end{figure}

\begin{figure}[th]
\centering
\includegraphics[width=0.8\textwidth]{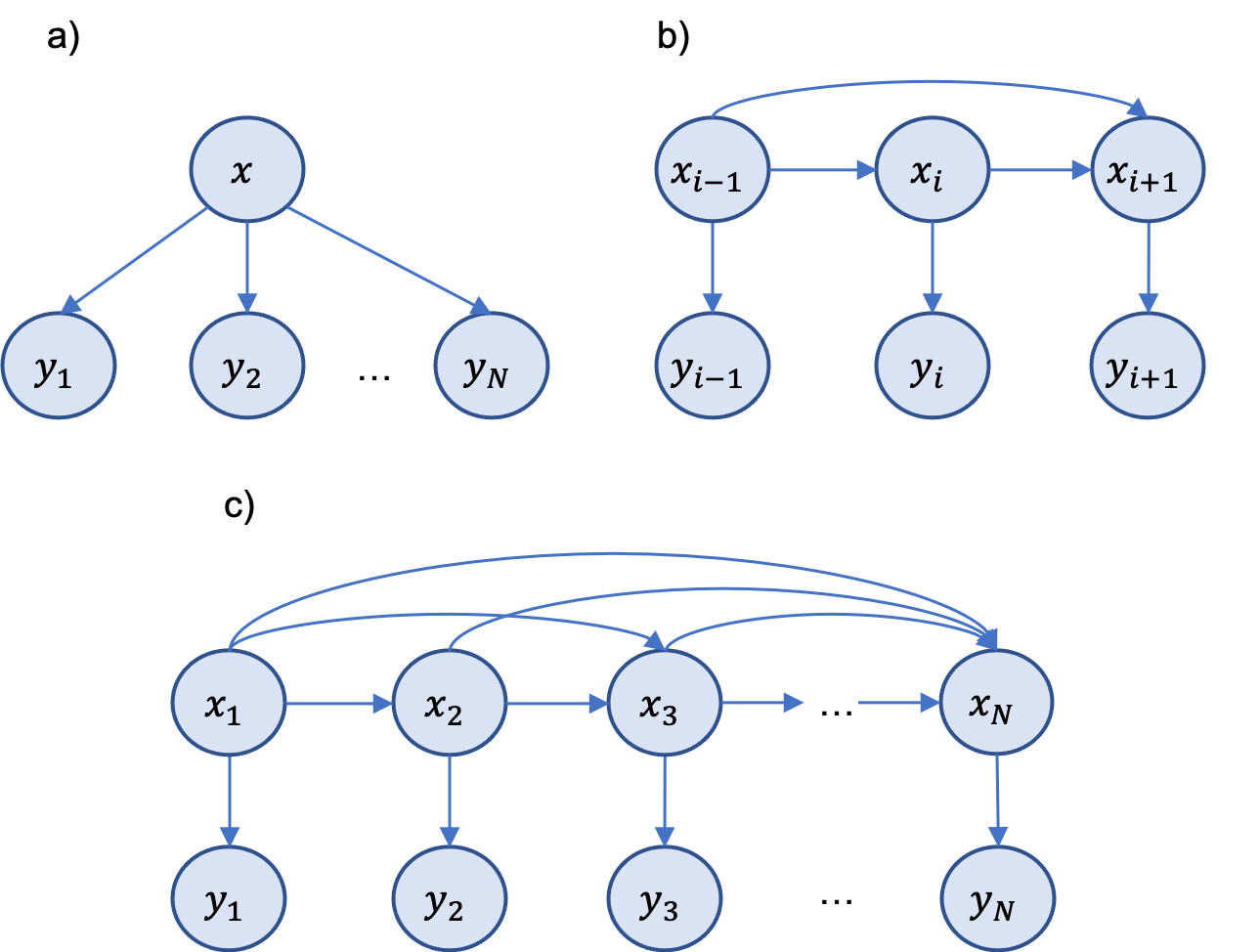}
\decoRule
\caption[Graphical models for different probabilistic model architectures: a) Naive Bayes - a single latent variable $x$ estimated from multiple independent observations $\{y_1,y_2,...,y_N\}$, b) Hidden Markov Model - a latent variable $x_i$ conditioned on the local context of two neighboring variables $x_{i-1}$ and $x_{i+1}$, and c) Gaussian Process - a model with infinite number of latent variables $\{x_1,x_2,...,x_N\}$ conditioned on independent observations $\{y_1,y_2,...,y_N\}$]{Graphical models for different probabilistic model architectures: a) Naive Bayes - a single latent variable $x$ estimated from multiple independent observations $\{y_1,y_2,...,y_N\}$, b) Hidden Markov Model - a latent variable $x_i$ conditioned on the local context of two neighboring variables $x_{i-1}$ and $x_{i+1}$, and c) Gaussian Process - a model with infinite number of latent variables $\{x_1,x_2,...,x_N\}$ conditioned on independent observations $\{y_1,y_2,...,y_N\}$.}
\label{fig:probml_graphical_model_architectures}
\end{figure}

Figure \ref{fig:probml_posterior_plots_for_pgms} shows the estimated (predicted) mean temperature values for four different probabilistic model architectures. The mean values are accompanied by the corresponding 95\% confidence intervals. Probabilistic machine learning gives confidence intervals 'for free' by modeling latent variables with probability distributions. Probabilistic distributions are usually parametrized by the mean $\mu$ and variance $\sigma_2$ parameters that can be converted to confidence intervals using the corresponding CDF function \cite{bishop2006pattern}. The corresponding probabilistic model architectures, Naive Bayes, Hidden Markov Model, and two variants of Non-parametric Gaussian Process, are presented in Figure \ref{fig:probml_graphical_model_architectures}. All models represent the latent temperature variables with the Gaussian distribution $p(x_i) = \mathcal{N}(\mu_{x_i}, \sigma_{x_i}^2)$. The noisy temperature observations follow the conditional Gaussian distribution $p(y_i|x_i) = \mathcal{N}(x_i, \sigma_{y_i}^2)$. However, these models differ in how the latent variable $x_t$ changes and correlates across time locations, leading to different abilities in interpolation and extrapolation tasks.
 
\subsubsection{Naive Bayes}

The simplest model, known as Naive Bayes \cite{murphy2012machine}, has one latent variable $x$ across all observations $y_i$  (Figure \ref{fig:probml_graphical_model_architectures}a). Consequently, the model estimates a single temperature value across all time locations, as shown in Figure \ref{fig:probml_posterior_plots_for_pgms}a).

\bigskip

\textbf{Posterior estimation}

\bigskip

The posterior value of the variable $x$ is defined by:  
 \begin{equation}
p(x|y_1,y_2,..,y_N) \propto p(x)\prod_{i=1}^{N}{p(y_i|x)}
\label{eqn:probml_singlelatentvariable_posterior}
\end{equation}
with the prior and conditional probability distributions $p(x) $ and $p(y_i|x)$ defined by:

\begin{equation}
p(x) = \mathcal{N}(\mu_{x}, \sigma_{x}^2)
\label{eqn:probml_singlelatentvariable_prior}
\end{equation}
\begin{equation}
p(y_i|x) = \mathcal{N}(x, \sigma_{y}^2)
\label{eqn:probml_singlelatentvariable_likelihood}
\end{equation}

Since all the terms in Eq. \ref{eqn:probml_singlelatentvariable_posterior} follow Gaussian distributions, the posterior over $x$ can be estimated analytically:

\begin{equation}
p(x|y_1,y_2,..,y_N)= \mathcal{N}(\tilde{\mu}_{x}, \tilde{\sigma}_{x}^2)
\end{equation}

, where:
\begin{equation}
\tilde{\mu}_x = \frac{\sigma_{y}^2}{N\sigma_{x}^2 + \sigma_{y}^2}\mu_{x} + \frac{N\sigma_{x}^2 }{N\sigma_{x}^2  + \sigma_{y}^2}\mu_{y_{ml}}
\end{equation}
\begin{equation}
\tilde{\sigma_x}^2 = \frac{1}{\frac{1}{\sigma_{x}^2} + \frac{N}{\sigma_{y}^2}}
\end{equation}

, where $\mu_{y_{ml}}$ is estimated with the maximum likelihood approach \cite{bishop2006pattern} given by:
 
 \begin{equation}
 \mu_{y_{ml}}=\frac{1}{N}\sum_{n=1}^Ny_i
\end{equation}

\bigskip

\textbf{Model training}

\bigskip

The Naive Bayes model presented in Figure \ref{fig:probml_graphical_model_architectures}a is parametrized with the parameters of the prior and conditional probability distributions (Equations \ref{eqn:probml_singlelatentvariable_prior} and \ref{eqn:probml_singlelatentvariable_likelihood}). The model parameters can be estimated in two ways. First, by extending Eq. \ref{eqn:probml_singlelatentvariable_posterior} to include the prior variables over the parameters of the model. For example, to learn the mean $\mu_x$ parameter of the prior distribution $p(x)$, a latent variable $p(\mu_x)$ can be added to the equation:

\begin{equation}
p(x|y_1,y_2,..,y_N) \propto \int p(\mu_x)p(x|\mu_x)\prod_{i=1}^{N}{p(y_i|x)}d\mu_x
\label{eqn:probml_singlelatentvariable_posterior_with_mu_prior}
\end{equation}
Interestingly, it can be seen that there is not much difference between inferring (estimating) the value of the latent variable of interest $x$ (Eq.  \ref{eqn:probml_singlelatentvariable_posterior}) and inferring (learning) and then integrating out the parameters of the model (Eq. \ref{eqn:probml_singlelatentvariable_posterior_with_mu_prior}). Both tasks, estimating and learning, use the same basic rules of the probability theory: the sum rule, product rule, and Bayes rule.

The second way to estimate the parameters of the model is to use a standard optimization technique such as gradient descent \cite{bishop2006pattern}, in which the parameters of the model $\theta=\{\theta_x,\theta_y\}$ are estimated by finding the maximum of the likelihood function $\arg\max_{\theta}\mathcal{L}(\theta)$. The likelihood function is defined by:

\begin{equation}
\mathcal{L}(\theta) = p(y_1,y_2,..,y_N|\theta) \propto \int p(x|\theta_x)\prod_{i=1}^{N}{p(y_i|x,\theta_y)}dx 
\end{equation}
Due to numerical instabilities, the log-likelihood function is minimized in practice. The likelihood function can be computed analytically for probabilistic models with both prior and conditional probability distributions represented by the Gaussian distribution \cite{bishop2006pattern}. For other distributions, approximation techniques such as Monte Carlo sampling \cite{koller2009probabilistic} and Variational Inference \cite{bishop2006pattern} are often used.

\bigskip

\textbf{Summary of the Naive Bayes model}

\bigskip

The example of the Naive Bayes model recalled above illustrates the general mechanism of using the probability theory to design probabilistic machine learning models. Probabilistic models differ in architecture. In some cases, the inference process is analytically tractable, but in others, optimization-based techniques are used. Some models have more, and some have fewer random variables. However, regardless of the model architecture, all models can be derived using the same probability theory. The following sections present more advanced probabilistic models for the temperature estimation task, for which the inference process has no analytical solution and requires optimization-based techniques. 

\subsubsection{Hidden markov model}

The Naive Bayes model described in the previous section estimates only the average temperature value across all time locations. The model introduced in this section, known as Hidden Markov Model (HMM) \cite{bishop2006pattern}, addresses this limitation by modeling local time dependencies between latent variables $x_i$, $x_{i-1}$, and $x_{i+1}$. Note that the vanilla HMM model only includes a dependency on the past variable $p(x_i|x_{i-1})$. Here, a slightly modified version of the model is presented, which takes into account both the past and future time dependencies $p(x_i|x_{i-1},x{_{i+1}})$. The model architecture is presented in Figure \ref{fig:probml_graphical_model_architectures}b. The estimated temperature values by the model are shown in Figure \ref{fig:probml_posterior_plots_for_pgms}b. The model interpolates well but is not capable of reasoning beyond the range of the training data. Modeling only the local context does not capture long-term dependencies in the data, which results in poor performance in the extrapolation task.

Probabilistic models based on local context dependencies have long been studied \cite{sarkka2013bayesian}. Most often, these models belong to the class of models known as Markov Chains \cite{bishop2006pattern}. The Markov Chain, or Markov Process, is a stochastic process in which the state of the latent variable $x_i$ depends only on the state of the latent variable $x_{i-1}$ at the previous time. In other words, the future and the past are independent of each other given the current state is known.  Kalman Filter and Exponential Moving Average (EMA) are two examples of Markov Chain-based models.

\bigskip

\textbf{Posterior estimation}

\bigskip

The posterior of the $i^{th}$ temperature latent variable is defined by:
 
\begin{equation}
p(x_i|y_1,y_2,..,y_N) \propto \int p(x_1)\prod_{i=2}^N{p(x_i|x_{i-1},x_{i+1}})\prod_{i=1}^{N}{p(y_i|x_i)}dx_{1..N \backslash i}
\label{eqn:probml_threelatentvariable_posterior}
\end{equation}
Similarly to the Naive Bayes model in the previous section, a posterior variable $p(x_i|y_1,y_2,..,y_N) $ can be calculated analytically for certain forms of conditional probability distributions, such as Gaussian. However, this process is computationally expensive for long sequences.

Belief propagation, also known as `message passing', is a popular algorithm that can efficiently compute posterior values for multiple latent variables $x_i$ \cite{koller2009probabilistic,bishop2006pattern}. In a nutshell, posteriors for latent variables $x_i$ are computed iteratively using the current best posterior estimates of the other dependent variables. Once the posterior value for one variable is estimated,  its state is sent as a message to other dependent variables in the PGM graph. Hence, the name of this algorithm is `message passing'. 

Let us consider a simplified model of three latent variables defined by $p(x_0,x_1,x_2)=p(x_0)p(x_1|x_0)p(x_2|x_1)$. To estimate the posterior value of the variable $x_1$ conditioned on the observed variable $x_2$, two incoming messages are needed from both neighboring variables, $m_{0->1}$ and $m_{2->1}$. The message $m_{0->1}$ is defined by:

\begin{equation}
m_{0->1} = \int p(x_0)p(x_1|x_0)dx_0
\end{equation}

whereas the message $m_{2->1}$ is defined by:

\begin{equation}
m_{2->1} = p(x_2|x_1)
\end{equation}

then the posterior of $x_1$ is defined as the product of both messages:

\begin{equation}
p(x_1|x_2) = m_{0->1}m_{2->1}  
\end{equation}
Messages are sent between the variables of the PGM graph till convergence, i.e., the delta between two consecutive posterior estimates is lower than a certain threshold. If the PGM graph is a tree, i.e., there are no loops between the variables and all messages can exactly be computed, i.e., no approximations are used to estimate any message, then the messages in the graph need to be passed only twice. This variant of Belief propagation is known as the forward-backward message passing algorithm \cite{koller2009probabilistic}. If the PGM is a graph, i.e., there are loops between the variables, and all messages are computed exactly,  the messages in the graph usually have to be passed more than twice to reach the convergence point. This variant is called Loopy Belief Propagation \cite{koller2009probabilistic}. In addition, if the messages are based on approximated probability distributions, then Loopy Belief Propagation is known as the Expectation Propagation algorithm \cite{minka2013expectation}.

\bigskip

\textbf{Model training}

\bigskip

Conceptually, the HMM model can be trained in the same way as the simpler Naive Bayes model from the previous section. That is, either by introducing latent variables representing the parameters of the model $\theta=\{\theta_x,\theta_y\}$ or by directly optimizing the likelihood function of the data. However, due to the complicated forms of the posterior distribution (Eq. \ref{eqn:probml_threelatentvariable_posterior}) and the likelihood function (Eq. \ref{eqn:probml_threelatentvariable_likelihood}), these techniques are often computationally intractable.

\begin{equation}
\mathcal{L}(\theta) = p(y_1,y_2,..,y_N|\theta) \\
\propto \int p(x_1|\theta_x)\prod_{i=2}^N{p(x_i|x_{i-1},x_{i+1}},\theta_x)\prod_{i=1}^{N}{p(y_i|x_i,\theta_y)}dx
\label{eqn:probml_threelatentvariable_likelihood}
\end{equation}

Expectation Maximization (EM) is an iterative algorithm that enables the training of complex probabilistic models \cite {moon1996expectation}. The Baum-Welch algorithm is a popular variant of EM-based methods of estimating the parameters of latent variables for more advanced probabilistic models \cite{welch2003hidden}. The algorithm decomposes a complex task of computing and optimizing the likelihood function into two simpler steps.

 The Maximization step maximizes the likelihood function in Eq. \ref{eqn:probml_threelatentvariable_likelihood} with respect to the model parameters. The calculation of the likelihood function is complicated due to the latent variables that have to be integrated out. If there were no latent variables, the likelihood function could be factorized into the product of independent likelihood terms and be much easier to estimate:
\begin{equation}
\mathcal{L_{EM}}(\theta) = p(y_1,y_2,..,y_N|\theta)
\propto \prod_{i=1}^{N}{p(y_i|x_i,\theta)}
\label{eqn:probml_threelatentvariable_likelihood_em}
\end{equation}
The redefined likelihood function $L_{EM}(\theta)$ is called the expected likelihood function because it depends on the estimates (expectations) of the latent variables. However, the model latent variables $x_1,..,x_N$ are not observed. To overcome this problem, the posteriors of the latent variables are computed based on the current best estimates of the model $\theta$ parameters using Eq. \ref{eqn:probml_threelatentvariable_posterior} - this is the Expectation step. 

The EM algorithm is a chicken and egg problem. To compute and maximize the likelihood function  $L_{EM}(\theta)$ in the Maximization step, the posteriors of the latent variables  $x_1,..,x_N$ have to be known in advance. To estimate the latent variables during the Expectation step, the model parameters $\theta$ are needed. The EM algorithm interchangeably iterates between the Expectation and Maximization steps till the model converges, that is, until the posteriors of the latent variables and the model parameters fall below a certain threshold. 

\bigskip

\textbf{Summary of the HMM model}

\bigskip

While discussing the HMM model, two important concepts were introduced. First, it has been shown that the exact estimation of the posteriors of latent variables in more complex probabilistic models is not always feasible. In theory, probabilistic machine learning attracts with the beauty of its basic principles based on the probability theory. However, in practice, optimization-based algorithms such as message-passing, Belief Propagation, and Expectation Propagation are required to compute the posteriors of latent variables.
	
The EM algorithm is another important concept introduced in this section.  There are many machine learning algorithms that have their roots in the EM method, such as $k$-means clustering, EM clustering, Auto-Encoders, Variational-Auto-Encoders, and Variational Inference \cite{bishop2006pattern,goodfellow2016deep}. Studying the similarities between different algorithms strengthens understanding of machine learning in general and makes it easier to invent new machine learning techniques to solve new problems.

\subsubsection{Non-parametric Gaussian processes}

The two previously described Naive Bayes and HMM models show that adding more latent variables increases the accuracy of the temperature estimation. While this is generally true, it comes at the cost of increasing the complexity of the model, making it more likely to overfit the training data. 

Interesting things happen when the model complexity grows to the point where there are an infinite number of latent variables and dependencies between them. Suddenly, the prior over the latent variables and their conditional dependencies can be computed using a relatively simple function parametrized with a few parameters only. Such a model is capable of representing complex distributions without overfitting to the training data. An example of such a model is the Gaussian Process (GP) model \cite{williams2006gaussian}.

\bigskip

\textbf{Definition of the Gaussian Process model}

\bigskip

In general form, GP is simply a multivariate Gaussian distribution over latent variables $\textbf{x}=\{x_1,...,x_N\}$ conditioned on observations $\textbf{y}=\{y_1,..,y_N\}$:

\begin{equation}
p(\textbf{x},\textbf{y}) = p(\textbf{x})p(\textbf{y}|\textbf{x})
\end{equation}where the latent variable $\textbf{x}$ follows the Multivariate Normal distribution $
\mathcal{N}(\boldsymbol{\mu},\boldsymbol{\Sigma})$ parametrized with the mean $\boldsymbol{\mu}$ and the covariance matrix $\boldsymbol{\Sigma}$ parameters. The covariance matrix $\boldsymbol{\Sigma}$ is computed using the covariance function, also known as the kernel function, or just the kernel. The $ij$-th element of the covariance matrix $\boldsymbol{\Sigma}$  is defined by:

\begin{equation}
\boldsymbol{\Sigma}_{ij}=cov(\mathbf{x}_i,\mathbf{x}_j)
\end{equation}
where $\mathbf{x}_i$ and $\mathbf{x}_j$ are the $i$-th and $j$-th elements of the latent variable $\mathbf{x}$. 

The covariance function defines the form of the function that can be modeled by the latent variable $\mathbf{x}$. For example, the underlying function can be smooth, periodic, linear, or it can model both global and local temporal dependencies. 

The likelihood function $p(\textbf{y}|\textbf{x})$ is also a Gaussian function conditioned on the latent variable $\mathbf{x}$. In its basic form, the likelihood function assumes that an individual observation $y_i$  is conditioned only on the corresponding latent variable $x_i$, defined by:
\begin{equation}
p( y_i|\mathbf{x}_i) = \mathcal{N}(\mathbf{x}_i,\sigma_y^2)
\label{eqn:probml_gp_likelihood_function}
\end{equation}where $\sigma_y^2$ is the noise related to imperfect observation $y_i$ of the latent variable $\mathbf{x}_i$.

\bigskip

\textbf{Gaussian Process for the temperature estimation problem}

\bigskip

A graphical representation of the GP model for the temperature estimation problem is shown in Figure \ref{fig:probml_graphical_model_architectures}c. The temperature observations are represented by the variable $\mathbf{y}$, whereas the latent variable $\mathbf{x}$ represents the temperature values over time that are estimated from noisy observations $\mathbf{y}$. The estimated temperature values for the GP model in both interpolation and extrapolation tasks are presented in Figures \ref{fig:probml_posterior_plots_for_pgms}c and\ref{fig:probml_posterior_plots_for_pgms}d for the Radial Basis Function (RBF) kernel and the linear kernel respectively.

The RBF kernel \cite{duvenaud2014kernel}, also known as Gaussian or Squared Exponential Kernel, is defined by:

\begin{equation}
cov_{rbf}(\mathbf{x}_i,\mathbf{x}_j)= \sigma^2\exp (-\frac{|\mathbf{f(x)}_i-\mathbf{f(x)}_j|^2}{2l^2})
\end{equation}
The RBF kernel imposes the constraint that the represented function is smooth, which means that points close to each other have more similar values than points falling more apart. The function $f(\mathbf{x})$ returns the feature vector for the latent variable $\mathbf{x}$. In the case of the temperature problem, the feature vector corresponds to time information, e.g., the number of seconds since 1970-01-01 00:00:00, but it can contain any multi-dimensional data that are supported by Euclidean distance. The variance parameter $\sigma_2$ tells how much the function values can differ from the mean value of the function. The length scale $l^2$ parameter indicates how many different variables $\mathbf{x}_i$ depend on each other over time. The higher the value, the stronger the temporal dependency. 

The RBF kernel performs well for both interpolation and extrapolation tasks. Its behavior in the extrapolation task is especially noteworthy. The function estimated with the RBF kernel can maintain its trend outside the regions of the training data, as shown in Figure \ref{fig:probml_posterior_plots_for_pgms}c, while using the confidence score to reflect the increasing uncertainty of the estimated values.

The Kernel Cookbook \cite{duvenaud2014kernel} presents different types of kernels such as Rational Quadratic Kernel, Periodic Kernel, Locally Periodic Kernel, and Linear Kernel. Different kernels can be combined to form new kernels by using the multiplication or addition functions. For example, Linear times Periodic kernel or RBF plus Linear kernel. 

\begin{figure}[t!h]
\centering
\includegraphics[width=0.9\textwidth]{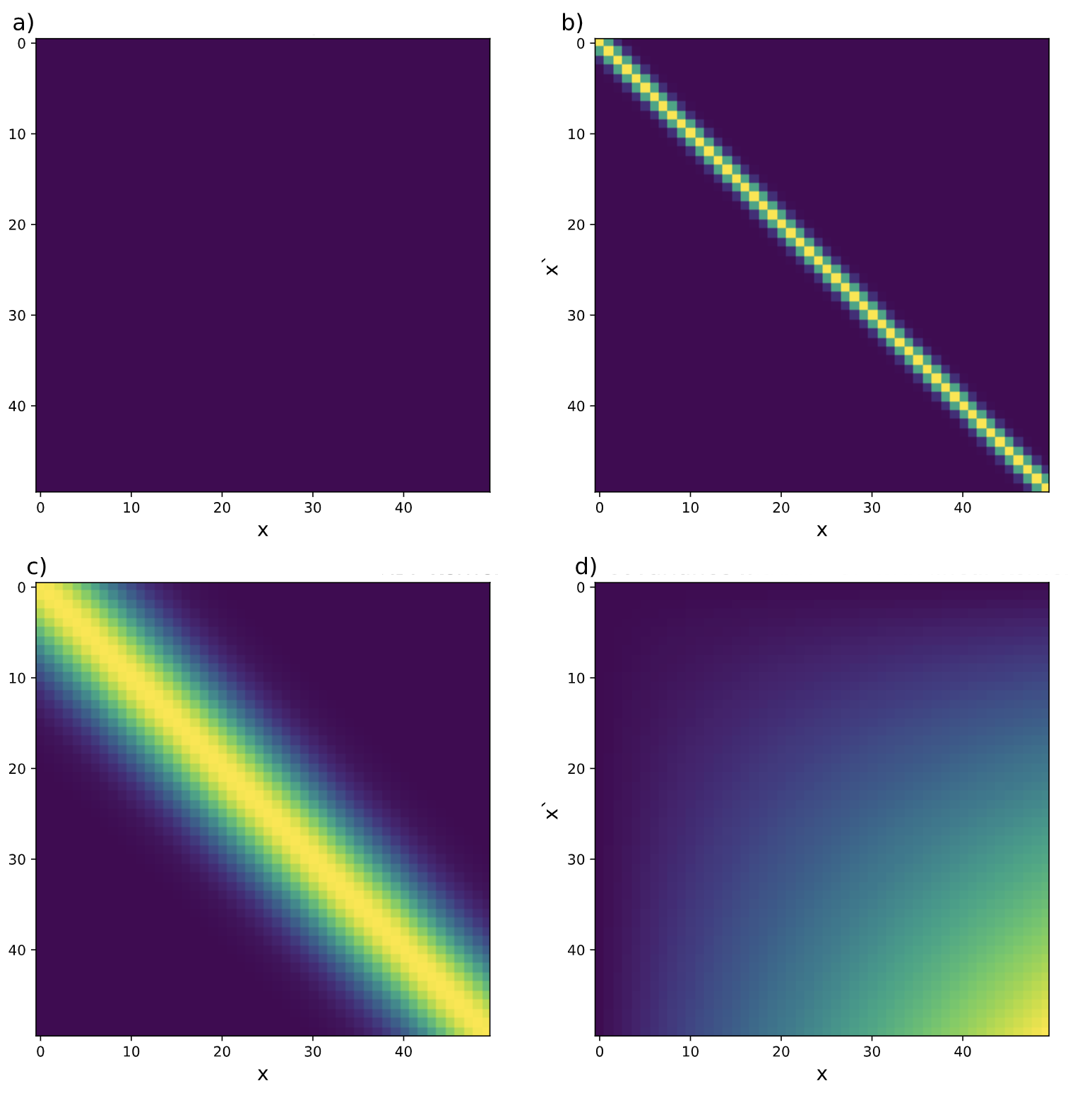}
\decoRule
\caption[Covariance plots for different probabilistic model architectures from Figure \ref{fig:probml_graphical_model_architectures}: a) Naive Bayes - a single latent variable $x$ estimated from multiple independent observations $\{y_1,y_2,...,y_n\}$, b) Hidden Markov Model - a latent variable $x_i$ conditioned on the local context of two neighboring variables $x_{i-1}$ and $x_{i+1}$, c) Gaussian Process with the RBF kernel - a model with infinite number of latent variables $\{x_1,x_2,...,x_n\}$ conditioned on independent observations $\{y_1,y_2,...,y_n\}$ , and d) Gaussian Process with the Linear kernel - a model with infinite number of latent variables. The covariance function, also known as a kernel or covariance matrix, is computed with $cov(x,x`)$ for all possible combinations of latent variables $\{x_1,x_2,...,x_n\}$. The form of a $cov()$ function depends on the probabilistic model architecture]{Covariance plots for different probabilistic model architectures from Figure \ref{fig:probml_graphical_model_architectures}: a) Naive Bayes - a single latent variable $x$ estimated from multiple independent observations $\{y_1,y_2,...,y_n\}$, b) Hidden Markov Model - a latent variable $x_i$ conditioned on the local context of two neighboring variables $x_{i-1}$ and $x_{i+1}$, c) Gaussian Process with the RBF kernel - a model with infinite number of latent variables $\{x_1,x_2,...,x_n\}$ conditioned on independent observations $\{y_1,y_2,...,y_n\}$ , and d) Gaussian Process with the Linear kernel - a model with infinite number of latent variables. The covariance function, also known as a kernel or covariance matrix, is computed with $cov(x,x`)$ for all possible combinations of latent variables $\{x_1,x_2,...,x_n\}$. The form of a $cov()$ function depends on the probabilistic model architecture.}
\label{fig:probml_covariance_plots_for_pgms}
\end{figure}

To get a better intuition on how different kernels perform in the temperature estimation problem, the GP model with a linear kernel is evaluated, with the results presented in Figure \ref{fig:probml_posterior_plots_for_pgms}d. The linear kernel (Eq. \ref{eqn:probml_linear_kernel}) corresponds to Bayesian linear regression \cite{williams2006gaussian}, having the ability to model only linear functions with respect to the feature vector $f(\mathbf{x})$ \cite{williams2006gaussian}.

\begin{equation}
cov_{linear}(\mathbf{x}_i,\mathbf{x}_j)= \sigma_b^2 +\sigma_v^2(f(\mathbf{x})_i-c)(f(\mathbf{x})_j-c)
\label{eqn:probml_linear_kernel}
\end{equation}
GP covariance matrices can be presented graphically, providing some insights into how the latent variables $\mathbf{x}_i$ are correlated with each other. Figures \ref{fig:probml_covariance_plots_for_pgms}c and \ref{fig:probml_covariance_plots_for_pgms}d show the covariance matrices for the RBF and linear kernels, respectively. Interestingly, two previously described models, the Naive Bayes and HMM, can be seen as special cases of GP with particular forms of the kernel function. Figure \ref{fig:probml_covariance_plots_for_pgms}a shows the covariance matrix for the Naive Bayes model, whereas the HMM model is presented in Figure \ref{fig:probml_covariance_plots_for_pgms}b.

\bigskip

\textbf{Posterior estimation}

\bigskip

Consider the task of estimating the temperature value $\tilde{x}_*\sim \mathcal{N}(\tilde{\mu}_*, \tilde{\sigma}_*^2)$ at the location $x_*$. The variable $\mathbf{y}$ represents the observed temperature values, and $\mathbf{x}$ denotes the corresponding latent variable. The posterior mean of  $x_*$ is defined by: 

\begin{equation}
\tilde{\mu}_*=k(x_*,\mathbf{x})(k(\mathbf{x},\mathbf{x}) + \sigma_y^2I)^{-1}\mathbf{y}
\end{equation}whereas the posterior variance is given by:
\begin{equation}
\tilde{\sigma}_*^2=k(x_*,x_*) - k(x_*,\mathbf{x})(k(\mathbf{x},\mathbf{x})+\sigma_y^2I)^{-1}k(\mathbf{x},x_*)
\end{equation}
$k(\mathbf{x},\mathbf{x})$ is a shortcut for the covariance function $cov(\mathbf{x},\mathbf{x})$. The variance $\sigma_y^2$ is the independent Gaussian noise of the likelihood function from Eq. \ref{eqn:probml_gp_likelihood_function}.

The equations for $\tilde{\mu}_*$ and $\tilde{\sigma}_*^2$ are computationally expensive, with cubic runtime complexity $O(N^3)$ and quadratic space complexity $O(N^2)$, where N is the number of observations (temperature measurement) in the training data. The key operation is to compute the inverse of the covariance function $(k(\mathbf{x},\mathbf{x})+\sigma_y^2I)^{-1}$, where the dimensionality of $\mathbf{x}$ is N. One way to overcome high computational complexity is to use inducing points, which will lower the dimensionality of the covariance matrix from $N x N$ to $N x M$, where M is the number of inducing points \cite{williams2006gaussian}. The inducing points can be selected directly from the training data by random selection, clustering the training data into clusters, or creating `virtual' inducing points during optimization of the likelihood function.

\bigskip
 
\textbf{Model training}

\bigskip

The model is trained with a gradient decent-based algorithm by optimizing the marginal likelihood function defined in Equations \ref{eqn:probml_gp_marginal_likelhood} and \ref{eqn:probml_gp_marginal_likelhood_details}. The runtime and space complexity are the same as for the case of posterior estimation presented in the previous section: $O(N^3)$ and $O(N^2)$, respectively. A similar technique based on inducing points can be used to scale training to larger datasets.

\begin{equation}
p(\mathbf{y}|\boldsymbol{\theta}) = \int p(\mathbf{y}|\mathbf{x},\boldsymbol{\theta})p(\mathbf{x},\boldsymbol{\theta})d\mathbf{x}
\label{eqn:probml_gp_marginal_likelhood}
\end{equation}

\begin{equation}
log\mathcal{L}(\boldsymbol{\theta}) = log p(\mathbf{y}|\boldsymbol{\theta}) = -0.5\mathbf{y}^T(k(\mathbf{x},\mathbf{x}) + \sigma_n^2I)^{-1}\mathbf{y} - 0.5log|k(\mathbf{x},\mathbf{x}) + \sigma_y^2I| - \frac{n}{2}log2\pi
\label{eqn:probml_gp_marginal_likelhood_details}
\end{equation}
where $\boldsymbol{\theta}$ represents trainable model parameters.

\bigskip

\textbf{Summary of Gaussian Processes}

\bigskip

Gaussian Processes (GPs) provide a powerful framework for creating probabilistic machine learning models. With the use of a covariance function, many model architectures can be created, each taking into account different prior assumptions. Depending on the choice of the covariance function, GPs  can capture both short-term and long-term temporal dependencies in the training data. GPs perform very well when the model has to make decisions under uncertainty with relatively little training data available.

GPs have some weaknesses, despite their solid mathematical foundations and the ability to generalize to multiple different modeling use cases. First, GPs are computationally expensive, and it is difficult to scale this method to millions of training examples. Second, GP is a shallow machine learning model, which means that it cannot easily discover deep dependencies in the data - something that deep neural networks \cite{goodfellow2016deep} and decision trees \cite{ali2012random} can do. There is a deep learning model called Deep Gaussian Processes \cite{damianou2013deep} that can include multiple GP layers stacked on top of each other, but this model is computationally expensive. Finally, GP models make Gaussian assumptions about the prior probability distribution and the likelihood function, which can lead to less accurate posterior estimates in applications such as vision and speech.

\subsubsection{Summary of probabilistic machine learning}

Probabilistic machine learning models provide an elegant framework for creating generative models that can reason under uncertainty. However, probabilistic models make strong assumptions about the generative process behind the training data, often modeling latent variables with the Gaussian distribution. The Gaussian distribution is used not because it represents the underlying process well, but because the mathematics behind it becomes simpler.  One alternative to probabilistic models are deep learning techniques such as deep neural networks.  Deep neural networks can more accurately represent the underlying generative process without making Gaussian assumptions, leading to more precise models. In addition, deep neural networks can incorporate elements of probabilistic machine learning to create models that are both precise and can reason under uncertainty. The following two sections present deep neural networks and their probabilistic perspective in more detail.

\subsection{Deep learning}

Deep learning generally refers to any machine learning model that can learn data representation at multiple levels. Such models consist of multiple layers processing the input signal through a series of transformations to generate the output signal. Each layer can take inputs from multiple layers and generate new data that represent specific signal characteristics. Deep Neural Networks (DNN), the most popular class of deep learning, the task is to estimate the variable $y=f(x)$, where the output $y$ and the input  $x$ variables can be scalars, vectors, or tensors, and the dependencies between the variables are represented by computational blocks such as Feed-forward Layer \cite{goodfellow2016deep}, Convolutional Neural Network (CNN) \cite{gu2018recent}, and Recurrent Neural Network (RNN) \cite{sutskever2014sequence}.

One of the first commercially deployed deep learning models is the speaker verification system based on multi-layer neural networks \cite{heck2000robustness}. Deep learning is commonly identified with neural networks, but there are other types of deep learning models, such as Deep Gaussian Processes \cite{damianou2013deep}. This section focuses on deep learning techniques that are used in the thesis to create various models for detecting pronunciation errors in non-native speech.

\subsubsection{Perceptron, dense layer and multi-Layer perceptron}

\begin{figure}[th]
\centering
\includegraphics[width=1\textwidth]{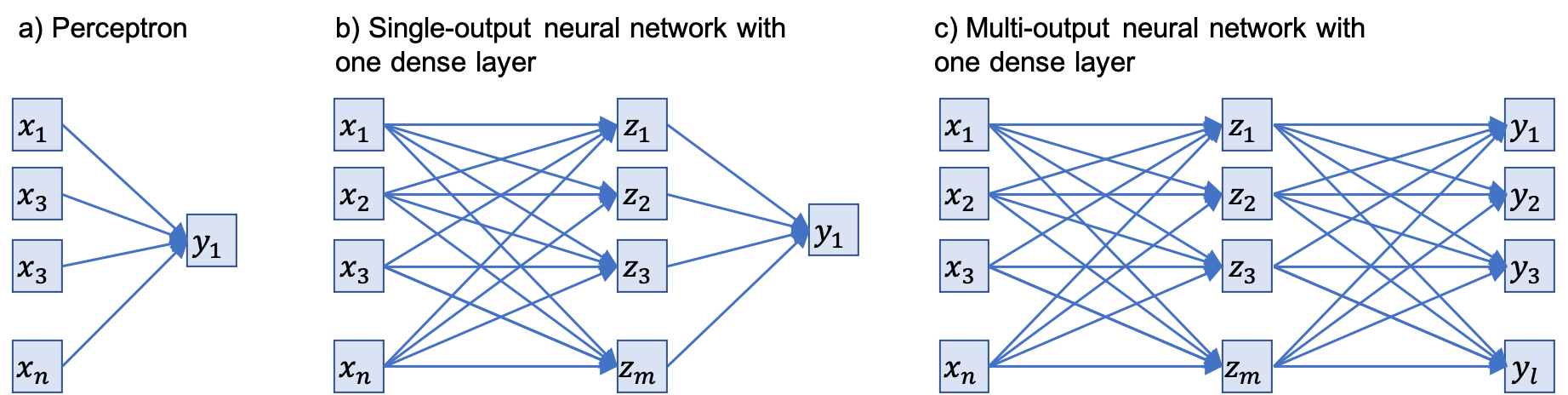}
\decoRule
\caption[Neural network architectures based on the perceptron and a dense layer components: a) neural network with input vector $\mathbf{x}$ and scalar output $y_1$, known as the perceptron, b) neural network with input vector $\mathbf{x}$, one dense layer $\mathbf{z}$, and scalar output $y_1$, c) neural network with input vector $\mathbf{x}$, one dense layer $\mathbf{z}$, and vector-based output $\mathbf{y}$]{Neural network architectures based on the perceptron and a dense layer components: a) neural network with input vector $\mathbf{x}$ and scalar output $y_1$, known as the perceptron, b) neural network with input vector $\mathbf{x}$, one dense layer $\mathbf{z}$, and scalar output $y_1$, c) neural network with input vector $\mathbf{x}$, one dense layer $\mathbf{z}$, and vector-based output $\mathbf{y}$.
}
\label{fig:deep_learning_dense}
\end{figure}

The perceptron is a basic building block of deep neural networks \cite{rosenblatt1960perceptron}. Let $\mathbf{x}$ be a $1 \times n$ input vector, $\mathbf{w}$ be $1 \times n$ vector of trainable parameters, $w_0$ be a trainable scalar parameter, and $\kappa$ be a non-linear transform function. The output scalar value $y_1$ is computed as follows:

\begin{equation}
y_1 = \kappa(\mathbf{xw^T} + w_0)
\end{equation}
A graphical representation of the perceptron  is shown in Figure \ref{fig:deep_learning_dense}a. The non-linear transform $\kappa$ is known as the activation function. Popular variants of the activation function include the sigmoid, TanH and ReLU (Rectified Linear Unit)  functions\cite{goodfellow2016deep}. The perceptron can be used as a binary classification model, but only for patterns that can be linearly separated. Exclusive OR (XOR) is a classic non-linear function $f: X \mapsto  y$, where $X\in \mathcal{R}^{2}$ and $y \in \{0,1\}$, which cannot be separated linearly into two binary categories.

The perceptron can be generalized by stacking multiple layers, also known as dense layers, on top of each other. Such a model is called a Multi-Layer Perceptron \cite {goodfellow2016deep}. MLP is shown in Figure \ref{fig:deep_learning_dense}b. By stacking multiple layers, the model is able to separate non-linear multi-dimensional spaces such as the XOR function, but only if the $\kappa$ activation function is non-linear.  Stacking multiple layers followed by linear activation functions does not make the model non-linear. In addition, MLP can support multi-output functions by producing a vector-based output $\mathbf{y}$ as shown in Figure \ref{fig:deep_learning_dense}c.

\subsubsection{Convolutional neural networks}

Convolution Neural Networks (CNN) \cite{goodfellow2016deep} are designed to detect patterns in highly-dimensional unstructured data such as images, video, and speech. The basic idea is based on the observation that the same processing block can be applied to different parts of the input signal. With this approach, fewer trainable parameters are needed and the network is less likely to overfit. Compared to CNN, the MLP network requires orders of magnitude more network parameters because of having to map between all elements of the input and output layers. 

\begin{figure}[th]
\centering
\includegraphics[width=0.6\textwidth]{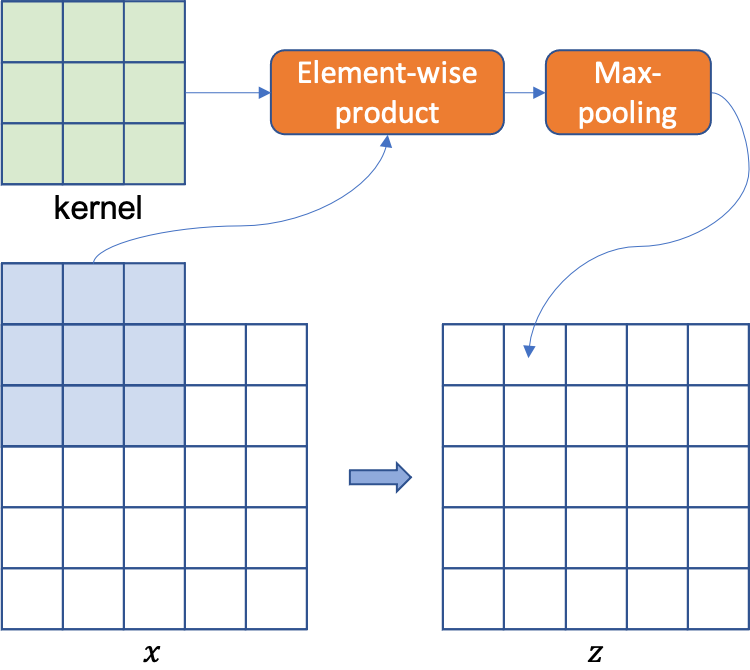}
\decoRule
\caption[An operation in a convolutional neural block that maps between a single $z_{ij}$ value in the $\mathbf{z}$ output tensor (layer) and the $\mathbf{x}$ input layer. A 3x3 convolutional kernel (filter) is multiplied element-wise by the corresponding region of the $\mathbf{x}$ input layer, followed by the max-pooling operation]{An operation in a convolutional neural block that maps between a single $z_{ij}$ value in the $\mathbf{z}$ output tensor (layer) and the $\mathbf{x}$ input layer. A 3x3 convolutional kernel (filter) is multiplied element-wise by the corresponding region of the $\mathbf{x}$ input layer, followed by the max-pooling operation.}
\label{fig:deep_learning_convolution}
\end{figure}

Let $\mathbf{x}$ be $n \times m$ dimensional input tensor and $\mathbf{z}$ be $n \times m$ dimensional output tensor. Let $i$ and $j$ be the indices of a single cell in the tensor, e.g. $z_{01}$, where $i=0$ and $j=1$, corresponds to the second element in the first row of the tensor $\mathbf{z}$ as shown in Figure \ref{fig:deep_learning_convolution}. The value of a single $z_{ij}$ element is calculated by multiplying (element-wise) the kernel tensor $\mathbf{k}$, for brevity called `kernel', by the corresponding region $\mathbf{x_k}$ of the input tensor $\mathbf{x}$. The result of the element-wise multiplication is passed through the $\max$ function, producing a single value $z_{ij}$. The complete operation to compute $z_{ij}$ is defined as follows:

\begin{equation}
z_{ij}= \max{(\mathbf{x_k} \odot \mathbf{k} )}
\end{equation}

In a generic case, multiple kernels can be applied to the input tensor $\mathbf{x}$, which results in the output tensor $\mathbf{z}$ of shape $n \times m \times l$, where $l$ is the number of kernels. Multiple convolutional blocks can be stacked on top of each other to extract features at different levels of abstraction. The dimensionality of the input $\mathbf{x}$ and output $\mathbf{z}$  kernels do not need to match, and the $\max$ function can be replaced with other options such as the $average$ function. 

\subsubsection{Recurrent neural networks}

\begin{figure}[th]
\centering
\includegraphics[width=0.7\textwidth]{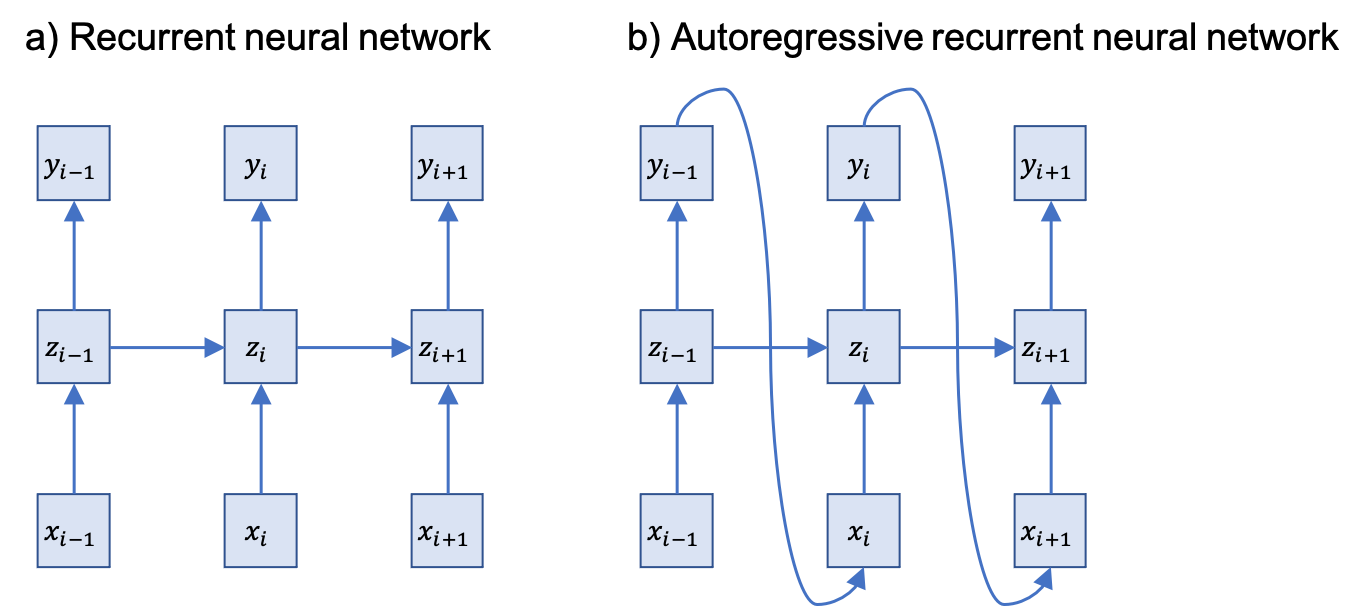}
\decoRule
\caption[Recurrent neural network architectures. a) Recurrent network without autoregressive loop. All $x_i$ inputs must be available in advance to the model. b) Autoregressive recurrent neural network. Only the first $x_0$ element must be available to the model. In general, the $x_i$ element is computed based on the value of the previous output $y_{i-1}$]{Recurrent neural network architectures. a) Recurrent network without autoregressive loop. All $x_i$ inputs must be available in advance to the model. b) Autoregressive recurrent neural network. Only the first $x_0$ element must be available to the model. In general, the $x_i$ element is computed based on the value of the previous output $y_{i-1}$.}
\label{fig:deep_learning_rnn}
\end{figure}

Recurrent Neural Networks (RNNs)  \cite{goodfellow2016deep} are suitable for modeling sequential data, such as a speech signal, where future values depend on past values. RNNs compute and maintain the $z_i$ latent state by sequentially processing the $x_i$ elements of the $\mathbf{x}$ input sequence to generate the $\mathbf{y}$ output sequence. RNNs can be used to process a signal known in advance to the model, such as recorded speech, as shown in Figure \ref{fig:deep_learning_rnn}a. Alternatively, RNNs can generate new sequential data, such as a speech signal. In this scenario, the value of $x_i$ input depends on the value of the $y_{i-1}$ output that was generated previously. 

Gated Recurrent Unit (GRU) and Long Short-Term Memory (LSTM) are the most popular variants of the blocks that compute the $z_i$ latent space \cite{goodfellow2016deep}. In a nutshell, the GRU and LSTM blocks track the latent state $z_i$ based on  previously processed inputs $x_{i-1}$ and $z_{i-1}$, and they can update the $z_i$ with new information or forget its state.

\subsubsection{Attention}

The attention mechanism \cite{vaswani2017attention} maps the $\mathbf{x}$ input sequence to the $\mathbf{y}$ output sequence. Each $y_i$ element in the output sequence is computed from all elements of the input sequence, with the attention mechanism, telling which elements of the input sequence should be used when computing the output value. In other words, to which elements of the input sequence the $y_i$ element should attend to. Hence, the name of this mechanism is attention.

\begin{figure}[th]
\centering
\includegraphics[width=0.5\textwidth]{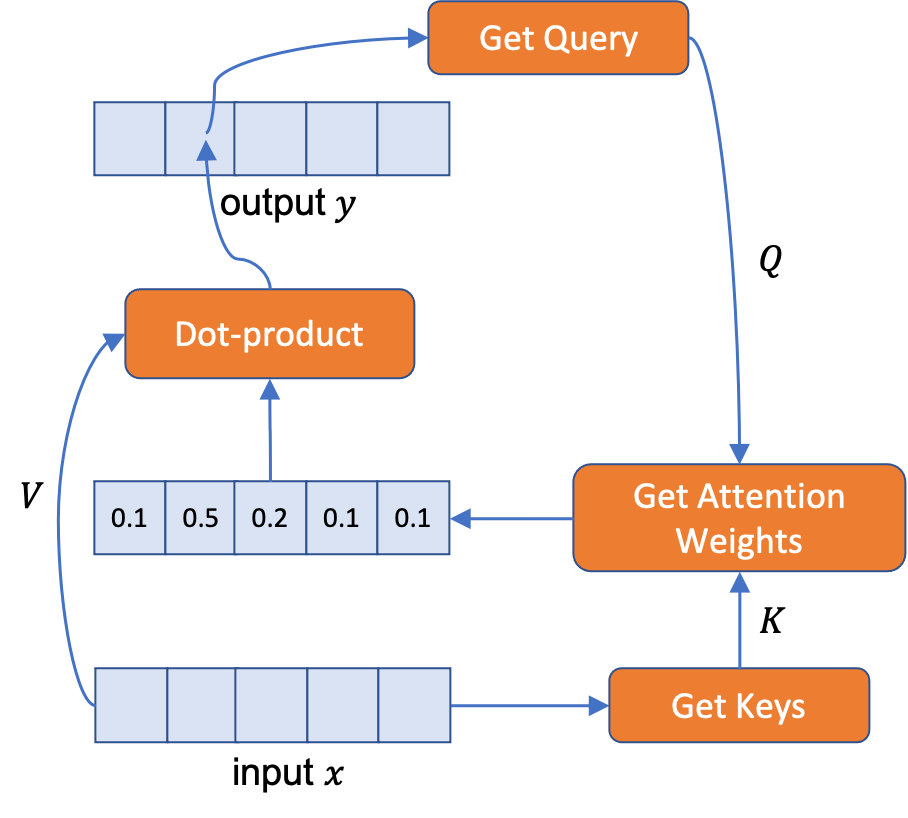}
\decoRule
\caption[The attention mechanism illustrated by the example of computing a single element of the output sequence $\mathbf{y}$ from the input sequence $\mathbf{x}$. $Q$ - query, $K$ - keys, $V$ - values]{The attention mechanism illustrated by the example of computing a single element of the output sequence $\mathbf{y}$ from the input sequence $\mathbf{x}$. $Q$ - query, $K$ - keys, $V$ - values.}
\label{fig:deep_learning_attention}
\end{figure}

The attention mechanism has three inputs: query $\mathbf{Q}$, values $\mathbf{V}$, and keys $\mathbf{K}$, as illustrated in Figure \ref{fig:deep_learning_attention}. The values $\mathbf{V}$ represent the $\mathbf{x}$ input sequence. The query $\mathbf{Q}$ corresponds to the element $y_i$ in the output sequence $\mathbf{y}$. The keys $\mathbf{K}$ are derived from the $\mathbf{x}$ input sequence, which tells how much each $x_i$ element should be included in the computation of $y_i$. The softmax function of the dot-product of the query $\mathbf{Q}$ and the keys $\mathbf{K}$ results in the vector of attention weights (probabilities). The dot-product between the attention weights and the values $\mathbf{V}$ returns the $y_i$ output. The attention equation is defined by:
\begin{equation}
Attention(\mathbf{Q}, \mathbf{K}, \mathbf{V}) = softmax(\dfrac{\mathbf{Q}\mathbf{K}^t}{\sqrt{d_k}})\mathbf{V}
\label{eq_att}
\end{equation}
In Eq. \ref{eq_att}, the dot-product between the query $\mathbf{Q}$ and the keys $\mathbf{K}$ is used to calculate attention weights, but there are other options available. Almost any type of neural network can be used to compute attention weights. Chaudhari et al. present a comprehensive review of various attention mechanisms \cite{chaudhari2021attentive}. The attention mechanism is suitable for tracking very long dependencies because it can attend to all elements in the input data. 

\subsection{Deep learning -- probabilistic perspective}
\label{subsec:prob_perspective_on_deep_learning}

Understanding the probability theory and the Bayesian rule concept is essential in getting to the origins of various neural network architectures. Many neural networks and other machine learning models have probabilistic counterparts. Linear regression, one of the simplest regression models, can be implemented as a probabilistic model known as Bayesian linear regression. Linear regression can be generalized  as the Gaussian Process, and the Gaussian Process can be implemented as a neural network with one hidden layer with an infinite number of layers. Dropout and L2 regularization in neural networks are related to the concept of a prior variable in Bayesian networks. There are endless examples of machine learning models with neural networks and probabilistic counterparts, many of which are presented in two excellent books on probabilistic machine learning by Christopher Bishop \cite{bishop2006pattern} and Kevin Murphy \cite{murphy2012machine}. 

\begin{figure}[th]
\centering
\includegraphics[width=0.8\textwidth]{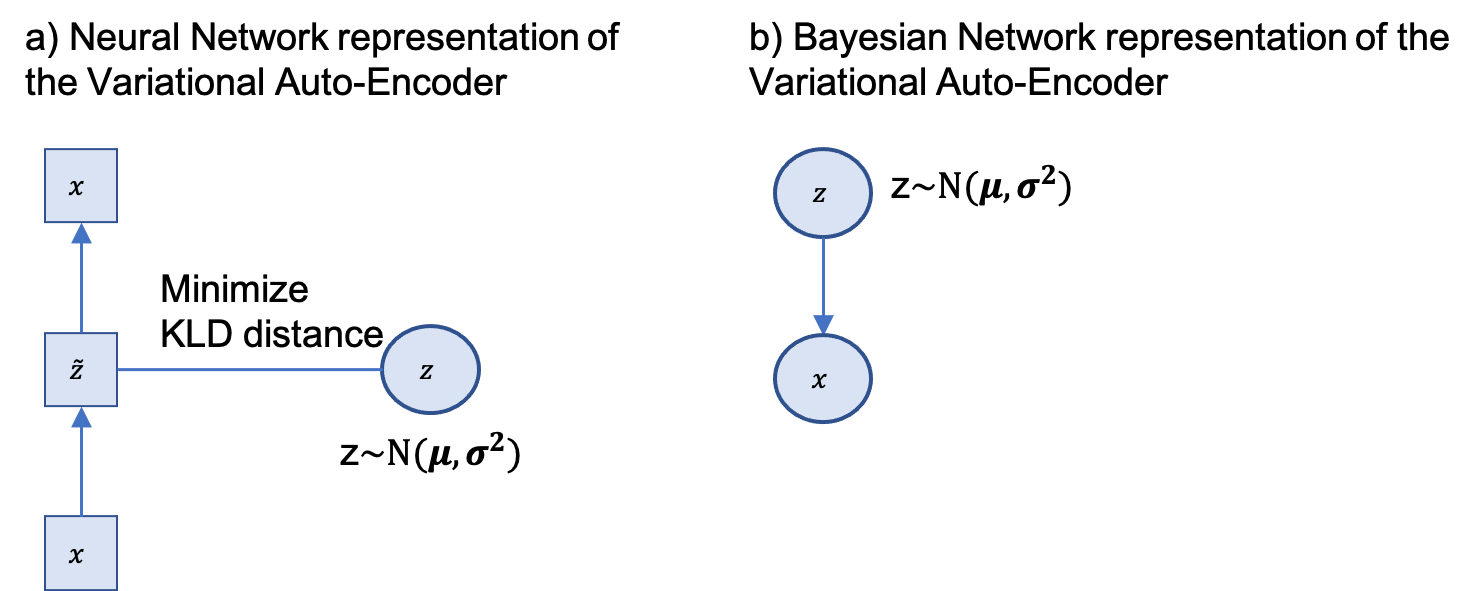}
\decoRule
\caption[Architecture of the Variational Auto-Encoder (VAE) model. a) Neural network representation of the VAE model, b) Bayesian network representation of the VAE model]{Architecture of the Variational Auto-Encoder (VAE) model. a) Neural network representation of the VAE model, b) Bayesian network representation of the VAE model.}

\label{fig:deep_learning_vae}
\end{figure}

To illustrate the relationship between the probability theory and neural networks, this section explains how the Variational Auto-Encoder (VAE) neural network can be derived with the use of the probability theory. VAE is an auto-encoder neural network that maps from the $\mathbf{x}$ input to the $\mathbf{x}$ output via the $\mathbf{\tilde{z}}$ bottleneck layer, as shown in Figure \ref{fig:deep_learning_vae}a. 

During training, the sum of the two loses is minimized:

\begin{equation}
log{p(\mathbf{x})}= log p(\mathbf{x} |\mathbf{z}) + D_{KL}(p(\mathbf{z}|\mathbf{x}) || p(\mathbf{z}))
\label{eqn:vae_neural_network}
\end{equation}
where $\mathbf{\tilde{z}}= p(\mathbf{z}|\mathbf{x})$ is the posterior probability of the variable $\mathbf{z}$. The first term is the reconstruction loss that minimizes the distance between the $\mathbf{x}$ input and the $\mathbf{x}$ output variables. The second term is the Kullback–Leibler Divergence (KLD) distance between the $\mathbf{\tilde{z}}$ bottleneck layer and the $\mathbf{z}$ Gaussian prior variable.

At first sight, the motivation for adding the KLD loss is difficult to explain, but it becomes more apparent when we consider the probabilistic variant of the model. Consider a Bayesian network shown in Figure \ref{fig:deep_learning_vae}b with two variables $\mathbf{x}$ and  $\mathbf{z}$. This network takes into account the prior belief that the observed $\mathbf{x}$ variable depends on the variable $\mathbf{z}$ that is unobserved (latent). To train this model, the latent variable has to be integrated out:

\begin{equation}
 log{ p(\mathbf{x})} = log\int p(\mathbf{x},\mathbf{z})d\mathbf{z} 
 \label{eqn:vae_simple}
\end{equation}The integral in Eq. \ref{eqn:vae_simple} can be approximated using the framework of variational inference \cite{jordan1999introduction} as shown in Eq. \ref{eqn:vae_full}:
\begin{align}
    log{ p(\mathbf{x})} = log\int p(\mathbf{x},\mathbf{z})d\mathbf{z} \\
    = log\int p(\mathbf{z}|\mathbf{x}) \frac{p(\mathbf{x},\mathbf{z})}{p(\mathbf{z}|\mathbf{x})}d\mathbf{z}  \\
    \geq E_{p(z|x)}[log\frac{p(\mathbf{x},\mathbf{z})}{p(\mathbf{z}|\mathbf{x})}] \\
    = E_{p(z|x)}[log\frac{p(\mathbf{x} |\mathbf{z})p(\mathbf{z})}{p(\mathbf{z}|\mathbf{x})}]  \\
    = E_{p(z|x)}[log p(\mathbf{x} |\mathbf{z}) + log\frac{p(\mathbf{z})}{p(\mathbf{z}|\mathbf{x})}]  \\
    = E_{p(z|x)}[log p(\mathbf{x} |\mathbf{z})] + E_{p(z|x)}[log\frac{p(\mathbf{z})}{p(\mathbf{z}|\mathbf{x})}]  \\
    = E_{p(z|x)}[log p(\mathbf{x} |\mathbf{z})] + D_{KL}(p(\mathbf{z}|\mathbf{x}) || p(\mathbf{z})) \\
    \label{eqn:vae_full}
\end{align}

The final derivation is as follows:
\begin{equation}
log{ p(\mathbf{x})} \geq E_{p(z|x)}[log p(\mathbf{x} |\mathbf{z})] + D_{KL}(p(\mathbf{z}|\mathbf{x}) || p(\mathbf{z}))
\end{equation}The first term is the VAE neural network reconstruction loss described earlier in Eq. \ref{eqn:vae_neural_network}, while the second term is the KLD loss. Both VAE representations based on neural networks and Bayesian networks are equivalent. The Bayesian representation made it possible to derive the VAE neural network architecture using an elegant mathematical framework of the probability theory. A similar approach can be used to derive other neural network architectures.

%% file: Chapters/ResearchMethodology/PerformanceMetrics.tex
\section{Performance metrics}
\label{sec:metrics}

In machine learning, performance metrics are used to evaluate different models to select the one that performs the best in the real-world scenario \cite{hossin2015review, botchkarev2018performance}. Generally, a performance metric is defined by a function that takes two arguments: the ground-truth value for a target variable and the estimated (predicted) value from a machine learning model. The metric function usually outputs a real-value number that indicates the  overall performance of the model averaged out over all examples in the test data.

As an intuitive example, let us consider a binary classification problem of classifying images into two classes, e.g., apples and oranges. One possible performance metric is `accuracy', defined as the ratio of correctly classified images. However, there are other possible options such as precision, recall, AUC, log-likelihood \cite{hossin2015review, sofaer2019area}. The choice depends on the machine learning task. 

In this section, a review of performance metrics used in the Ph.D. thesis is given, and the choices compared to other possible options are justified. This discussion is divided into two parts dedicated to different types of machine learning problems that require different types of metrics: 
\begin{itemize}
\item Detection of pronunciation errors (mispronounced phones and  incorrect lexical stress errors) - this is a classification machine learning problem in which the task is to estimate the probability of a speech error at the word or the syllable level.
\item Generation of synthetic pronunciation errors in non-native speech and reconstruction of dysarthric speech - this is a regression problem with a goal of generating speech of desired characteristics such as including mispronunciations (non-native speech) or improving the intelligibility of speech (dysarthric speech).
\end{itemize}
\subsection{Metrics for the detection of pronunciation errors}
\label{sec:metrics_pronunciation_errors}
Performance metrics for detecting pronunciation errors are designed to ensure the optimal user experience of using a CAPT tool. Foremost, the tool should correctly identify mispronunciations. A user might get demotivated and eventually abandon using CAPT if the tool often provides incorrect feedback. Second, even if the tool is always correct while providing feedback, it should not miss too many mispronunciations made by the user. Otherwise, the user will be consolidating bad pronunciation habits and language learning will be less efficient. To summarize, a good CAPT tool should aim to: 1) not provide incorrect feedback, 2) not miss mispronunciations. 

\subsubsection{Key metrics}

There are three key metrics to address the user experience requirements: precision, recall, and Area Under the Curve (AUC) \cite{hossin2015review, sofaer2019area}.

The precision metric reflects the requirement `do not provide incorrect feedback'. It is defined as the proportion of raised mispronunciations that are identified correctly:

\begin{equation}
precision  = \frac{TP}{TP+FP} 
\label{eqn:precision}
\end{equation}
where $TP$ (true positives) is the number of correctly detected mispronunciations and $FP$ (false positives) is the number of incorrectly detected mispronunciations.

\bigskip

The recall metric addresses the requirement `do not miss mispronunciations', and it is defined as the proportion of all mispronunciations that are identified correctly:

\begin{equation}
recall  = \frac{TP}{TP+FN} 
\label{eqn:recall}
\end{equation}

where $FN$ (false negatives) is the number of missed mispronunciations. 

\bigskip

In addition to the statistics $TP$, $FP$, and $FN$, there is also the $TN$ (true negatives) quantity, which is the number of correctly identified good pronunciations. All four statistics, when summed up, give the total number of speech segments, e.g., words, for which the pronunciation error detection model is evaluated for. They serve as basic information for other more high-level metrics such as precision, recall, and AUC.

To compute the statistics $TP$, $FP$, $TN$, and $FN$, the test data with spoken sentences are first annotated to provide ground-truth information. Human listeners skilled in English listen to spoken sentences and label speech segments, e.g., words, with a binary label $e_g\in \{0,1\}$, where the value of 1 means that the speech segment is mispronounced. The ground-truth label $e_g$ is compared with the corresponding output of the pronunciation error detection model  $\tilde{e} \in \{0,1\}$. There are four possible combinations of each pair $\{e_g,\tilde{e}\}$, contributing to one of the statistics $TP$, $FP$, $TN$, and $FN$. For example, $\{e_g=0,\tilde{e}=1\}$ adds to the total number of FP.

Instead of directly producing a binary label $\tilde{e} \in \{0,1\}$, the pronunciation error detection models proposed in the Ph.D. thesis estimate the probability of mispronunciation denoted as $e$. The variable $e$ is modeled as a conditional Bernoulli distribution  $e \sim p(e|speech+context)$, conditioned on the speech signal and additional context such as pronunciation of a native speaker. However,  to compute the statistics $TP$, $FP$, $TN$, and $FN$, a binary output from the model is needed. To convert the probability of mispronunciation to a binary output, a threshold $t$ is used as follows:
\begin{equation}
\label{eqn:pron_error_e_thres}
\tilde{e} = 
\left\{ 
  \begin{array}{rl}
     1 &\mbox{ if $p(e)>t$} \\
     0 &\mbox{ otherwise}
   \end{array} 
 \right.
\end{equation}

Changing the threshold $t$ value allows for different trade-offs between precision and recall metrics.  Increasing $t$, increases precision and decreases recall. Decreasing $t$, has the opposite effect. However, this controllability makes it difficult to estimate precision and recall metrics because it is unclear which threshold $t$ value should be used. AUC metric overcomes the need for selecting the value of threshold $t$ \cite{sofaer2019area}. Intuitively, AUC summarizes precision and recall metrics across all possible thresholds, producing a single score between 0 and 1. The value of 0 indicates that pronunciation errors are always detected incorrectly, and the value of 1 means the opposite. The value of 0.5 represents a model that detects pronunciation errors by random, assuming 50\% of all speech segments are mispronounced. The AUC metric is defined as follows:
\begin{equation}
AUC  = \int_0^1 precision(recall^{-1}(x)) dx
\label{eqn:auc}
\end{equation}
where $recall^{-1}(x)$ returns the threshold $t$ value for the recall value $x$. This function is the inverse of $x=recall(t)$ that returns the recall value for a given threshold. Graphically, the AUC metric can be visualized as the area under the curve on a precision-recall plot, with precision placed on the y-axis and recall on the x-axis. Precision-recall plots provide an intuitive view of how precision and recall change across different values of  threshold $t$. For illustration, the examples of precision-recall plots with the corresponding AUC values are presented in Section \ref{sec:weaklys_pron_model_sota}.

To summarize, there are three key metrics used for the evaluation of pronunciation error detection: precision, recall, and AUC. Precision and recall reflect the two user experience requirements:  `do not provide incorrect feedback' and `do not miss mispronunciations', respectively. The AUC metric provides a single-number performance metric, accounting for all possible trade-offs between precision and recall. 

\subsubsection{Discussion}

The metrics of our choice, precision and recall, are already used in the field of pronunciation error detection \cite{leung2019cnn, zhang2021text, yan2021end}. They are especially useful when the data are imbalanced, with fewer positive (incorrect pronunciation) than negative (correct pronunciation) examples. Precision and recall do not depend on the statistic $TN$ (the number of correctly identified good pronunciations), and therefore, they are unlikely to underestimate the negative impact of either missing mispronunciation or raising a false alarm.

FPR (False Positive Rate), also known as False Rejection Rate (FRR), is another popular metric \cite{li2016mispronunciation, leung2019cnn, zhang2021text}. FPR is the ratio of good pronunciations that were incorrectly raised as mispronunciations, and in such a sense, it is similar to precision.
\begin{equation}
FPR  = \frac{FP}{FP+TN} 
\label{eqn:fpr}
\end{equation}
However, contrary to precision, FPR may underestimate the negative effect of raising false pronunciation alarms. In the denominator of the FPR formula, there is the number of correctly identified good pronunciations ($TN$), which may outweigh the number of incorrectly raised mispronunciations ($FP$).

The recall metric is closely related to the False Negative Rate (FNR), also known as the False Acceptance Rate (FAR) \cite{li2016mispronunciation, leung2019cnn, zhang2021text}. FNR is defined as the ratio of all mispronunciations that are identified as good pronunciations. There is no difference between using both metrics, except that recall should be maximized, and FNR minimized.  
\begin{equation}
FNR = \frac{FN}{FN+TP}  =1-recall
\end{equation}
It is somewhat difficult to compare the different pronunciation error detection models using precision and recall metrics. One model may have higher precision, whereas the other model may be better in recall. AUC metric mitigates this problem by providing a single score based on precision and recall values (Eq. \ref{eqn:auc}). F1-score is another single-score metric based on precision and recall, and it is widely used in other works on pronunciation error detection \cite{leung2019cnn, zhang2021text, yan2021end}:
\begin{equation}
f_1=2 \cdot \frac{precision \cdot recall}{ precision+recall}
\end{equation}
Contrary to AUC, F1-score depends on precision and recall values computed for a specific value of threshold $t$ (Eq. \ref{eqn:pron_error_e_thres}). This threshold is applied to the probability of mispronunciation used to compute the precision and recall values. Different pronunciation error detection models might perform differently for the same threshold, and it is hard to decide on its value in order to compare different models. AUC metric averages out over all possible values of threshold $t$, making it easier for model comparison.

In two works, the accuracy metric is used \cite{leung2019cnn, zhang2021text}, defined as the ratio of correctly classified speech segments, either as mispronunciations or good pronunciations:
\begin{equation}
accuracy = \frac{TP+TN}{TP+TN+FP+FN}
\end{equation}
However, this metric is not used in the Ph.D. thesis because it does not work well with imbalanced data. For example, for the data set with 10\% of mispronunciations, the model that never raises any mispronunciations would have an accuracy of 90\%, which does not sound correct. On the other hand, both precision and recall values would equal 0, correctly indicating poor model performance.

Many discussed metrics have multiple names, making it harder to review and compare different models in the field. A good example is the recall metric, also known as True Positive Rate (TPR), Sensitivity, and Hit rate. In the Ph.D. thesis, the naming convention from the machine learning field is used with names, such as precision, recall, TPR, FPR and FNR.

\subsection{Metrics for the generation of speech}
There are two types of machine learning models for speech generation discussed in the Ph.D. thesis. First, the generation of synthetic pronunciation errors helps to improve the accuracy of detecting pronunciation errors in non-native speech. Thanks to improved accuracy, a person learning a foreign language receives a better user experience of using a CAPT tool. The second machine learning model performs the reconstruction of dysarthric speech that helps people with dysarthria disorder to better communicate with other people.  

Both models are different in the way they influence the user experience. Synthetic pronunciation errors generated by the first model are not visible to language learners; they are used only to increase the size of the training data, improving accuracy of machine learning models. This is an example of an indirect impact on the user experience. Besides,  speech reconstruction performed by the second model directly influences the user experience. Poor reconstruction may negatively influence the intelligibility and fluency of speech perceived by humans. The second model impacts the user experience directly. The difference between the direct and indirect impact on the user experience suggests that dedicated approaches to performance metrics should be used.

\subsubsection{Metrics for the generation of synthetic pronunciation errors}

Synthetic mispronounced speech is added to the training data to improve accuracy of pronunciation error detection. Intuitively, to help achieve better accuracy, synthetic speech should simulate as closely as possible real speech of non-native speakers. This intuition suggests that a good performance metric should reflect relevant aspects of a synthetic speech signal, such as the signal quality and the similarity to the mispronounced speech of human speakers. However, what really matters to CAPT users are not the characteristics of a synthetic speech signal, but whether using synthetic pronunciation errors improves the accuracy of pronunciation error detection. Therefore, to measure the benefits of using synthetic pronunciation errors, the same performance metrics as for the detection of pronunciation errors are used (see Section \ref{sec:metrics_pronunciation_errors}). 

To measure the effect of adding synthetic speech errors to the training data, two models for the detection of pronunciation errors are evaluated and compared with each other. For the first model, synthetic speech errors are added to the training data, whereas for the second model, they are not. Precision, recall, and AUC metrics are computed for both models, and their deltas are analyzed. Such investigation in which one aspect of the model is removed to understand its contribution to the overall model performance is known as an ablation study \cite{meyes2019ablation}.

\subsubsection{Metrics for speech reconstruction}

The goal of speech reconstruction is to make it easier for people with speech disorders to communicate with other people. Performance metrics should reflect human opinions about reconstructed speech. In a perceptual speech test, human listeners listen to multiple samples of speech and answer various questions, for example, 'please rate the naturalness of speech on the scale from 0 (the least natural) to 100 (the most natural)'. Ratings obtained from multiple listeners are aggregated into performance metrics, such as Mean Opinion Score (MOS) and MUltiple Stimuli with Hidden Reference and Anchor (MUSHRA), reflecting human opinions on certain aspect of speech \cite{merritt2018comprehensive, wagner2019speech}. By varying questions asked to listeners, multiple characteristics of speech may be assessed, such as naturalness, fluency, intelligibility, and similarity to other speech. Perceptual speech tests performed by human listeners are also known as subjective evaluation tests, because they reflect personal human opinions.

Human perceptual tests are laborious. They usually engage between 20 and 50 human listeners who have to listen to each audio sample and score it carefully. Automated perceptual evaluation tests are designed to simulate human perception and complement human-based evaluation \cite{valizada2021development, wagner2019speech}. Some automated tests attempt to mimic directly human listeners, such as AutoMOS \cite{patton2016automos} that estimates the naturalness of speech on a scale from 1 (the most natural) to 5 (the least natural). In comparison, other automated models produce less interpretable metrics, such as the distance between generated and reference speech samples. Mel Cepstral Distortion (MCD) is an example of such distance based metrics \cite{skerry2018towards, valizada2021development}. The AutoMOS model does not require providing a reference audio signal, whereas, in MCD, this signal is required. Reference-free methods are more flexible, as they can be used to assess any generated speech sample, even if the reference signal is not available. While working on new machine learning models for speech generation, multiple evaluations have to be conducted to assess the progress of work. Automated perceptual speech tests are often used in this research phase. Final evaluations of the speech generation models are usually conducted by human listeners. 

In this Ph.D. thesis, MUSHRA is used as the primary metric to assess the performance of speech reconstruction. MUSHRA has been initially designed to evaluate the quality of audio coders in telecommunications \cite{series2014method}, but in recent years it has been successfully adopted in the field of speech synthesis \cite{rosenberg2017bias, merritt2018comprehensive, wagner2019speech, mu2021review}, and music \cite{hines2015visqolaudio}. In the MUSHRA test, listeners evaluate multiple systems, for example, different machine learning models for speech reconstruction. Various aspects of speech may be evaluated, such as signal quality, naturalness, and intelligibility. The goal of the test depends only on how the question is formulated, for example, `please rate the naturalness of speech'. A listener is presented with audio samples, one sample for each system, and rates them on a scale from 0 (the lowest performance) to 100 (the highest performance). There are multiple rounds (screens) in which a listener listens to audio samples and scores them. Collected scores are aggregated across listeners into multiple statistics such as the mean, median, and rank values, and then statistical tests such as $p$-value and $t$-test are conducted to conclude the final outcome of the MUSHRA test.
 
Original MUSHRA specification created by International Telecomm. Union – Radio communication Sector (ITU-R) makes a few additional recommendations for the MUSHRA test construction \cite{series2014method}. On each MUSHRA screen, listeners are asked to rank one system with a score of 100 (upper anchor) and one system with a score of 0 (lower anchor). These anchors help calibrate the evaluated system on the 0-100 scale. In the field of speech synthesis, sometimes, only the upper anchor is employed, and the user is not forced to score one system as 100 \cite{merritt2018comprehensive}.
 
Merrit et al. suggest using 50 listeners and assigning 40 screens to each listener to achieve repeatable and statistically significant results \cite{merritt2018comprehensive}. However, measuring statistical significance in perceptual tests is a complex problem. In MUSHRA, standard $p$-value-based statistical tests are common. These tests cannot be reliably used because they rely on the assumption that listener responses are independently and identically distributed (iid), but this is not guaranteed. For example, one tester can strongly prefer audio samples generated by one system, whereas the second listener can have a strong preference for the second system. In this case, all scores within a listener will be correlated more than the scores between different listeners \cite{bishop2006pattern}. Effectively, in such situations, $p$-value-based tests provide an over-optimistic estimate of statistical significance. Due to violating the iid assumption, selecting the number of listeners, the number of unique texts for which audio samples are generated, and the number of screens per tester is often a trial and error process.

MOS is another popular metric for synthetic speech evaluation \cite{rosenberg2017bias, wang2017tacotron}. Listeners listen to audio samples for multiple systems one sample at a time and rate them on a scale between 1 and 5, sometimes between 1 and 7. The average scores are computed for all systems and compared against each other. The statistical significance of the results is computed with the paired $t$-test. Calculations of the mean and $p$-value statistics used in the $t$-test assume that the input data are normally distributed; however, the MOS scale is ordinal, which violates this assumption \cite {rosenberg2017bias}. On the other hand, the MUSHRA scale is more granular (0-100), making the scale closer to the continuous nature of the normal distribution. It must be noted that there exist statistical tests and statistics that do not require the data to be normally distributed, such as median \cite{lee2008introduction} and Wilcoxon signed-rank test \cite{woolson2007wilcoxon}. Another difference between both tests is that in MUSHRA, a listener is presented with audio samples for all systems at once and then rates them, whereas in MOS, a listener listens to audio samples and rates them one at a time.  Thanks to presenting multiple systems at once, listeners can calibrate between different systems before providing their scores. Therefore, MUSHRA obtains statistically significant results faster than MOS  \cite{wagner2019speech}.

A preference test \cite{mu2021review, gabrys2021improving} is similar to a MUSHRA test. A listener listens to multiple systems in parallel and then rates them. One difference is that only two systems are evaluated in the preference test. Second, contrary to the fine-grained 0-100 scale in MUSHRA, a listener selects from the limited set of choices: system A is better, system B is better, and both systems are the same. Sometimes, the scale is extended with two additional options: system A or B is significantly better. The preference test is also known as the AB test. There exists a variant of the AB test called the ABX test \cite{mu2021review}. A listener is presented with the reference audio signal X and has to decide which of the A and B systems is closer to the reference signal. Statistical significance of AB tests is conducted with the Binomial test \cite {abdi2007binomial}. This test provides the $p$-value score that gives the probability that systems A and B are the same based on provided preference scores. If the $p$-value is low, e.g., <0.01, then it means that one of the systems has been scored higher by listeners; otherwise, it is assumed that the difference between the two systems is due to random sampling. The Bernoulli test assumes that all individual scores provided by listeners are iid. Because this assumption does not always hold, the $p$-value tends to be over-estimated (lower than it should be).

In conclusion, the MUSHRA test is used in the doctoral dissertation for the evaluation of speech reconstruction. MUSHRA enables listeners to listen to speech samples from multiple systems simultaneously and score them on a continuous scale from 0 to 100, providing more precise results on the quality of the speech being assessed. 

%% file: Chapters/PronunciationErrorDetection/PronunciationErrorDetection.tex

\chapter{Pronunciation error detection} 

\label{chapter:pron_error_detection} 



\section{Introduction}

This chapter constitutes the main scientific part of the doctoral dissertation. The aim is to explore the key research thesis to create new deep learning models for pronunciation error detection:

\begin{center}
\textbf{It is possible to improve the accuracy of deep learning methods for detecting pronunciation errors in non-native English by employing synthetic speech generation and end-to-end modeling techniques that reduce the need for phonetically transcribed mispronounced speech.}
\end{center}

The results of this research have been published in scientific publications at major international speech conferences and scientific journals \cite{korzekwa22_speechcomm, korzekwa21b_interspeech, korzekwa2021mispronunciation, korzekwa21_interspeech, korzekwa2019deep}. These publications address the following challenges of the existing methods for detecting pronunciation errors in non-native speech, with respect to the research thesis. The research background on these challenges was presented in Section \ref{sec:research_theses}.

\begin{enumerate}
\item Transcription of non-native speech is a difficult and costly process

Section \ref{sec:weaklys_model} describes a new approach to pronunciation error detection that does not require phonetic transcriptions of non-native speech.

\item Aligning canonical and recognized phonemes accurately is challenging

Section \ref{sec:weaklys_model} describes an end-2-end model for detecting pronunciation errors that does not need to align between canonical and recognized phonemes.

\item Not all pronunciation errors are the same

Section \ref{sec:weaklys_model} describes a new approach to categorizing pronunciation errors by severity to further improve the accuracy of detecting pronunciation errors.

\item A sentence can be pronounced correctly in multiple different ways

Section \ref{sec:uncertainty_modeling} describes a probabilistic model that reduces the number of false mispronunciation alarms by accounting for multiple correct pronunciations of the same sentence.

\item Practicing lexical stress is an important part of CAPT

Section \ref{sec:lexical_stress} describes a new method for the detection of lexical stress errors based on synthetically generated lexical errors and the attention mechanism.

\item The availability of non-native speech with pronunciation errors is limited

Section \ref{sec:speech_synthesis} describes a new approach to pronunciation error detection that reformulates the problem of detecting pronunciation errors as a speech generation task.

\item Multi-task learning as an approach to tackling overfitting in deep learning

Section \ref{sec:weaklys_model} presents the model that includes a phoneme recognizer as a secondary task to regularize the primary task of computing the probability of a pronunciation error at the word level.

\end{enumerate}
\input{Chapters/PronunciationErrorDetection/WeaklySupervisedModel}

\input{Chapters/PronunciationErrorDetection/UncertaintyModeling}

\input{Chapters/PronunciationErrorDetection/LexicalStressErrorDetection}

\input{Chapters/PronunciationErrorDetection/SpeechSynthesis}

%% file: Chapters/PronunciationErrorDetection/WeaklySupervisedModel.tex
\section{Weakly-supervised word-level pronunciation error detection in non-native English speech}
\label{sec:weaklys_model}

\begin{center}
\textit{Daniel Korzekwa, Jaime Lorenzo-Trueba, Thomas Drugman, Shira Calamaro, Bozena Kostek, Weakly-supervised word-level pronunciation error detection in non-native English speech, Interspeech, 2021}
\end{center}

\bigskip

\textbf{Abstract}

\bigskip

We propose a weakly-supervised model for word-level mispronunciation detection in non-native (L2) English speech. To train this model, phonetically transcribed L2 speech is not required and we only need to mark mispronounced words. The lack of phonetic transcriptions for L2 speech means that the model has to learn only from a weak signal of word-level mispronunciations. Because of that and due to the limited amount of mispronounced L2 speech, the model is more likely to overfit. To limit this risk, we train it in a multi-task setup. In the first task, we estimate the probabilities of word-level mispronunciation. For the second task, we use a phoneme recognizer trained on phonetically transcribed L1 speech that is easily accessible and can be automatically annotated. Compared to state-of-the-art approaches, we improve the accuracy of detecting word-level pronunciation errors in AUC metric by 30\% on the GUT Isle Corpus of L2 Polish speakers, and by 21.5\% on the Isle Corpus of L2 German and Italian speakers.

\subsection{Introduction}

\begin{figure}[th]
\centering
\centerline{
\includegraphics[width=1.25\textwidth]{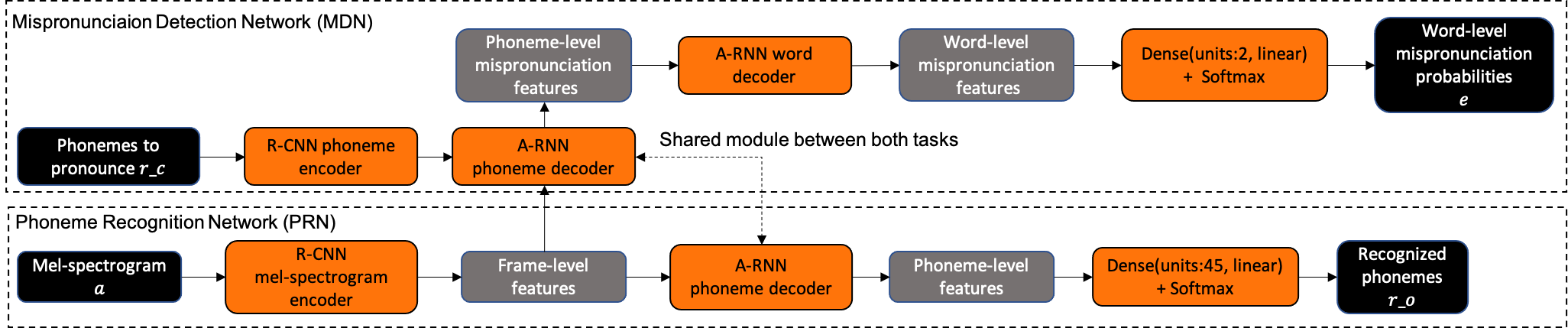}}
\decoRule
\caption[Neural network architecture of the WEAKLY-S model for word-level pronunciation error detection]{Neural network architecture of the WEAKLY-S model for word-level pronunciation error detection.}
\label{fig:neural_network}
\end{figure}

\begin{figure}[th]
\centering
\centerline{
\includegraphics[width=1.25\textwidth]{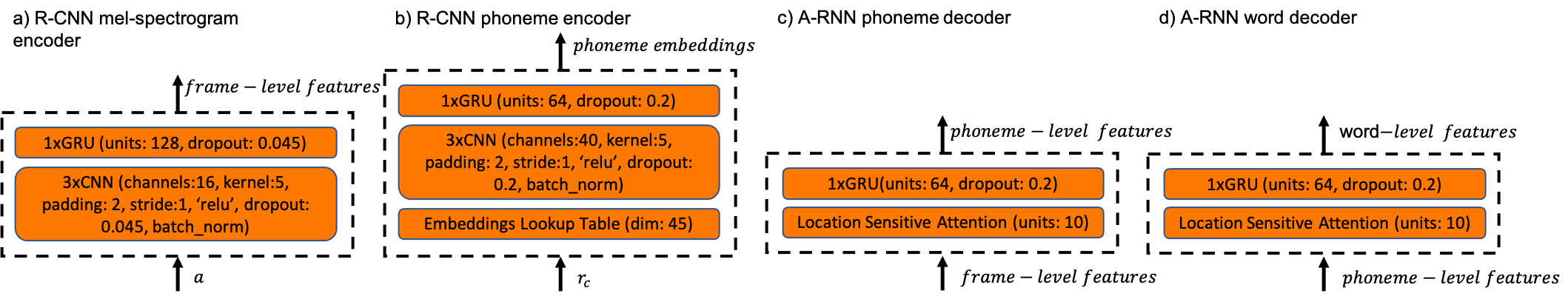}}
\decoRule
\caption[Details of the neural network architecture of the WEAKLY-S model for word-level pronunciation error detection]{Details of the neural network architecture of the WEAKLY-S model for word-level pronunciation error detection.}
\label{fig:implementation_details}
\end{figure}

It has been shown that Computer-Assisted Pronunciation Training (CAPT) helps people practice and improve pronunciation skills \cite{neri2008effectiveness,tejedor2020assessing}. Despite significant progress over the last two decades, standard methods are still unable to detect mispronunciations with high accuracy. These methods can detect phoneme-level mispronunciations at about 60\% precision and 40\%-80\% recall \cite{leung2019cnn, korzekwa2021mispronunciation, zhang2021text}. By further raising precision we can lower the risk of providing incorrect feedback, whereas with higher recall, we can detect more mispronunciation errors.

Standard methods aim at recognizing the phonemes pronounced by a speaker and compare them with expected (canonical) pronunciation of correctly pronounced speech. Any mismatch between recognized and canonical phonemes yields a pronunciation error at the phoneme level. Phoneme recognition-based approaches rely on phonetically transcribed speech labeled by human listeners. Human-based transcription is a laborious task, especially, in the case of L2 speech where listeners have to identify mispronunciations. Sometimes, it might be even impossible to transcribe L2 speech because different languages have different phoneme sets and it is unclear which phonemes were pronounced by the speaker.

Phoneme recognition-based approaches generally fall into two categories. The first category uses forced-alignment techniques \cite{Hongyan2011, li2016mispronunciation, sudhakara2019improved, cheng2020asr} based on the work by Franco et al. \cite{franco1997automatic} and the Goodness of Pronunciation (GOP) method \cite{witt2000phone}. The GOP uses Bayesian inference to find the most likely alignment between canonical phonemes and the corresponding audio signal (forced alignment). Then, the GOP uses the likelihoods of the aligned audio signal as an indicator for mispronounced phonemes. In the second category there are methods that recognize phonemes pronounced by a speaker purely from a speech signal, and only then align them with canonical phonemes \cite{Minematsu2004PronunciationAB, harrison2009implementation, Lee2013PronunciationAV, plantinga2019towards,Sudhakara2019NoiseRG}. Techniques falling into both categories can be complemented with the use of a reference signal obtained either from a database of speech \cite{xiao2018paired, nicolao2015automatic, wang2019child} or generated from phonetic representation \cite{korzekwa2021mispronunciation, qian2010capturing}.

There are two challenges for the phoneme recognition approaches. First, phonemes pronounced by a speaker have to be recognized accurately, which has been shown to be difficult \cite{zhang2021text, chorowski2014end, chorowski2015attention, bahdanau2016end}. Second, standard approaches expect only a single canonical pronunciation of a given text, but this assumption does not always hold true due to phonetic variability of speech. In \cite{korzekwa2021mispronunciation}, we addressed these problems by modeling uncertainty in the model by incorporating a pronunciation model of L1 speech. Nonetheless, this approach still relies on phonetically transcribed L2 speech.

In this paper, we introduce a novel model (noted as WEAKLY-S) for the detection of word-level pronunciation errors that does not require phonetically transcribed L2 speech. The model produces the probabilities of mispronunciation for all words, conditioned on a spoken sentence and canonical phonemes. Mispronunciation error types include any of phoneme replacement, addition, deletion or unknown speech sound. During training, the model is weakly supervised, in the sense that we only mark mispronounced words in L2 speech and the data do not have to be phonetically transcribed. Due to the limited availability of L2 speech and the fact it is not phonetically transcribed, the model is more likely to overfit. To solve this problem, we train the model in a multi-task setup. In addition to a primary task of word-level mispronunciation detection, we use a phoneme recognizer trained on automatically transcribed L1 speech for the secondary task. Both tasks share common parts of the model, which makes the primary task less likely to overfit. Additionally, we address the overfitting problem with synthetically generated  pronunciation errors that are derived from L1 speech.

Leung et al. \cite{leung2019cnn} used a phoneme recognizer based on Connectionist Temporal Classification (CTC) for pronunciation error detection. Instead, we use an attention-based phoneme recognizer following Chorowski et al. \cite{chorowski2015attention} so that we can regularize the model by both tasks sharing a common component (attention). With a CTC-based phoneme recognizer it would not be possible because this technique does not use attention that could be shared between both tasks. Zhang et al. \cite{zhang2021text} employed a multi-task model for pronunciation assessment, but with two important differences. First, they use a Needleman-Wunsch algorithm \cite{needleman1970general} for aligning canonical and recognized sequences of phonemes, but this algorithm cannot be tuned towards sequences of phonemes. We use an attention mechanism that automatically maps the speech signal to the sequence of word-level pronunciation errors. Second, Zhang et al. detect pronunciation errors at the phoneme level and they expect L2 speech to be phonetically transcribed. This differs from our method of recognizing pronunciation errors at the word level with no need for phonetic transcriptions of L2 speech. To the best of our knowledge, this is the first approach to train word-level pronunciation error detection model that does not require phonetically transcribed L2 speech and can be optimized directly towards word-level mispronunciation detection.

\subsection{Proposed model}
\label{sec:proposed_model}

\subsubsection{Model definition}

The model is made of two sub-networks: \emph{i)} a word-level Mispronunciations Detection Network (MDN) detects word-level pronunciation errors $\mathbf{e}$ from the audio signal $\mathbf{a}$ and canonical phonemes $\mathbf{r_c}$, \emph{ii)} a Phoneme Recognition Network (PRN) recognizes phonemes $\mathbf{r_o}$ pronounced by a speaker from the audio signal $\mathbf{a}$ (Fig. \ref{fig:neural_network}). 

More formally, let us define the following variables: $\mathbf{a}$ - speech signal represented by a mel-spectrogram, $\mathbf{r_c}$ - canonical phonemes that the speaker was expected to pronounce, $\mathbf{r_o}$ -   phonemes pronounced, and $\mathbf{e}$ -  the probabilities of mispronouncing words in the spoken sentence. The model outputs the probabilities of word-level mispronunciation, denoted as $\mathbf{e} \sim  p(\mathbf{e}|\mathbf{a},\mathbf{r_c},\bm{\theta})$, where $\bm{\theta}$ represent parameters of the model.

We train the WEAKLY-S model in a multi-task setup. In addition to the primary task $\mathbf{e}$, we use a phoneme recognizer denoted as $\mathbf{r_o} \sim  p(\mathbf{r_o}|\mathbf{a},\bm{\theta})$ for the secondary task. The parameters $\bm{\theta}$  are shared between both tasks, which makes the MDN less likely to overfit. We define the loss function as the sum of two losses: a word-level mispronunciation loss and a phoneme recognition loss. Its formulation for the \textit{ith} training example is presented in Eq. \ref{eq:loss_function}. We train the model using two types of training data: phonetically transcribed L1 speech (both losses are used) and untranscribed L2 speech (only the mispronunciation loss is used). Having a separate loss for word-level mispronunciation lets us train the model from speech data that are not phonetically transcribed.

\begin{equation}
\mathcal{L(\bm{\theta})}=log(p(\mathbf{e}|\mathbf{a},\mathbf{r_c},\bm{\theta})) + log(p(\mathbf{r_o}|\mathbf{a},\bm{\theta}))
\label{eq:loss_function}
\end{equation}

\subsubsection{Neural network details}

\begin{figure}[th]
\centering
\centerline{
\includegraphics[width=1.25\textwidth]{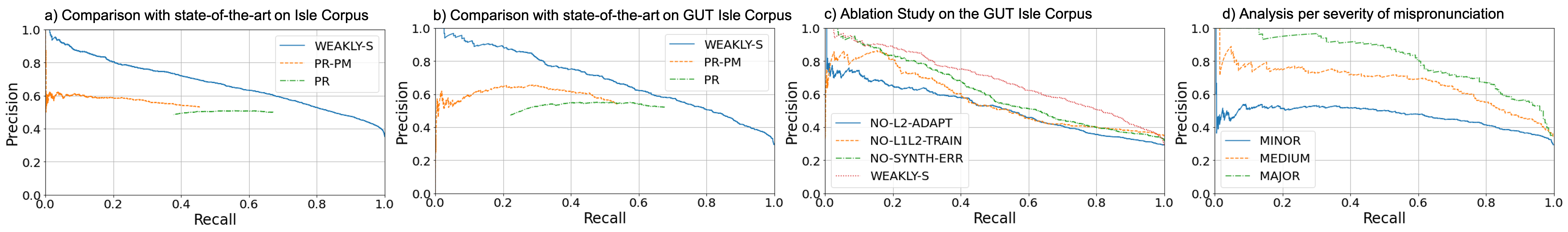}}
\decoRule
\caption[Precision-recall curves for the WEAKLY-S and baseline models, PR-PM and PR, (a) tested on Isle Corpus of German and Italian speakers and (b) GUT Isle Corpus of Polish speakers. (c) Ablation study on the GUT Isle corpus. (d) Analysis of mispronunciation severity levels]{Precision-recall curves for the WEAKLY-S and baseline models, PR-PM and PR, (a) tested on Isle Corpus of German and Italian speakers and (b) GUT Isle Corpus of Polish speakers. (c) Ablation study on the GUT Isle corpus. (d) Analysis of mispronunciation severity levels.}
\label{fig:precision_recall_plots}
\end{figure}

Following Sutskever et al. \cite{sutskever2014sequence}, the MDN network encodes the mel-spectrogram $\mathbf{a}$ and the canonical phonemes  $\mathbf{r_c}$ with Recurrent Convolutional Neural Network (RCNN) encoders (Fig. \ref{fig:implementation_details}a and Fig. \ref{fig:implementation_details}b). These encoded representations are passed into an attention-based \cite{vaswani2017attention} Recurrent Neural Network (A-RNN) decoder (Fig. \ref{fig:implementation_details}c) that generates phoneme-level mispronunciation features. Phoneme-level features are transformed into word-level features (Fig. \ref{fig:implementation_details}d) based on an attention mechanism and these finally are used for computing word-level mispronunciation probabilities $\mathbf{e}$.

The PRN recognizes phonemes $\mathbf{r_o}$ pronounced by the speaker. It is similar to the attention-based phoneme recognizer by Chorowski et al. \cite{chorowski2015attention}. To generate phoneme-level features, it uses the same RCNN mel-spectrogram encoder and A-RNN decoder as the MDN. The only difference is that the A-RNN decoder is not conditioned on canonical phonemes. Phoneme-level features are transformed to the probabilities of pronounced phonemes. We added a phoneme recognition task due to the limited amount of L2 speech annotated with word-level mispronunciations. Without it, the MDN would be prone to overfitting if it was trained only on its own. By sharing common parts between both models, the PRN acts as a backbone for the MDN and makes it more robust.

The model was implemented in MxNet framework \cite{chen2015mxnet} and tuned for hyper-parameters with AutoGluon Bayesian optimization framework \cite{erickson2020autogluon}. The model was first pretrained on L1 and L2 speech corpora and then the MDN part was fine-tuned only on L2 speech data. We used the Adam optimizer with learning rate 0.001 and gradient clipping 5. Training data were segmented into buckets with batch size 32, using GluonCV \cite{guo2020gluoncv}. The A-RNN phoneme and word decoders are based on Location Sensitive Attention by Chorowski et al. \cite{chorowski2015attention}.

\subsection{Experiments}
\label{sec:experiments}

We present three experiments. We start with comparing our model against state-of-the-art approaches in the task of word-level mispronunciation detection. In an ablation study we analyze which elements of the model contribute the most to its performance. Finally, we analyze how the severity of pronunciation error affects the accuracy of the model.

\subsubsection{Speech corpora and metrics}

In our experiments, we use a combination of L1 and L2 English speech. L1 speech is obtained from TIMIT \cite{garofolo1993darpa} and LibriTTS \cite{Zen2019} corpora. L2 data come from the Isle \cite{atwell2003isle} corpus (German and Italian speakers) and the GUT Isle \cite{Weber2020} corpus (Polish speakers). In total, we collected 102,812 utterances, summarized in Table \ref{tab:weakly_speech_corpora}. We split the data into training and test sets, holding out 28 L2 speakers (11 German, 11 Italian, and 6 Polish) only for testing the performance of the model.

The L2 corpus of Polish speakers was annotated for word-level pronunciation errors by 5 native English speakers. Annotators marked mispronounced words and indicated their severity levels using one of the three possible values: 1 - MINOR, 2 - MEDIUM, 3 - MAJOR. The Isle corpus of German and Italian speakers comes with phoneme level mispronunciations. Words with at least one mispronounced phoneme were automatically marked as mispronounced. The Isle corpus is not mapped to severity levels of mispronunciations. In total, there are 35,555 L2 words, including 8035 mispronounced words. All data were re-sampled to 16 kHz.

We extended the train set with 292,242 utterances of L1 speech with synthetically generated pronunciation errors. We use a simple approach of perturbing phonetic transcription for the corresponding speech audio.  First, we sample these utterances with replacement from L1 corpora of human speech. Then, for each utterance, we replace phonemes with random phonemes with a probability of 0.2. In \cite{korzekwa21_interspeech} we found that generating incorrectly stressed speech using Text-To-Speech (TTS) improves the accuracy of detecting lexical stress errors in L2 speech. Although, as opposed to using TTS, we create pronunciation errors by perturbing the text, we expect this simpler approach should still help recognizing word-level pronunciation errors.

\begin{table}[htb]
\caption{Summary of speech corpora used in experiments. * - audiobooks read by volunteers from all over the world \cite{Zen2019} }
\label{tab:weakly_speech_corpora}
\centering
\begin{tabular}{lll}
\toprule  
\tabhead{Native Language} & \tabhead{Hours} & \tabhead{Speakers} \\
\midrule
English & 90.47 & 640\\ 
Unknown* & 19.91 & 285\\  
German and Italian & 13.41 & 46\\  
Polish & 1.49 & 12\\ 
\bottomrule\\
\end{tabular}
\end{table}

To evaluate our model, we use three standard metrics: Area Under Curve (AUC), precision and recall. The AUC metric provides an overall performance of the model accounting for all possible trade offs between precision and recall. Precision-recall plots illustrate relations between both metrics. Complementary, to analyze precision, in all our experiments we consistently fix recall at the value of 0.4 to be comparable with two baseline models that do not cover the whole range of recall values (see Section \ref{sec:weaklys_pron_model_sota}).

\subsubsection{Comparison with state-of-the-art}\label{sec:weaklys_pron_model_sota}

We compare our proposed WEAKLY-S model against two state-of-the-art baselines. The phoneme recognizer (PR) model by Leung et al. \cite{leung2019cnn} is our first baseline. The PR is based on CTC loss \cite{graves2012connectionist} and it outperforms multiple alternative approaches for pronunciation assessment. The original CTC-based model uses a hard likelihood threshold applied to recognized phonemes. To compare it with two other models, following our work in \cite{korzekwa2021mispronunciation}, we replaced hard likelihood threshold with a soft threshold. The second baseline is the PR extended by a pronunciation model (PR-PM model \cite{korzekwa2021mispronunciation}). The pronunciation model accounts for phonetic variability of speech produced by native speakers, which results in higher precision of detecting pronunciation errors.

The results are presented in Fig. \ref{fig:precision_recall_plots}a, Fig. \ref{fig:precision_recall_plots}b and Table \ref{tab:accuracy_metrics}. The WEAKLY-S model turns out to outperform the second best model in AUC by 30\% from 52.8 to 68.63 and in precision by 23\% from 61.21 to 75.25 on the GUT Isle Corpus of Polish speakers. We observe similar improvements on the Isle Corpus of German and Italian speakers.

\begin{table}[htb]
  \caption{Accuracy metrics of detecting word-level pronunciation errors. WEAKLY-S vs baseline models.}
  \label{tab:accuracy_metrics}
  \centering
   \begin{tabular}{llll}
    \toprule
     \tabhead{Model} & \tabhead{AUC [\%]} & \tabhead{Precision [\%,95\%CI]} & \tabhead{Recall [\%,95\%CI]} \\
    \midrule
    \multicolumn{4}{c}{\textbf{Isle corpus (German and Italian)}} \\
    PR & 55.52 & 49.39 (47.59-51.19) & 40.20 (38.62-41.81)\\ 
    PR-PM & 48.00 & 54.20 (52.32-56.08) & 40.20 (38.62-41.81)\\  
    WEAKLY-S & \textbf{67.47} & 71.94 (69.96, 73.87) & 40.14 (38.56, 41.75) \\  
	
	 \multicolumn{4}{c}{\textbf{GUT Isle corpus (Polish)}} \\
     PR & 52.8 & 54.91 (50.53-59.24) & 40.29 (36.66-44.02)\\ 
     PR-PM & 50.50 & 61.21 (56.63-65.65) & 40.15 (36.51-43.87)\\  
     WEAKLY-S & \textbf{68.63} & 75.25 (71.67-78.59) & 40.38 (37.52-43.29)\\    
    \bottomrule\\
  \end{tabular}
\end{table}

One difference between our model and the two baselines is that they both use the Needleman-Wunsch algorithm \cite{needleman1970general} for aligning canonical and recognized sequences of phonemes. This is a dynamic programming-based algorithm for comparing biological sequences and cannot be optimized for mispronunciation errors. Our model automatically finds the mapping between regions in the speech signal and the corresponding canonical phonemes, and then identifies word-level mispronunciation errors. In this way, we eliminate the Needleman-Wunsch algorithm as a possible source of error. 

The second difference is the use of phonetic transcriptions for L2 speech. Both baselines use automatic transcriptions provided by an Amazon-proprietary grapheme-to-phoneme model. In \cite{korzekwa2021mispronunciation} we found that for the PR and PR-PM models it is better to use automatically transcribed L2 speech for training a phoneme recognizer than not use L2 speech at all. Note that these automatic transcriptions will include phoneme mistakes for mispronounced speech. Our model does not use transcriptions of L2 speech, and instead it is guided by the word-level pronunciation errors of L2 speech in a weakly-supervised fashion.

\subsubsection{Ablation study}\label{subsec:ablation_study}

We now investigate which elements of our new model contribute the most to its performance. Along with the WEAKLY-S model, we trained three additional variants, each with a certain feature removed. The NO-L2-ADAPT variant does not fine-tune the model on L2 speech, though it is still exposed to L2 speech while it is trained on a combined corpus of L1 and L2 speech. The NO-L1L2-TRAIN model is not trained on L1/L2 speech, and fine-tuning on L2 speech starts from scratch. It means that the model will not use a large amount of phonetically transcribed L1 speech data and ultimately the secondary task of the phoneme recognizer will not be used. In the NO-SYNTH-ERR model, we exclude synthetic samples of mispronounced L1 speech. It significantly reduces the amount of incorrectly pronounced words used during training from 1,129,839 to only 5,273 L2 words.

L2 Fine-tuning (NO-L2-ADAPT) is the most important factor that contributes to the performance of the model (Fig. \ref{fig:precision_recall_plots}c and Table \ref{tab:ablation_study}), with an AUC of 51.72\% compared to 68.63\% for the full model. Training the model on both L2 and L1 speech together is not sufficient. We think it is because L2 speech accounts for less than 1\% of the training data and the model naturally leans towards L1 speech. The second most important feature is training the model on a combined set of L1 and L2 speech (NO-L1L2-TRAIN), with AUC of 56.46\%. L1 speech accounts for more than 99\% of the training data. These data are also phonetically transcribed, and therefore can be used for the phoneme recognition task. The phoneme recognition task acts as a 'backbone' and reduces the effect of overfitting in the main task of detecting word pronunciation errors. Finally, excluding synthetically generated pronunciation errors (NO-SYNTH-ERR) reduces the AUC from 68.63\% to 61.54\%.

\begin{table}[htb]
  \caption{Ablation study for the GUT Isle corpus.}
  \label{tab:ablation_study}
  \centering
   \begin{tabular}{llll}
    \toprule
      \tabhead{Model} &  \tabhead{AUC [\%]} &  \tabhead{Precision [\%]} &  \tabhead{Recall [\%]} \\
    \midrule
     NO-L2-ADAPT & 51.72 & 57.89 & 40.11 \\ 
     NO-L1L2-TRAIN & 56.46 & 59.73 & 40.20 \\  
     NO-SYNTH-ERR & 61.54 & 67.22 & 40.38 \\  
     WEAKLY-S & \textbf{68.63} & 75.25  & 40.38 \\    
    \bottomrule\\
  \end{tabular}
\end{table}

\subsubsection{Severity of mispronunciation}\label{sec:severity}

When providing feedback to the L2 speaker about mispronounced words, we want to reflect the severity of mispronunciation, in order to focus on more severe errors and not report them all at once. We segment pronunciation errors into three categories: LOW, MEDIUM and HIGH, based on an inter-tester agreement of annotating sentences for word-level mispronunciations. Mispronounced words with less than 40\% inter-tester agreement belong to the LOW category, between 40\% and 80\% to MIDDLE, and over 80\% to HIGH. We validated that the proposed inter-tester agreement bands are well correlated with explicit listener opinions on the severity of mispronunciation, as shown in Table \ref{tab:inter-tester-agreement-by-severity}. This result shows that data on mispronunciation severity can be derived automatically, without the need to collect it.

\begin{table}[th]
  \caption{Severity of mispronunciation by inter-tester agreement for the GUT Isle Corpus. 1 - MINOR, 2 - MEDIUM, 3 - MAJOR.}
  \label{tab:inter-tester-agreement-by-severity}
  \centering
   \begin{tabular}{ll}
    \toprule
      \tabhead{Inter-tester agreement} &  \tabhead{Severity [mean and 95\% CI ]}  \\
    \midrule
     LOW (Less than 40\%) & 1.32 (1.28-1.35)  \\ 
     MEDIUM (Between 40\% and 80\%) & 1.58 (1.54-1.62)  \\  
     HIGH( Higher than 80\%) & 2.08 (2.03-2.13)  \\    
    \bottomrule\\
  \end{tabular}
\end{table}

We aim at detecting the words of HIGH inter-tester agreement with higher precision to provide more relevant feedback to L2 speakers. To make AUC, precision, and recall metrics comparable between different levels of inter-tester agreement, we enforce the ratio of mispronounced words across all categories to the same level of 29.2\% by randomly down-sampling correctly pronounced words. This value is the proportion of mispronounced words across all inter-tester agreement levels in the GUT Isle Corpus. We observe that we can detect pronunciation errors of HIGH inter-tester agreement with 91.67\% precision at 40.38\% recall (Fig. \ref{fig:precision_recall_plots}d and Table \ref{tab:severity_of_mispronunciation}). By segmenting pronunciation errors into three difference bands, we can report to a language learner only the errors of HIGH inter-tester agreement, and improve their learning experience.

\begin{table}[th]
  \caption{Accuracy metrics for different severity levels of mispronunciation for the GUT Isle Corpus.}
  \label{tab:severity_of_mispronunciation}
  \centering
   \begin{tabular}{llll}
    \toprule
      \tabhead{Inter-test agreement} &  \tabhead{AUC [\%]} &  \tabhead{Precision [\%]} &  \tabhead{Recall [\%]} \\
    \midrule
     LOW & 46.99 & 51.84 & 40.48 \\ 
     MEDIUM & 66.90 & 71.89 & 40.80 \\  
     HIGH & 81.48 & 91.67  & 40.31 \\    
    \bottomrule\\
  \end{tabular}
\end{table}

\subsection{Conclusions and future work}
\label{sec:conclusions}

We proposed a model for detecting pronunciation errors in English that can be trained from L2 speech labeled only for word-level mispronunciations. The data do not have to be phonetically transcribed. The model outperforms state-of-the-art models in AUC metric on the GUT Isle Corpus of Polish speakers and the Isle Corpus of German and Italian speakers. The limited amount of L2 speech and the lack of phonetically transcribed speech makes this model prone to overfitting. We overcame this issue by proposing a multi-task training with two tasks: a word-level pronunciation error detector trained on L1 and L2 speech, and a phoneme recognizer trained on L1  speech. The most important factors that contribute to the model accuracy are:  \emph{i)} fine-tuning on L2 speech,  \emph{ii)} pre-training on a joined corpus of L1 and L2 speech, and  \emph{iii)} use of synthetically generated pronunciation errors.  

The level of inter-tester agreement in annotating pronunciation errors correlates with explicit human opinions about the severity of mispronunciation. By detecting pronunciation errors only for high inter-tester agreement, we may significantly lower the number of false positives reported to a language learner. 

In the future, we will experiment with discrete representation of the latent phoneme space such as Vector-Quantized Variational-Auto-Encoder (VQ-VAE) \cite{chorowski2019unsupervised, van2017neural}, which should fit better to discrete nature of phonemes. We plan to generate synthetic mispronounced speech, which is motivated by our recent work on using speech synthesis for generating speech errors in the related task of lexical stress error detection \cite{korzekwa21_interspeech}.

%% file: Chapters/PronunciationErrorDetection/UncertaintyModeling.tex
\section{The role of uncertainty modeling}
\label{sec:uncertainty_modeling}

\begin{center}
    \textit{Daniel Korzekwa, Jaime Lorenzo-Trueba, Szymon Zaporowski, Shira Calamaro, Thomas Drugman, Bozena Kostek, Mispronunciation Detection in Non-Native (L2) English with Uncertainty Modeling, IEEE International Conference on Acoustics, Speech and Signal Processing (ICASSP), 2021}
\end{center}

\bigskip

\textbf{Abstract}

A common approach to the automatic detection of mispronunciation in language learning is to recognize the phonemes produced by a student and compare it to the expected pronunciation of a native speaker. This approach makes two simplifying assumptions:  a) phonemes can be recognized from speech with high accuracy, b) there is a single correct way for a sentence to be pronounced. These assumptions do not always hold, which can result in a significant amount of false mispronunciation alarms. We propose a novel approach to overcome this problem based on two principles: a) taking into account uncertainty in the automatic phoneme recognition step, b) accounting for the fact that there may be multiple valid pronunciations. We evaluate the model on non-native (L2) English speech of German, Italian and Polish speakers, where it is shown to increase the precision of detecting mispronunciations by up to 18\% (relative) compared to the common approach.

\bigskip

\subsection{Introduction}

In Computer Assisted Pronunciation Training (CAPT), students are presented with a text and asked to read it aloud. A computer informs students on mispronunciations in their speech, so that they can repeat it and improve. CAPT has been found to be an effective tool that helps non-native (L2) speakers of English to improve their pronunciation skills \cite{neri2008effectiveness,tejedor2020assessing}.

A common approach to CAPT is based on recognizing the phonemes produced by a student and comparing them with the expected (canonical) phonemes that a native speaker would pronounce \cite{witt2000phone,li2016mispronunciation,sudhakara2019improved,leung2019cnn}.  It makes two simplifying assumptions. First, it assumes that phonemes can be automatically recognized from speech with high accuracy. However, even in native (L1) speech, it is difficult to get the Phoneme Error Rate (PER) below 15\% \cite{chorowski2015attention}. Second, this approach assumes that this is the only `correct' way for a sentence to be pronounced, but due to phonetic variability this is not always true. For example, the word `enough' can be pronounced by native speakers in multiple correct ways:  /ih n ah f/ or /ax n ah f/ (short `i' or `schwa' phoneme at the beginning). These assumptions do not always hold which can result in a significant amount of false mispronunciation alarms and making students confused when it happens.

We propose a novel approach that results in fewer false mispronunciation alarms, by formalizing the intuition that we will not be able to recognize exactly what a student has pronounced or say precisely how a native speaker would pronounce it. First, the model estimates a belief over the phonemes produced by the student, intuitively representing the uncertainty in the student's pronunciation. Then, the model converts this belief into the probabilities that a native speaker would pronounce it, accounting for phonetic variability. Finally, the model makes a decision on which words were mispronounced in the sentence by processing three pieces of information: a) what the student pronounced, b) how likely a native speaker would pronounce it that way, and c) what the student was expected to pronounce. 

In Section \ref{subsec:uncertainty_modeling_related_work}, we review the related work. In Section \ref{subsec:uncertainty_modeling_proposed_model}, we describe the proposed model. In Section \ref{subsec:uncertainty_modeling_experiments}, we present the experiments, and we conclude in Section \ref{subsec:uncertainty_modeling_conclusion}.

\subsection{Related work}
\label{subsec:uncertainty_modeling_related_work}

In 2000, Witt et al. coined the term Goodness of Pronunciation (GoP) \cite{witt2000phone}. GoP starts by aligning the canonical phonemes with the speech signal using a forced-alignment technique. This technique aims to find the most likely mapping between phonemes and the regions of a corresponding speech signal. In the next step, GoP computes the ratio between the likelihoods of the canonical and the most likely pronounced phonemes. Finally, it detects a mispronunciation if the ratio falls below a given threshold. GoP was further extended with Deep Neural Networks (DNNs), replacing Hidden Markov Model (HMM) and Gaussian Mixture Model (GMM) techniques for acoustic modeling \cite{li2016mispronunciation,sudhakara2019improved}. Cheng et al. \cite{cheng2020asr} improved the performance of GoP with the latent representation of speech extracted in an unsupervised way.

As opposed to GoP, we do not use forced-alignment that requires both speech and phoneme inputs. Following the work of Leung et al. \cite{leung2019cnn}, we use a phoneme recognizer, which recognizes phonemes from only the speech signal. The phoneme recognizer is based on a Convolutional Neural Network (CNN), a Gated Recurrent Unit (GRU), and Connectionist Temporal Classification (CTC) loss. Leung et al. report that it outperforms other forced-alignment \cite{li2016mispronunciation} and forced-alignment-free \cite{harrison2009implementation} techniques on the task of detecting phoneme-level mispronunciations in L2 English. Contrary to Leung et al., who rely only on a single recognized sequence of phonemes, we obtain top $N$ decoded sequences of phonemes, along with the phoneme-level posterior probabilities.

It is common in pronunciation assessment to employ the speech signal of a reference speaker. Xiao et al. use a pair of speech signals from a student and a native speaker to classify native and non-native speech \cite{xiao2018paired}. Mauro et al. incorporate the speech of a reference speaker to detect mispronunciations at the phoneme level \cite{nicolao2015automatic}. Wang et al. use siamese networks for modeling discrepancy between normal and distorted children's speech \cite{wang2019child}. We take a similar approach but we do not need a database of reference speech. Instead, we train a statistical model to estimate the probability of pronouncing a sentence by a native speaker. Qian et al. propose a statistical pronunciation model as well \cite{qian2010capturing}. Unlike our work, in which we create a model of `correct` pronunciation, they build a model that generates hypotheses of mispronounced speech.

\subsection{Proposed model}
\label{subsec:uncertainty_modeling_proposed_model}

\begin{figure}[th]
\centering
\centerline{
\includegraphics[width=1.25\textwidth]{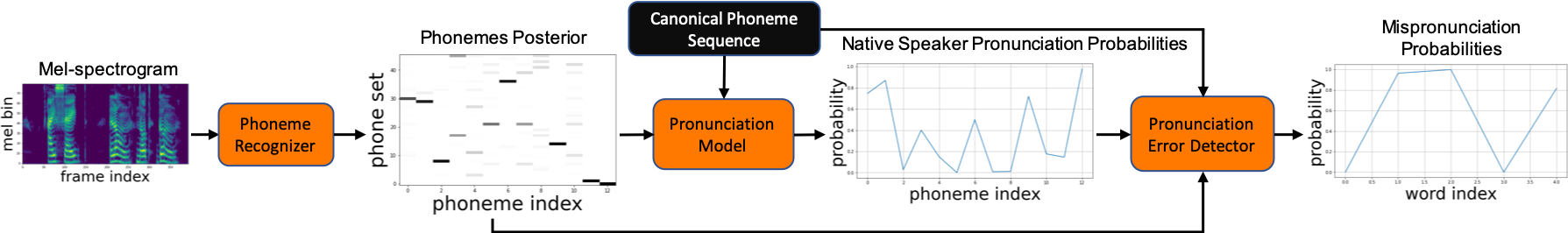}}
\decoRule
\caption[Architecture of the system for detecting mispronounced words in a spoken sentence]{Architecture of the system for detecting mispronounced words in a spoken sentence.}
\label{fig:uncertainty_model_architecture}
\end{figure}

The design consists of three subsystems: a Phoneme Recognizer (PR), a Pronunciation Model (PM), and a Pronunciation Error Detector (PED), illustrated in Figure \ref{fig:uncertainty_model_architecture}. The PR recognizes phonemes spoken by a student. The PM estimates the probabilities of having been pronounced by a native speaker. Finally, the PED computes word-level mispronunciation probabilities. In Figure \ref{fig:nn_architecture}, we present detailed architectures of the PR, PM, and PED.

For example, considering the text: `I said alone not gone' with the canonical representation of /ay - s eh d - ax l ow n - n aa t - g aa n/. Polish L2 speakers of English often mispronounce the /eh/ phoneme in the second word as /ey/. The PM would identify the /ey/ as having a low probability of being pronounced by a native speaker in the middle of the word `said’, which the PED would translate into a high probability of mispronunciation.

\subsubsection{Phoneme recognizer}
\label{subsubsec:uncertainty_modeling_phoneme_recognizer}

\begin{figure}[th]
\centering
\centerline{
\includegraphics[width=1.25\textwidth]{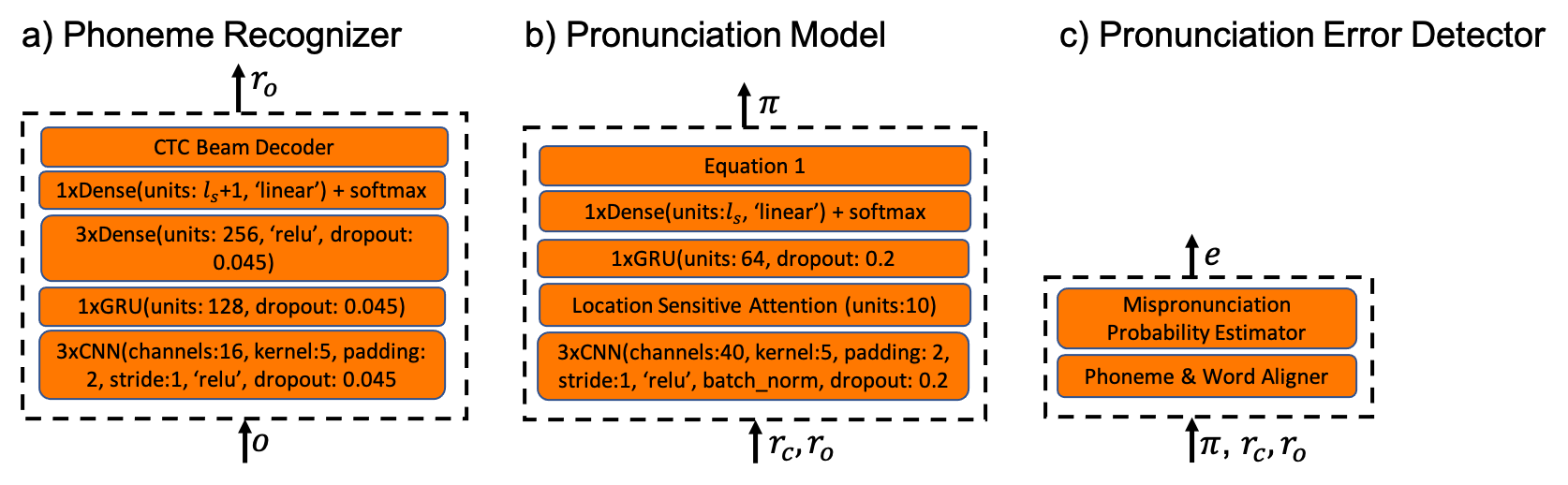}}
\decoRule
\caption[Architecture of the PR, PM, and PED subsystems. $l_s$ - the size of the phoneme set]{Architecture of the PR, PM, and PED subsystems. $l_s$ - the size of the phoneme set.}
\label{fig:nn_architecture}
\end{figure}

The PR (Figure \ref{fig:nn_architecture}a) uses beam decoding \cite{graves2013speech} to estimate $N$ hypotheses of the most likely sequences of phonemes that are recognized in the speech signal $\mathbf{o}$. A single hypothesis is denoted as $\mathbf{r_o} \sim p(\mathbf{r_o}|\mathbf{o})$. The speech signal $\mathbf{o}$ is represented by a mel-spectrogram with $f$ frames and 80 mel-bins. Each sequence of phonemes $\mathbf{r_o}$ is accompanied by the posterior phoneme probabilities of shape: $(l_{r_o},l_s+1)$. $l_{r_o}$ is the length of the sequence and $l_s$ is the size of the phoneme set (45 phonemes including `pause', `end of sentence (eos)', and a `blank' label required by the CTC-based model).

\subsubsection{Pronunciation model}
\label{subsubsec:uncertainty_modeling_proonunciation_model}

The PM (Figure \ref{fig:nn_architecture}b) is an encoder-decoder neural network following Sutskever et al. \cite{sutskever2014sequence}. Instead of building a text-to-text translation system between two languages, we use it for phoneme-to-phoneme conversion. The sequence of phonemes $\mathbf{r_c}$ that a native speaker was expected to pronounce is converted into the sequence of phonemes  $\mathbf{r}$ they had pronounced, denoted as  $\mathbf{r} \sim p(\mathbf{r}|\mathbf{r_c})$. Once trained, the PM acts as a probability mass function, computing the likelihood sequence $\bm{\pi}$ of the phonemes $\mathbf{r_o}$ pronounced by a student conditioned on the expected (canonical) phonemes $\mathbf{r_c}$. The PM is denoted in Eq. \ref{eq_proposed_model}, which we implemented in MxNet \cite{chen2015mxnet} using `sum' and `element-wise multiply' linear-algebra operations.

\begin{equation}
\bm{\pi}=\sum_{\mathbf{r_o}} p(\mathbf{r_o}|\mathbf{o})p(\mathbf{r}=\mathbf{r_o}|\mathbf{r_c})
\label{eq_proposed_model}
\end{equation}

The model is trained on phoneme-to-phoneme speech data created automatically by passing the speech of the native speakers through the PR. By annotating the data with the PR, we can make the PM model more resistant to possible phoneme recognition inaccuracies of the PR at testing time.

\subsubsection{Pronunciation error detector}
\label{subsubsec:uncertainty_modeling_pron_error_detector}

The PED (Figure \ref{fig:nn_architecture}c) computes the probabilities of mispronunciations $\mathbf{e}$ at the word level, denoted as $\mathbf{e} \sim p(\mathbf{e}|\mathbf{r_o},\bm{\pi},\mathbf{r_c})$. The PED is conditioned on three inputs: the phonemes $\mathbf{r_o}$ recognized by the PR, the corresponding pronunciation likelihoods $\bm{\pi}$ from the PM, and the canonical phonemes $\mathbf{r_c}$. The model starts with aligning the canonical and recognized sequences of phonemes. We adopted a dynamic programming algorithm for aligning biological sequences developed by Needleman-Wunsch \cite{needleman1970general}. Then, the probability of mispronunciation for a given word is computed with Eq. \ref{eq_ped}, $k$ denotes the word index, and $j$ is the phoneme index in the word with the lowest probability of pronunciation.

\begin{equation}
p(\mathbf{e}_k) =
    \begin{cases}
      0 & \text{if aligned phonemes match,}\\
       1-\bm{\pi}_{k,j}  & \text{otherwise.}
    \end{cases} 
 \label{eq_ped}
\end{equation}

We compute the probabilities of mispronunciation for $N$ phoneme recognition hypotheses from the PR. Mispronunciation for a given word is detected if the probability of mispronunciation falls below a given threshold for all hypotheses. The hyper-parameter $N=4$ was manually tuned on a single L2 speaker from the testing set to optimize the PED in the precision metric.

\subsection{Experiments and discussion}
\label{subsec:uncertainty_modeling_experiments}

We want to understand the effect of accounting for uncertainty in the PR-PM system presented in Section \ref{sec:proposed_model}. To do this, we compare it with two other variants, PR-LIK and PR-NOLIK, and analyze precision and recall metrics. The PR-LIK system helps us understand how important is it to account for the phonetic variability in the PM. To switch the PM off, we modify it so that it considers only a single way for a sentence to be pronounced correctly. 

The PR-NOLIK variant corresponds to the CTC-based mispronunciation detection model proposed by Leung et al. \cite{leung2019cnn}. To reflect this, we make two modifications compared to the PR-PM system. First, we switch the PM off in the same way we did it in the PR-LIK system. Second, we set the posterior probabilities of recognized phonemes in the PR to 100\%, which means that the PR is always certain about the phonemes produced by a speaker. There are some slight implementation differences between Leung's model and PR-NOLIK, for example, regarding the number of units in the neural network layers. We use our configuration to make a consistent comparison with PR-PM and PR-LIK systems. One can hence consider PR-NOLIK as a fair state-of-the-art baseline \cite{leung2019cnn}.

\subsubsection{Model details}
\label{subsubsec:uncertainty_modeling_model_details}

For extracting mel-spectrograms, we used a time step of 10 ms and a window size of 40 ms. The PR was trained with CTC Loss and Adam Optimizer (batch size: 32, learning rate: 0.001, gradient clipping: 5). We tuned the following hyper-parameters of the PR with Bayesian Optimization: dropout, CNN channels, GRU, and dense units. The PM was trained with the cross-entropy loss and AdaDelta optimizer (batch size: 20, learning rate: 0.01, gradient clipping: 5). The location-sensitive attention in the PM follows the work by Chorowski et al. \cite{chorowski2015attention}. The PR and PM models were implemented in MxNet Deep Learning framework.

\subsubsection{Speech corpora}
\label{subsubsec:uncertainty_modeling_speech_corpora}

For training and testing the PR and PM, we used 125.28 hours of L1 and L2 English speech from 983 speakers segmented into 102812 sentences, sourced from multiple speech corpora: TIMIT \cite{garofolo1993darpa}, LibriTTS \cite{Zen2019}, Isle \cite{atwell2003isle} and GUT Isle \cite{Weber2020}. We summarize it in Table \ref{tab:speech_corpora}. All speech data were downsampled to 16 kHz. Both L1 and L2 speech were phonetically transcribed using Amazon proprietary grapheme-to-phoneme model and used by the PR. Automatic transcriptions of L2 speech do not capture pronunciation errors, but we found it is still worth including automatically transcribed L2 speech in the PR. L2 corpora were also annotated by 5 native speakers of American English for word-level pronunciation errors. There are 3624 mispronounced words out of 13191 in the Isle Corpus and 1046 mispronounced words out of 5064 in the GUT Isle Corpus.

From the collected speech, we held out 28 L2 speakers and used them only to assess the performance of the systems in the mispronunciation detection task. It includes 11 Italian and 11 German speakers from the Isle corpus \cite{atwell2003isle}, and 6 Polish speakers from the GUT Isle corpus \cite{Weber2020}.

\begin{table}[htb]
\caption{The summary of speech corpora used by the PR.}
  \label{tab:speech_corpora}
  \centering
  \begin{tabular}{lll}
    \toprule  
    \tabhead{Native Language} &  \tabhead{Hours} &  \tabhead{Speakers} \\
    \midrule
    English & 90.47 & 640\\ 
    Unknown & 19.91 & 285\\  
    German and Italian & 13.41 & 46\\  
    Polish & 1.49 & 12\\ 
    \bottomrule\\
  \end{tabular}
\end{table}

\subsubsection{Experimental results}
\label{subsubsec:uncertainty_modeling_experiment_results}

The PR-NOLIK detects mispronounced words based on the difference between the canonical and recognized phonemes. Therefore, this system does not offer any flexibility in optimizing the model for higher precision. 

\begin{figure}[th]
\centering
\includegraphics[width=1\textwidth]{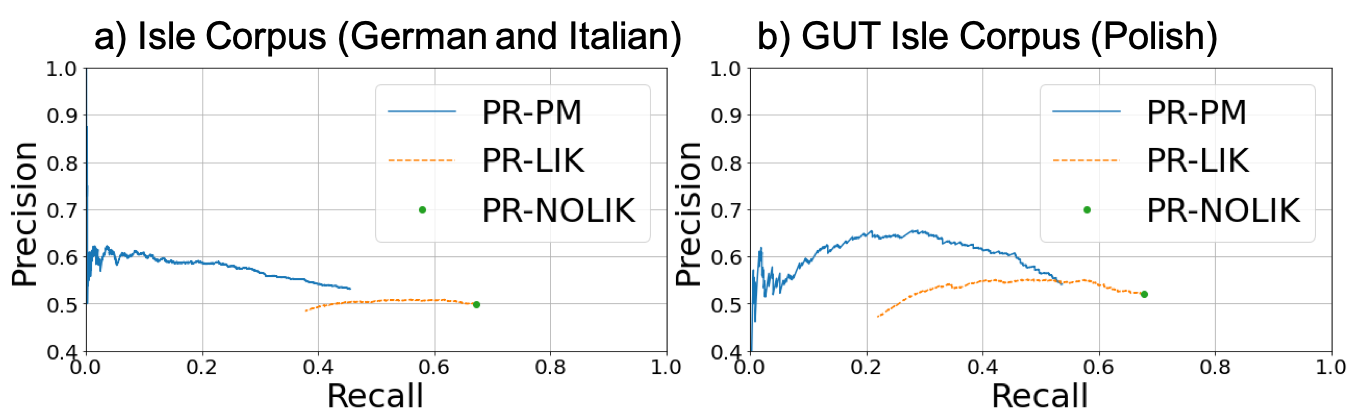}
\decoRule
\caption[Precision-recall curves for the evaluated systems]{Precision-recall curves for the evaluated systems.}
\label{fig:uncertainty_precision_recall_plot}
\end{figure}

The PR-LIK system incorporates posterior probabilities of recognized phonemes. It means that we can tune this system towards higher precision, as illustrated in Figure \ref{fig:uncertainty_precision_recall_plot}. Accounting for uncertainty in the PR helps when there is more than one likely sequence of phonemes that could have been uttered by a user, and the PR model is uncertain which one it is. For example, the PR reports two likely pronunciations for the text `I said' /ay s eh d/. The first one, /s eh d/ with /ay/ phoneme missing at the beginning and the alternative one  /ay s eh d/ with the /ay/ phoneme present. If the PR considered only the mostly likely sequence of phonemes, like PR-NOLIK does, it would incorrectly raise a pronunciation error. In the second example, a student read the text `six' /s ih k s/ mispronouncing the first phoneme /s/ as /t/. The likelihood of the recognized phoneme is only 34\%. It suggests that the PR model is quite uncertain on what phoneme was pronounced. However, sometimes even in such cases, we can be confident that the word was mispronounced. It is because the PM computes the probability of pronunciation based on the posterior probability from the PR model. In this particular case, other phoneme candidates that account for the remaining 66\% of uncertainty are also unlikely to be pronounced by a native speaker. The PM can take it into account and correctly detect a mispronunciation.

However, we found that the effect of accounting for uncertainty in the PR is quite limited. Compared to the PR-NOLIK system, the PR-LIK raises precision on the GUT Isle corpus only by 6\% (55\% divided by 52\%), at the cost of dropping recall by about 23\%. We can observe a much stronger effect when we account for uncertainty in the PM model. Compared to the PR-LIK system, the PR-PM system further increases precision between 11\% and 18\%, depending on the decrease in recall between 20\% to 40\%. One example where the PM helps is illustrated by the word `enough' that can be pronounced in two similar ways: /ih n ah f/ or /ax n ah f/ (short `i' or `schwa' phoneme at the beginning.) The PM can account for phonetic variability and recognize both versions as pronounced correctly. Another example is word linking \cite{hieke1984linking}. Native speakers tend to merge phonemes of neighboring words. For example, in the text `her arrange' /hh er - er ey n jh/, two neighboring phonemes /er/ can be pronounced as a single phoneme: /hh er ey n jh/. The PM model can correctly recognize multiple variations of such pronunciations.

Complementary to precision-recall curve showed in Figure \ref{fig:uncertainty_precision_recall_plot}, we present in Table \ref{tab:pron_error_detection} one configuration of the precision and recall scores for the PR-LIK and PR-PM systems. This configuration is selected in such a way that:  a) recall for both systems is close to the same value, b) to illustrate that the PR-PM model has a much bigger potential of increasing precision than the PR-LIK system. A similar conclusion can be made by inspecting multiple different precision and recall configurations in the precision and recall plots for both Isle and GUT Isle corpora.

\begin{table}[thb]
\caption{Precision and recall of detecting word-level mispronunciations. CI - Confidence Interval.}
  \label{tab:pron_error_detection}
  \centering
  \begin{tabular}{lll}
    \toprule  
    \tabhead{Model}  &  \tabhead{Precision [\%,95\%CI]} &  \tabhead{Recall [\%,95\%CI]} \\
    \midrule
    \multicolumn{3}{c}{\textbf{Isle corpus (German and Italian)}} \\
    PR-LIK & 49.39 (47.59-51.19) & 40.20 (38.62-41.81)\\ 
    PR-PM & 54.20 (52.32-56.08) & 40.20 (38.62-41.81)\\  
  
     \multicolumn{3}{c}{\textbf{GUT Isle corpus (Polish)}} \\
    PR-LIK & 54.91 (50.53-59.24) & 40.29 (36.66-44.02)\\ 
    PR-PM & 61.21 (56.63-65.65) & 40.15 (36.51-43.87)\\  
    
    \bottomrule\\
  \end{tabular}
\end{table}
 
\subsection{Conclusion and future work}
\label{subsec:uncertainty_modeling_conclusion}

To report fewer false pronunciation alarms, it is important to move away from the two simplifying assumptions that are usually made by common methods for pronunciation assessment: a) phonemes can be recognized with high accuracy, b) a sentence can be read in a single correct way. We acknowledged that these assumptions do not always hold. Instead, we designed a model that: a) accounts for the uncertainty in phoneme recognition and b) accounts for multiple ways a sentence can be pronounced correctly due to phonetic variability. We found that to optimize precision, it is more important to account for the phonetic variability of speech than accounting for uncertainty in phoneme recognition. We showed that the proposed model can raise the precision of detecting mispronounced words by up to 18\% compared to the common methods.

In the future, we plan to adapt the PM model to correctly pronounced L2 speech to account for phonetic variability of non-native speakers. We plan to combine the PR, PM, and PED modules and train the model jointly to eliminate accumulation of statistical errors coming from disjoint training of the system.

%% file: Chapters/PronunciationErrorDetection/LexicalStressErrorDetection.tex
\section{Detection of Lexical Stress Errors in Non-Native (L2) English with Data Augmentation and Attention}
\label{sec:lexical_stress}

\begin{center}
\textit{Daniel Korzekwa, Roberto Barra-Chicote, Szymon Zaporowski, Grzegorz Beringer, Jaime Lorenzo-Trueba, Alicja Serafinowicz, Jasha Droppo, Thomas Drugman, Bozena Kostek, Detection of Lexical Stress Errors in Non-Native (L2) English with Data Augmentation and Attention, Interspeech, 2021}
\end{center}

\bigskip

\textbf{Abstract}

This paper describes two novel complementary techniques that improve the detection of lexical stress errors in non-native (L2) English speech: attention-based feature extraction and data augmentation based on Neural Text-To-Speech (TTS). In a classical approach, audio features are usually extracted from fixed regions of speech such as the syllable nucleus. We propose an attention-based deep learning model that automatically derives optimal syllable-level representation from frame-level and phoneme-level audio features. Training this model is challenging because of the limited amount of incorrect stress patterns. To solve this problem, we propose to augment the training set with incorrectly stressed words generated with Neural TTS. Combining both techniques achieves 94.8\% precision and 49.2\% recall for the detection of incorrectly stressed words in L2 English speech of Slavic and Baltic speakers.

\subsection{Introduction}

Computer Assisted Pronunciation Training (CAPT) usually focuses on practicing pronunciation of phonemes \cite{witt2000phone, leung2019cnn, korzekwa2021mispronunciation}, while there is evidence in non-native (L2) English speakers that practicing lexical stress improves speech intelligibility \cite{field2005intelligibility,lepage2014intelligibility}. Lexical stress is a syllable-level phonological feature. It is a part of the phonological rules that define how words should be spoken in a given language. Stressed syllables are usually longer, louder, and expressed with a higher pitch than their unstressed counterparts \cite{jung2018acoustic}. Lexical stress is inter-connected with phonemic representation. For example, placing lexical stress on a different syllable of a word may lead to different phonemic realizations known as `vowel reduction' \cite{bergem1991acoustic}.

The focal point of our work is the detection of words with incorrect stress patterns. The training data with human speech is usually highly imbalanced, with few training examples of incorrectly stressed words. It makes training machine learning models for this task challenging. We address this problem by augmenting the training set with synthetic speech that is generated with Neural Text-To-Speech (TTS) \cite{latorre2019effect}. Neural TTS allows us generating words with both correct and incorrect stress patterns.

Most of the existing approaches for automated lexical stress assessment are based on carefully designed features that are extracted from fixed regions of speech signal such as the syllable nucleus \cite{ferrer2015classification,shahin2016automatic, chen2010automatic_2}. We introduce attention mechanism \cite{vaswani2017attention} to automatically learn optimal syllable-level representation. 
Attention-based approach originates from the intuition of how people detect specific patterns in high dimensional and unstructured data such as visual and speech signals \cite{posner1990attention}. For example, we might focus our attention on the duration ratio between nuclei of two neighboring syllables, incidentally, an important predictor of lexical stress. The syllable-level representation is derived from frame-level (F0, intensity) and phoneme-level (duration) audio features and the corresponding phonetic representation of a word. We do not indicate precisely the regions of the audio signal that are important for the detection of lexical stress errors. The attention mechanism does it automatically.

To the best of our knowledge, this paper is the first attempt, for the task of lexical stress error detection, to: \emph{i)} augment the training data with Neural TTS, \emph{ii)} use attention mechanisms to automatically extract syllable-level features for lexical stress error detection. Ruan et al. \cite{ruan2019end} used attention-based architecture of transformers for lexical stress detection. However, their paper concerns recognizing stressed and unstressed phonemes. They do not detect lexical stress errors, which is crucial in CAPT applications.

The paper is structured as follows. In Section \ref{sec:lexical_stress_related_work}, we review the related work. Section \ref{sec:lexical_stress_proposed_model} describes the proposed model. Section \ref{sec:lexical_stress_speech_corpora} reviews human and synthetic speech corpora. In Section \ref{sec:lexical_stress_experiments}, we present our experiments, and Section \ref{sec:lexical_stress_conclusion} concludes the paper.

\subsection{Related work}
\label{sec:lexical_stress_related_work}

\begin{figure}[th]
\centering
\centerline{
\includegraphics[width=1.25\textwidth]{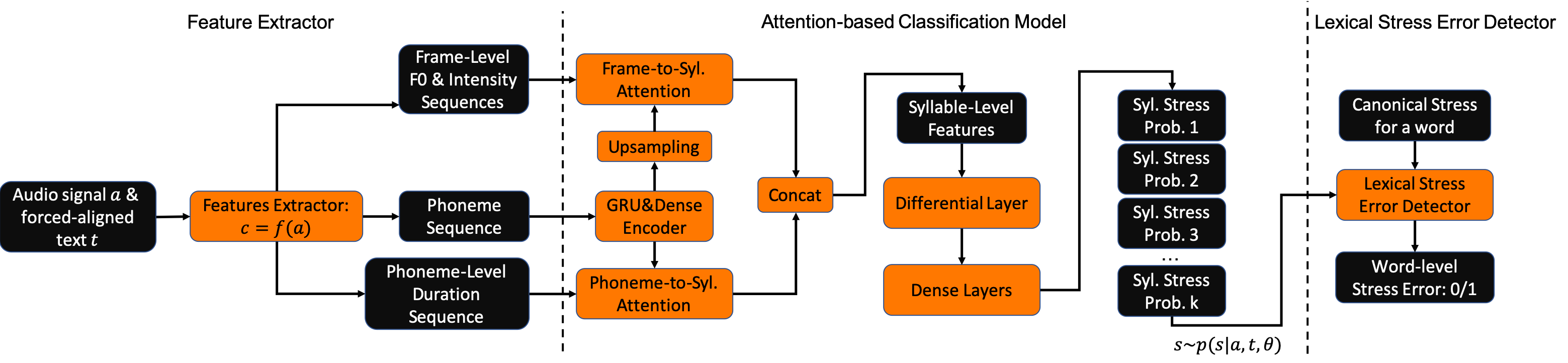}}
\decoRule
\caption[Attention-based Deep Learning model for the detection of lexical stress errors]{Attention-based Deep Learning model for the detection of lexical stress errors.}
\label{fig:lexical_stress_model_architecture}
\end{figure}

The existing work focuses on the supervised classification of lexical stress using Neural Networks \cite{li2018automatic, shahin2016automatic}, Support Vector Machines \cite{chen2010automatic_2,zhao2011automatic} and Fisher’s linear
discriminant \cite{chen2007using}. There are two popular variants: a) discriminating syllables between primary stress/no stress \cite{ferrer2015classification}, and b) classifying between primary stress/secondary stress/no stress \cite{li2013lexical,li2018automatic}. Ramanathi et al. \cite{ramanathi2019asr} have followed an alternative unsupervised way of classifying lexical stress, which is based on computing the likelihood of an acoustic signal for a number of possible lexical stress representations of a word.

Accuracy is the most commonly used performance metric, and it indicates the ratio of correctly classified stress patterns on a syllable \cite{li2013lexical} or word level \cite{chen2010automatic_2}. On the contrary, following Ferrer et al. \cite{ferrer2015classification}, we analyze precision and recall metrics because we aim to detect lexical stress errors and not just classify them.

Existing approaches for the classification and detection of lexical stress errors are based on carefully designed features. They start with aligning a speech signal with phonetic transcription, performed via forced-alignment \cite{shahin2016automatic, chen2010automatic_2}. Alternatively, Automatic Speech Recognition (ASR) can provide both phonetic transcription and its alignment with a speech signal \cite{li2013lexical}. Then, prosodic features such as duration, energy and pitch \cite{chen2010automatic_2} and cepstral features such as MFCC and Mel-Spectrogram \cite{ferrer2015classification,shahin2016automatic} are extracted. These features can be extracted on the syllable \cite{shahin2016automatic} or syllable nucleus \cite{ferrer2015classification,chen2010automatic_2} level. 

Shahin et al. \cite{shahin2016automatic} computed features of neighboring vowels, and Li et al. \cite{li2013lexical} included the features for two preceding and two following syllables in the model. The features are often preprocessed and normalized to avoid potential confounding variables \cite{ferrer2015classification}, and to achieve better model generalization by normalizing the duration and pitch on a word level \cite{ferrer2015classification,chen2007using}. Li et al. \cite{li2018automatic} added canonical lexical stress to input features, which improves the accuracy of the model. 

In our approach, we use attention mechanisms to derive automatically regions of the audio signal that are important for the detection of lexical stress errors. We also use data augmentation through the generation of artificial data with Neural TTS.

\subsection{Proposed model}
\label{sec:lexical_stress_proposed_model}

The proposed model consists of three subsystems: Feature Extractor, Attention-based Classification Model, and Lexical Stress Error Detector. It is illustrated in Figure \ref{fig:lexical_stress_model_architecture}.

\subsubsection{Feature extractor}
\label{subsec:lexical_stress_feature_extractor}

The Feature Extractor extracts prosodic features and phonemes from speech signal $\mathbf{a}$ and forced-aligned text $\mathbf{t}$. To obtain forced-alignment, we used Montreal toolkit \cite{mcauliffe2017montreal} along with an acoustic model pretrained on LibriSpeech ASR corpus \cite{panayotov2015librispeech}. The prosodic features $\mathbf{c}=f(\mathbf{a})$ are formed by: F0, intensity [dB SPL] and phoneme-level durations. The F0 and intensity features are computed at the frame level using Praat library \cite{boersma2006praat} (time step: 10 ms, window size: 40 ms). The F0 contour is linearly interpolated in unvoiced regions. These raw features will be further transformed by the attention-based model to the syllable-level representation.

\subsubsection{Attention-based classification model}
\label{subsec:lexical_stress_classification_model}

The Attention-based Classification Model maps frame-level and phoneme-level features to the syllable-level representation. Then, it produces a lexical stress pattern $\mathbf{s}$, modeled as a sequence of Bernoulli random variables $\mathbf{s}=\{s_1,..,s_k\}$ (stressed/unstressed) over $K$ syllables of a multi-syllable word, conditioned on audio $\mathbf{a}$ and text $\mathbf{t}$ representations. Let us define it as a conditional probability distribution $\mathbf{s} \sim p(\mathbf{s}|\mathbf{a},\mathbf{t},\bm{\theta})$, where $\bm{\theta}$ are the parameters of the model.

\begin{figure}[th]
\centering
\centerline{
\includegraphics[width=1.2\textwidth]{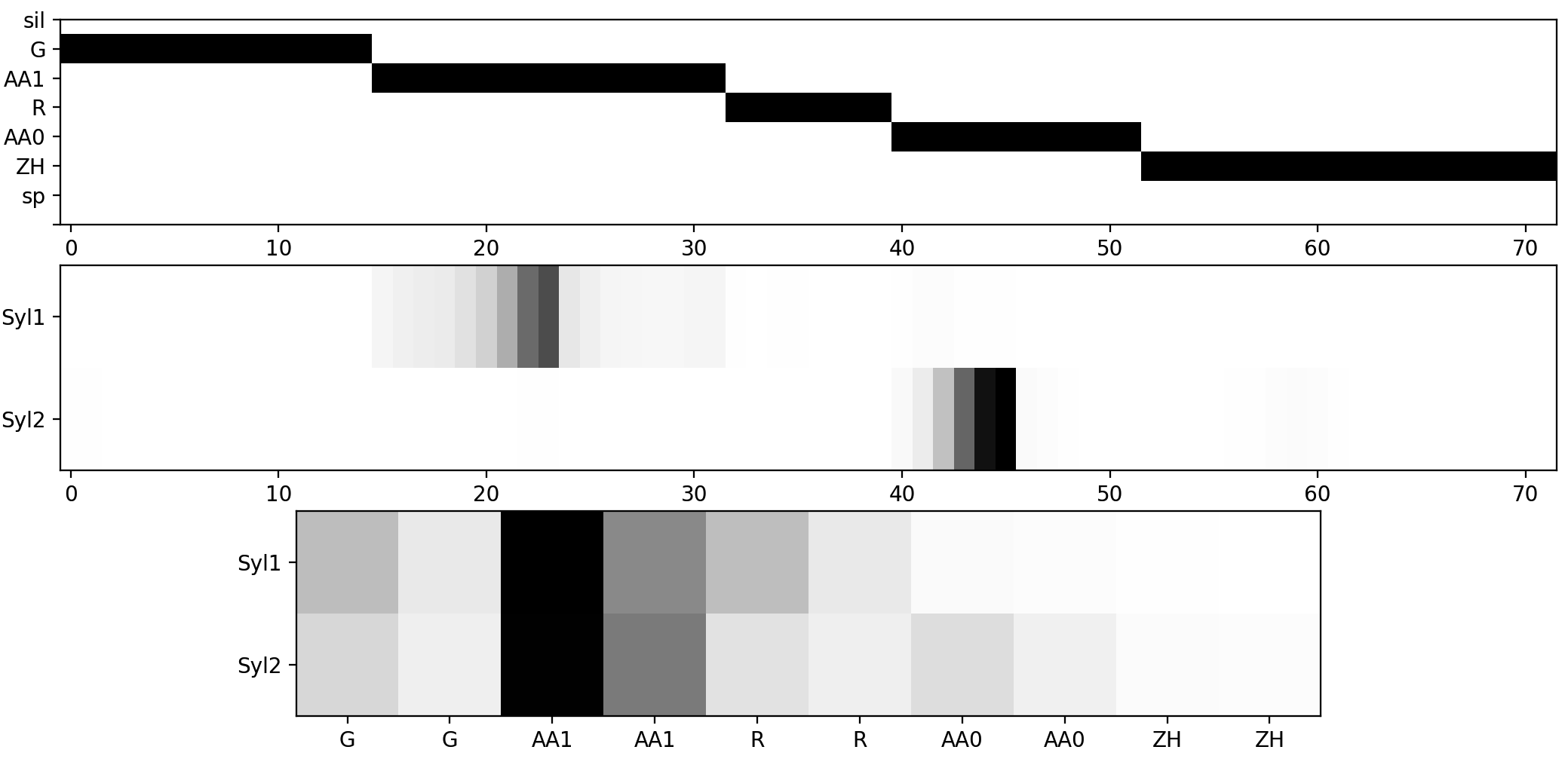}}
\decoRule
\caption[Top: forced-alignment mapping between phonemes and frames for the word 'garage'. Middle: Frame-to-syllable attention weights matrix. Bottom: (Sub)Phoneme-to-syllable attention weights matrix]{Top: forced-alignment mapping between phonemes and frames for the word 'garage'. Middle: Frame-to-syllable attention weights matrix. Bottom: (Sub)Phoneme-to-syllable attention weights matrix.}
\label{fig:frame_phoneme_level_att}
\end{figure}

To extract syllable-level features, we use two dot-product attentions operating on the frame and phoneme levels. To build better intuition on what these two attention do, in Figure \ref{fig:frame_phoneme_level_att} we show the frame-level and phoneme-level attention plots for the word 'garage' pronounced by a Polish speaker and incorrectly stressed on the first syllable in reference to American English. This word has a similar pronunciation but different lexical stress in Polish and American English languages (`G AA1 R AA0 ZH' vs `G ER0 AA1 ZH'). Both attentions find the most relevant regions of the frame-level and phoneme-level features.

The dot-product attention is presented in Eq. \ref{lexical_stress_eq_att}, and it follows the notation proposed by Vaswani et al. \cite{vaswani2017attention}. It is based on three inputs: Query ($\mathbf{Q}$), Keys ($\mathbf{K}$) and Values ($\mathbf{V}$), where $d_k$ is the dimensionality of $\mathbf{K}$.

\begin{equation}
Attention(\mathbf{Q}, \mathbf{K}, \mathbf{V}) = softmax(\dfrac{\mathbf{Q}\mathbf{K}^t}{\sqrt{d_k}})\mathbf{V}
\label{lexical_stress_eq_att}
\end{equation}

The attention inputs are represented as follows.  Query refers to the syllable positional embeddings defined by one-hot syllable index encodings. Keys represents a sequence of sub-phonemes. Each sub-phoneme is represented by a set of features: $phoneme\_id$, $syllable\_index$, $is\_vowel$, $left\_or\_right\_sub\_phoneme$. All features are one-hot encoded and processed with a Gated Recurrent Unit (GRU) layer \cite{cho2014learning} (units:4, dropout: 0.24). In the end, encoded sub-phoneme sequence is passed through linear dense layers. In the case of the frame-level attention, the encoded sub-phoneme sequence is upsampled to the frame level using phoneme durations from forced-alignment. In upsampling, we simply replicate phonemes across aligned frames of audio signal. Similar phoneme-to-frame upsampling has been recently adopted in Text-To-Speech \cite{elias2020parallel}. Finally, Values are the $F0/intensity$ and $duration$ features for frame-level and phoneme-level attentions respectively.

To model relative prominence, we introduce a differential bi-directional layer that computes the ratios of syllable-level acoustic features for each syllable and its two neighbors (Figure \ref{fig:lexical_stress_model_architecture}). The bi-directional layer is implemented as a simple `division' math operation and it does not contain any trainable parameters. The output of the differential layer is further processed by three dense layers (units: 4, activation: tanh, dropout: 0.24), followed by a linear dense layer (units: 2, dropout: 0.24) that produces a two-dimensional output for each syllable. It is then squeezed by a softmax function to generate lexical stress probabilities.

\subsubsection{Training of the classification model}
\label{subsec:lexical_stress_training_procedure}

We train the model on a set of $N$ triplets that contains 1) human recorded words and 2) synthetic words generated using Neural TTS. A single triplet is represented by $\{\mathbf{s_n},\mathbf{a_n},\mathbf{t_n}\}$, where $n=1..N$ is the index of a training example.

The concept of data augmentation can be explained using a framework of Bayesian Inference. Consider three random variables, lexical stress $\mathbf{s_n}$, audio signal $\mathbf{a_n}$ and text $\mathbf{t_n}$. All variables are observed for the training examples of human speech. However, for the synthetic speech, we only observe the lexical stress and text variables. The audio signal is unobserved (hidden) because we have to generate it.

To train this model, we derive a negative log-likelihood loss over a joint probability distribution of lexical stress $\mathbf{s}$ and audio $\mathbf{a}$ random variables, as depicted in Eq. \ref{eq1}. The loss is further approximated with the variational lower bound \cite{jordan1999introduction}, as presented in Eq. \ref{eq2} (we omit $\bm{\theta}$ for brevity). For the training examples of synthetic speech, the conditional probability distribution over the audio signal $\mathbf{a_n} \sim p(\mathbf{a_n}|\mathbf{s_n},\mathbf{t_n})$ is estimated with Neural TTS, and for human recorded words, it is given explicitly.

\begin{equation}
\mathcal{L(\bm{\theta})} = -\sum_n^N log \int p(\mathbf{s_n},\mathbf{a_n}|\mathbf{t_n},\bm{\theta})d\mathbf{a_n}
  \label{eq1}
\end{equation}

\begin{equation}
 log \int p(\mathbf{s_n},\mathbf{a_n}|\mathbf{t_n})\mathbf{da_n}\approx E_{\mathbf{a_n}\sim p(\mathbf{a_n}|\mathbf{t_n},\mathbf{s_n})}[logp(\mathbf{s_n}|\mathbf{a_n},\mathbf{t_n})]
  \label{eq2}
\end{equation}

The model was implemented in MxNet \cite{chen2015mxnet}, trained with Stochastic Gradient Descent optimizer (learning rate: 0.1, batch size: 20) and tuned with Bayesian optimization \cite{paleyes2019emulation}. Training data were split into buckets based on the number of frames in an audio signal, using Gluon-NLP package \cite{guo2020gluoncv}. A single bucket contains words with the same number of syllables with zero-padded acoustic and sub-phoneme sequences.

\subsubsection{Lexical stress error detector}
\label{subsec:lexical_stress_error_detector}

The Lexical Stress Error Detector reports on lexical stress error if the expected (canonical) and estimated lexical stress for a given syllable do not match and the corresponding probability is higher than a given threshold.

\subsection{Speech corpus}
\label{sec:lexical_stress_speech_corpora}

Our speech corpus consists of human and synthetic speech. The data were split into training and testing sets with disjointed speakers ascribed to each set. Human speech contains L1 and L2 speakers of English. Synthetic data were generated with Neural TTS and are included only in the training set. All audio files were downsampled to a 16 kHz sampling rate. The data are summarized in Table \ref{tab:speech_corpora_traintest}, and we provide more details in the following subsections.

\begin{table}[!ht]
  \caption{Train and test sets details.}
  \label{tab:speech_corpora_traintest}
  \centering
  \begin{tabular}{llll}
    \toprule  
    \tabhead{Data set}  &  \tabhead{\makecell[l]{Speakers \\ (L2)}} &  \tabhead{\makecell[l]{Words \\ (unique)}} &   \tabhead{\makecell[l]{Stress \\ Errors}} \\
    \midrule
    Train set (human) & 473 (10) & 8223 (1528) & 425\\
    Train set (TTS) & 1 (0) & 3937 (1983) & 2005\\
    Test set (human) & 176 (21) & 2108 (378) & 189\\
    \bottomrule\\
  \end{tabular}

\end{table}

\subsubsection{Human speech}
\label{subsec:lexical_stress_human_speech}

Due to the limited availability of L2 corpora, we recorded our own L2-English corpus of Slavic and Baltic speakers. It also allows us to evaluate the model during interactive English learning sessions with our students. The corpus contains speech from 25 speakers (23 Polish, 1 Ukrainian and 1 Lithuanian): 7 females and 18 males, all between 24 and 40 years old. All speakers read a list of two hundred words. One hundred words were prepared by a professional English teacher, including frequently mispronounced words by Slavic and Baltic students. The second half consists of the most common words that were obtained from Google's Trillion Word Corpus \cite{michel2011quantitative} based on n-gram frequency analysis. We excluded abbreviations and one-syllable words.

Additionally, L1 and L2 English speech was collected from publicly available speech data sets, including TIMIT \cite{garofolo1993darpa}, Arctic \cite{kominek2004cmu}, L2-Arctic \cite{zhao2018l2} and Porzuczek \cite{porzuczek2017english}. 

\subsubsection{Synthetic speech}
\label{subsec:lexical_stress_synthetic_speech}

Complementary to human recordings, synthetic speech was generated with Neural TTS by Latorre et al. \cite{latorre2019effect}. The Neural TTS consists of two modules. Context-generation module is an attention-based encoder-decoder neural network that generates a mel-spectrogram from a sequence of phonemes. Then, a Neural Vocoder converts it to the speech signal. The Neural Vocoder is a neural network of architecture similar to the work by \cite{oord2018parallel}. The Neural TTS was trained using speech of a professional American voice talent. To generate words with different lexical stress patterns, we modify lexical stress markers associated with the vowels in the phonemic transcription of a word. For example, with the input of /r iy1 m ay0 n d/ we can place lexical stress on the first syllable of the word `remind'. 1980 popular English words were synthesized with correct and incorrect stress patterns.

\subsubsection{Lexical stress annotations}
\label{subsec:lexical_stress_lexical_stress_annotations}

L1 corpora were segmented into words and annotated automatically using a proprietary Amazon American English Lexicon, taking into account the syntactic context of the word. Neural TTS speech and the speech of L2 speakers were annotated by 5 American English linguists into `primary' and `no stress' categories, keeping the words for which a minimum of 4 out of 5 linguists agreed on the stress pattern. Annotators were not able to distinguish between primary and secondary lexical stress.
81.5\% of synthesized words matched the intended stress patterns with a minimum of 4 annotators' agreement. It shows that Neural TTS can be used to generate incorrectly stressed speech.

\subsection{Experiments}
\label{sec:lexical_stress_experiments}

\begin{figure}[th]
\centering
\includegraphics[width=0.8\textwidth]{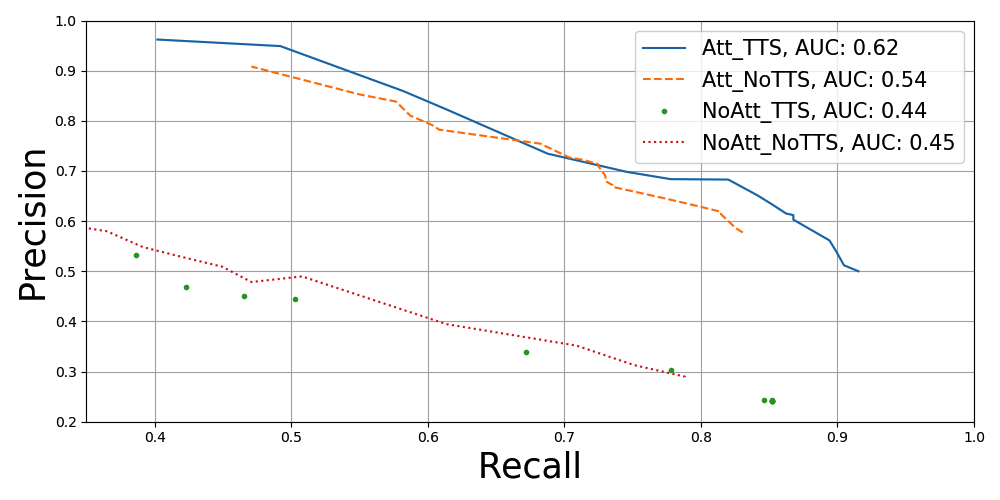}
\decoRule
\caption[Precision-recall curves for evaluated systems]{Precision-recall curves for evaluated systems.}
\label{fig:lexical_stress_precision_recall_plot}
\end{figure}

The proposed model (Att\_TTS) from Section \ref{sec:lexical_stress_proposed_model}  is compared to three baseline models that are designed to measure the impact of the Neural TTS data augmentation and the attention mechanism. To compare these models, we plotted their precision-recall curves and gave their corresponding area under a curve (AUC) along with our results, see Figure \ref{fig:lexical_stress_precision_recall_plot}.

The Att\_NoTTS model has the same architecture as the Att\_TTS, but the synthetic speech is excluded from the `training set'. The NoAtt\_TTS model uses the same training set as the Att\_TTS, but it has no attention mechanism. Instead, as a syllable-level representation, it uses mean values of acoustic features for the corresponding syllable nucleus. The NoAtt\_NoTTS model has no attention, and it does not use Neural TTS data augmentation.

As a state-of-the-art baseline, we use the work by Ferrer et al. \cite{ferrer2015classification}. However, a direct comparison is not possible. In their test corpus, there were 46.4\% (191 out of 411) of incorrectly stressed words, far more than 9.4\% (189 out of 2109) words in our experiment. The fewer lexical stress errors are made by users, the more challenging it is to detect it. They also used proprietary L2 English of Japanese speakers. Due to the lack of available benchmark and standard speech corpora for the task of lexical stress assessment, we could not make a fairer comparison with the state-of-the-art.

\subsubsection{Experimental results}
\label{subsec:lexical_stress_experiments_results}
First, we compare Att\_NoTTS and NoAtt\_NoTTS models. Using the attention mechanism for automatic extraction of syllable-level features significantly improves the detection of lexical stress errors. It is illustrated by precision-recall curves and AUC metric in Figure \ref{fig:lexical_stress_precision_recall_plot}. To be comparable with the study by Ferrer et al., we fix recall to around 50\% and compare the models using precision as shown in Table \ref{tab:precision_recall_acc}.

The Att\_NoTTS attention-based can be further improved. Augmenting the training set with incorrectly stressed words (Att\_TTS) boosts precision from 87.85\% to 94.8\%, at a recall level of 50\%. Data augmentation helps because it increases the number of words with incorrect stress patterns in the training set. It prevents the model from exploiting a strong correlation between phonemes and lexical stress in correctly stressed words. Using data augmentation in the simpler no-attention-based model (NoAtt\_TTS) does not help. It is because NoAtt\_TTS uses only prosodic features for fixed regions of speech, so this model cannot overfit to phonetic input.

\begin{table}[!ht]
 \caption{Precision and recall [\%, 95\% Confidence Interval] of detecting lexical stress errors, at around 50\% recall. * - Ferrer et al. model has been evaluated on the data with 46.4\% of lexical stress errors, compared to 9.4\% of errors on our data set. This data point indicates that our proposed model AttTTS should outperform Ferrr et al. model if both were evaluated exactly in the same conditions.}
  \label{tab:precision_recall_acc}
  \centering
  \begin{tabular}{lll}
    \toprule  
    \tabhead{Model} &  \tabhead{Precision} &  \tabhead{Recall}  \\
    \midrule
    AttTTS & 94.8 (89.18-98.03) & 49.2 (42.13-56.3) \\
    AttNoTTS & 87.85 (80.67-93.02) & 49.74 (42.66-56.82)\\
    NoAttTTS & 44.39 (37.85-51.09) & 50.26 (43.18-57.34)  \\
    NoAttNoTTS & 48.98 (42.04-55.95) & 50.79 (43.70-57.86)  \\
    Ferrer et al. \cite{ferrer2015classification} * & 95.00 (na-na) & 48.3 (na-na)  \\
    \bottomrule\\
  \end{tabular}
\end{table}

Ferrer et al. \cite{ferrer2015classification} reported on a similar performance to our Att\_TTS model with a precision of 95\% and a recall of 48.3\% on L2 English speech of Japanese speakers. However, in their testing data, the proportion of incorrectly stressed words is much larger, which makes it easier to detect lexical stress errors.

\subsection{Conclusion and future work}
\label{sec:lexical_stress_conclusion}

Using an attention-based neural network for the automatic extraction of syllable-level features significantly improves the detection of lexical stress errors in L2 English speech, compared to baseline models. However, this model has a tendency to classify lexical stress based on highly-correlated phonemes. We can counteract this effect by augmenting the training set with incorrectly stressed words generated with Neural TTS. It boosts the performance of the attention-based model by 14.8\% in the AUC metric and by 7.9\% in precision, while maintaining recall at a level close to 50\%. Data Augmentation, however, does not help when applied to a simpler model without an attention mechanism.

We found that the current word-level model is not able to correctly classify lexical stress when two words are linked \cite{hieke1984linking} and stress shift may occur  \cite{shattuck1994stress}. For example, two neighboring phonemes /er/ in the text `her arrange' /hh er - er ey n jh/ are pronounced as a single phoneme. Therefore, in future, we plan to move away from the assessment of isolated words and extend the current model to detect lexical stress errors at the sentence level. 
We plan to replace a single-speaker TTS model to generate synthetic lexical stress errors with a multi-speaker model. We plan to analyze the accuracy of detecting lexical stress errors for speakers with different proficiency levels of English.

%% file: Chapters/PronunciationErrorDetection/SpeechSynthesis.tex
\section{Speech synthesis is almost all you need}
\label{sec:speech_synthesis}

\begin{center}
\textit{Daniel Korzekwa, Jaime Lorenzo-Trueba, Thomas Drugman, Bozena Kostek, Computer-assisted Pronunciation Training - Speech synthesis is almost all you need, accepted for publication in Speech Communication Journal on June 17 ‘2022, in print}
\end{center}

\bigskip

\textbf{Abstract}

\bigskip

The research community has long studied computer-assisted pronunciation training (CAPT) methods in non-native speech. Researchers focused on studying various model architectures, such as Bayesian networks and deep learning methods, as well as on the analysis of different representations of the speech signal. Despite significant progress in recent years, existing CAPT methods are not able to detect pronunciation errors with high accuracy (only 60\% precision at 40\%-80\% recall). One of the key problems is the low availability of mispronounced speech that is needed for the reliable training of pronunciation error detection models. If we had a generative model that could mimic non-native speech and produce any amount of training data, then the task of detecting pronunciation errors would be much easier. We present three innovative techniques based on phoneme-to-phoneme (P2P), text-to-speech (T2S), and speech-to-speech (S2S) conversion to generate correctly pronounced and mispronounced synthetic speech. We show that these techniques not only improve the accuracy of three machine learning models for detecting pronunciation errors but also help establish a new state-of-the-art in the field. Earlier studies have used simple speech generation techniques such as P2P conversion, but only as an additional mechanism to improve the accuracy of pronunciation error detection. We, on the other hand, consider speech generation to be the first-class method of detecting pronunciation errors. The effectiveness of these techniques is assessed in the tasks of detecting pronunciation and lexical stress errors. Non-native English speech corpora of German, Italian, and Polish speakers are used in the evaluations. The best proposed S2S technique improves the accuracy of detecting pronunciation errors in AUC metric by 41\% from 0.528 to 0.749 compared to the state-of-the-art approach.

\subsection{Introduction}

Language plays a key role in online education, giving people access to large amounts of information contained in articles, books, and video lectures. Thanks to spoken language and other forms of communication, such as a sign-language, people can participate in interactive discussions with teachers and take part in lively brainstorming with other people. Unfortunately, education is not available to everybody. According to the UNESCO report, 40\% of the global population do not have access to education in the language they understand \cite{unesco2016if}.  ‘If you don’t understand, how can you learn?' the report says. English is the leading language on the Internet, representing 25.9\% of the world's population \cite{statista2020}. Regrettably, research by EF (Education First) \cite{EPI2020} shows a large disproportion in English proficiency across countries and continents. People from regions of 'very low' language proficiency, such as the Middle East, are unable to navigate through English-based websites or communicate with people from an English-speaking country.

Computer-Assisted Language Learning (CALL) helps to improve the English language proficiency of people in different regions \cite{levy2013call}. CALL relies on computerized self-service  tools that are used by students to practice a language, usually a foreign language, also known as a non-native (L2) language. Students can practice multiple aspects of the language, including grammar, vocabulary, writing, reading, and speaking. Computer-based tools can also be used to measure student's language skills and their learning potential by using Computerized Dynamic Assessment (C-DA) test \cite{mehri2019diagnosing}. CALL can complement traditional language learning provided by teachers. It also has a chance to make second language learning more accessible in scenarios where traditional ways of learning languages are not possible due to the cost of learning or the lack of access to foreign language teachers.

Computer-Assisted Pronunciation Training (CAPT) is a part of CALL responsible for learning pronunciation skills. It has been shown to help people practice and improve their pronunciation skills \cite{neri2008effectiveness,golonka2014technologies, tejedor2020assessing}. CAPT consists of two components: an automated pronunciation evaluation component \cite{leung2019cnn,zhang2021text,korzekwa2021mispronunciation} and a feedback component \cite{ai2015automatic}. The automated pronunciation evaluation component is responsible for detecting pronunciation errors in spoken speech, for example, for detecting words pronounced incorrectly by the speaker. The feedback component informs the speaker about mispronounced words and advises  how to pronounce them correctly. This article is devoted to the topic of automated detection of pronunciation errors in non-native speech. This area of CAPT can take advantage of technological advances in machine learning and bring us closer to creating a fully automated assistant based on artificial intelligence for language learning.

The research community has long studied the automated detection of pronunciation errors in non-native speech. Existing work has focused on various tasks such as detecting mispronounced phonemes \cite{leung2019cnn} and lexical stress errors \cite{ferrer2015classification}. Researcher have given most attention to studying various machine learning models such as Bayesian networks \cite{witt2000phone, Hongyan2011} and deep learning methods \cite{leung2019cnn, zhang2021text}, as well as analyzing different representations of the speech signal such as prosodic features (duration, energy and pitch) \cite{chen2010automatic_2}, and cepstral/spectral features \cite{ferrer2015classification, shahin2016automatic, leung2019cnn}. Despite significant progress in recent years, existing CAPT methods detect pronunciation errors with relatively low accuracy of 60\% precision at 40\%-80\% recall \cite{leung2019cnn,korzekwa2021mispronunciation,zhang2021text}.  Highlighting correctly pronounced words as pronunciation errors by the CAPT tool can demotivate students and lower the confidence in the tool. Likewise, missing pronunciation errors can slow down the learning process.

One of the main challenges with the existing CAPT methods is poor availability of mispronounced speech, which is required for the reliable training of pronunciation error detection models. We propose a reformulation of the problem of pronunciation error detection as a task of synthetic speech generation. Intuitively, if we had a generative model that could mimic mispronounced speech and produce any amount of training data, then the task of detecting pronunciation errors would be much easier. The probability of pronunciation errors for all the words in a sentence can then be calculated using the Bayes rule \cite{bishop2006pattern}. In this new formulation, we move the complexity to learning the speech generation process that is well suited to the problem of limited speech availability \cite{huybrechts2021low, shah21_ssw, fazel21_interspeech}. The proposed method outperforms the state-of-the-art model \cite{leung2019cnn} in detecting pronunciation errors in AUC metric by 41\% from 0.528 to 0.749 on the GUT Isle Corpus of L2 Polish speakers. 

To put the new formulation of the problem into action, we propose three innovative techniques based on phoneme-to-phoneme (P2P), text-to-speech (T2S), and speech-to-speech (S2S) conversion to generate correctly pronounced and mispronounced synthetic speech. We show that these techniques not only improve the accuracy of three machine learning models for detecting pronunciation errors but also help establish a new state-of-the-art in the field. The effectiveness of these techniques is assessed in two tasks: detecting mispronounced words (replacing, adding, removing phonemes, or pronouncing an unknown speech sound) and detecting lexical stress errors. The results presented in this study are the culmination of our recent work on speech generation in pronunciation error detection task \cite{korzekwa2021mispronunciation,korzekwa21b_interspeech,korzekwa21_interspeech}, including a new S2S technique.

In short, the contributions of the paper are as follows:
\begin{itemize}
\item A new paradigm for the automated detection of pronunciation errors is proposed, reformulating the problem as a task of generating synthetic speech.
\item A unified probabilistic view on P2P, T2S, and S2S techniques is presented in the context of detecting pronunciation errors.
\item A new S2S method to generate synthetic speech is proposed, which outperforms the state-of-the-art model \cite{leung2019cnn} in detecting pronunciation errors.
\item Comprehensive experiments are described to demonstrate the effectiveness of speech generation in the tasks of pronunciation and lexical stress error detection.
\end{itemize}

The outline of the rest of this paper is: Section \ref{sec:related_work} presents related work. Section \ref{sec:synth_speech_generation_methods} describes the proposed methods of generating synthetic speech for automatic detection of pronunciation errors. Section \ref{sec:human_speech_corpora} describes the human speech corpora used to train the pronunciation error detection models in the experiments. Section \ref{sec:experiments_pronunciation} presents experiments demonstrating the effectiveness of various synthetic speech generation methods in improving the accuracy of the detection of pronunciation and lexical stress errors. Finally, conclusions and future work are presented in Section \ref{sec:speech_synth_conclusions}.

\subsection{Related work}
\label{sec:related_work}

\subsubsection{Pronunciation error detection}

\bigskip

\textbf{Phoneme recognition approaches} 

\bigskip

Most existing CAPT methods are designed to recognize the phonemes pronounced by the speaker and compare them with the expected (canonical) pronunciation of correctly pronounced speech \cite{witt2000phone,li2016mispronunciation,sudhakara2019improved,leung2019cnn}. Any discrepancy between the recognized and canonical phonemes results in a pronunciation error at the phoneme level. Phoneme recognition approaches generally fall into two categories: methods that align a speech signal with phonemes (forced-alignment techniques) and methods that first recognize the phonemes in the speech signal and then align the recognized and canonical phoneme sequences. Aside these two categories, CAPT methods can be split into multiple other categories:

Forced-alignment techniques \cite{Hongyan2011, li2016mispronunciation, sudhakara2019improved, cheng2020asr} are based on the work of Franco et al. \cite{franco1997automatic} and the Goodness of Pronunciation (GoP) method \cite{witt2000phone}. In the first step, GoP uses Bayesian inference to find the most likely alignment between canonical phonemes and the corresponding audio signal (forced alignment).  In the next step, GoP calculates the ratio between the likelihoods of the canonical and the most likely pronounced phonemes. Finally, it detects mispronunciation if the ratio drops below a certain threshold.  GoP has been further extended with Deep Neural Networks (DNNs), replacing the Hidden Markov Model (HMM) and Gaussian Mixture Model (GMM) techniques for acoustic modeling \cite{li2016mispronunciation,sudhakara2019improved}. Cheng et al. \cite{cheng2020asr} improves GoP performance with the hidden representation of speech extracted in an unsupervised way. This model can detect pronunciation errors based on the input speech signal and the reference canonical speech signal, without using any linguistic information such as text and phonemes. 

The methods that do not use forced-alignment recognize the phonemes pronounced by the speaker purely from the speech signal and only then align them with the canonical phonemes \cite{Minematsu2004PronunciationAB, harrison2009implementation, Lee2013PronunciationAV, plantinga2019towards,Sudhakara2019NoiseRG,zhang2020end}. Leung et al. \cite{leung2019cnn} use a phoneme recognizer that recognizes phonemes only from the speech signal. The phoneme recognizer is based on Convolutional Neural Network (CNN), a Gated Recurrent Unit (GRU), and Connectionist Temporal Classification (CTC) loss. Leung et al. report that it outperforms other forced-alignment \cite{li2016mispronunciation} and forced-alignment-free \cite{harrison2009implementation} techniques in the task of detecting mispronunciations at the phoneme-level in L2 English. 

There are two challenges with presented approaches for pronunciation error detection. First, phonemes pronounced by the speaker must be recognized accurately, which has been proved difficult \cite{zhang2021text, chorowski2014end, chorowski2015attention, bahdanau2016end}. Phoneme recognition is difficult, especially in non-native speech, as different languages have different phoneme spaces. Second, standard approaches assume only one canonical pronunciation of a given text, but this assumption is not always true due to the phonetic variability of speech, e.g., differences between regional accents. For example, the word `enough' can be pronounced by native speakers in multiple ways:  /ih n ah f/ or /ax n ah f/ (short `i' or `schwa' phoneme at the beginning). In our previous work, we solve these problems by creating a native speech pronunciation model that returns the probability of the sentence to be spoken by a native speaker \cite{korzekwa2021mispronunciation}.

Techniques based on phoneme recognition can be supplemented by a reference speech signal obtained  from the speech database \cite{xiao2018paired, nicolao2015automatic, wang2019child} or generated from the phonetic representation \cite{korzekwa2021mispronunciation, qian2010capturing}. Xiao et al. \cite{xiao2018paired} use a pair of speech signals from a student and a native speaker to classify native and non-native speech. Mauro et al. \cite{nicolao2015automatic} use the speech of the reference speaker to detect mispronunciation errors at the phoneme level. Wang et al. \cite{wang2019child} use Siamese networks to model the discrepancy between normal and distorted children's speech. Qian et al. \cite{qian2010capturing} propose a statistical  model of pronunciation in which they build a model that generates hypotheses of mispronounced speech. 

In this work, we use the end-to-end method to detect pronunciation errors directly, without having to recognize phonemes as an intermediate step. The end-to-end approach is discussed in more detail in the next section.

\bigskip

\textbf{End-to-end methods}

\bigskip

The phoneme recognition approaches presented so far rely on phonetically transcribed speech labeled by human listeners. Phonetic transcriptions are needed to train a phoneme recognition model. Human-based transcription is a time-consuming task, especially with L2 speech, where listeners need to recognize mispronunciation errors. Sometimes L2 speech transcription may be even impossible because different languages have different phoneme sets, and it is unclear which phonemes were pronounced by the speaker. In our recent work, we have introduced a novel model (known as WEAKLY-S,  i.e., weakly supervised) for detecting pronunciation errors at the world level that does not require phonetically transcribed L2 speech \cite{korzekwa21b_interspeech}. During training, the model is weakly supervised, in the sense that in L2 speech, only mispronounced words are marked, and the data do not need to be phonetically transcribed. In addition to the primary task of detecting mispronunciation errors at the world level, the second task uses a phoneme recognizer trained on automatically transcribed L1 speech. 
Zhang et al. \cite{zhang2021text} employ a multi-task model with two tasks: phoneme-recognition and pronunciation error detection tasks. Unlike our WEAKLY-S model, they use the Needleman-Wunsch algorithm \cite{needleman1970general} from  bioinformatics to align the canonical and recognized phoneme sequences, but this algorithm cannot be tuned to detect pronunciation errors. The WEAKLY-S model automatically learns the alignment, thus eliminating a potential source of inaccuracy. The alignment is learned through an attention mechanism that automatically maps the speech signal to a sequence of pronunciation errors at the word level. Tong et al. [39] propose to use a multi-task framework in which a neural network model is used to learn the joint space between the acoustic characteristics of adults and children. Additionally, Duan et al. \cite{duan2019cross} propose a multi-task model for acoustical modeling with two tasks for native and non-native speech respectively.

The work of Zhang et al. \cite{zhang2021text} and our recent work \cite{korzekwa21b_interspeech} are end-to-end methods of direct estimation of pronunciation errors, setting up a new trend in the field of automated pronunciation assessment. In this article, we use the end-to-end method as well, but we extend it by the S2S method of generating mispronounced speech. 

\bigskip

\textbf{Other trends}

\bigskip

All the works presented so far treat pronunciation errors as discrete categories, at best producing the probability of mispronunciation. In  contrast, Bi-Cheng et al. \cite{yan20_interspeech} propose a model capable of identifying phoneme distortions, giving the user more detailed feedback on mispronunciation. In our recent work, we provide more fine-grained feedback by indicating the severity level of mispronunciation \cite{korzekwa21b_interspeech}.

Active research is conducted not only on modeling techniques but also on speech representation. Xu et al. \cite{xu21k_interspeech} and Peng et al. \cite{peng21e_interspeech} use the Wav2vec 2.0 speech representation that is created in an unsupervised way. They report that it outperforms existing methods and requires three times less speech training data. Lin et al. \cite{lin21j_interspeech} use transfer learning by taking advantage of deep latent features extracted from the Automated Speech Recognition (ASR) acoustic model and report improvements over the classic GOP-based method.

In this work, we use a mel-spectrogram as a speech representation in the pronunciation error detection model. We also use a mel-spectrogram to represent the speech signal in the T2S and S2S methods of generating mispronounced speech.

\subsubsection{Lexical stress error detection} 

CAPT usually focuses on practicing the pronunciation of phonemes \cite{witt2000phone, leung2019cnn, korzekwa2021mispronunciation}. However, there is evidence that practicing lexical stress improves the intelligibility of non-native English speech \cite{field2005intelligibility,lepage2014intelligibility}. Lexical stress is a phonological feature of a syllable. It is part of the phonological rules that govern how words should be pronounced in a given language. Stressed syllables are usually longer, louder, and expressed with a higher pitch than their unstressed counterparts \cite{jung2018acoustic}. The lexical stress is related to the phonemic representation. For example, placing lexical stress on a different syllable of a word can lead to various phonemic realizations known as `vowel reduction' \cite{bergem1991acoustic}. Students should be able to practice both pronunciation and lexical stress in spoken language. We study both topics to better understand the potential of using speech generation methods in CAPT.

The existing works focus on the supervised classification of lexical stress using Neural Networks \cite{li2018automatic, shahin2016automatic}, Support Vector Machines \cite{chen2010automatic_2, zhao2011automatic}, and Fisher’s linear
discriminant \cite{chen2007using}. There are two popular variants: a) discriminating syllables between primary stress/no stress \cite{ferrer2015classification}, and b) classifying between primary stress/secondary stress/no stress \cite{li2013lexical,li2018automatic}. Ramanathi et al. \cite{ramanathi2019asr} have followed an alternative unsupervised way of classifying lexical stress, which is based on computing the likelihood of an acoustic signal for a number of possible lexical stress representations of a word.  

Accuracy is the most commonly used performance metric, and it indicates the ratio of correctly classified stress patterns on a syllable \cite{li2013lexical} or word level \cite{chen2010automatic_2}. On the contrary, Ferrer et al. \cite{ferrer2015classification}, analyzed the precision and recall metrics to detect lexical stress errors and not just classify them.

Most existing approaches for the classification and detection of lexical stress errors are based on carefully designed features. They start with aligning a speech signal with phonetic transcription, performed via forced-alignment \cite{shahin2016automatic, chen2010automatic_2}. Alternatively, ASR can provide both phonetic transcription and its alignment with a speech signal \cite{li2013lexical}. Then, prosodic features such as duration, energy and pitch \cite{chen2010automatic_2} and cepstral features such as Mel Frequency Cepstral Coefficients (MFCC) and Mel-Spectrogram \cite{ferrer2015classification,shahin2016automatic} are extracted. These features can be extracted on the syllable \cite{shahin2016automatic} or syllable nucleus \cite{ferrer2015classification,chen2010automatic_2} level. Shahin et al. \cite{shahin2016automatic} computes features of neighboring vowels, and Li et al. \cite{li2013lexical} includes the features for two preceding and two following syllables in the model. The features are often preprocessed and normalized to avoid potential confounding variables \cite{ferrer2015classification}, and to achieve better model generalization by normalizing the duration and pitch on a word level \cite{ferrer2015classification,chen2007using}. Li et al. \cite{li2018automatic} adds canonical lexical stress to input features, which improves the accuracy of the model. 

In our recent work, we use attention mechanisms to automatically derive areas of the audio signal that are important for the detection of lexical stress errors  \cite{korzekwa21_interspeech}. In this work, we use the T2S method to generate synthetic lexical stress errors to improve the accuracy of detecting lexical stress errors.

\subsubsection{Synthetic speech generation for pronunciation error detection}

Existing synthetic speech generation techniques for detecting pronunciation errors can be divided into two categories: data augmentation and data generation.  

Data augmentation techniques are designed to generate new training examples for existing mispronunciation labels. Badenhorst et al. \cite{badenhorst2017limitations} simulate new speakers by adjusting the speed of raw audio signals. Eklund \cite {eklund2019data} generates additional training data by adding background noise and convolving the audio signal with the impulse responses of the microphone of a mobile device and a room.

Data generation techniques are designed to generate new training data with new labels of both correctly pronounced and mispronounced speech. Most existing works are based on the P2P technique to generate mispronounced speech by perturbing the phoneme sequence of the corresponding audio using a variety of strategies \cite{lee2016language, komatsu2019speech, fu2021full, yan2021maximum,korzekwa2021mispronunciation}. In addition to P2P techniques, in our recent work, we use T2S to generate synthetic lexical stress errors \cite{korzekwa21b_interspeech}. Qian et al. \cite{qian2010capturing} introduce a generative model to create hypotheses of mispronounced speech and use it as a reference speech signal to detect pronunciation errors. Recently, we proposed a similar technique to create a pronunciation model of native speech to account for many ways of correctly pronouncing a sentence by a native speaker \cite{korzekwa2021mispronunciation}.

Synthetic speech generation techniques have recently gained attention in other related fields. Fazel et al. \cite{fazel21_interspeech} use synthetic speech generated with T2S to improve accuracy in ASR. Huang et al. \cite{huang2016machine} use a machine translation technique to generate text to train an ASR language model in a low-resource language. At the same time, Shah et al. \cite{shah21_ssw} and Huybrechts et al. \cite{huybrechts2021low} employ S2S voice conversion to improve the quality of speech synthesis in the data reduction scenario.

All the presented works on the detection of pronunciation errors treat synthetic speech generation as a secondary contribution. In this article, we present a unified perspective of synthetic speech generation methods for detecting pronunciation errors. This article extends our previous work \cite{korzekwa2021mispronunciation,korzekwa21_interspeech,korzekwa21b_interspeech} and introduces a new S2S method to detect pronunciation errors. To the best of our knowledge, there are no papers devoted to generating pronunciation errors with the S2S technique and using it in the detection of pronunciation errors.

\subsection{Methods of generating pronunciation errors}
\label{sec:synth_speech_generation_methods}

To detect pronunciation errors, first, the spoken language must be separated from other factors in the signal and then incorrectly pronounced speech sounds have to be identified. Separating speech into multiple factors is difficult, as speech is a complex signal. It consists of prosody (F0, duration, energy), timbre of the voice, and the representation of the spoken language. Spoken language is defined by the sounds (phones) perceived by people. Phones are the realizations of phonemes - a human abstract representation of how to pronounce a word/sentence. Speech may also present variability due to the recording channel and environmental effects such as noise and reverberation. Detecting pronunciation errors is very challenging, also because of the limited amount of recordings with mispronounced speech. To address these challenges, we reformulate the problem of pronunciation error detection as the task of synthetic speech generation.

Let $\mathbf{s}$ be the speech signal, $\mathbf{r}$ be the sequence of phonemes that the user is trying to pronounce (canonical pronunciation), and $\mathbf{e}$ be the sequence of probabilities of mispronunciation at the phoneme or word level. The original task of detecting pronunciation errors is defined by:

\begin{equation}
 \mathbf{e} \sim p(\mathbf{e}|\mathbf{s},\mathbf{r})
\end{equation}

where the formulation of the problem as the task of synthetic speech generation is defined as follows:

\begin{equation}
 \mathbf{s} \sim p(\mathbf{s}|\mathbf{e},\mathbf{r})
\end{equation} 

The probability of pronunciation errors for all the words in a sentence can then be calculated using the Bayes rule \cite{bishop2006pattern}:

\begin{equation}
p(\mathbf{e}|\mathbf{s},\mathbf{r}) = \frac{p(\mathbf{e}|\mathbf{r})p(\mathbf{s}|\mathbf{e},\mathbf{r})}{p(\mathbf{s}|\mathbf{r})}
\label{eqn:pron_error_detection_bayes_rule}
\end{equation} 

From Eq. \ref{eqn:pron_error_detection_bayes_rule}, one can see that there is no need to directly learn the probability of pronunciation errors $p(\mathbf{e}|\mathbf{s},\mathbf{r})$, since the complexity of the problem has now been transferred to learning the speech generation process $p(\mathbf{s}|\mathbf{e}, \mathbf{r})$. Such a formulation of the problem opens the way to the inclusion of additional prior knowledge into the model:

\begin{enumerate}
\item Replacing the phoneme in a word while preserving the original speech signal results in a pronunciation error (P2P method).
\item Changing the speech signal while retaining the original pronunciation results in a pronunciation error (T2S method).
\item There are many variations of mispronounced speech that differ in terms of the voice timbre and the prosodic aspects of speech (S2S method).
\end{enumerate}

To solve Eq. \ref{eqn:pron_error_detection_bayes_rule}, we use Markov Chain Monte Carlo Sampling (MCMC) \cite{koller2009probabilistic}. In this way, the prior knowledge can be incorporated by generating $N$ training examples $\{\mathbf{e_i},\mathbf{s_i}, \mathbf{r_i}\}$ for $i=1..N$ with the use of P2P (prior knowledge 1), T2S (prior knowledge 2), and S2S (prior knowledge 3) methods. Accounting for the prior knowledge, intuitively corresponds to an increase in the amount of training data, which contributes to outperforming state-of-the-art models for detecting pronunciation errors, as presented in Section \ref{sec:experiments_pronunciation}. Eq. \ref{eqn:pron_error_detection_bayes_rule} can then be optimized with standard gradient-based optimization techniques. In the following subsections, we present the P2P conversion, T2S, and S2S methods of generating correctly and incorrectly pronounced speech in details. 

\subsubsection{P2P method}
\label{sec:p2p_method}

\begin{figure}[th]
\centering
\includegraphics[width=1\textwidth]{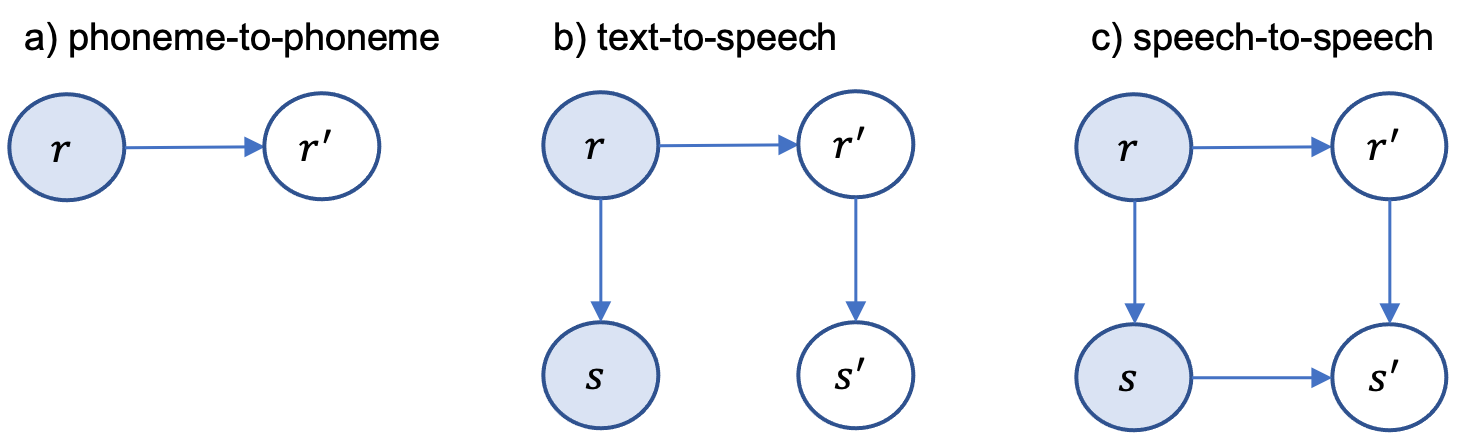}
\decoRule
\caption[Probabilistic graphical models for three methods to generate pronunciation errors: P2P, T2S and S2S. Empty circles represent hidden (latent) variables, while filled (blue) circles represent observed variables.  $\mathbf{s}$ - the speech signal, $\mathbf{r}$ - the sequence of phonemes that the user is trying to pronounce (canonical pronunciation), the superscript $\mathbf{'}$ represents a variable with generated mispronunciations]{Probabilistic graphical models for three methods to generate pronunciation errors: P2P, T2S and S2S. Empty circles represent hidden (latent) variables, while filled (blue) circles represent observed variables.  $\mathbf{s}$ - the speech signal, $\mathbf{r}$ - the sequence of phonemes that the user is trying to pronounce (canonical pronunciation), the superscript $\mathbf{'}$ represents a variable with generated mispronunciations.}
\label{fig:pgms}
\end{figure}

To generate synthetic mispronounced speech, it is enough to start with correctly pronounced speech and modify the corresponding sequence of phonemes. This simple idea does not even require generating the speech signal itself. It can be observed that the probability of mispronunciations depends on the discrepancy between the speech signal and the corresponding canonical pronunciation. This leads to the P2P conversion model shown in Figure \ref{fig:pgms}a.

Let $\{\mathbf{e_{noerr}}, \mathbf{s},\mathbf{r}\}$ be a single training example containing: the sequence of $0s$ denoting correctly pronounced phonemes, the speech signal, and the sequence of phonemes representing the canonical pronunciation.  Let $\mathbf{r^{'}}$ be the sequence of phonemes with injected mispronunciations such as phoneme replacements, insertions, and deletions:
\begin{equation}
\mathbf{r^{'}} \sim p(\mathbf{r^{'}}|\mathbf{r})
\end{equation}
then the probability of mispronunciation for the $j^{th}$ phoneme is defined by:

\begin{equation}
e_j^{'} = 
\left\{ 
  \begin{array}{rl}
     1 &\mbox{ if $r_j^{'} != r_j $} \\
     0 &\mbox{ otherwise}
   \end{array} 
 \right.
 \label{eqn:ptp_pron_error}
\end{equation}

The probabilities of mispronunciation can be projected from the level of phonemes to the level of words. A word is treated as mispronounced if at least one pair of phonemes in the word $\{r_j^{'},r_j\}$ does not match. At the end of this process, a new training example is created with artificially introduced pronunciation errors: $\{\mathbf{e_{err}}, \mathbf{s},\mathbf{r^{'}}\}$. Note that the speech signal $\mathbf{s}$ in the new training example is unchanged from the original training example and only phoneme transcription is manipulated.

\bigskip

\textbf{Implementation}

\bigskip

To generate synthetic pronunciation errors, we use a simple approach of perturbing phonetic transcription for the corresponding speech audio.  First, we sample these utterances with replacement from the input corpora of human speech. Then, for each utterance, we replace the phonemes with random phonemes with a given probability.

\subsubsection{T2S method}
\label{sec:t2s_method}

The T2S method expands on P2P by making it possible to create speech signals that match the synthetic mispronunciations. The T2S method for generating mispronounced speech is a generalization of the P2P method, as can be seen by the comparison of the two methods shown in Figures \ref{fig:pgms}a and \ref{fig:pgms}b.

One problem with the P2P method is that it cannot generate a speech signal for the newly created sequence of phonemes $\mathbf{r^{'}}$. As a result, pronunciation errors will dominate in the training data containing new sequences of phonemes $\mathbf{r^{'}}$. Therefore, it will be possible to detect pronunciation errors only from the canonical representation $\mathbf{r^{'}}$, ignoring information contained in the speech signal. To mitigate this issue, there should be two training examples for the phonemes $\mathbf{r^{'}}$, one representing mispronounced speech: $\{\mathbf{e_{err}}, \mathbf{s},\mathbf{r^{'}}\}$, and the second one for correct pronunciation: $\{\mathbf{e_{noerr}}, \mathbf{s^{'}},\mathbf{r^{'}}\}$, where: 
\begin{equation}
 \mathbf{s^{'}} \sim p(\mathbf{s^{'}}|\mathbf{e_{noerr}},\mathbf{r^{'}})
\end{equation}

Because we now have the speech signal $\mathbf{s^{'}}$, another training example can be created as: $\{\mathbf{e_{err}}, \mathbf{s^{'}},\mathbf{r}\}$. In summary, T2S method extends a single training example of correctly pronounced speech to four combinations of correctly and incorrect pronunciations: 
\begin{itemize}
\item $\{\mathbf{e_{noerr}}, \mathbf{s},\mathbf{r}\}$ -- correctly pronounced input speech

\item $\{\mathbf{e_{err}}, \mathbf{s},\mathbf{r^{'}}\}$ -- mispronounced speech generated by the P2P method

\item $\{\mathbf{e_{noerr}}, \mathbf{s^{'}},\mathbf{r^{'}}\}$ -- correctly pronounced speech generated by the T2S method

\item $\{\mathbf{e_{err}}, \mathbf{s^{'}},\mathbf{r}\}$ --  mispronounced speech generated by the T2S method
\end{itemize}

\bigskip

\textbf{Implementation}

\bigskip

The synthetic speech is generated with the Neural TTS described by Latorre et al. \cite{latorre2019effect}. The Neural TTS consists of two modules. The context-generation module is an attention-based encoder-decoder neural network that generates a mel-spectrogram from a sequence of phonemes. The Neural Vocoder then converts it into a speech signal. The Neural Vocoder is a neural network of architecture similar to Parallel Wavenet \cite{oord2018parallel}. The Neural TTS is trained using the speech of a single native speaker. To generate words with different lexical stress patterns, we modify the lexical stress markers associated with the vowels in the phonetic transcription of the word. For example, with the input of /r iy1 m ay0 n d/ we can place lexical stress on the first syllable of the word `remind'.

\subsubsection{S2S method}
\label{sec:s2s_method}

The S2S method is designed to simulate the diverse nature of speech, as there are many ways to correctly pronounce a sentence. The prosodic aspects of speech, such as pitch, duration, and energy, can vary. Similarly, phonemes can be pronounced differently. To mimic human speech, speech generation techniques should allow a similar level of variability. The T2S method outlined in the previous section always produces the same output for the same phoneme input sequence. The S2S method is designed to overcome this limitation.

S2S converts the input speech signal $\mathbf{s}$ in a way to change the pronounced phonemes (phoneme replacements, insertions, and deletions) from the input phonemes $\mathbf{r}$ to target phonemes $\mathbf{r^{'}}$ while preserving other aspects of speech, including voice timbre and prosody (Eq. \ref{eqn:sts_equation_for_s} and Figure \ref{fig:pgms}c). In this way, the natural variability of human speech is preserved, resulting in generating many variations of incorrectly pronounced speech. The prosody will differ in various versions of the sentence of the same speaker, while the same sentence spoken by many speakers will differ in the voice timbre. 
\begin{equation}
 \mathbf{s^{'}} \sim p(\mathbf{s^{'}}|\mathbf{e_{noerr}},\mathbf{r^{'}}, \mathbf{s})
 \label{eqn:sts_equation_for_s}
\end{equation}
Similarly to the T2S method, the S2S method outputs four types of speech pronounced correctly and incorrectly: $\{\mathbf{e_{noerr}}, \mathbf{s},\mathbf{r}\}$, $\{\mathbf{e_{err}}, \mathbf{s},\mathbf{r^{'}}\}$, $\{\mathbf{e_{noerr}}, \mathbf{s^{'}},\mathbf{r^{'}}\}$, and $\{\mathbf{e_{err}}, \mathbf{s^{'}},\mathbf{r}\}$.

\bigskip

\textbf{Implementation}

\bigskip

\begin{figure}[th]
\centering
\centerline{
\includegraphics[width=1.20\textwidth]{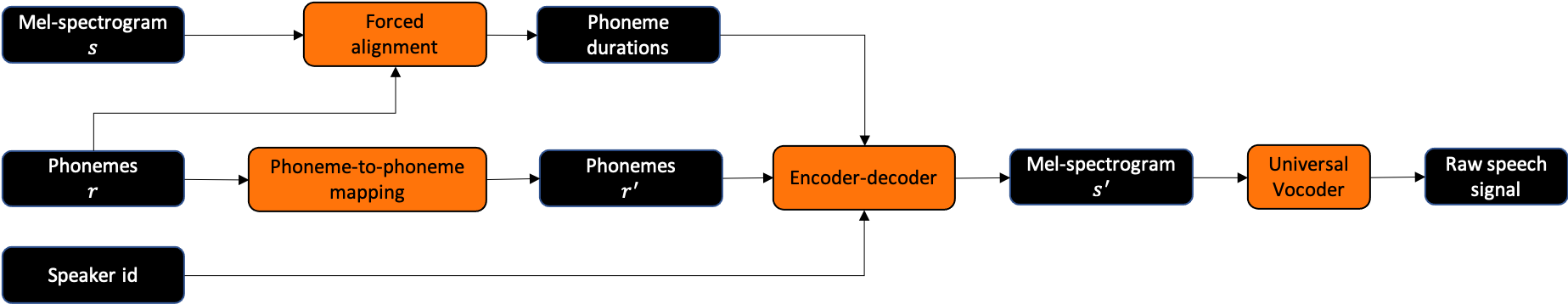}}
\decoRule
\caption[Architecture of the S2S model to generate mispronounced synthetic speech while maintaining prosody and voice timbre of the input speech. The black rectangles represent the data (tensors) and the orange boxes represent processing blocks. This color notation is used in all  machine learning model diagrams throughout the article]{Architecture of the S2S model to generate mispronounced synthetic speech while maintaining prosody and voice timbre of the input speech. The black rectangles represent the data (tensors) and the orange boxes represent processing blocks. This color notation is used in all  machine learning model diagrams throughout the article.}
\label{fig:speech_to_speech_architecture}
\end{figure}

Synthetic speech is generated by introducing mispronunciations into the input speech, while preserving the duration of the phonemes and timbre of the voice. The architecture of the S2S model is shown in Figure \ref{fig:speech_to_speech_architecture}. The mel-spectrogram of the input speech signal $\mathbf{s}$ is forced-aligned with the corresponding canonical phonemes $\mathbf{r}$ to get the duration of the phonemes. The speaker id has to be provided together with the input speech to enable the source speaker's voice to be maintained. Mispronunciations are introduced into the canonical phonemes $\mathbf{r}$ according to the P2P method described in Section \ref{sec:p2p_method}. Mispronounced phonemes $\mathbf{r^{'}}$ along with phonemes duration and speaker id are processed by the encoder-decoder, which generates the mel-spectrogram $\mathbf{s^{'}}$. The encoder-decoder transforms the phoneme-level representation into frame-level features and then generates all mel-spectrogram frames in parallel. The mel-spectrogram is converted to an audio signal with Universal Vocoder \cite{jiao2021universal}. Without the Universal Vocoder, it would not be possible to generate the raw audio signal for hundreds of speakers included in the LibriTTS corpus. Details of the S2S method are shown in the works of Shah et al. \cite{shah21_ssw} and Jiao et al. \cite{jiao2021universal}. The main difference between these two models and our S2S model is the use of the P2P mapping to introduce pronunciation errors.

\subsubsection{Summary of mispronounced speech generation}

Generation of synthetic mispronounced speech and detection of pronunciation errors were presented from the probabilistic perspective of the Bayes-rule. With this formulation, we can better understand the relationship between P2P, T2S and S2S methods, and see that the S2S method generalizes two simpler methods. Following this reasoning, we can argue that using the Bayes rule gives us a nice mathematical framework to potentially further generalize the S2S method, e.g. by adding a language variable to the model to support multilingual pronunciation error detection. There is another advantage of modeling pronunciation error detection from the probabilistic perspective - it paves the way for joint training of mispronounced speech generation and pronunciation error detection models. In the present work, we are training separate machine learning models for both tasks, but it should be possible to train both models jointly using the framework of Variational Inference \cite{jordan1999introduction} instead of MCMC to infer the probability of mispronunciation in Eq. \ref{eqn:pron_error_detection_bayes_rule}.

\subsection{Speech corpora}
\label{sec:human_speech_corpora}

\subsubsection{Corpora of continuous speech}

Speech corpora of recorded sentences is a combination of L1 and L2 English speech. L1 speech is obtained from the TIMIT \cite{garofolo1993darpa} and the LibriTTS \cite{Zen2019} corpora. L2 speech comes from the Isle \cite{atwell2003isle} corpus (German and Italian speakers) and the GUT Isle \cite{Weber2020} corpus (Polish speakers). In total, we used 125.28 hours of L1 and L2 English speech from 983 speakers segmented into 102812 sentences. A summary of the speech corpora is presented in Table \ref{tab:weaklys_speech_corpora}, whereas the details are presented in our recent work \cite{korzekwa21b_interspeech}.

The speech data are used in all the pronunciation error detection experiments presented in Section \ref{sec:experiments_pronunciation}. From the collected speech, we held out 28 L2 speakers and used them only to assess the performance of the systems in the mispronunciation detection task. It includes 11 Italian and 11 German speakers from the Isle corpus \cite{atwell2003isle}, and 6 Polish speakers from the GUT Isle corpus \cite{Weber2020}. The human speech training data is extended with synthetic pronunciation errors generated by the methods presented in Section \ref{sec:synth_speech_generation_methods}.

\begin{table}[htb]
\caption{Summary of human speech corpora used in the pronunciation error detection experiments. * - audiobooks read by volunteers from all over the world \cite{Zen2019} }
\label{tab:weaklys_speech_corpora}
\centering
 \begin{tabular}{lll}
    \toprule  
   \tabhead{Native Language} &  \tabhead{Hours} &  \tabhead{Speakers} \\
    \midrule
    English & 90.47 & 640\\ 
    Unknown* & 19.91 & 285\\  
    German and Italian & 13.41 & 46\\  
    Polish & 1.49 & 12\\ 
\bottomrule\\
\end{tabular}
\end{table}

\subsubsection{Corpora of isolated words}
\label{sec:speech_corpora_words}

The speech corpora consist of human and synthetic speech. The data were divided into training and testing sets with separate speakers assigned to each set. Human speech includes native (L1) and non-native (L2) English speech. L1 speech corpora are made of TIMIT \cite{garofolo1993darpa} and Arctic \cite{kominek2004cmu}. L2 corpora contain speech from L2-Arctic [32], Porzuczek \cite{porzuczek2017english}, and our own recordings of 25 speakers (23 Polish, 1 Ukrainian and 1 Lithuanian). The synthetic data were generated using the T2S method and are only included in the training set. The data are summarized in Table \ref{tab:lexical_stress_speech_corpora_traintest}. For a more detailed description of speech corpora, see Section 4 of our recent work \cite{korzekwa21_interspeech}. The speech corpora of isolated words are used in the lexical stress error detection experiment presented in Section \ref{sec:experiments_lexical_stress}.

\begin{table}
 \caption{Details of the training and test sets for the lexical stress error detection model.}
   \label{tab:lexical_stress_speech_corpora_traintest}
   \centering
\begin{tabular}{llll}
    \toprule  
     \tabhead{Data set}  &  \tabhead{Speakers  (L2)} &  \tabhead{Words (unique)} &   \tabhead{Stress Errors} \\
    \midrule
    Train set (human) & 473 (10) & 8223 (1528) & 425\\
    Train set (TTS) & 1 (0) & 3937 (1983) & 2005\\
    Test set (human) & 176 (21) & 2108 (378) & 189\\
    \bottomrule\\
  \end{tabular}
\end{table}

\subsection{Experiments}
\label{sec:experiments_pronunciation}

\subsubsection{Generation of mispronounced speech}
\label{sec:p2p_gen_of_incorrectly_pron_speech}

\bigskip

\textbf{Experimental setup}

\bigskip

The effect of using synthetic pronunciation errors based on the P2P, T2S and S2S methods is evaluated in the task of detecting pronunciation errors in spoken sentences at the word level. First, we analyze the P2P method by comparing it with the state-of-the-art techniques and measure the effect of adding synthetic pronunciation errors to the training data. We then compare P2P with T2S and S2S to assess the benefits of using more complex methods of generating pronunciation errors. The accuracy of detecting pronunciation errors is reported in standard Area Under the Curve (AUC), precision and recall metrics.

\bigskip

\textbf{Overview of our WEAKLY-S model}

\bigskip

We use the pronunciation error detection model (WEAKLY-S) recently proposed by us \cite{korzekwa21b_interspeech}. To train the model, the human speech training set is extended with 292,242 utterances of L1 speech with synthetically generated pronunciation errors. To generate pronunciation errors, the P2P, T2S, and S2S methods described in Section \ref{sec:synth_speech_generation_methods} are used.

The WEAKLY-S model produces probabilities of mispronunciation for all words, conditioned by the spoken sentence and canonical phonemes. Mispronunciation errors include phoneme replacement, addition, deletion, or an unknown speech sound. During training, the model is weakly supervised, in the sense that only mispronounced words in L2 speech are marked by listeners and the data do not have to be phonetically transcribed. Due to the limited availability of L2 speech and the fact that it is not phonetically transcribed, the model is more likely to overfit. To solve this problem, the model is trained in a multi-task setup. In addition to the primary task of detecting mispronunciation error at the word level, the second task uses a phoneme recognizer which is trained on automatically transcribed L1 speech. Both tasks share components of the model, which makes the primary task less likely to overfit.

\begin{figure}[th]
\centering
\centerline{
\includegraphics[width=1.2\textwidth]{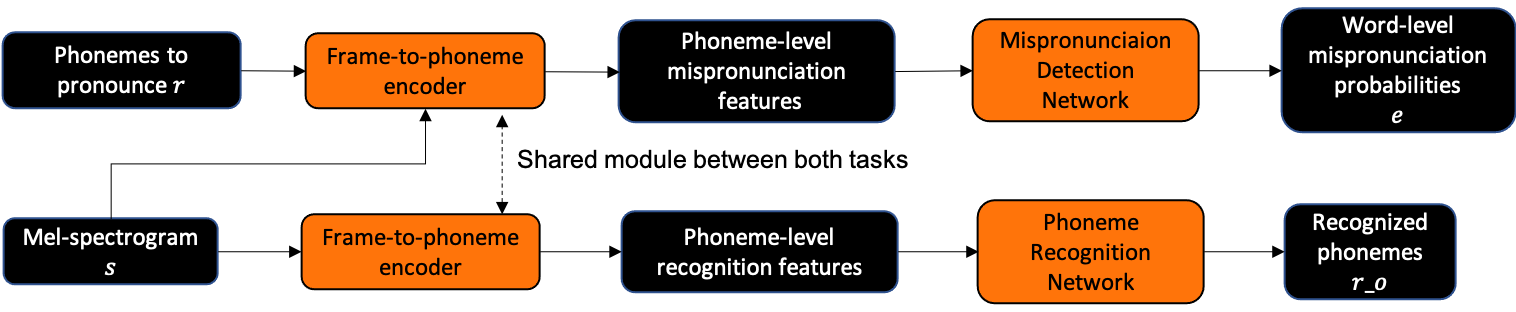}}
\decoRule
\caption[Architecture of the WEAKLY-S model for word-level pronunciation error detection trained in the multi-task setup. Task 1 - to detect pronunciation errors $e$. Task 2 - to recognize phonemes $r_o$]{Architecture of the WEAKLY-S model for word-level pronunciation error detection trained in the multi-task setup. Task 1 - to detect pronunciation errors $e$. Task 2 - to recognize phonemes $r_o$.}
\label{fig:weakly_architecture}
\end{figure}

The architecture of the pronunciation error detection model is shown in Figure \ref{fig:weakly_architecture}. The model consists of two sub-networks. The Mispronunciations Detection Network (MDN) detects word-level pronunciation errors $\mathbf{e}$ from the audio signal $\mathbf{s}$ and canonical phonemes $\mathbf{r}$, while the Phoneme Recognition Network (PRN) recognizes phonemes $\mathbf{r_o}$ pronounced by a speaker from the audio signal $\mathbf{s}$. The detailed model architecture is presented in Section 2 of our recent work \cite{korzekwa21b_interspeech}.

\bigskip

\textbf{Results - P2P method}

\bigskip

We conducted an ablation study to measure the effect of removing synthetic pronunciation errors from the training data. We trained four variants of the WEAKLY-S model to measure the effect of using synthetic data against other elements of the model. WEAKLY-S is a complete model that also includes synthetic data during training. In the NO-SYNTH-ERR model, we exclude synthetic samples of mispronounced L1 speech, significantly reducing the number of mispronounced words seen during training from 1,129,839 to just 5,273 L2 words. The NO-L2-ADAPT variant does not fine-tune the model on L2 speech, although it is still exposed to L2 speech while being trained on a combined corpus of L1 and L2 speech. The NO-L1L2-TRAIN model is not trained on L1/L2 speech, and fine-tuning on L2 speech starts from scratch. This means that this model will not use a large amount of phonetically transcribed L1 speech data and ultimately no secondary phoneme recognition task will be used. 

\begin{figure}[th]
\centering
\includegraphics[width=0.8\textwidth]{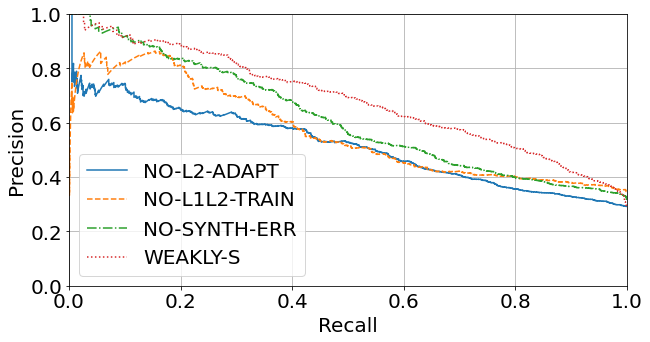}
\decoRule
\caption[Precision-recall curve for the ablation study on the GUT Isle corpus, illustrating the effect of using synthetic pronunciation errors generated by the P2P method]{Precision-recall curve for the ablation study on the GUT Isle corpus, illustrating the effect of using synthetic pronunciation errors generated by the P2P method.}
\label{fig:weaklys_ablation_precision_recall_plots}
\end{figure}

L2 fine-tuning (NO-L2-ADAPT) is the most important factor influencing the performance of the model (Fig. \ref{fig:weaklys_ablation_precision_recall_plots} and Table \ref{tab:weaklys_ablation_study}), with an AUC of 0.517 compared to 0.686 for the full model. Training the model on both L2 and L1 human speech together is not enough. This is because L2 speech accounts for less than 1\% of the training data and the model naturally leans towards L1 speech. The second most important feature is training the model on a combined set of L1 and L2 speech (NO-L1L2-TRAIN), with an AUC of 0.565. L1 speech accounts for over 99\% of training data. These data are also phonetically transcribed, and therefore can be used for the phoneme recognition task. The phoneme recognition task acts as a 'backbone' and reduces the effect of overfitting in the main task of detecting errors in the pronunciation of words. Finally, excluding synthetically generated pronunciation errors (NO-SYNTH-ERR) reduces an AUC from 0.686 to 0.615. Although, the synthetic data provides the least improvement to the model, it still increases the accuracy of the model by 11.5\% in AUC, contributing to setting up a new state-of-the-art. 

\begin{table}
  \caption{Ablation study for the GUT Isle corpus to show the effect of using synthetic data and other elements of the WEAKLY-S model. Pr. - Precision, Re. - Recall}
  \label{tab:weaklys_ablation_study}
  \centering
   \begin{tabular}{lllll}
    \toprule
     \tabhead{Model} & \tabhead{Description} & \tabhead{AUC} & \tabhead{Pr. [\%]} & \tabhead{Re.[\%]} \\
    \midrule
     NO-L2-ADAPT & No fine-tuning on L2 speech & 0.517 & 57.89 & 40.11 \\ 
     NO-L1L2-TRAIN & No pretraining on L1\&L2 speech & 0.565 & 59.73 & 40.20 \\  
     NO-SYNTH-ERR & \makecell[l]{ No synthetically \\generated pronunciation \\ errors in the training data}& 0.615 & 67.22 & 40.38 \\  
     WEAKLY-S & Complete model & \textbf{0.686} & 75.25  & 40.38 \\    
    \bottomrule\\
  \end{tabular}
\end{table}

We compare the WEAKLY-S model with two state-of-the-art baselines. The Phoneme Recognizer (PR) model by Leung et al. \cite{leung2019cnn} is our first baseline. The PR is based on the CTC loss \cite{graves2012connectionist} and outperforms multiple alternative approaches of pronunciation assessment. The original CTC-based model uses a hard likelihood threshold applied to the recognized phonemes. To compare it with two other models, following our recent work \cite{korzekwa2021mispronunciation}, we have replaced the hard likelihood threshold with a soft threshold. The second baseline is  PR extended by the pronunciation model (PR-PM model \cite{korzekwa2021mispronunciation}). The pronunciation model takes into account the phonetic variability of the speech spoken by native speakers, which results in greater precision in detecting pronunciation errors. The results are shown in Table \ref{tab:weaklys_accuracy_metrics}. It turns out that the WEAKLY-S model outperforms the second-best model in terms of an AUC by 30\% from 0.528 to 0.686 and precision by 23\% from 0.612 to 0.752 on the GUT Isle Corpus of Polish speakers. We are seeing similar improvements on the Isle Corpus of German and Italian speakers. The use of synthetic data is an important contribution to the performance of the WEAKLY-S model.

\begin{table}
  \caption{Accuracy metrics of detecting word-level pronunciation errors. WEAKLY-S vs. baseline models.}
     \label{tab:weaklys_accuracy_metrics}
  \centering
   \begin{tabular}{llll}
    \toprule
      \tabhead{Model} &  \tabhead{AUC} &  \tabhead{Precision [\%,95\%CI]} &  \tabhead{Recall [\%,95\%CI]} \\
    \midrule
    \multicolumn{4}{c}{\textbf{Isle corpus (German and Italian)}} \\
    PR & 0.555 & 49.39 (47.59-51.19) & 40.20 (38.62-41.81)\\ 
    PR-PM & 0.480 & 54.20 (52.32-56.08) & 40.20 (38.62-41.81)\\  
    WEAKLY-S & \textbf{0.678} & 71.94 (69.96, 73.87) & 40.14 (38.56, 41.75) \\  
	
	 \multicolumn{4}{c}{\textbf{GUT Isle corpus (Polish)}} \\
     PR & 0.528 & 54.91 (50.53-59.24) & 40.29 (36.66-44.02)\\ 
     PR-PM & 0.505 & 61.21 (56.63-65.65) & 40.15 (36.51-43.87)\\  
     WEAKLY-S & \textbf{0.686} & 75.25 (71.67-78.59) & 40.38 (37.52-43.29)\\    
    \bottomrule\\
  \end{tabular}
\end{table}

\bigskip

\textbf{Results - T2S and S2S methods}

\bigskip

The main limitation of the P2P method is that it does not generate a new speech signal. The method introduces mispronunciations by operating only on the sequence of phonemes for the corresponding speech. In this experiment, we demonstrate the T2S and S2S methods that can directly generate a speech signal to overcome this limitation. The S2S method introduces mispronunciations into the input native speech while preserving the prosody (phoneme durations) and timbre of the voice. Preserving speech attributes other than pronunciation increases speech variability during training and makes the pronunciation error detection model more reliable during testing. The T2S method can be considered as a simplified variant of the S2S method, in which there is only text as input.

The T2S and S2S methods are compared with the P2P method. Three WEAKLY-S models are trained, differing in the technique of generating mispronounced speech contained in the training data. The S2S method outperforms the P2P method by increasing an AUC score by 9\% from 0.686 to 0.749 in the Gut Isle corpus of Polish speakers (Table \ref{tab:speech_to_speech_auc_results}). Additionally, an AUC increases from 0.815 to 0.834 for major pronunciation errors (Table \ref{tab:speech_to_speech_auc_results_high_sev_real_speech}), according to a similar experiment presented in Section 3.4 of \cite{korzekwa21b_interspeech}. Interestingly, the T2S method is only slightly better than the P2P method, which suggests that the variability of the generated mispronounced speech provided by the S2S method is really important. The presented experiments show the potential of the S2S method in improving the accuracy of detecting pronunciation errors. The S2S method is able to control voice timbre, phoneme duration, and pronunciation, opening the door to transplanting all three properties from non-native speech and potentially further improving the accuracy of the model. 

One downside of the S2S method is its complexity. Compared to the straightforward P2P method, the 9\% improvement in an AUC is associated with high costs. The method involves training a complex multi-speaker S2S model to convert between input and output mel-spectrograms and requires training a Universal Vocoder model to convert a mel-spectrogram into a raw speech signal.

To better understand what prevents the model from achieving higher accuracy, we measure the performance of the model on synthetic pronunciation errors. We divide all synthetic pronunciation errors into four categories to reflect the severity of pronunciation errors. The `low' category includes mispronounced words with only one mismatched phoneme between the canonical and pronounced phonemes of the word. The `medium' category  includes two mispronounced phonemes. The `high' category gets three, and the 'very high' category includes four mispronounced errors. The AUC across different severity levels varies from 0.928 (low severity) to 1.00 (very high severity) as shown in Table \ref{tab:speech_to_speech_auc_synth_speech}. These AUC values are significantly higher than the results for non-native human  speech, suggesting that making synthetic speech errors more similar to non-native speech may improve the accuracy of detecting pronunciation errors.

\begin{table}
  \caption{Comparison of the P2P, T2S and S2S methods in the task of pronunciation error detection assessed on the GUT Isle corpus.}
    \label{tab:speech_to_speech_auc_results}
  \centering
   \begin{tabular}{llll}
    \toprule
     \tabhead{Model} & \tabhead{AUC} & \tabhead{Precision [\%]} & \tabhead{Recall [\%]} \\
    \midrule
     P2P & 0.686 &  75.25 (71.67-78.59)  & 40.38 (37.52-43.29) \\  
     T2S & 0.695 & 76.15 (72.59-79.36) & 40.25 (37.44-43.22) \\ 
     S2S & 0.749 & 80.45 (76.94-83.47)  & 40.12 (37.12-43.02) \\  
    \bottomrule\\
  \end{tabular}
\end{table}

\begin{table}
  \caption{Comparison of the P2P, T2S and S2S methods in the task of pronunciation error detection assessed on the GUT Isle corpus only for major pronunciation errors.}
  \label{tab:speech_to_speech_auc_results_high_sev_real_speech}
  \centering
   \begin{tabular}{llll}
    \toprule
     \tabhead{Model} & \tabhead{AUC} & \tabhead{Precision [\%]} & \tabhead{Recall [\%]} \\
    \midrule
     P2P & 0.815 &  91.67 (88.55-94.45) & 40.31 (37.43-43.23) \\
     T2S & 0.819 & 92.11 (89.09-94.83) & 40.21 (36.81-43.31)\\  
     S2S & 0.834 & 93.54 (90.53-96.23)  & 40.15 (37.26-43.11)\\  
    \bottomrule\\
  \end{tabular}
\end{table}

\begin{table}
  \caption{Accuracy (AUC) in  detecting  pronunciation errors assessed in synthetic speech at different severity levels of mispronunciation for the best S2S method.}
    \label{tab:speech_to_speech_auc_synth_speech}
  \centering
   \begin{tabular}{ll}
    \toprule
     \tabhead{Severity} & \tabhead{AUC} \\
    \midrule
    Low (phoneme distance=1) & 0.928\\   
    Medium (phoneme distance=2) & 0.974 \\  
    High (phoneme distance=3) & 0.993 \\  
    Very High (phoneme distance=4) & 1.00 \\ 
    \bottomrule\\
  \end{tabular}
\end{table}

\subsubsection{Model of native speech pronunciation}

\bigskip

\textbf{Experimental setup}

\bigskip

The P2P, T2S, and S2S are generative models that provide the probability of generating a particular output sequence. This probability can be used directly to detect pronunciation errors without generating the mispronounced speech and adding it to the training data. In this experiment, we show how to apply this approach in practice.

One of the challenges in detecting pronunciation errors is that a native speaker can pronounce a sentence correctly in many ways. The classic approach for detecting pronunciation errors is based on identifying the difference between pronounced and canonical phonemes. All pronunciations that do not correspond precisely to the canonical pronunciation will result in false pronunciation errors. One way to solve this problem is to use the P2P technique to create a native speech Pronunciation Model (PM) that determines the probability that a sentence is pronounced by a native speaker. A low likelihood value indicates a high probability of mispronunciation. 

To evaluate the performance of the PM model, the pronunciation error detection model has been designed such that the PM model can be turned on and off. To disable the PM, we are modifying  it so that it only takes into account  one way of correctly pronouncing a sentence. In an ablation study, we measure  whether the PM model improves the accuracy in detecting pronunciation errors at the word level. Note that in this experiment, synthetically generated pronunciation errors are not used explicitly. Instead, the native speech pronunciation model is used to implicitly represent the generative speech process. 

\bigskip

\textbf{Overview of the pronunciation error detection model}

\bigskip

\begin{figure}[th]
\centering
\centerline{
\includegraphics[width=1.25\textwidth]{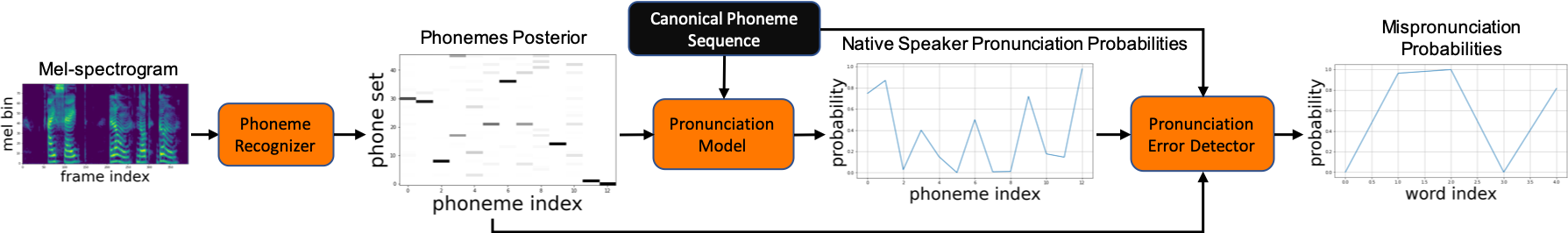}}
\decoRule
\caption[Architecture of the system for detecting mispronounced words in a spoken sentence based on the native speech pronunciation model]{Architecture of the system for detecting mispronounced words in a spoken sentence based on the native speech pronunciation model.}
\label{fig:uncertainty_modeling_architecture}
\end{figure}

The design of the pronunciation error detection model consists of three subsystems: a Phoneme Recognizer (PR), a Pronunciation Model (PM), and a Pronunciation Error Detector (PED), shown in Figure \ref{fig:uncertainty_modeling_architecture}. First, the PR model estimates a belief over the phonemes produced by the student, intuitively representing the uncertainty in the student's pronunciation. The PM model transforms this belief into a probability that a native speaker would pronounce the sentence this way, given the phonetic variability. Finally, the PED model decides which words were mispronounced in the sentence by processing three pieces of information: a) what the student pronounced, b) how likely it is that the native speaker would pronounce it that way, and c) what the student was supposed to pronounce. Details of the entire model of pronunciation error detection are presented in Section 3 of our recent work \cite{korzekwa2021mispronunciation}. We will now only show the details of the PM model that are relevant to this experiment. 

\bigskip

\textbf{Overview of the native speech pronunciation model}

\bigskip

PM is an encoder-decoder neural network following Sutskever et al. \cite{sutskever2014sequence}. Instead of building a text-to-text translation system between two languages, we use it for the P2P conversion. The sequence of phonemes $\mathbf{r}$ that the native speaker was supposed to pronounce is converted to the sequence of phonemes  $\mathbf{r^{'}}$ they had pronounced, denoted as  $\mathbf{r^{'}} \sim p(\mathbf{r^{'}}|\mathbf{r})$. Once trained,  PM acts as a probability mass function, computing the probability sequence $\boldsymbol{\pi}$ of the recognized phonemes $\mathbf{r_o}$ pronounced by the student conditioned by the expected (canonical) phonemes $\mathbf{r}$.  PM is denoted as in Eq. \ref{eqn:speech_synth_proposed_model}.

\begin{equation}
\boldsymbol{\pi}=\sum_{\mathbf{r_o}} p(\mathbf{r_o}|\mathbf{o})p(\mathbf{r^{'}}=\mathbf{r_o}|\mathbf{r})
\label{eqn:speech_synth_proposed_model}
\end{equation}
The PM model is trained on P2P speech data generated automatically by passing the speech of the native speakers through the PR. By using PR to annotate the data, we can make the PM model more robust against possible phoneme recognition inaccuracies in  PR at the time of testing.

\bigskip

\textbf{Results}

\bigskip

The complete model with PM enabled is called  PR-PM that stands for a Phoneme Recognizer + Pronunciation Model. The model with PM turned off is called PR-LIK that stands for Phoneme Recognizer outputting the likelihoods of recognized phonemes. PR-LIK is an extension of the PR-NOLIK model --  the mispronunciation detection model proposed by Leung et al. \cite{leung2019cnn} that only returns the most likely recognized phonemes and does not use phoneme likelihoods to detect pronunciation errors.  PR-NOLIK detects mispronounced words based on the difference between the canonical and recognized phonemes. Therefore, this system does not offer any flexibility in optimizing the model for higher precision by fine-tuning the threshold applied to the phoneme recognition probabilities.

\begin{figure}[th]
\centering
\includegraphics[width=1\textwidth]{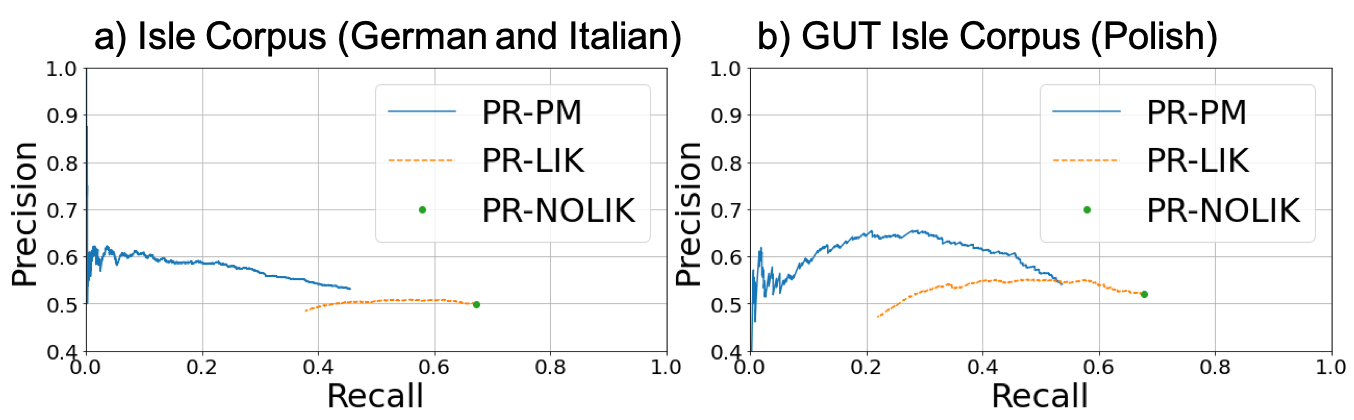}
\decoRule
\caption[Precision-recall curves for the evaluated systems to measure the effect of using the PM model in detecting pronunciation errors. PR-PM - full model with the PM enabled. PR-LIK - the PR-PM model with the PM disabled. PR-NOLIK - non-probabilistic variant of the PR-LIK  model proposed by Leung et al. \cite{leung2019cnn}]{Precision-recall curves for the evaluated systems to measure the effect of using the PM model in detecting pronunciation errors. PR-PM - full model with the PM enabled. PR-LIK - the PR-PM model with the PM disabled. PR-NOLIK - non-probabilistic variant of the PR-LIK  model proposed by Leung et al. \cite{leung2019cnn}.}
\label{fig:pr_uncertainty_modeling_precision_recall_plot}
\end{figure}

Turning off PM reduces the precision between 11\% and 18\%, depending on the decrease in recall between 20\% to 40\%,  as shown in Figure \ref{fig:pr_uncertainty_modeling_precision_recall_plot}. One example where the PM helps is the word `enough' that can be pronounced in two similar ways: /ih n ah f/ or /ax n ah f/ (short `i' or `schwa' phoneme at the beginning.) The PM can take into account the phonetic variability and recognize both versions as correctly pronounced. Another example is coarticulation \cite{hieke1984linking}. Native speakers tend to merge phonemes of adjacent words. For example, in the text `her arrange' /hh er - er ey n jh/, two adjacent phonemes /er/ can be pronounced as one phoneme: /hh er ey n jh/. The PM model can correctly recognize multiple variations of such pronunciations.

Complementary to the precision-recall curve shown in Figure \ref{fig:pr_uncertainty_modeling_precision_recall_plot}, we present in Table \ref{tab:pr_pm_results} one configuration of the precision and recall scores for the PR-LIK and PR-PM systems. This configuration is chosen in a way to: a) make the recall for both systems close to the same value, and b) to illustrate that the PR-PM model has much greater potential to increase precision than the PR-LIK system. A similar conclusion can be drawn by checking various different precision and recall configurations in the precision and recall plots for both Isle and GUT Isle corpora.

\begin{table}
\caption{Precision and recall of detecting word-level mispronunciations. CI - Confidence Interval. PR-PM - full model with the PM enabled. PR-LIK - the PR-PM model with the PM disabled.}
  \label{tab:pr_pm_results}
  \centering
  \begin{tabular}{lll}
    \toprule  
  \tabhead{ Model}  & \tabhead{Precision [\%,95\%CI]} & \tabhead{Recall [\%,95\%CI]} \\
    \midrule
    \multicolumn{3}{c}{\textbf{Isle corpus (German and Italian)}} \\
    PR-LIK & 49.39 (47.59-51.19) & 40.20 (38.62-41.81)\\ 
    PR-PM & 54.20 (52.32-56.08) & 40.20 (38.62-41.81)\\  
     \multicolumn{3}{c}{\textbf{GUT Isle corpus (Polish)}} \\
    PR-LIK & 54.91 (50.53-59.24) & 40.29 (36.66-44.02)\\ 
    PR-PM & 61.21 (56.63-65.65) & 40.15 (36.51-43.87)\\  
    \bottomrule\\
  \end{tabular}
\end{table}

\subsubsection{Lexical stress error detection}
\label{sec:experiments_lexical_stress}

\bigskip

\textbf{Experimental setup}

\bigskip

The full CAPT learning experience includes both the detection of pronunciation and lexical stress errors. To investigate the potential of speech generation in the lexical stress error detection task, we evaluate the T2S method, which is a simpler version of the S2S method evaluated in Section \ref{sec:p2p_gen_of_incorrectly_pron_speech}. 

The lexical stress error detection model is trained to measure the benefits of employing synthetic mispronounced speech. The first model, denoted as Att\_TTS is based on an attention mechanism and is trained on both human and synthetic speech with pronunciation errors. In this model, 1980 the most popular English words \cite{michel2011quantitative} were synthesized with correct and incorrect stress patterns using the method outlined in Section \ref{sec:t2s_method}, and added to the speech corpora of isolated words presented in  Section \ref{sec:speech_corpora_words}. The Att\_NoTTS model is trained only on human speech. Each of the two models presented has its simpler version without the attention mechanism, marked as NoAtt\_TTS and NoAtt\_NoTTS. Both models will help to understand whether the benefits of using synthetic pronunciation errors depend on the model capacity.

The accuracy of detecting lexical stress errors is measured in terms of an AUC metric. To be comparable to the study by Ferrer et al. \cite{ferrer2015classification}, we use precision as an additional metric, while setting recall to 50\%.

\bigskip

\textbf{Overview of the lexical stress detection model}

\bigskip

\begin{figure}[th]
\centering
\centerline{
\includegraphics[width=1.25\textwidth]{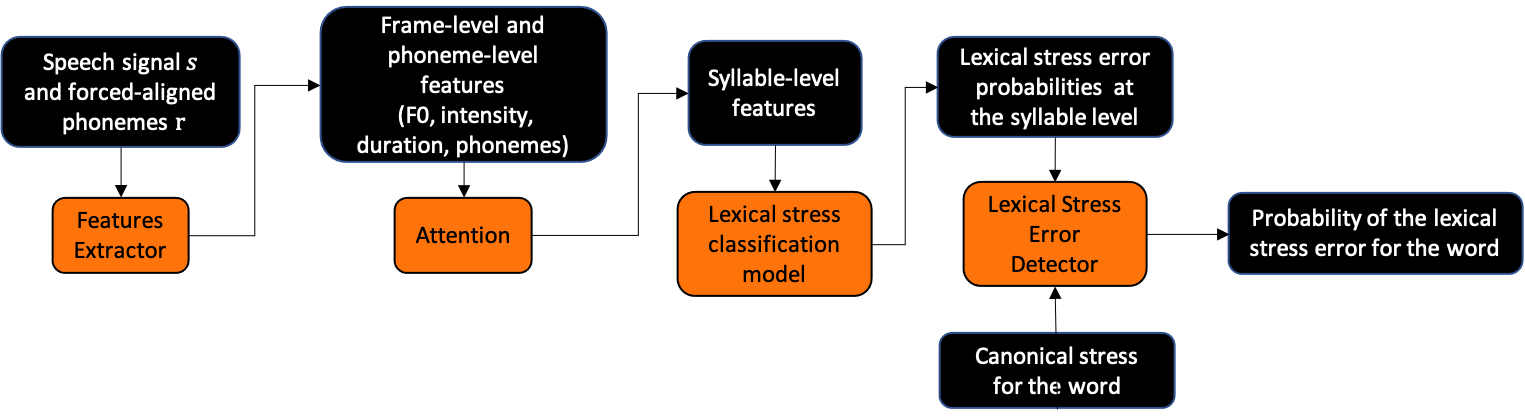}}
\decoRule
\caption[Attention-based model for the detection of lexical stress errors]{Attention-based model for the detection of lexical stress errors.}
\label{fig:lexical_stress_architecture}
\end{figure}

As shown in Figure \ref{fig:lexical_stress_architecture}, the lexical stress error detection model consists of three subsystems: Feature Extractor, Attention-based Classification Model, and Lexical Stress Error Detector. The Feature Extractor extracts prosodic features and phonemes from the speech signal $\mathbf{s}$ and the forced-aligned canonical phonemes $\mathbf{r}$. Prosodic features include: F0, intensity [dB SPL] and duration of phonemes. The F0 and intensity features are computed at the frame level. The Attention-based Classification Model uses the attention mechanism \cite{vaswani2017attention} to map frame-level and phoneme-level features to a syllable-level representation. It then produces lexical stress error probabilities at the syllable level. The Lexical Stress Error Detector reports a lexical stress error if the expected (canonical) and estimated lexical stress for a given syllable do not match and the corresponding probability is higher than the specified threshold. The detailed architecture of the model is presented in Section 3 of our recent work \cite{korzekwa21_interspeech}.

The NoAtt\_TTS and NoAtt\_NoTTS models do not have the attention mechanism. Instead, as a representation at the syllable level, they use the average acoustic feature values for the corresponding syllable nucleus. The hypothesis is that synthetic data will not be beneficial to a simpler model due to its limited capacity. 

\bigskip

\textbf{Results}

\bigskip

\begin{figure}[th]
\centering
\includegraphics[width=0.8\textwidth]{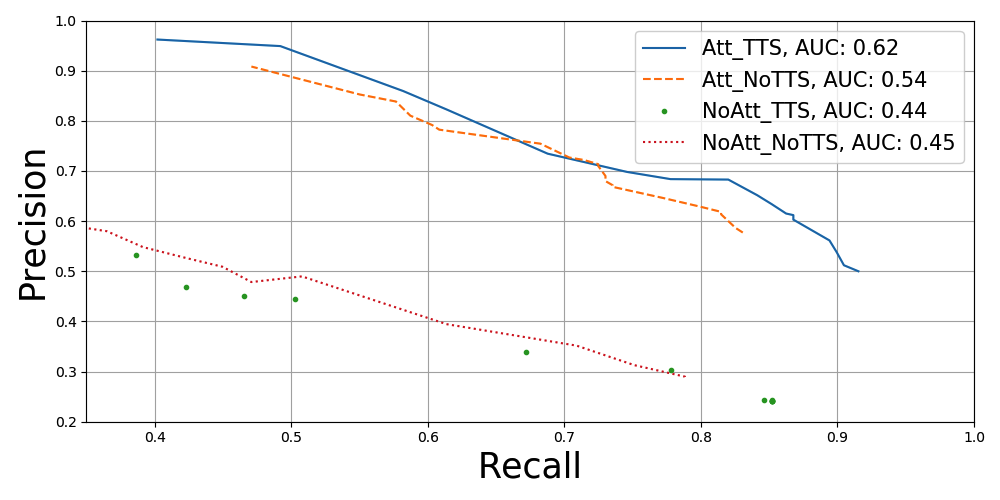}
\decoRule
\caption[Precision-recall curves for lexical stress error detection models]{Precision-recall curves for lexical stress error detection models.}
\label{fig:lexical_stress_error_detection_precision_recall_plot}
\end{figure}

Enriching the training set with the incorrectly stressed words increases an AUC score from 0.54 to 0.62 (Att\_TTS vs. Att\_NoTTS in Figure \ref{fig:lexical_stress_error_detection_precision_recall_plot} and Table \ref{tab:lexical_stress_precision_recall_acc}). Data augmentation helps because it increases the number of words with incorrect stress patterns in the training set. This prevents the model from using the strong correlation between phonemes and lexical stress in the correctly stressed words. Using data augmentation in the simpler model without the attention mechanism slightly reduced an AUC score from 0.45 to 0.44 (NoAtt\_NoTTS vs NoAtt\_TTS). The NoAtt\_TTS model has limited capacity due to not using the attention mechanism to model prosodic features, and thus is unable to benefit from synthetic speech.

We compare our results with the work of Ferrer et al. \cite{ferrer2015classification}. There were 46.4\% (191 out of 411) of incorrectly stressed words in their corpus, well over 9.4\% (189 out of 2109) words in our experiment. The fewer lexical stress errors that users make, the more difficult it is to detect them. Under these conditions, we can state that our lexical stress detection model based on T2S generated synthetic speech achieves higher scores in precision and recall compared to the work of Ferrer et al. \cite{ferrer2015classification}.

\begin{table}
 \caption{AUC, precision and recall [\%, 95\% Confidence Interval] metrics for lexical stress error detection models. Att. - Model with attention. Syn. - Synthetic mispronunciations.}
   \label{tab:lexical_stress_precision_recall_acc}
  \centering
  \begin{tabular}{llllll}
    \toprule  
   \tabhead{Model} & \tabhead{\makecell[l]{Att.}} & \tabhead{\makecell[l]{Syn.}} & \tabhead{AUC} & \tabhead{Precision [\%]} & \tabhead{Recall[\%]}  \\
    \midrule
    Att\_TTS & yes & yes & 0.62 & 94.8 (89.18-98.03) & 49.2 (42.13-56.3) \\
    Att\_NoTTS & yes & no & 0.54 & 87.85 (80.67-93.02) & 49.74 (42.66-56.82)\\
    NoAtt\_TTS & no & yes & 0.44 & 44.39 (37.85-51.09) & 50.26 (43.18-57.34)  \\
    NoAtt\_NoTTS & no & no & 0.45 & 48.98 (42.04-55.95) & 50.79 (43.70-57.86)  \\
    \cite{ferrer2015classification} & na & na & na & 95.00 (na-na) & 48.3 (na-na)  \\
    \bottomrule\\
  \end{tabular}
\end{table}

\subsection{Conclusions}
\label{sec:speech_synth_conclusions}
 
We propose a new paradigm for detecting pronunciation errors in non-native speech. Rather than focusing on detecting  pronunciation errors directly, we reformulate the detection problem as a speech generation task. This approach is based on the assumption that it is easier to generate speech with specific characteristics than to detect those characteristics in speech with limited availability. In this way, we address one of the main problems of the existing CAPT methods, which is the low availability of mispronounced speech for reliable training of pronunciation error detection models.

We present a unified look at three different speech generation techniques for detecting pronunciation errors based on P2P, T2S and S2S conversion. The P2P, T2S, and S2S methods improve the accuracy of detecting pronunciation and lexical stress errors. The methods outperform strong baseline models and establish a new state-of-the-art. The best S2S method outperforms the baseline method \cite{leung2019cnn} by improving the accuracy of detecting pronunciation errors in AUC metric by 41\% from 0.528 to 0.749. The S2S method has the ability to control many properties of speech, such as voice timbre, prosody (duration), and pronunciation. This opens the door to the generation of mispronounced speech that can mimic certain aspects of non-native speech, such as voice timbre. The S2S method can be seen as a generalization of the simpler methods, T2S and P2P, providing a general framework for building a first-class models of pronunciation assessment. For better reproducibility, in addition to using publicly available speech corpora, we recorded the GUT Isle corpus of non-native English speech  \cite{Weber2020}. The corpus is available to other researchers in the field.

In the future, we plan to extend the S2S method in order to generate synthetic speech as close as possible to non-native speech: a) we will extract the voice timbre from the speech of non-native speakers and transfer it to native speech, following the paper of Merritt et al. on text-free voice conversion \cite{merritt22_icassp_vc}, and b) we will mimic the distribution of pronunciation errors in non-native speech. We expect both changes to increase the accuracy of detecting pronunciation errors in non-native speech. In the long run, we hope to demonstrate that ''synthetic speech is all you need'' by training the model with synthetic speech only and achieving  state-of-the-art results in the pronunciation error detection task. This may revolutionize computer-assisted English L2 learning and CAPT. Moreover, such a paradigm may be transferred to the whole domain of computer-assisted foreign language learning.

%% file: Chapters/RelatedApplications.tex

\chapter{Generalization of deep learning methods for pronunciation error detection} 

\label{chapter:dysarthric_speech} 


In this section, we explore the generalization capabilities of deep learning methods for pronunciation error detection. For this purpose, the following secondary research thesis has been formulated:

\begin{center}
\textbf{Deep learning methods for the detection of pronunciation errors in non-native speech are transferable to the related tasks of detection and reconstruction of dysarthric speech.}
\end{center}

The first task related to pronunciation error detection is the detection of dysarthric speech. For this purpose, generalization capabilities of the attention mechanism and the multi-task deep learning techniques are investigated. 

The reconstruction of dysarthric speech was selected for the second related task. Reconstructing dysarthric speech and generating synthetic pronunciation errors  are examples of speech-to-speech deep learning methods, therefore, similar deep learning techniques may perform well in both scenarios.

The research on both topics, detection and reconstruction of dysarthric speech, resulted in a publication at the Interspeech 2019 conference, which is presented in this chapter.

\begin{center}
\textit{Daniel Korzekwa, Roberto Barra-Chicote, Bozena Kostek, Thomas Drugman, Mateusz Lajszczak, Interpretable deep learning model for the detection and reconstruction of dysarthric speech, Interspeech, 2019}
\end{center}

\bigskip

\textbf{Abstract}

\bigskip

We present a novel deep learning model for the detection and reconstruction of dysarthric speech. We train the model with a multi-task learning technique to jointly solve dysarthria detection and speech reconstruction tasks. The model key feature is a low-dimensional latent space that is meant to encode the properties of dysarthric speech. It is commonly believed that neural networks are “black boxes” that solve problems but do not provide interpretable outputs. On the contrary, we show that this latent space successfully encodes interpretable characteristics of dysarthria, is effective at detecting dysarthria, and that manipulation of the latent space allows the model to reconstruct healthy speech from dysarthric speech. This work can help patients and speech pathologists to improve their understanding of the condition, lead to more accurate diagnoses and aid in reconstructing healthy speech for afflicted patients.

\section{Introduction}

Dysarthria is a motor speech disorder manifesting itself by a weakness
of muscles controlled by the brain and nervous system that are used in the process of speech production, such as lips, jaw and throat \cite{ASHA2018}. Patients with dysarthria produce harsh and breathy speech with abnormal prosodic
patterns, such as very low speech rate or flat intonation, which makes
their speech unnatural and difficult to comprehend. Damage to the nervous system is the main cause of dysarthria \cite{ASHA2018}.
It can happen as an effect of multiple possible neurological disorders
such as cerebral palsy, brain stroke, dementia or brain cyst \cite{Cuny2017,Banovic2018}.

Early onset detection of dysarthria may improve the quality of life
for people affected by these neurological disorders. According to Alzheimer's Research UK2015 \cite{Alzheimersresearchuk2015}, 1 out of 3 people
in the UK born in 2015 will develop dementia in their life. 
Manual detection of dysarthria conducted 
in clinical conditions by speech pathologists is costly, time-consuming and can lead to an incorrect diagnosis \cite{2012E121001,Carmichael2008}. With an automated analysis of speech, we can detect an early onset of
dysarthria and recommend further health checks with a clinician even when a human speech pathologist is not available. Speech reconstruction may help with better identification of the symptoms and
enable patients with severe dysarthria to communicate with other people.

Section 2 presents related work. In Section 3 we describe the proposed model for detection
and reconstruction of dysarthria. In Section 4 we demonstrate the
performance of the model with experiments on detection, interpretability, and
reconstruction of healthy speech from dysarthric speech. We conclude with our remarks.

\section{Related work}

\subsection{Dysarthria detection}

Deep neural networks can automatically detect dysarthric patterns
without any prior expert knowledge \cite{Krishna,DBLP:conf/interspeech/Vasquez-CorreaA18}.
Unfortunately, these models are difficult to interpret because
they are usually composed of multiple layers producing multidimensional outputs
with an arbitrary meaning and representation. Contrarily, statistical
models based on a fixed vector of handcrafted prosodic and spectral
features such as jitter, shimmer, Noise to Harmonic Ratio (NHR) or
Mel-Frequency Cepstral Coefficients (MFCC) offer good interpretability but require experts to manually design predictor features \cite{Falk2012,Sarria-Paja2012a,Gillespie2017,Lansford2014}.

The work of Tu Ming et al. on interpretable objective evaluation of dysarthria
\cite{DBLP:conf/interspeech/TuBL17} is the closest we found to our
proposal. The main difference is that our model not only provides interpretable characteristics of dysarthria but also reconstructs healthy speech. Their model is based
on feed-forward deep neural networks with a latent layer representing
four dimensions of dysarthria: nasality, vocal quality, articulatory
precision, and prosody. The final output of the network represents
general dysarthria severity on a scale from 1 to 7. The input to this
model is described by a 1201-dimensional vector of spectral
and cepstral features that capture various aspects of dysarthric
speech such as rhythm, glottal movement or formants. As opposed to
this work, we use only mel-spectrograms to present the input speech
to the model. Similarly to our approach, Vasquez-Correa et al. \cite{DBLP:conf/interspeech/Vasquez-CorreaA18}
uses a mel-spectrogram representation for dysarthria detection. However,
they use 160 ms long time windows at the transition points between
voiced and unvoiced speech segments, in contrast to using a full mel-spectrogram
in our approach.

\subsection{Speech reconstruction}

There are three different approaches to the reconstruction of dysarthric speech: voice banking, voice adaptation and voice reconstruction \cite{2012E121001}. Voice banking is a simple idea of collecting  a patient's speech samples before their speech becomes unintelligible and using it to build a personalized Text-To-Speech (TTS) voice.
It requires about 1800 utterances for a basic unit-selection TTS
technology \cite{Modeltalker} and more than 5K utterances for building a Neural TTS voice \cite{DBLP:journals/corr/abs-1811-06315}. Voice adaptation requires as little as
7 minutes of recordings. In this approach, we start with a TTS
model of an average speaker and adapt its acoustic and articulatory
parameters to the target speaker \cite{AhmadKhan2011}. 

Both voice banking and voice adaptation techniques rely on the availability of recordings for a healthy speaker. The voice reconstruction technique overcomes this shortcoming.
This technique aims at restoring damaged speech by tuning parameters representing the glottal source and the vocal tract filter \cite{rabiner_schafer78,Drugman2014}.
In our model, we
take a similar approach. However, instead of making assumptions on
what parameters should be restored, we let the model automatically learn
the best dimensions of the latent space that are responsible for dysarthric
speech. Reconstruction of healthy speech by manipulating the latent space of a dysarthric speech is a promising direction,
however, so far we only managed to successfully apply this technique in a single-speaker setup. 

Variational Auto-Encoder (VAE) \cite{doersch2016tutorial}
is a probabilistic latent space model that
has recently become popular for the reconstruction of various signals
such as text \cite{DBLP:journals/corr/HuYLSX17,DBLP:journals/corr/BowmanVVDJB15}
and speech \cite{DBLP:journals/corr/abs-1812-04342,DBLP:journals/corr/abs-1709-07902}.

\section{Proposed model}

\begin{figure}[th]
\centering
\centerline{
\includegraphics[width=1.2\textwidth]{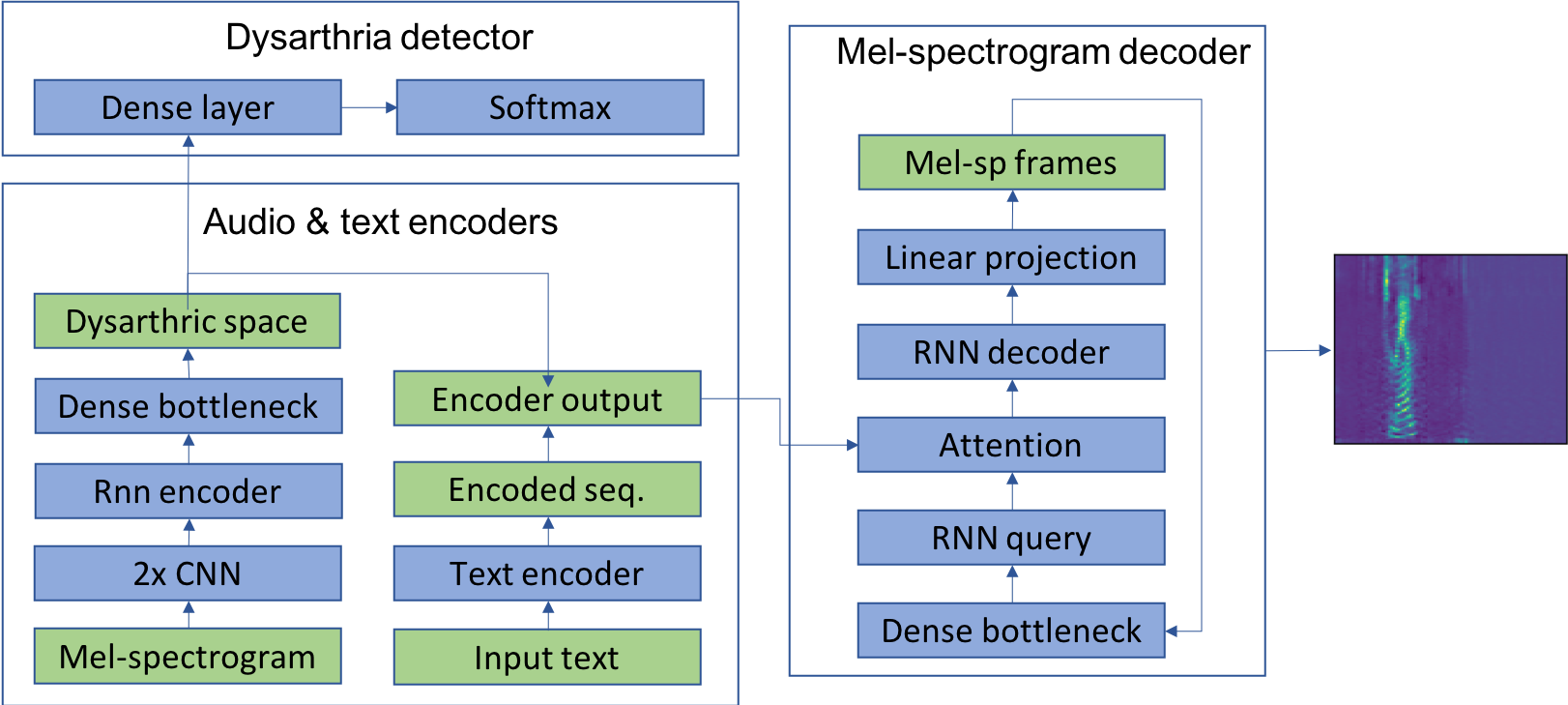}}
\decoRule
\caption[Architecture of deep learning model for detection and reconstruction
of dysarthric speech]{Architecture of deep learning model for detection and reconstruction
of dysarthric speech.}
\label{fig:Architecute-of-deep}
\end{figure}

The model consists of two output networks, jointly trained, with a
shared encoder as shown in Figure \ref{fig:Architecute-of-deep}. The audio and text encoders produce a low-dimensional dysarthric latent space and a sequential encoding of the input text.
The audio decoder reconstructs input mel-spectrogram from a dysarthric latent space and encoded text. Logistic classification model predicts the probability of dysarthric speech from the dysarthric latent space. In Table \ref{tab:Configuration-of-neural}
we present the details of various neural blocks used in the model.

Let us define a matrix \textbf{$X:[n_{mels},n_{f}]$} representing a mel-spectrogram (frame length=50ms and frame shift=12.5ms), where
 \emph{$n_{mels}=128$} is the number of mel-frequency
bands and \emph{$n_{f}$} is the number of frames. Let us define a matrix $T:[n_{c},n_{t}]$ representing a one-hot encoded input text, where $n_{c}$ is the number of unique characters in the
alphabet and $n_{t}$ is the number of characters in the input text.
The mel-spectrogram $X$ is encoded into 2-dimensional
dysarthria latent space $\mathbf{l}=\{l_{1},l_{2}\}$ and then used as a conditioning variable for estimating the probability 
of dysarthria $d \sim p(d|X,\theta)$ and reconstructing the mel-spectrogram $Y \sim p(Y|X,T,\theta)$. Limiting the latent space to 2 dimensions makes the model more resilient to overfitting. The theta is a vector of trainable parameters of the model.

Let us define a training set of $m$ tuples of $((X,T),y)$, where $y\in\{0,1\}$ is the label for normal/dysarthric
speech and $m$ is the number of speech mel-spectrograms for dysarthric
and normal speakers. We optimize a joint cost of the predicted probability
of dysarthria and mel-spectrogram reconstruction defined as a weighted
function:
\begin{equation}
\sum_{i=1}^{m}\alpha log(p(d_{i}|X_{i},\theta))+(1-\alpha)log(p(Y_{i}|X_{i},T_{i},\theta))\label{eq:}
\end{equation}
where $log(p(d_{i}|X_{i},\theta))$ is the cross-entropy between the predicted
and actual labels of dysarthria, and $log(p(Y_{i}|X_{i},T_{i},\theta))$ is the log-likelihood
of a Gaussian distribution for the predicted mel-spectrogram with a unit
variance, a.k.a L2 loss. We used backpropagation and mini-batch stochastic gradient
descent with a learning rate of 0.03 and a batch size of 50.
The whole model is initialized with Xavier’s method \cite{journals/jmlr/GlorotB10} using the magnitude
value of 2.24. Hyper-parameters of the model presented in Table \ref{tab:Configuration-of-neural} were tuned with a  grid search optimization. We used MxNet framework for implementing the model \cite{DBLP:journals/corr/ChenLLLWWXXZZ15}.

\begin{table}[t]
  \caption{Configuration of the neural network blocks.}
  \label{tab:Configuration-of-neural}
  \centering
  \begin{tabular}{ll}
    \toprule  
    \tabhead{Neural block}  & \tabhead{Config}           \\
    \midrule
    \multicolumn{2}{c}{\textbf{Audio encoder}} \\
    2x CNN & 20 channels, 5x5 kernel, RELU, VALID \\
    GRU & 20 hidden states, 1 layer \\
    Dense & 20 units, tanh \\
    Dysarthric space & 2 units, linear \\
    \multicolumn{2}{c}{\textbf{Text encoder}} \\
    3x CNN & 40 channels, 5x5 kernel, RELU, SAME \\
    GRU & 27 hidden states, 1 layer \\
    \multicolumn{2}{c}{\textbf{Audio decoder}} \\
    Dense bottleneck & 96 units, RELU \\
    GRU query & 29 hidden states, 1 layer \\
    GRU decoder & 128 hidden states, 1 layer \\
    Linear projection & frames\_num x melsp bins units, linear \\
    \bottomrule\\
  \end{tabular}
\end{table}

\subsection{Mel-spectrogram and text encoders}

For the spectrogram encoder, we use a Recurrent Convolutional Neural Network
model (RCNN) \cite{DBLP:journals/corr/abs-1803-09047}. The convolutional layers,
each followed by a max-pooling layer, extract local and time-invariant patterns of
the glottal source and the vocal tract. The GRU layer
models temporal patterns of dysarthric speech \cite{DBLP:journals/corr/ChoMGBSB14}. The last
state of the GRU layer is processed by two dense layers. Dropout \cite{JMLR:v15:srivastava14a} with probability of 0.5 is applied to the output of the activations for both
CNN layers, GRU layer, and the dense layer.

Text encoder encodes the input text using one-hot encoding, followed by
three CNN layers and one GRU layer. Outputs of both audio and text encoders are concatenated via matrix broadcasting, producing a matrix $E:[n_{c}+n_{l},n_{t}]$, where $n_{l}$ is dimensionality 
of the dysarthria latent space.

\subsection{Spectrogram decoder and dysarthria detector }

For decoding a mel-spectrogram, similarly to Wang et al. \cite{DBLP:journals/corr/WangSSWWJYXCBLA17}, we use a Recurrent Neural Network (RNN) model with attention. The
dot-product attention mechanism \cite{DBLP:journals/corr/VaswaniSPUJGKP17} plays a crucial role.
It informs to which elements of the encoder output the decoder should pay attention at every decoder step. The RNN network that produces a query vector for the attention, takes as input $r$ predicted mel-spectrogram frames from the previous time-step. 
The output of the RNN decoder is projected via a linear dense
layer into $r$ number of mel-spectrogram frames. Similarly to Wang
et al. \cite{DBLP:journals/corr/WangSSWWJYXCBLA17}, we found that
it is important to preprocess the mel-spectrogram with a dense layer and
dropout regularization to improve the overall generalization of the model.

The dysarthria detector is created from a 2-dimensional dense layer. It
uses a tanh activation followed by a softmax function that represents
the probability of dysarthric speech.

\section{Experiments}

\subsection{Dysarthric speech database\label{subsec:Dysarthric-speech-database}}

There is no well-established benchmark in the literature to compare
different models for detecting dysarthria. Aside from the most popular
dysarthric corpora, UA-Speech \cite{Kim2008} and TORGO \cite{Rudzicz2012}, there are multiple speech databases created for the purpose of a
specific study, for example, corpora of 57 dysarthric speakers \cite{Lansford2014}
and Enderby Frenchay Assessment dataset \cite{Carmichael2008}.
Many corpora, including TORGO and HomeService \cite{Nicolao2016}, are
available under non-commercial license.

In our experiments we use the UA-Speech database from the University
of Illinois \cite{Kim2008}. It contains 11 male and 4 female dysarthric
speakers of different dysarthria severity levels and 13 control speakers.
455 isolated words are recorded for each speaker with 1 to 3 repetitions.
Every word is recorded through a 7-channel microphone array, producing a separate wav file of 16 kHz sampling rate
for every channel. It contains 9.4 hours of speech for dysarthric speakers
and 4.85 hours for control speakers. UA-Speech corpus comes with intelligibility scores that are obtained from a transcription task performed by 5 naive listeners.

To control variabilities in recording conditions, we normalized mel-spectrograms for every recorded
word independently with a z-score normalization. We considered removing the initial period of silence at the beginning
of recorded words but we decided against it. We found that for dysarthric
speakers of high speech intelligibility, the average length of the initial
silence period that lasts 0.569sec +- 0.04674 (99\% CI) is comparable with
healthy speakers with the length of 0.532sec +- 0.055. Because we can predict unvoiced periods with merely 85\% of accuracy \cite{Johnston:2012:WAR:2432294}, removing the periods of silence for dysarthric speakers with poor intelligibility is very inaccurate.

\subsection{Automatic detection of dysarthria\label{subsec:Automatic-detection-of}}

To define the training and test sets, we use a Leave-One-Subject-Out (LOSO) cross-validation scheme. For each
training, we include all speakers but one that is left out to measure
the prediction accuracy on unseen examples. The accuracy, precision and recall metrics are computed at a speaker level (the average dysarthria probability
of all the words produced by the speaker is compared to a target speaker dysarthria
label $\in\{0,1\}$), and a word level (comparing target dysarthria
label with predicted dysarthria probability for all words independently).

As a baseline, we use the Gillespie's et al. model that is based on Support Vector Machine classifier \cite{Gillespie2017}.
It uses 1595 low-level predictor features processed with a global z-score normalization.
It reports a 75.3 and 92.9 accuracy in the dysarthria detection
task at the word and speaker levels respectively, following LOSO cross-validation. However, Gillespie uses 336
words from the UA-Speech corpus with 12 words per speaker, whereas we
use all 455 words across all speakers. 

In our first model, only dysarthric labels are observed and we achieved an accuracy on the word and speaker
levels of 82\% and 93\% respectively. By training the multi-task model,
in which both targets, i.e. mel-spectrogram and dysarthric labels, are observed, the accuracy on the word level increased by 3 percents to the value of 85.3\% (Table \ref{tab:Accuracy-metrics-for}). We found that the UA-Speech database includes multiple recorded words for healthy speakers that contain intelligibility errors, different words than asked or background speech of other people. These
issues affect the accuracy of detecting dysarthric speech.

\begin{table}[ht]
  \caption{Accuracy of dysarthria detection including  95\% CI. Classifier task - target mel-spectrogram (ML) is not
observed during training. Multitask - both targets ML and dysarthric
labels are observed}
  \label{tab:Accuracy-metrics-for}
  \centering
  \begin{tabular}{llll}
    \toprule  
    \tabhead{System} & \tabhead{Accuracy} & \tabhead{Precision}  & \tabhead{Recall}\\
    \midrule
    \multicolumn{4}{c}{\textbf{Word level}} \\
    Multitask & 0.853 (0.849 - 0.857) & 0.831 & 0.911\\
    Classifier task & 0.820 (0.815 - 0.824) & 0.818 & 0.855 \\
    Gillespie et al.\cite{Gillespie2017} & 0.753 (na) & 0.823 & 0.728 \\
    \multicolumn{4}{c}{\textbf{Speaker level}} \\
    Multitask & 0.929 (0.790-0.984) & 1.000 & 0.867\\
    Classifier task & 0.929 (0.790-0.984) & 0.933 & 0.933 \\
    Gillespie et al.\cite{Gillespie2017} & 0.929 (na) & na & na \\
    \bottomrule\\
  \end{tabular}
\end{table}

Krishna reports a 97.5\% accuracy on UA-Corpus \cite{Krishna}. However,
after email clarification with the author, we found that they estimated the accuracy
taking into account only the speakers with a medium level of dysarthria.
Narendra et al. achieved 93.06\% utterance level accuracy on the
TORGO dysarthric speech database \cite{DBLP:conf/interspeech/NarendraA18}. As opposed to the related work, our model does not need any expert knowledge to design hand-crafted features and it can learn automatically using a low-dimensional latent space that encodes characteristics of dysarthria.

\subsection{Interpretable modeling of dysarthric patterns\label{subsec:Interpretable-modeling-of}}

We analyze the correlation between the dysarthric
latent space and the intelligibility of speakers. We look at 550 audio samples
of a single 'Command' word across the 15 dysarthric
speakers and 13 healthy speakers.

\begin{figure}[th]
\centering
\centerline{
\includegraphics[width=1.2\textwidth]{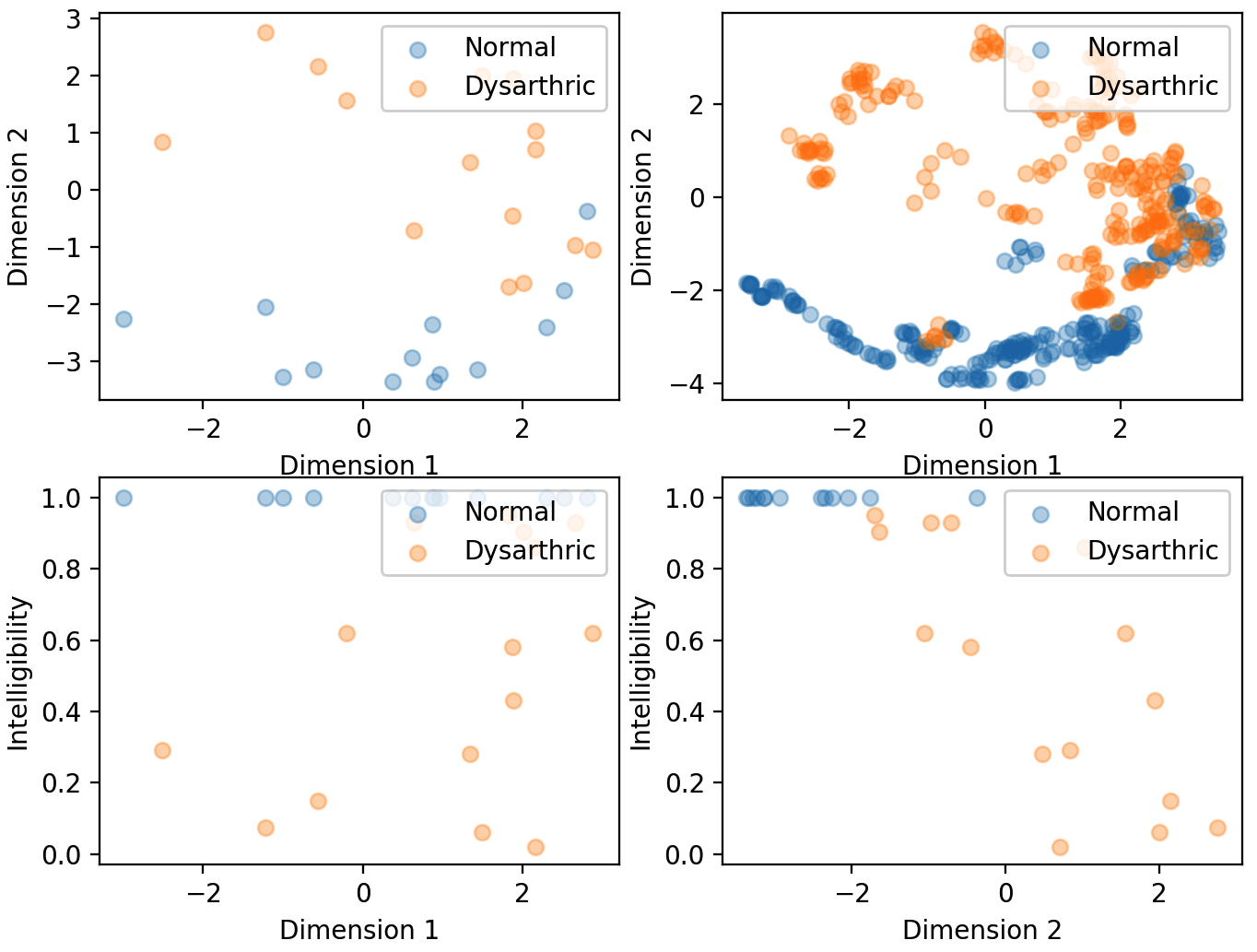}}
\decoRule
\caption[Unsupervised learning. Top row: Separation between dysarthric and control speakers in the latent space on a speaker (left) and word (right) level. Bottom row: Correlation between both dimensions of the latent space and the intelligibility scores]{Unsupervised learning. Top row: Separation between dysarthric and control speakers in the latent space on a speaker (left) and word (right) level. Bottom row: Correlation between both dimensions of the latent space and the intelligibility scores.}
\label{fig:Correlation-between-dysarthric}
\end{figure}

\begin{figure}[th]
\centering
\centerline{
\includegraphics[width=1.2\textwidth]{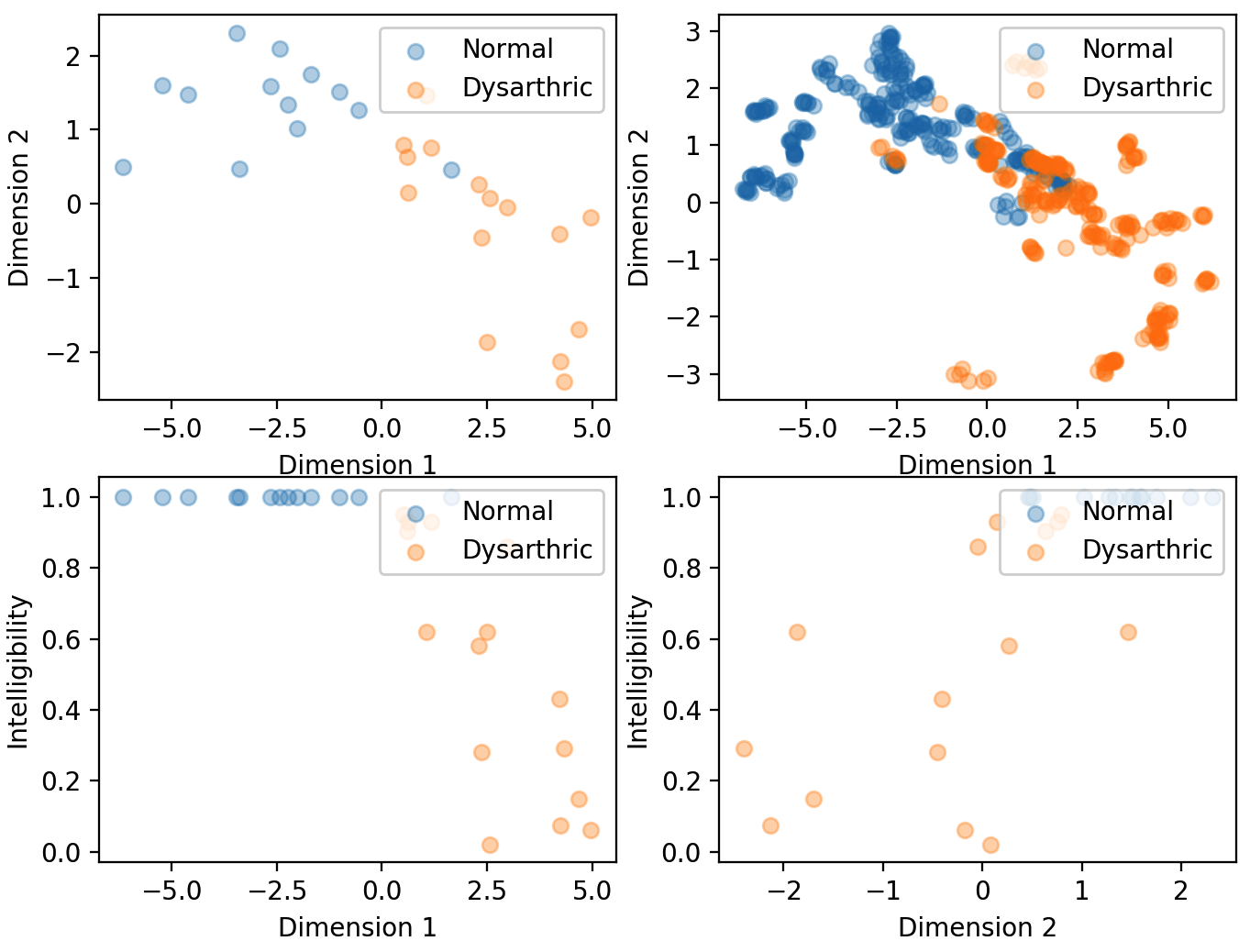}}
\decoRule
\caption[Supervised learning. As in Figure \ref{fig:Correlation-between-dysarthric}]{Supervised learning. As in Figure \ref{fig:Correlation-between-dysarthric}.}
\label{fig:Correlation-between-dysarthric-1}
\end{figure}

In an unsupervised training (Figure \ref{fig:Correlation-between-dysarthric}), target labels of dysarthric/normal speech are not presented
to the model. Dysarthric speakers are well separated from normal speakers and the
dimension 2 of the latent space is negatively correlated with the
intelligibility scores (Pearson correlation of -0.84, two-sided $p$-value \verb|<| 0.001).
In a supervised variant (Figure \ref{fig:Correlation-between-dysarthric-1}), we train the model jointly
with both reconstructed mel-spectrogram and the target dysarthria labels
observed. Both dimensions of the latent space are highly correlated with the
intelligibility scores (dimension 1 with correlation of -0.76 and
dimension 2 with correlation of 0.70, both with $p$-value \verb|<| 0.001). 

The sign of the correlation has no particular meaning. Retraining the model multiple times results in both positive and negative correlations between the latent space and the intelligibility of speech.
A high correlation 
between dysarthric latent space and intelligibility
scores suggests that by moving along the dimensions of the latent space,
we should be able to reconstruct speech of dysarthric speakers and
improve its intelligibility. We explore this in the next experiment.

\subsection{Reconstruction of dysarthric speech\label{subsec:Reconstruction-of-dysarthric}}

First we trained a supervised multi-speaker model with all dysarthric and control speakers but
we achieved poor reconstruction results with almost unintelligible speech. We think
this is due to a high variability of dysarthric speech across all speakers,
including various articulation, prosody and fluency problems. To better understand the potential for speech reconstruction, we narrowed the experiment down to two speakers, male speaker M05 and a corresponding control speaker. We have chosen M05 subject because their speech varies across different levels of fluency and we wanted to observe
this pattern when manipulating the latent space. For example, when pronouncing the word 'backspace', M05 uttered consonants 'b' and 's' multiple times, resulting in 'ba ba cs space'.

We analyzed a single category of 19 computer command words,
such as 'command' or 'backspace'. For every word uttered by M05,
we generated 5 different versions of speech, fixing dimension 2 of
the latent space to the value of -0.1, and using the values of {[}-0.5,
0, 0.5, 1, 1.5{]} for dimension 1. Audio samples of reconstructed
speech were obtained by converting predicted mel-spectrograms to waveforms
using the Griffin-Lim algorithm \cite{Griffin1984}.

We conducted MUSHRA perceptual test \cite{merritt2018comprehensive}.
Every listener was presented with 6 versions of a
given word at the same time, 5 reconstructions and one version of
recorded speech. We asked listeners to evaluate the fluency of speech
on a scale from 0 to 100. We used
10 US based listeners from the Amazon mTurk platform,
in total providing us with 1140 evaluated speech samples.

\begin{figure}[th]
\centering
\centerline{
\includegraphics[width=1.2\textwidth]{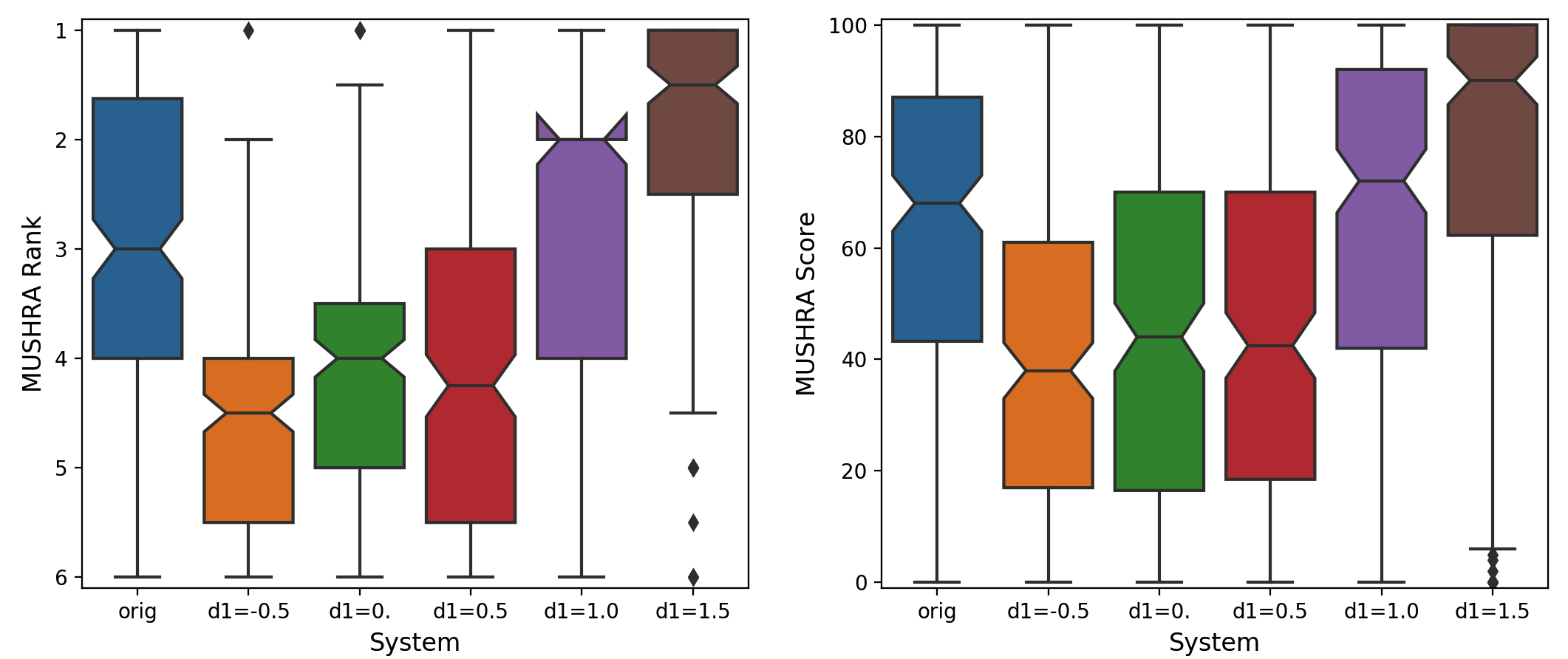}}
\decoRule
\caption[MUSHRA results for the fluency of speech for 5 reconstructions and one recorded speech. Rank order (left) and the median score on the scale from 0 to 100 (right)]{MUSHRA results for the fluency of speech for 5 reconstructions and one recorded speech. Rank order (left) and the median score on the scale from 0 to 100 (right).}
\label{fig:MUSHRA-results-for}
\end{figure}

\begin{figure}[th]
\centering
\centerline{
\includegraphics[width=1.2\textwidth]{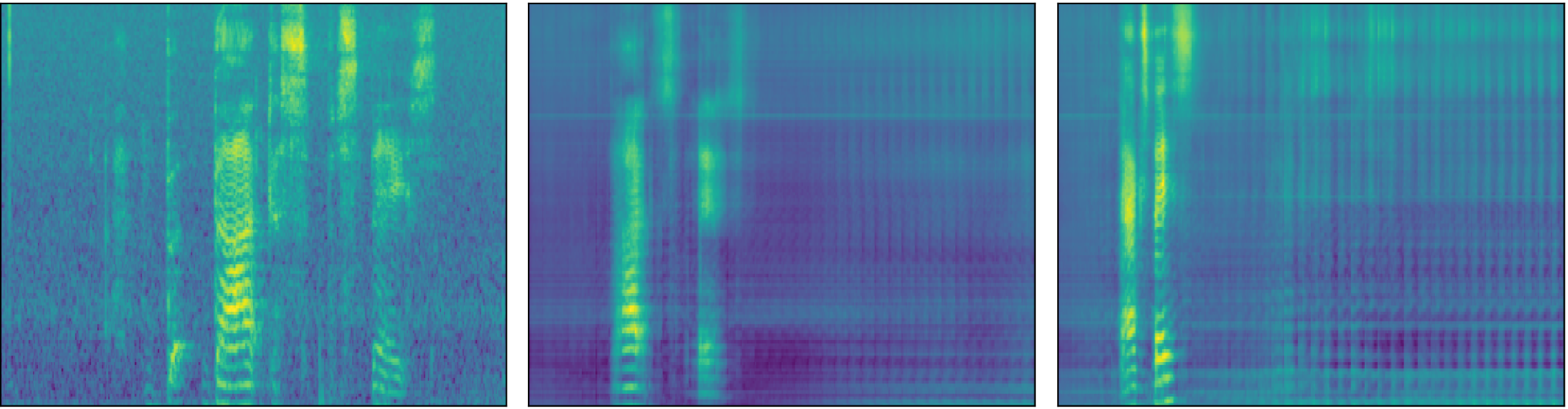}}
\decoRule
\caption[Reconstruction of dysarthric speech ('command' word)]{Reconstruction of dysarthric speech ('command' word). From left to right (MUSHRA scores of 51.8, 61.9 and 89.5): Recorded dysarthric speech. Reconstructed speech
with dimension 1 of 0.0 and 1.5 respectively.}
\label{fig:Reconstruction-of-dysarthric}
\end{figure}

As shown in Figure \ref{fig:MUSHRA-results-for}, by moving along dimension 1 of the latent space, we
can improve the fluency of speech, generating speech with levels of fluency not observed in the training data.
In the pairwise two-sided Wilcoxon signed-rank, all pairs of ranks are different
from each other with $p$-value \verb|<| 0.001, except of \{orig, d1=1.0\}, \{d1=-0.5,
d1=0.0\}, \{d1=-0.5, d1=0.5\}. Examples of original and reconstructed mel-spectrograms are shown in Figure \ref{fig:Reconstruction-of-dysarthric}. 

We found that manipulation of the latent space changes both the fluency of speech and the timbre of voice and it is possible that dysarthria is so tied up with speaker identify making it fruitless to disentangle them. We replaced a deterministic dysarthric latent space with a Gaussian variable and trained the model with an additional Kullback\verb|-|Leibler loss \cite{doersch2016tutorial,Mathieu2018} but we did not manage to separate the timbre of voice from dysarthria. Training the model with an additional discriminative cost to ensure that every dimension of the latent space is directly associated with a particular speech factor can potentially help with this problem \cite{DBLP:journals/corr/HuYLSX17}. 

\section{Conclusions}
 
This paper proposed a novel approach for the detection and reconstruction of dysarthric speech. The encoder-decoder model factorizes speech into a low-dimensional latent space and encoding of the input text. We showed that the latent space conveys interpretable characteristics of dysarthria, such as intelligibility and fluency of speech. MUSHRA perceptual test demonstrated that the adaptation of the latent space let the model generate speech of improved fluency. The multi-task supervised approach for predicting both the probability of dysarthric speech and the mel-spectrogram helps improve the detection of dysarthria with higher accuracy. This is thanks to a low-dimensional latent space of the auto-encoder as opposed to directly predicting dysarthria from a highly dimensional mel-spectrogram.
 
\section{Acknowledgements}

We would like to thank A. Nadolski, J. Droppo, J. Rohnke and V. Klimkov for insightful discussions on this work. 

%% file: Chapters/Conclusions.tex

\chapter{Conclusions} 

\label{chapter:conclusions} 


\section{Summary}

Within the research carried out in the framework of the Ph.D. work, novel deep learning methods were developed to detect pronunciation errors in non-native English speech automatically. Detecting pronunciation errors is part of CAPT that enables people to learn foreign languages without the assistance of a language teacher. As already mentioned, regarding the UNESCO report, 40\% of the world's population does not have access to education in a language they understand, so there is a great potential for the new CAPT methods to make education more accessible to people all over the world.

Existing CAPT methods based on deep learning cannot detect pronunciation errors with high accuracy. The best method proposed in this Ph.D. research improves the accuracy of detecting pronunciation errors in the AUC metric by 41\%, from 0.528 to 0.749, compared to the state-of-the-art approach \cite{korzekwa22_speechcomm}. This improvement corresponds to 80.45\% precision and 40.12\% recall. Taking into account only severe pronunciation errors, the AUC metric raises from 0.749 to 0.834, corresponding to 93.54\% precision and 40.15\% recall. These achievements successfully validate the primary research thesis:

\begin{center}
\textbf{It is possible to improve the accuracy of deep learning methods for detecting pronunciation errors in non-native English by employing synthetic speech generation and end-to-end modeling techniques that reduce the need for phonetically transcribed mispronounced speech.}
\end{center}

Extensive experiments have been conducted to measure the effectiveness of the proposed methods in CAPT. Deep learning models were developed and assessed to detect both pronunciation and lexical stress errors. Non-native speech of German, Italian and Polish speakers were used in the evaluations. As part of the doctoral research, two speech corpora of non-native Slavic and Baltic speakers have been recorded \cite{Weber2020}. 

To investigate generalization capabilities, the developed deep learning techniques for detecting pronunciation errors were successfully applied to the related areas of detection and reconstruction of dysarthric speech \cite {korzekwa2019interpretable}. The auto-encoder model was proposed to factorize dysarthric speech into a low-level latent representation. By controlling the latent representation, the fluency of the output speech can be improved, as shown in the MUSHRA perceptual speech test. In addition, the latent presentation can be used to detect dysarthric speech at the word level with 83.1\% precision and 91.1\% recall metrics. The new deep learning techniques applied to the topic of dysarthric speech successfully prove the secondary research thesis:

\begin{center}
\textbf{Deep learning methods for the detection of pronunciation errors in non-native speech are transferable to the related tasks of detection and reconstruction of dysarthric speech.}
\end{center}

\section{Novelty}

Many important observations have been made on existing state-of-the-art methods, which led to the development of novel techniques for detecting pronunciation errors.

Performing phonetic transcription of non-native speech is time-consuming, and sometimes, transcription is impossible due to differences between spoken languages. A new method of detecting pronunciation errors called WEAKLY-S (Weakly-supervised) has been proposed, which does not require phonetic transcriptions of non-native speech \cite{korzekwa21b_interspeech}. 

State-of-the-art methods align the canonical and recognized phoneme sequences to identify mispronounced speech segments such as phonemes and words. Any inaccuracies introduced in the alignment process would lower the accuracy of detecting pronunciation errors. As part of the WEAKLY-S model, a new end-2-end method has been proposed to directly detect pronunciation errors at the word level without having to align with canonical and recognized phoneme sequences \cite{korzekwa21b_interspeech}.  The WEAKLY-S model increases the accuracy of detecting pronunciation errors in the AUC metric by up to 30\% compared to the state-of-the-art approach.

There are two sources of variability and uncertainty that can affect the accuracy of detecting pronunciation errors. First, the same sentence can be pronounced in multiple correct ways, which should not trigger a pronunciation error. Second, it is challenging to recognize phonemes pronounced by the speaker accurately, and this ubiquitous uncertainty has to be accounted for. A new method has been proposed to this end, accounting for: i) multiple correct ways of pronouncing the same sentence and ii) uncertainty in phoneme recognition \cite{korzekwa2021mispronunciation}. The proposed method increases the precision of detecting mispronunciations by up to 18\% compared to the state-of-the-art approach.

Existing methods of detecting pronunciation errors often rely on hand-crafted speech features such as f0, energy, and phoneme alignment. A new method based on the attention mechanism has been proposed to automatically extract optimal speech features from a speech signal \cite{korzekwa21_interspeech}. The  method introduced plays a vital role in all proposed deep learning models in detecting pronunciation and lexical stress errors. 

The attention mechanism helps factorize a black-box deep learning model into multiple dependent components. Factorization leads to better interpretability of the model, e.g., visualizing the attention of the model for detecting lexical stress errors \cite{korzekwa21_interspeech}. Multi-task learning is a type of model factorization that can make a deep learning model more robust and less prone to overfitting \cite{korzekwa21b_interspeech}. Training the proposed multi-task WEAKLY-S pronunciation error detection model with two tasks, phoneme recognizer and pronunciation error detector, increase the accuracy of detecting pronunciation errors. Factorization can also take the form of an interpretable bottleneck layer that can be used to modify specific characteristics of the signal, e.g., make dysarthric speech more fluent and intelligible \cite{korzekwa2019interpretable}.

There is limited availability of non-native speech that is time-consuming to collect and difficult to annotate with phonetic transcriptions. Resorting to the probability theory and Bayes-rule, the problem of pronunciation error detection is reformulated as a speech generation task. Intuitively, if we had an unlimited amount of synthetic speech that could mimic non-native human speech, deep learning models for detecting pronunciation errors would be less prone to overfitting. The best proposed speech-to-speech generation method for generating mispronounced speech increases the accuracy of detecting pronunciation errors in the AUC metric by 41\%, from 0.528 to 0.749, compared to the state-of-the-art approach \cite{korzekwa22_speechcomm}. 

The experiments carried out to investigate the performance of the proposed approaches supported research theses no. 1 and no. 2 of this doctoral dissertation. In summary, the following major original contributions were introduced in this Ph.D. dissertation:
\begin{enumerate}
\item To reduce the need for phonetically transcribed non-native speech, the problem of pronunciation error detection has been reformulated as a speech generation task  \cite{korzekwa22_speechcomm}, which enables to generate synthetic mispronounced speech.
\item To eliminate the need to align canonical and recognized phoneme sequences and not rely on transcribed non-native speech, a novel end-to-end multi-task technique to directly detect pronunciation errors was proposed, called WEAKLY-S (Weakly-supervised) \cite{korzekwa21b_interspeech}.
\item To take into account the variability of pronunciation and the uncertainty in phoneme recognition while recognizing pronunciation errors, a new probabilistic deep learning architecture was proposed \cite{korzekwa2021mispronunciation}.
\item To automatically extract speech features in the pronunciation \cite{korzekwa21b_interspeech}  and lexical stress \cite{korzekwa21_interspeech} error detection tasks, the attention mechanism was proposed.
\item To enable the generation of mispronounced speech \cite{korzekwa21b_interspeech} and improve the fluency of disordered speech \cite{korzekwa2019interpretable}, controllable deep learning models were proposed.
\end{enumerate}

\section{Applicability}

The machine learning models created as part of the doctoral dissertation can be divided into two groups: models for automated pronunciation error detection and models of speech synthesis and voice conversion. Both types of models have been applied to real-world problems at Amazon.

The pronunciation error detection models were applied to automatically detect pronunciation errors in a synthetic speech in two scenarios: during inference and training of speech synthesis models. During inference, the goal is to automatically evaluate the quality of speech generated by speech synthesis models. After the speech synthesis model is trained, a large number of synthetic utterances are synthesized and automatically processed by the pronunciation error detection model. Automatically detecting pronunciation errors enables to evaluate synthetic voices on a large scale and greatly reduces the number of perceptual tests conducted by human listeners. During training, the pronunciation error detection model is used as a perceptual loss to ensure that the speech synthesis model will generate intelligible speech.

Speech synthesis and voice conversion pipelines consist of two steps, a context generation module that generates a mel-spectrogram from the input text and/or the input speech signal and a vocoder component that produces a raw speech signal based on the mel-spectrogram. Both components have been implemented into Alexa devices and serve millions of Amazon customers worldwide. In addition, synthetic speech generated by speech synthesis and voice conversion models improved the accuracy of the pronunciation error detection models in the synthetic speech evaluation task.

\section{Future work}
During the doctoral research, multiple interesting research directions emerged. The most forward-thinking idea is to continue the work from the Ph.D. research on reformulating the problem of pronunciation error detection as a speech generation task \cite{korzekwa22_speechcomm}. The proposed Speech-to-Speech (S2S) method can generate synthetic mispronounced speech but is not yet able to perfectly mimic non-native human speech. To improve the S2S method, a universal speech model should be created in order to generate any type of speech. The model should be able of transforming native speech into non-native speech, reflecting  the identity, prosody, speaking style, and pronunciation of the target speaker. This approach could make non-native human speech unnecessary, as the pronunciation error detection model will only be trained on synthetic speech data.  

Another interesting research direction is to explore unsupervised speech representations such as Wav2vec \cite {peng21e_interspeech}. A more compact speech representation might reduce the need for a large amount of speech data for training pronunciation error detection models. Multi-modal pronunciation error detection to benefit from audio-visual speech corpora is an attractive future direction as well \cite{czyzewski2017audio, oneata2022improving}.

So the vision is that future work will also focus on the development of a complete CAPT system with the goal of raising foreign language proficiency in the global population. An AI-based conversational agent will be created. The agent will consist of two elements:  a pronunciation error detection model and a feedback component. The pronunciation error detection model will be based on the results of this doctoral research, while the feedback component will require additional research. The CAPT system will only be controlled via the voice interface and the student will have a learning experience similar to the one provided by a human language teacher.

%% file: Appendices/Appendix_author_statements.tex
\chapter{Declaration of authorship} 
\label{sec:authorship} 
\includepdf[pages=-]{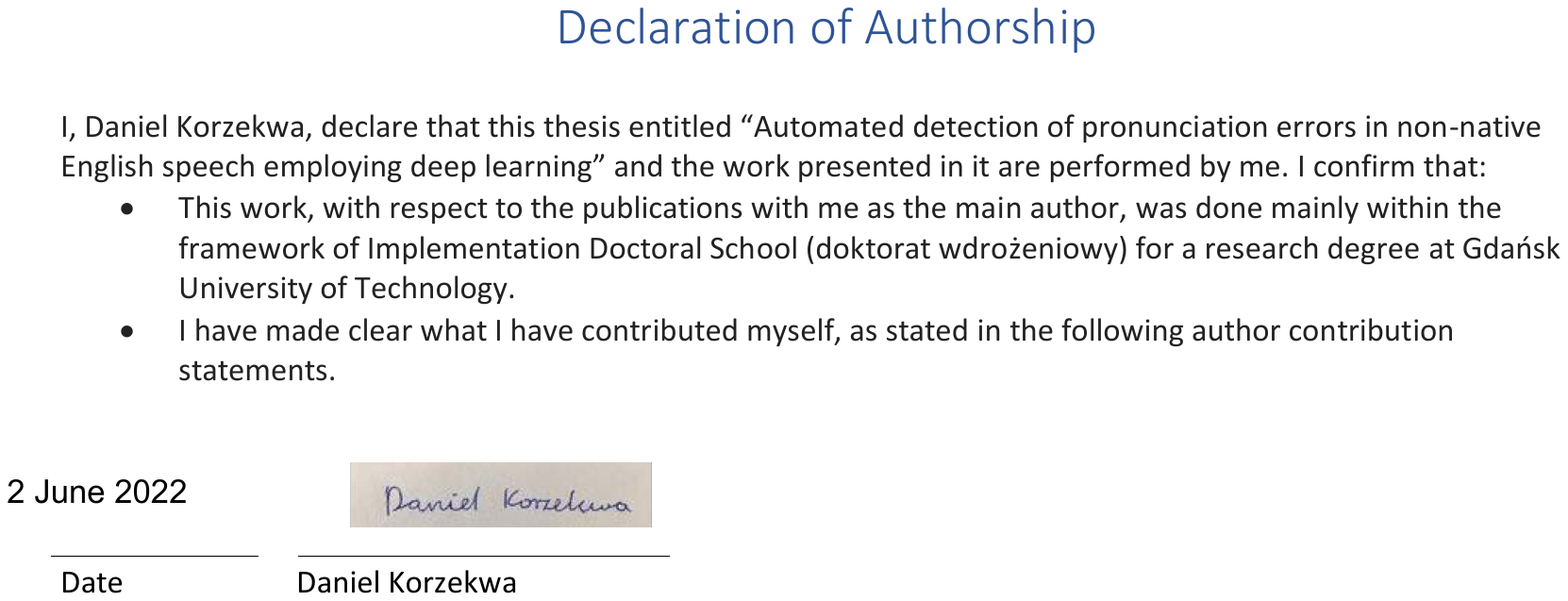}

\includepdf[pages=-]{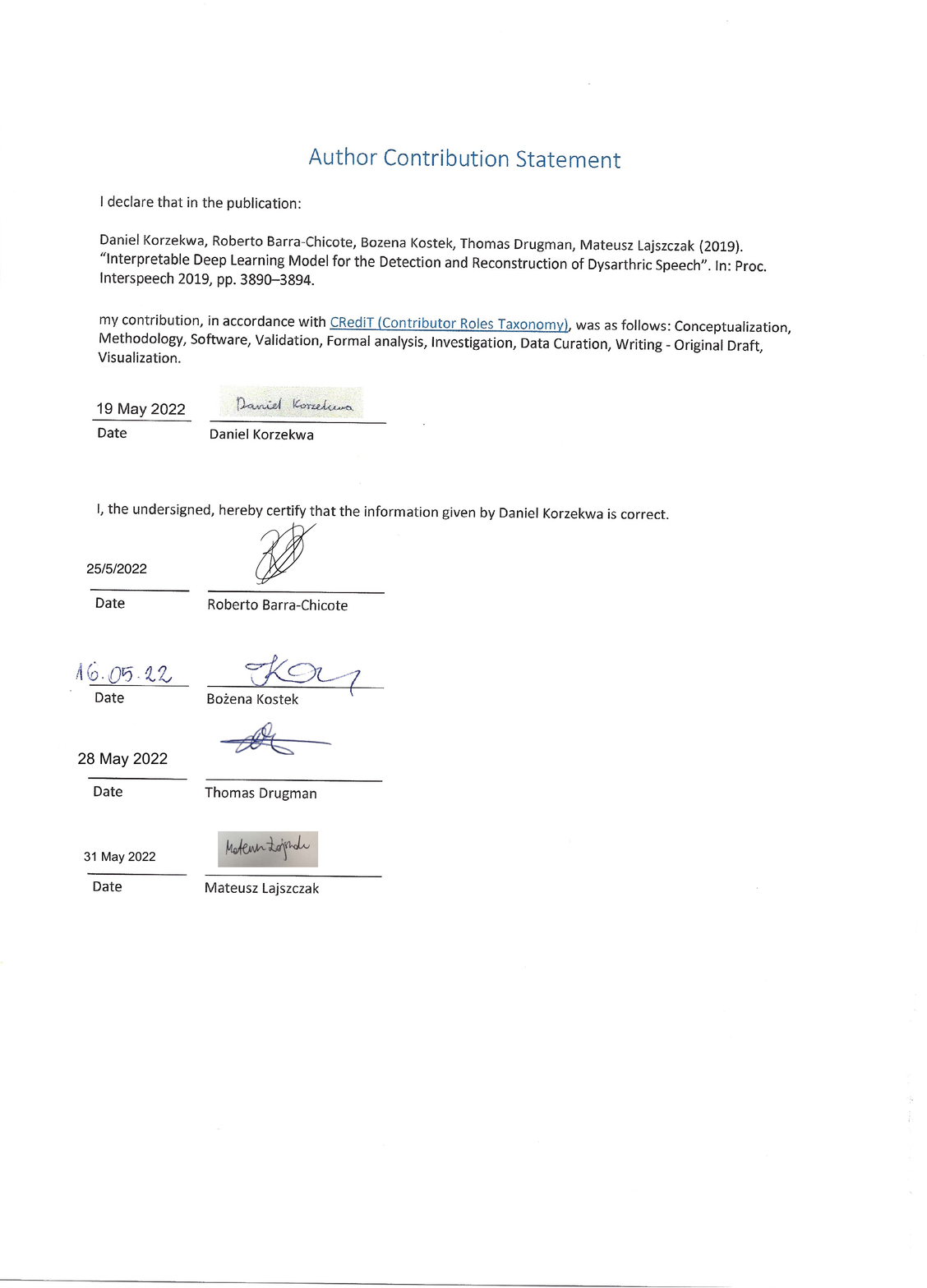}
\includepdf[pages=-]{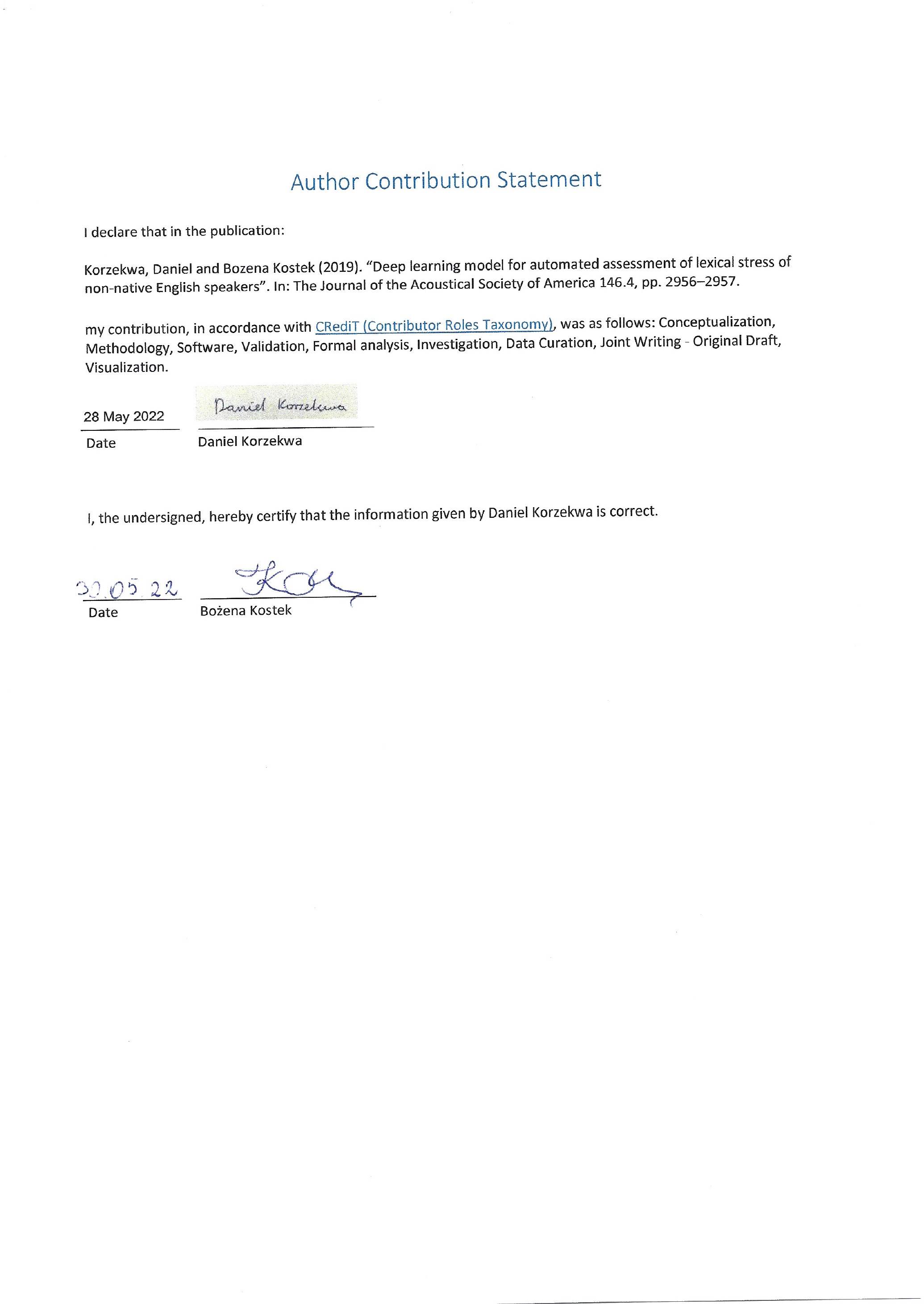}
\includepdf[pages=-]{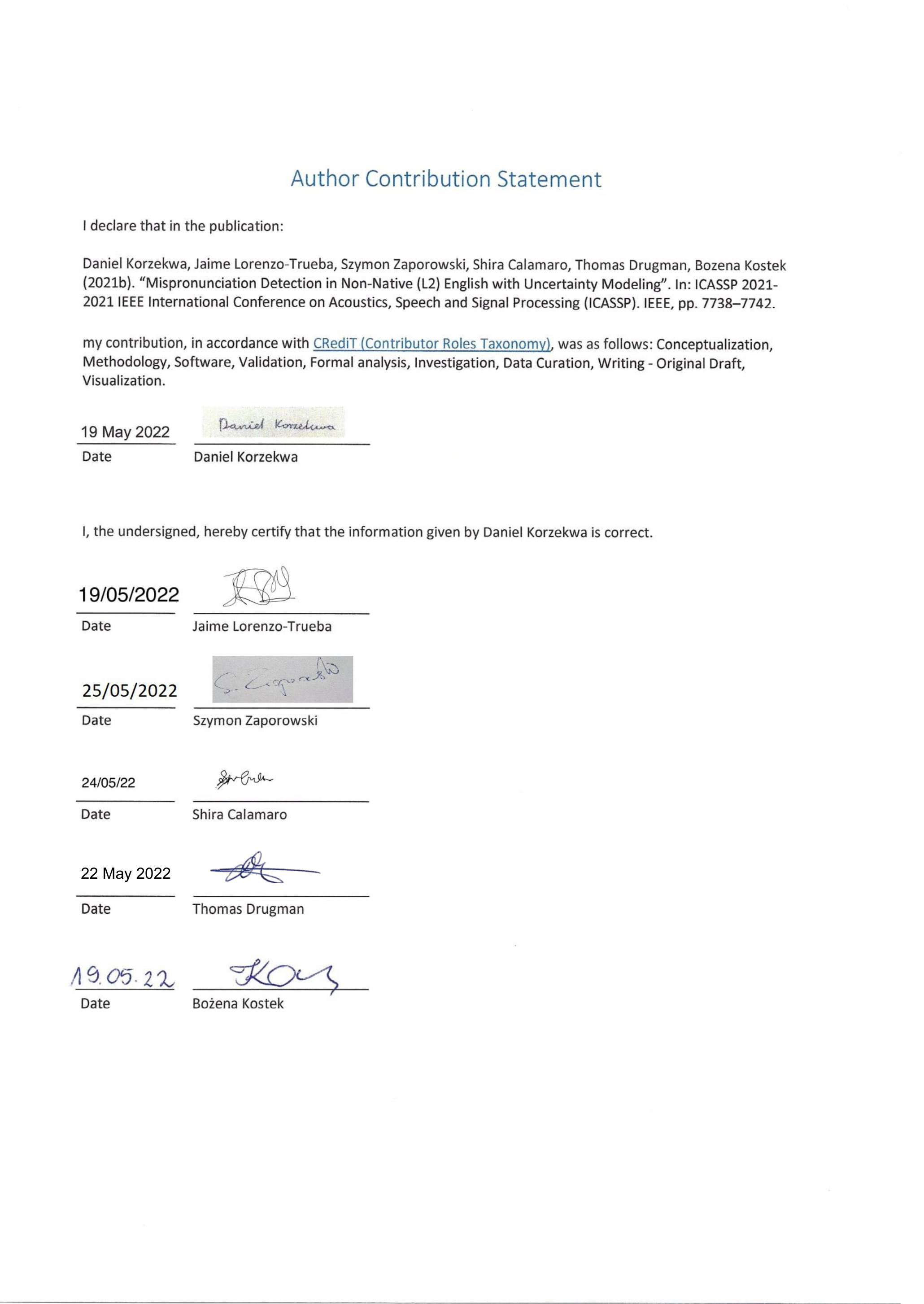}
\includepdf[pages=-]{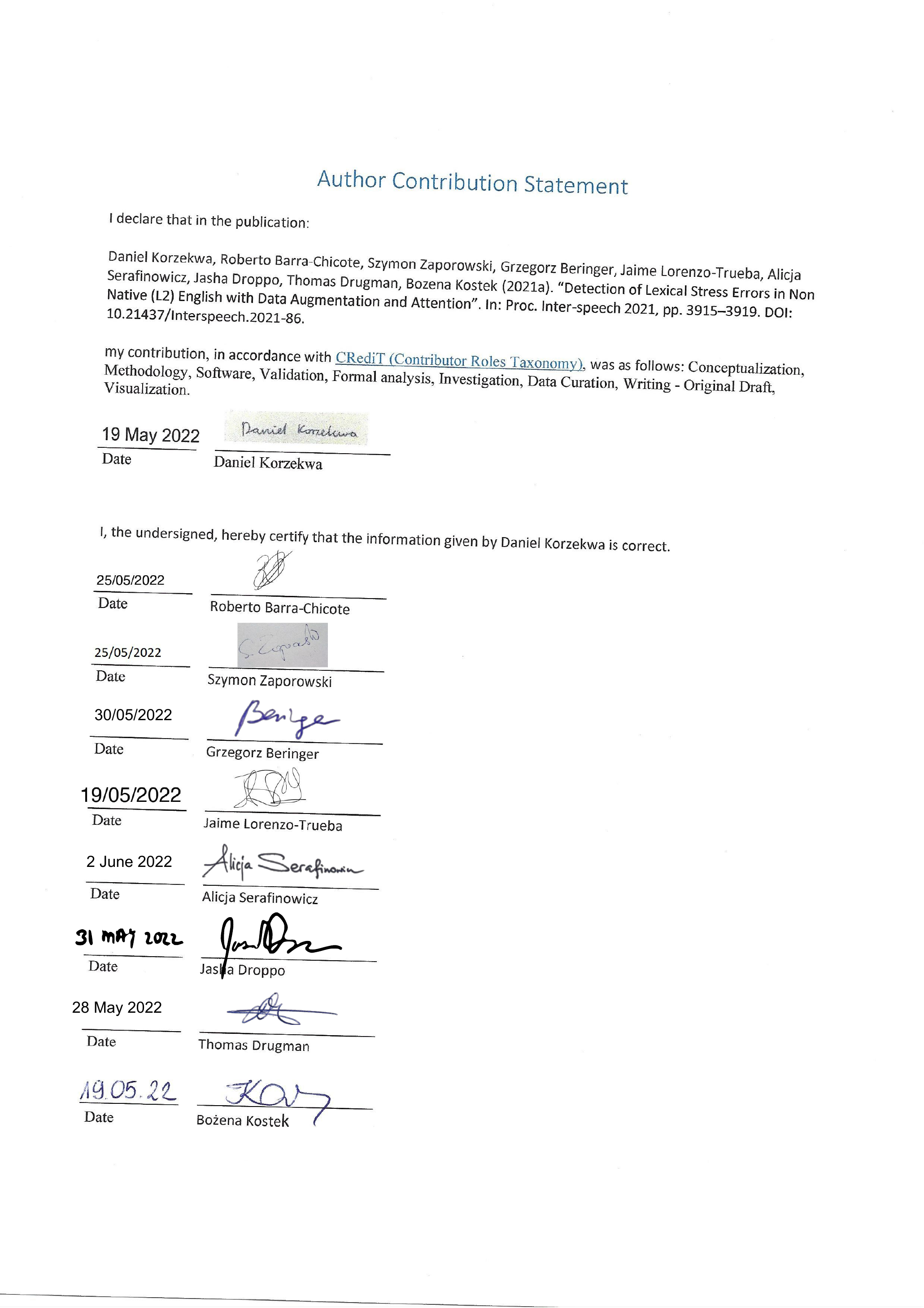}
\includepdf[pages=-]{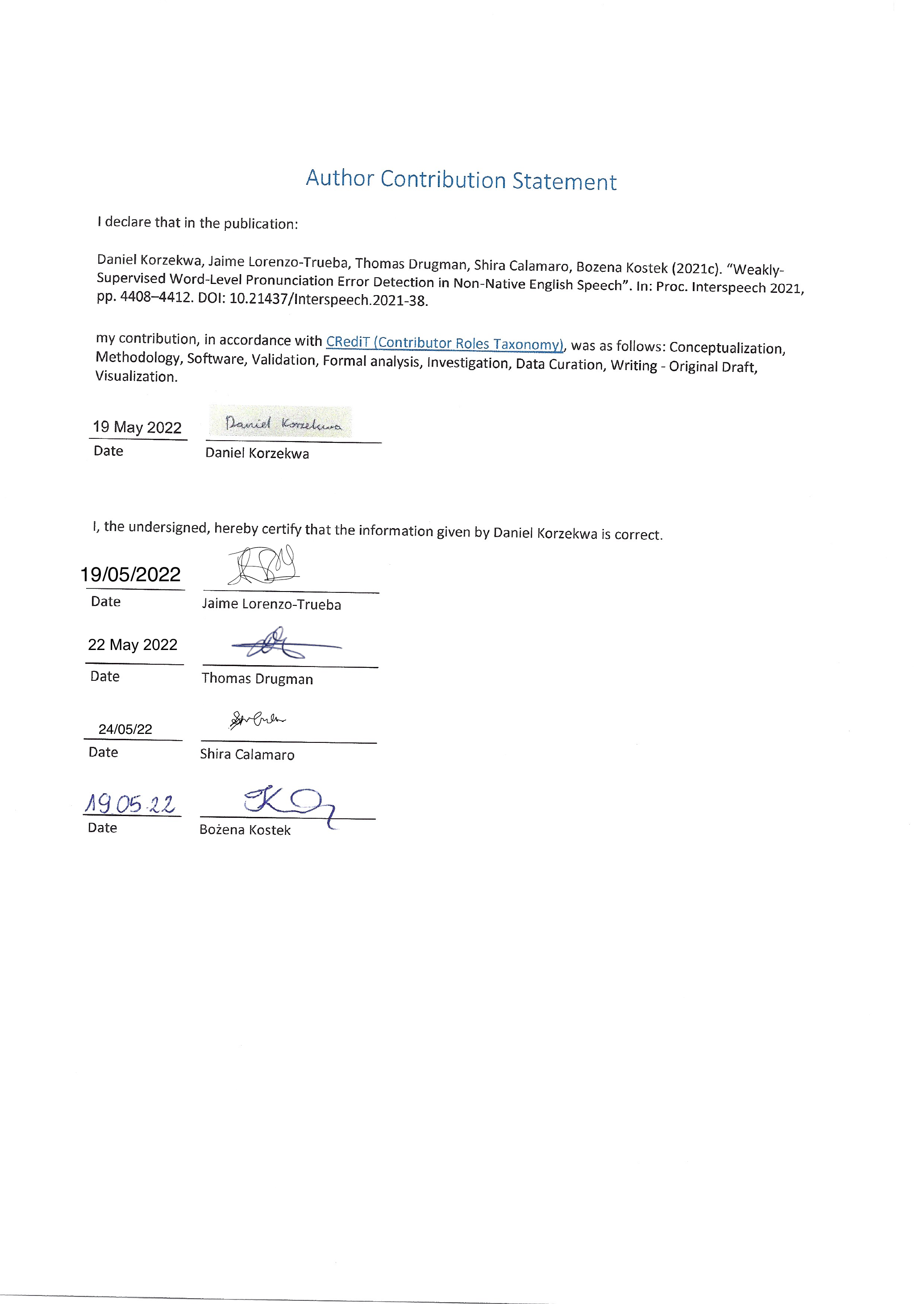}
\includepdf[pages=-]{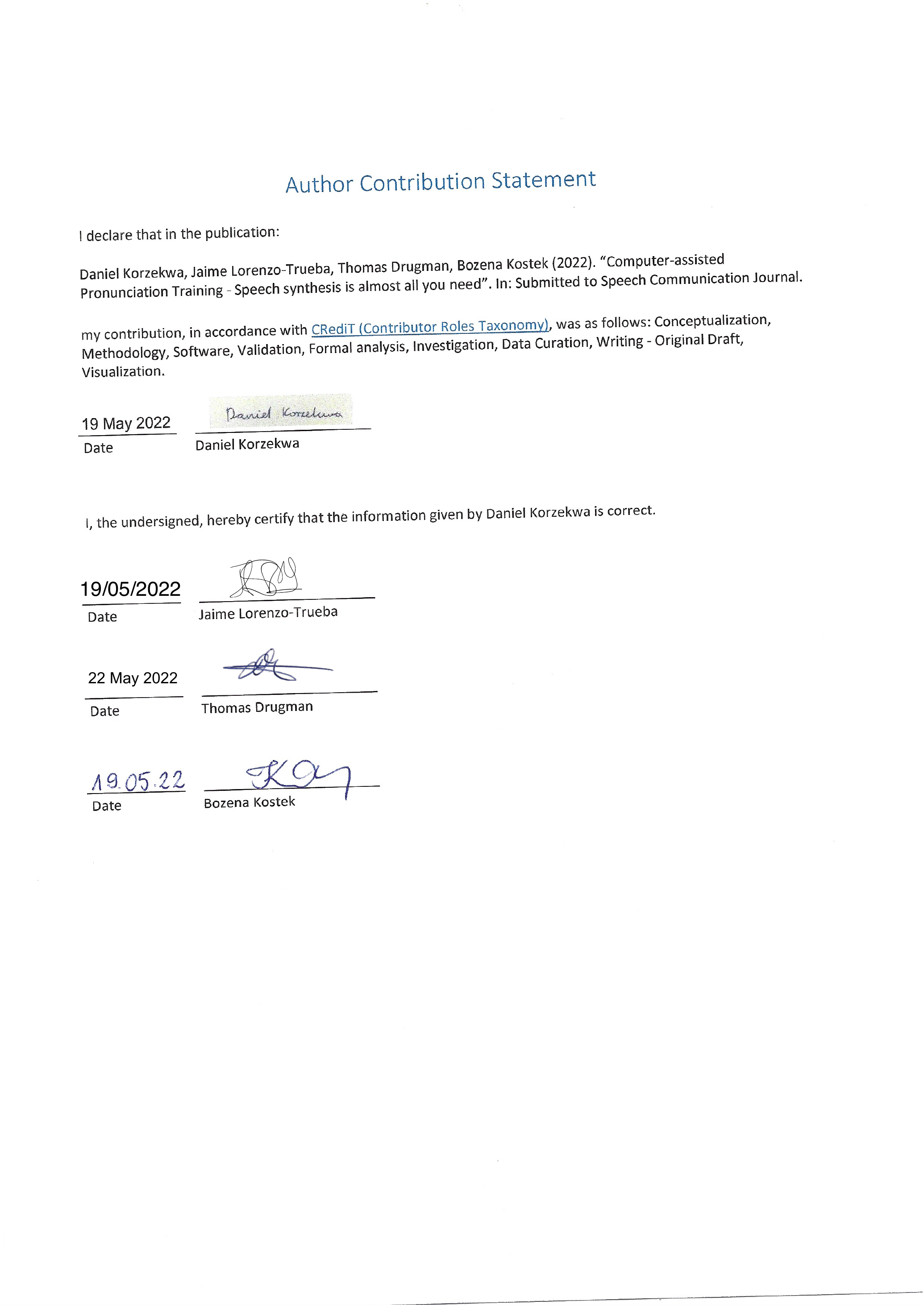}

%% file: Appendices/AppendixC.tex

\chapter{List of publications of the author of the doctoral dissertation} 

\label{AppendixC} 

The articles published or accepted for publication with Daniel Korzekwa as the primary author:

\begin{refcontext}[sorting=ydnt]
\printbibliography[heading=bibintoc, keyword=own,heading=none, title=none]
\end{refcontext}

Other articles co-authored by Daniel Korzekwa:

\begin{refcontext}[sorting=ydnt]
\printbibliography[heading=bibintoc, keyword=co-own,heading=none, title=none]
\end{refcontext}







%% file: Appendices/AppendixA.tex

\chapter{Primary author publications in the original format} 

\label{AppendixA} 

\includepdf[pages=-]{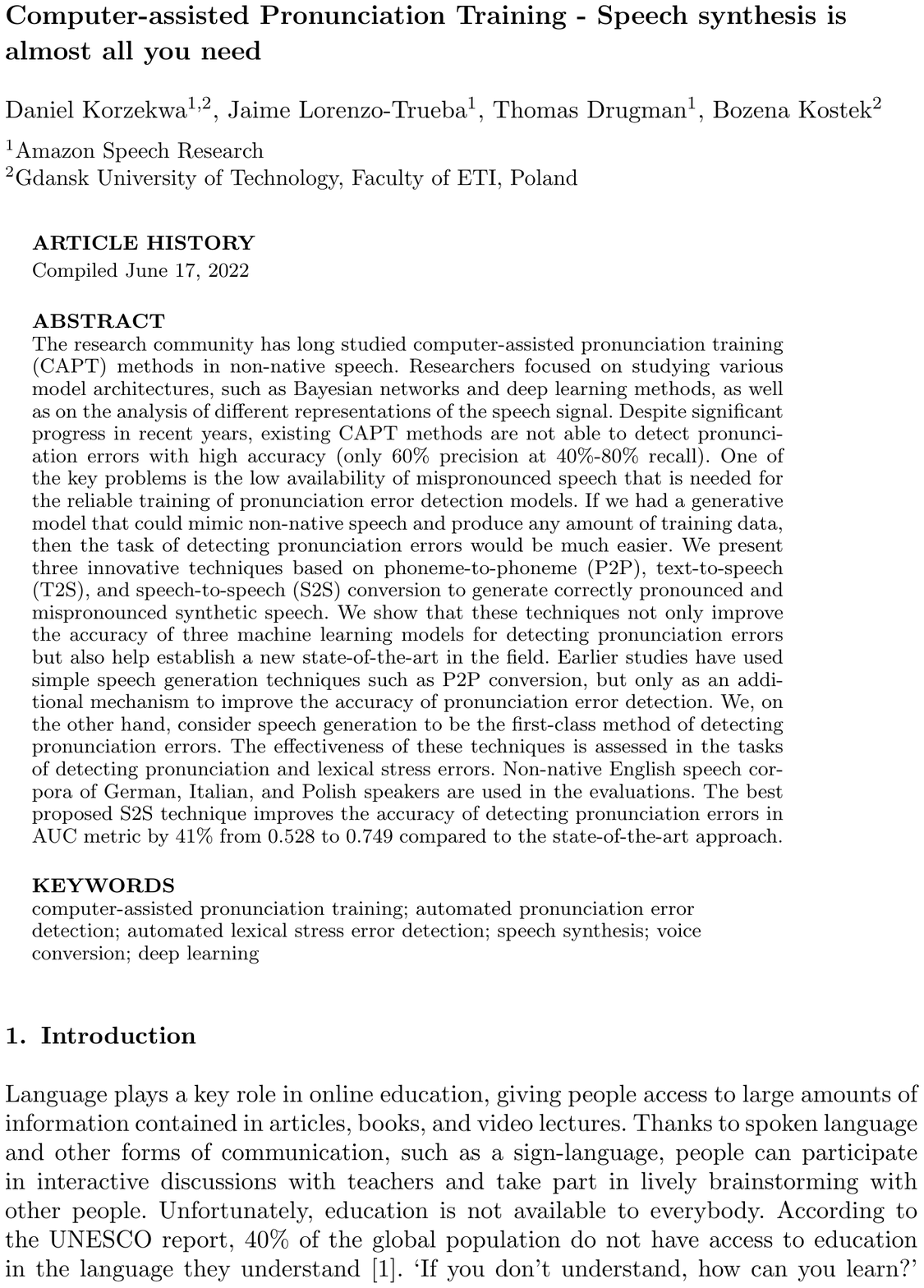}

\includepdf[pages=-]{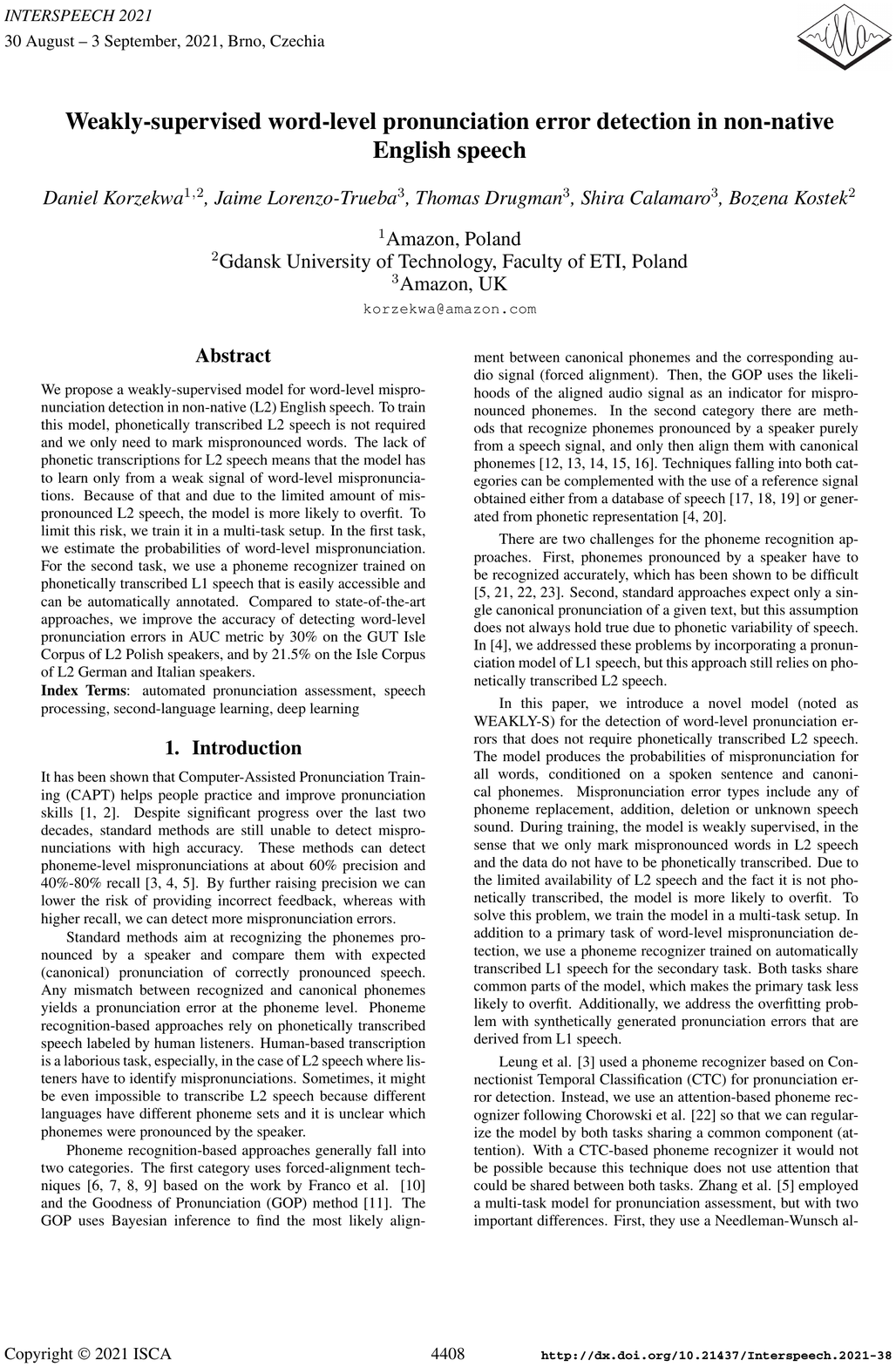}

\includepdf[pages=-]{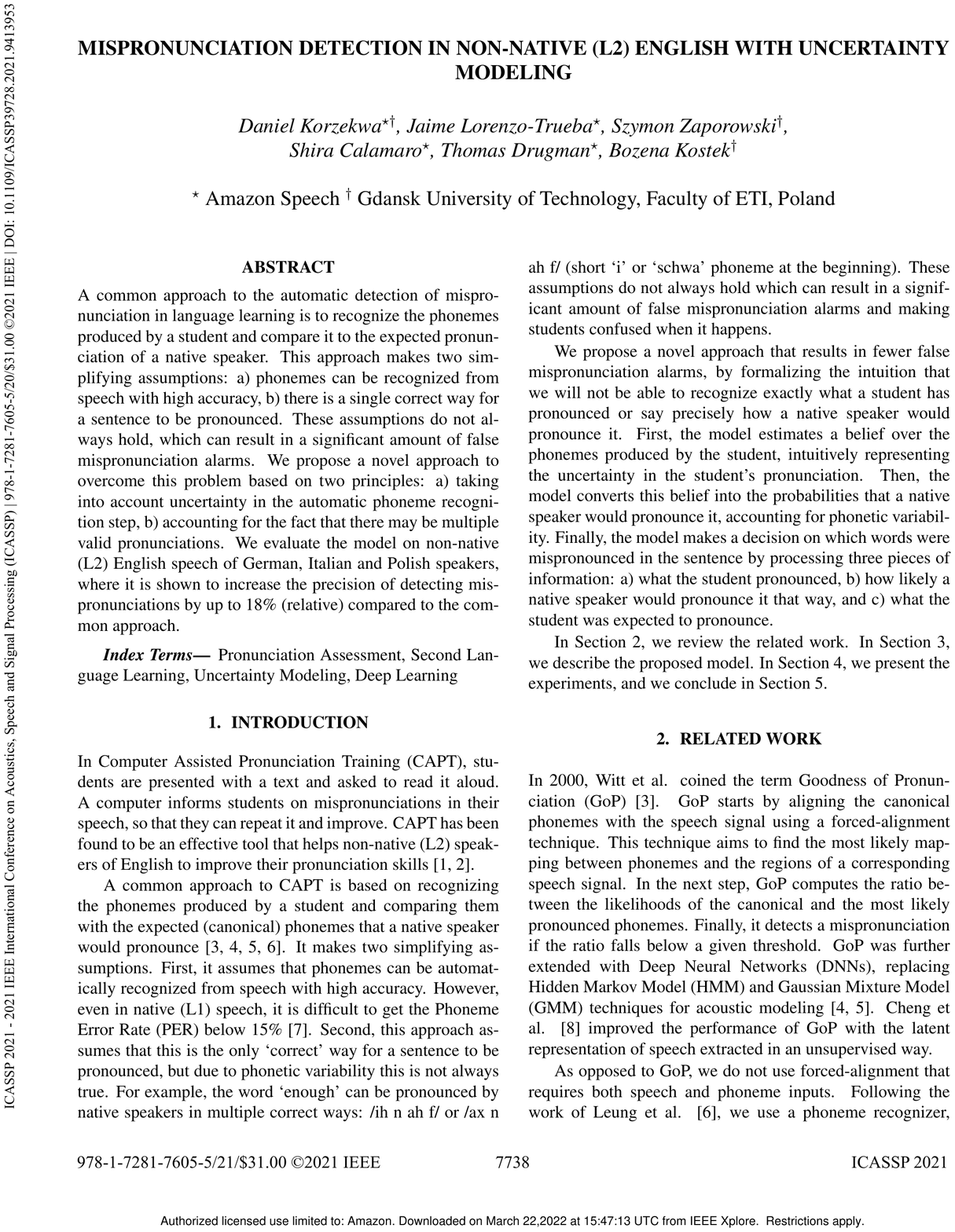}

\includepdf[pages=-]{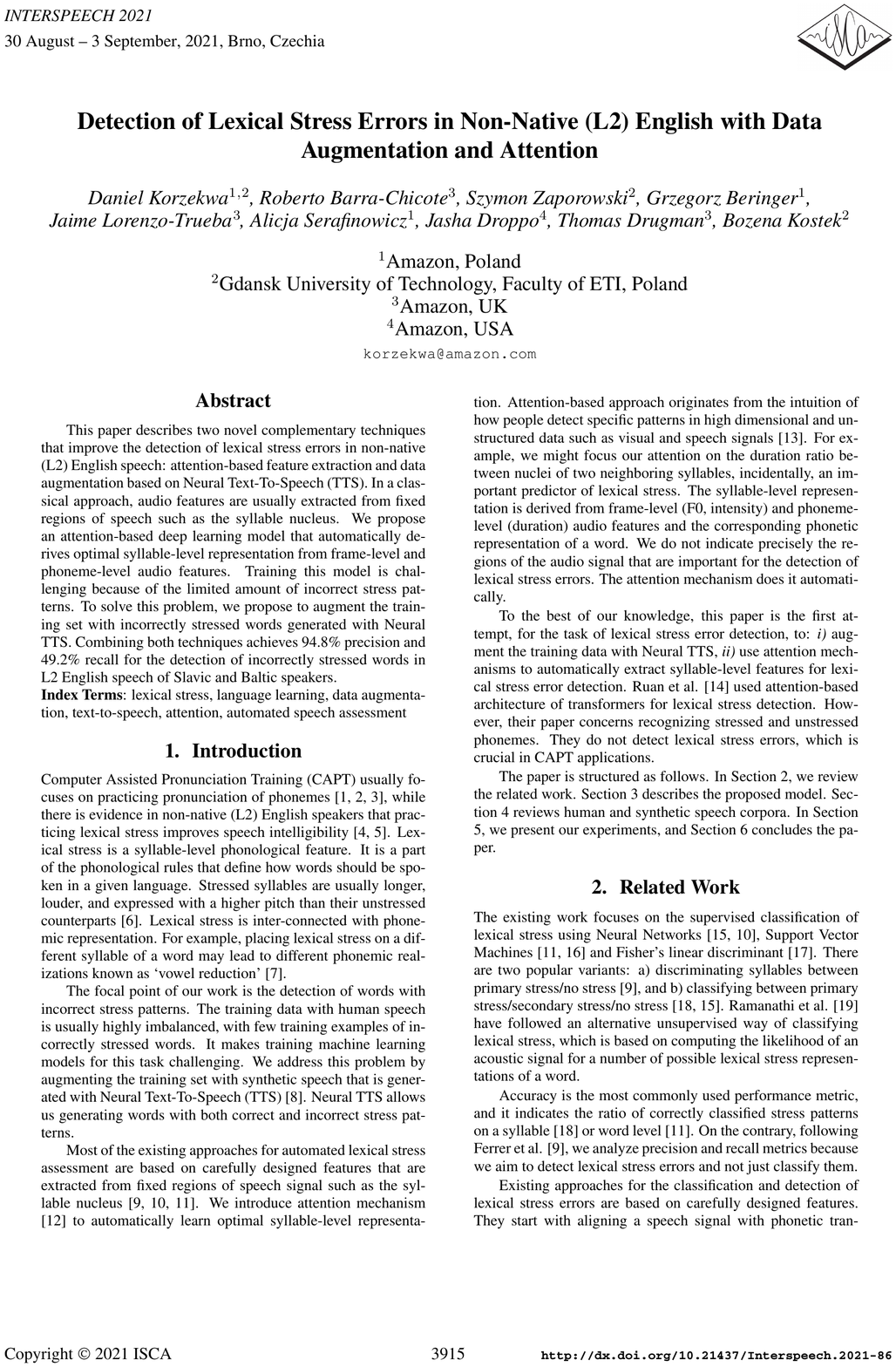}

\includepdf[pages=-]{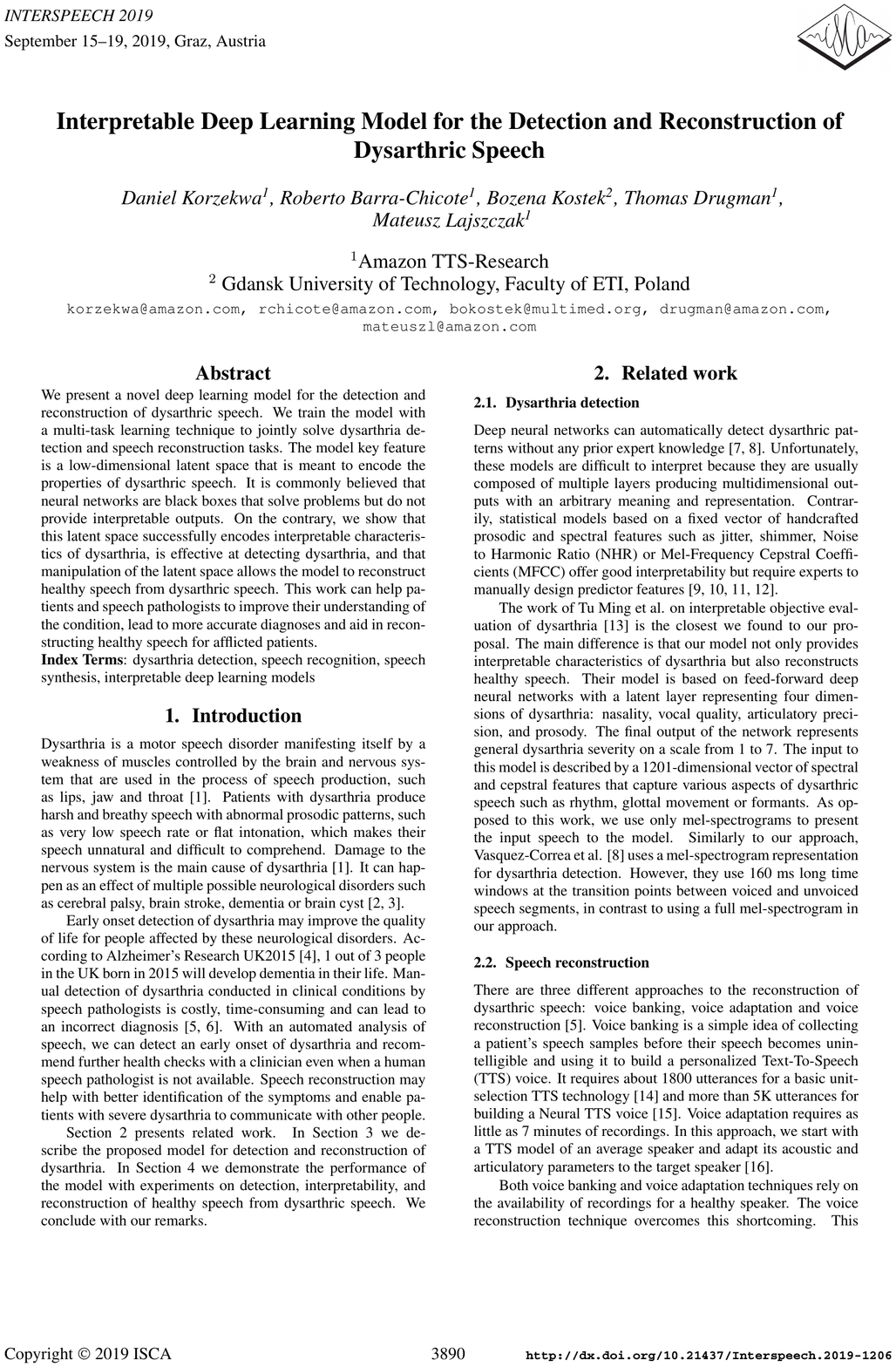}

%% file: Appendices/AppendixB.tex

\chapter{Co-authored publication on pronunciation error detection prior to Ph.D. research} 

\label{AppendixB} 

The first work on detecting pronunciation errors conducted by Daniel Korzekwa, preceding the doctorate, resulted in the co-authorship of the publication by Grzegorz Beringer. Grzegorz conducted a science internship on pronunciation assessment at Amazon, and Daniel Korzekwa was his mentor. The publication was presented at internal Amazon Machine Learning Conference (AMLC) in 2020, Seattle, United States.

\includepdf[pages=-]{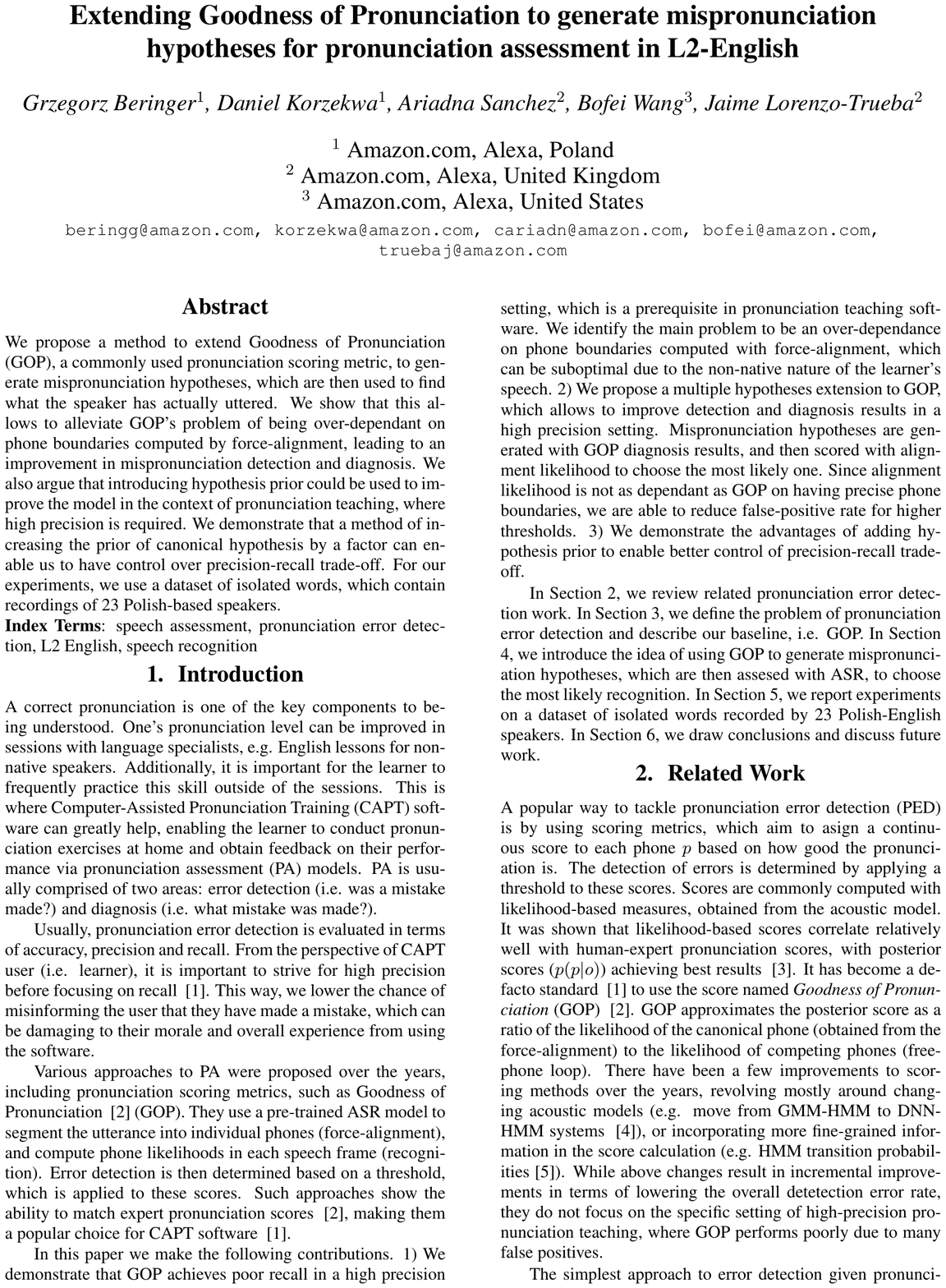}

%% file: dysarthric_speech.bib
@article{Banovic2018,
author = {Banovic, Silva and Zunic, Lejla and Sinanovic, Osman},
doi = {10.5455/msm.2018.30.221-224},
issn = {1512-7680},
journal = {Materia Socio Medica},
number = {2},
pages = {221},
title = {{Communication Difficulties as a Result of Dementia}},
url = {https://www.ejmanager.com/fulltextpdf.php?mno=302643414},
volume = {30},
year = {2018}
}

@article{ASHA2018,
author = {ASHA},
title = {{The American Speech-Language-Hearing Association (ASHA) - Dysarthria}},
@url = {https://www.asha.org/public/speech/disorders/dysarthria/},
year = {2018}
}

@article{Cuny2017,
author = {Cuny, M. L. and Pallone, M. and Piana, H. and Boddaert, N. and Sainte-Rose, C. and Vaivre-Douret, L. and Piolino, P. and Puget, S.},
doi = {10.1007/s00381-016-3285-x},
isbn = {0256-7040},
issn = {14330350},
journal = {Child's Nervous System},
pmid = {27832354},
title = {{Neuropsychological improvement after posterior fossa arachnoid cyst drainage}},
year = {2017}
}

@article{2012E121001,
abstract = {In this invited paper, we overview the clinical applications of speech synthesis technologies and explain a few selected researches. We also introduce the University of Edinburgh's new project ``Voice Banking and reconstruction'' for patients with degenerative diseases, such as motor neurone disease and Parkinson's disease and show how speech synthesis technologies can improve the quality of life for the patients.},
author = {Yamagishi, Junichi and Veaux, Christophe and King, Simon and Renals, Steve},
doi = {10.1250/ast.33.1},
journal = {Acoustical Science and Technology},
number = {1},
pages = {1--5},
title = {{Speech synthesis technologies for individuals with vocal disabilities: Voice banking and reconstruction}},
@url = {http://www.jstage.jst.go.jp/browse/ast/33/1/{\_}contents},
volume = {33},
year = {2012}
}

@article{DBLP:journals/corr/ChenLLLWWXXZZ15,
  author    = {Tianqi Chen and
               Mu Li and
               Yutian Li and
               Min Lin and
               Naiyan Wang and
               Minjie Wang and
               Tianjun Xiao and
               Bing Xu and
               Chiyuan Zhang and
               Zheng Zhang},
  title     = {MXNet: {A} Flexible and Efficient Machine Learning Library for Heterogeneous
               Distributed Systems},
  journal   = {CoRR},
  volume    = {abs/1512.01274},
  year      = {2015},
  url       = {http://arxiv.org/abs/1512.01274},
  archivePrefix = {arXiv},
  eprint    = {1512.01274},
  bibsource = {dblp computer science bibliography, https://dblp.org}
}

@article{Alzheimersresearchuk2015,
author = {Alzheimersresearchuk},
title = {{One in three people born in 2015 will develop dementia, new analysis shows}},
@url = {https://www.alzheimersresearchuk.org/one-in-three-2015-develop-dementia/},
year = {2015}
}

@inproceedings{Carmichael2008,
author = {Carmichael, James and Wan, Vincent and Green, Phil},
booktitle = {Proceedings of the Annual Conference of the International Speech Communication Association, INTERSPEECH},
issn = {19909772},
title = {{Combining neural network and rule-based systems for dysarthria diagnosis}},
year = {2008}
}

@article{Krishna,
author = {Krishna, Gurugubelli},
journal = {WSPD},
title = {{Excitation Source Analysis of Dysarthric Speech for Early Stage Detection of Dysarthria}},
year = {2018}
}

@inproceedings{DBLP:conf/interspeech/NarendraA18,
author = {Narendra, N P and Alku, Paavo},
booktitle = {Interspeech 2018, 19th Annual Conference of the International Speech Communication Association, Hyderabad, India, 2-6 September 2018.},
doi = {10.21437/Interspeech.2018-1059},
editor = {Yegnanarayana, B},
file = {:Users/korzekwa/Library/Application Support/Mendeley Desktop/Downloaded/Narendra, Alku - 2018 - Dysarthric Speech Classification Using Glottal Features Computed from Non-words, Words and Sentences.pdf:pdf},
pages = {3403--3407},
publisher = {ISCA},
title = {{Dysarthric Speech Classification Using Glottal Features Computed from Non-words, Words and Sentences}},
@url = {https://doi.org/10.21437/Interspeech.2018-1059},
year = {2018}
}

@inproceedings{DBLP:conf/interspeech/Vasquez-CorreaA18,
author = {V{\'{a}}squez-Correa, Juan Camilo and Arias-Vergara, Tomas and Orozco-Arroyave, Juan Rafael and N{\"{o}}th, Elmar},
booktitle = {Interspeech 2018, 19th Annual Conference of the International Speech Communication Association, Hyderabad, India, 2-6 September 2018.},
doi = {10.21437/Interspeech.2018-1988},
editor = {Yegnanarayana, B},
file = {:Users/korzekwa/Library/Application Support/Mendeley Desktop/Downloaded/V{\'{a}}squez-Correa et al. - 2018 - A Multitask Learning Approach to Assess the Dysarthria Severity in Patients with Parkinson's Disease.pdf:pdf},
pages = {456--460},
publisher = {ISCA},
title = {{A Multitask Learning Approach to Assess the Dysarthria Severity in Patients with Parkinson's Disease}},
@url = {https://doi.org/10.21437/Interspfeech.2018-1988},
year = {2018}
}

@article{Falk2012,
author = {Falk, Tiago H. and Chan, Wai Yip and Shein, Fraser},
doi = {10.1016/j.specom.2011.03.007},
isbn = {0167-6393},
issn = {01676393},
journal = {Speech Communication},
title = {{Characterization of atypical vocal source excitation, temporal dynamics and prosody for objective measurement of dysarthric word intelligibility}},
year = {2012}
}

@inproceedings{Sarria-Paja2012a,
author = {Sarria-Paja, Milton and Falk, Tiago},
booktitle = {Interspeech},
isbn = {9781622767595},
title = {{Automated Dysarthria Severity Classification for Improved Objective Intelligibility Assessment of Spastic Dysarthric Speech.}},
year = {2012}
}

@inproceedings{Gillespie2017,
author = {Gillespie, Stephanie and Logan, Yash Yee and Moore, Elliot and Laures-Gore, Jacqueline and Russell, Scott and Patel, Rupal},
booktitle = {Proceedings of the Annual Conference of the International Speech Communication Association, INTERSPEECH},
doi = {10.21437/Interspeech.2017-216},
issn = {19909772},
title = {{Cross-database models for the classification of dysarthria presence}},
year = {2017}
}

@article{Lansford2014,
archivePrefix = {arXiv},
arxivId = {NIHMS150003},
author = {Lansford, Kaitlin L. and Liss, Julie M.},
doi = {10.1044/1092-4388(2013/12-0262)},
eprint = {NIHMS150003},
isbn = {1092-4388},
issn = {1092-4388},
journal = {Journal of Speech Language and Hearing Research},
pmid = {24687467},
title = {{Vowel Acoustics in Dysarthria: Speech Disorder Diagnosis and Classification}},
year = {2014}
}

@inproceedings{DBLP:conf/interspeech/TuBL17,
author = {Tu, Ming and Berisha, Visar and Liss, Julie},
booktitle = {Interspeech 2017, 18th Annual Conference of the International Speech Communication Association, Stockholm, Sweden, August 20-24, 2017},
doi = {10.21437/Interspeech.2017},
editor = {Lacerda, Francisco},
file = {:Users/korzekwa/Library/Application Support/Mendeley Desktop/Downloaded/Tu, Berisha, Liss - 2017 - Interpretable Objective Assessment of Dysarthric Speech Based on Deep Neural Networks.pdf:pdf},
pages = {1849--1853},
publisher = {ISCA},
title = {{Interpretable Objective Assessment of Dysarthric Speech Based on Deep Neural Networks}},
@url = {http://www.isca-speech.org/archive/Interspeech{\_}2017/abstracts/1222.html},
year = {2017}
}

@misc{Modeltalker,
author = {Modeltalker},
title = {www.modeltalker.com}
}

@article{DBLP:journals/corr/abs-1811-06315,
archivePrefix = {arXiv},
arxivId = {1811.06315},
author = {Latorre, Javier and Lachowicz, Jakub and Lorenzo-Trueba, Jaime and Merritt, Thomas and Drugman, Thomas and Ronanki, Srikanth and Viacheslav, Klimkov},
eprint = {1811.06315},
journal = {CoRR},
title = {{Effect of data reduction on sequence-to-sequence neural {\{}TTS{\}}}},
@url = {http://arxiv.org/abs/1811.06315},
volume = {abs/1811.0},
year = {2018}
}

@misc{AhmadKhan2011,
author = {{Ahmad Khan}, Zahoor and Green, Phil and Creer, Sarah and Cunningham, Stuart},
booktitle = {AAC: Augmentative and Alternative Communication},
doi = {10.3109/07434618.2010.545078},
isbn = {1477-3848 (Electronic)$\backslash$r0743-4618 (Linking)},
issn = {07434618},
pmid = {21284563},
title = {{Reconstructing the voice of an individual following laryngectomy}},
year = {2011}
}

@book{rabiner_schafer78,
author = {Rabiner, L and Schafer, R},
publisher = {Englewood Cliffs: Prentice Hall},
title = {{Digital Processing of Speech Signals}},
year = {1978}
}

@article{Drugman2014,
author = {Drugman, Thomas and Alku, Paavo and Alwan, Abeer and Yegnanarayana, Bayya},
year = {2014},
month = {09},
pages = {},
title = {Glottal Source Processing: from Analysis to Applications},
volume = {28},
journal = {Computer Speech and Language},
doi = {10.1016/j.csl.2014.03.003}
}

@misc{doersch2016tutorial,
abstract = {In just three years, Variational Autoencoders (VAEs) have emerged as one of
the most popular approaches to unsupervised learning of complicated
distributions. VAEs are appealing because they are built on top of standard
function approximators (neural networks), and can be trained with stochastic
gradient descent. VAEs have already shown promise in generating many kinds of
complicated data, including handwritten digits, faces, house numbers, CIFAR
images, physical models of scenes, segmentation, and predicting the future from
static images. This tutorial introduces the intuitions behind VAEs, explains
the mathematics behind them, and describes some empirical behavior. No prior
knowledge of variational Bayesian methods is assumed.},
annote = {cite arxiv:1606.05908},
author = {Doersch, Carl},
keywords = {VAE sgd},
title = {{Tutorial on Variational Autoencoders}},
@url = {http://arxiv.org/abs/1606.05908},
year = {2016}
}

@article{DBLP:journals/corr/HuYLSX17,
archivePrefix = {arXiv},
arxivId = {1703.00955},
author = {Hu, Zhiting and Yang, Zichao and Liang, Xiaodan and Salakhutdinov, Ruslan and Xing, Eric P},
eprint = {1703.00955},
journal = {CoRR},
title = {{Controllable Text Generation}},
@url = {http://arxiv.org/abs/1703.00955},
volume = {abs/1703.0},
year = {2017}
}

@article{DBLP:journals/corr/BowmanVVDJB15,
archivePrefix = {arXiv},
arxivId = {1511.06349},
author = {Bowman, Samuel R and Vilnis, Luke and Vinyals, Oriol and Dai, Andrew M and J{\'{o}}zefowicz, Rafal and Bengio, Samy},
eprint = {1511.06349},
journal = {CoRR},
title = {{Generating Sentences from a Continuous Space}},
@url = {http://arxiv.org/abs/1511.06349},
volume = {abs/1511.0},
year = {2015}
}

@article{DBLP:journals/corr/abs-1812-04342,
archivePrefix = {arXiv},
arxivId = {1812.04342},
author = {Zhang, Ya-Jie and Pan, Shifeng and He, Lei and Ling, Zhen-Hua},
eprint = {1812.04342},
journal = {CoRR},
title = {{Learning latent representations for style control and transfer in end-to-end speech synthesis}},
@url = {http://arxiv.org/abs/1812.04342},
volume = {abs/1812.0},
year = {2018}
}

@article{DBLP:journals/corr/abs-1709-07902,
archivePrefix = {arXiv},
arxivId = {1709.07902},
author = {Hsu, Wei-Ning and Zhang, Yu and Glass, James R},
eprint = {1709.07902},
journal = {CoRR},
title = {{Unsupervised Learning of Disentangled and Interpretable Representations from Sequential Data}},
@url = {http://arxiv.org/abs/1709.07902},
volume = {abs/1709.0},
year = {2017}
}

@inproceedings{journals/jmlr/GlorotB10,
  author = {Glorot, Xavier and Bengio, Yoshua},
  booktitle = {AISTATS},
  editor = {Teh, Yee Whye and Titterington, D. Mike},
  pages = {249-256},
  publisher = {JMLR.org},
  series = {JMLR Proceedings},
  title = {Understanding the difficulty of training deep feedforward neural networks.},
  volume = 9,
  year = 2010
}

@article{DBLP:journals/corr/abs-1803-09047,
archivePrefix = {arXiv},
arxivId = {1803.09047},
author = {Skerry-Ryan, R J and Battenberg, Eric and Xiao, Ying and Wang, Yuxuan and Stanton, Daisy and Shor, Joel and Weiss, Ron J and Clark, Rob and Saurous, Rif A},
eprint = {1803.09047},
journal = {CoRR},
title = {{Towards End-to-End Prosody Transfer for Expressive Speech Synthesis with Tacotron}},
@url = {http://arxiv.org/abs/1803.09047},
volume = {abs/1803.0},
year = {2018}
}

@article{DBLP:journals/corr/ChoMGBSB14,
archivePrefix = {arXiv},
arxivId = {1406.1078},
author = {Cho, Kyunghyun and van Merrienboer, Bart and G{\"{u}}l{\c{c}}ehre, {\c{C}}aglar and Bougares, Fethi and Schwenk, Holger and Bengio, Yoshua},
eprint = {1406.1078},
journal = {CoRR},
title = {{Learning Phrase Representations using {\{}RNN{\}} Encoder-Decoder for Statistical Machine Translation}},
@url = {http://arxiv.org/abs/1406.1078},
volume = {abs/1406.1},
year = {2014}
}

@article{JMLR:v15:srivastava14a,
author = {Srivastava, Nitish and Hinton, Geoffrey and Krizhevsky, Alex and Sutskever, Ilya and Salakhutdinov, Ruslan},
journal = {Journal of Machine Learning Research},
pages = {1929--1958},
title = {{Dropout: A Simple Way to Prevent Neural Networks from Overfitting}},
@url = {http://jmlr.org/papers/v15/srivastava14a.html},
volume = {15},
year = {2014}
}

@article{DBLP:journals/corr/WangSSWWJYXCBLA17,
archivePrefix = {arXiv},
arxivId = {1703.10135},
author = {Wang, Yuxuan and Skerry-Ryan, R J and Stanton, Daisy and Wu, Yonghui and Weiss, Ron J and Jaitly, Navdeep and Yang, Zongheng and Xiao, Ying and Chen, Zhifeng and Bengio, Samy and Le, Quoc V and Agiomyrgiannakis, Yannis and Clark, Rob and Saurous, Rif A},
eprint = {1703.10135},
journal = {CoRR},
title = {{Tacotron: {\{}A{\}} Fully End-to-End Text-To-Speech Synthesis Model}},
@url = {http://arxiv.org/abs/1703.10135},
volume = {abs/1703.1},
year = {2017}
}

@article{DBLP:journals/corr/VaswaniSPUJGKP17,
archivePrefix = {arXiv},
arxivId = {1706.03762},
author = {Vaswani, Ashish and Shazeer, Noam and Parmar, Niki and Uszkoreit, Jakob and Jones, Llion and Gomez, Aidan N and Kaiser, Lukasz and Polosukhin, Illia},
eprint = {1706.03762},
journal = {CoRR},
title = {{Attention Is All You Need}},
@url = {http://arxiv.org/abs/1706.03762},
volume = {abs/1706.0},
year = {2017}
}

@article{Kim2008,
author = {Kim, Heejin and Hasegawa-Johnson, Mark and Perlman, Adrienne and Gunderson, Jon and Huang, Thomas and Watkin, Kenneth and Frame, Simone},
issn = {19909772},
journal = {INTERSPEECH},
title = {{Dysarthric Speech Database for Universal Access Research}},
year = {2008}
}

@book{Johnston:2012:WAR:2432294,
 author = {Johnston, Alan B. and Burnett, Daniel C.},
 title = {WebRTC: APIs and RTCWEB Protocols of the HTML5 Real-Time Web},
 year = {2012},
 isbn = {0985978805, 9780985978808},
 publisher = {Digital Codex LLC},
 address = {USA},
}

@article{Rudzicz2012,
archivePrefix = {arXiv},
arxivId = {arXiv:astro-ph/0507464v2},
author = {Rudzicz, Frank and Namasivayam, Aravind Kumar and Wolff, Talya},
doi = {10.1007/s10579-011-9145-0},
eprint = {0507464v2},
isbn = {1574020X},
issn = {1574020X},
journal = {Language Resources and Evaluation},
pmid = {12658535},
primaryClass = {arXiv:astro-ph},
title = {{The TORGO database of acoustic and articulatory speech from speakers with dysarthria}},
year = {2012}
}

@inproceedings{Nicolao2016,
author = {Nicolao, Mauro and Christensen, Heidi and Cunningham, Stuart and Green, Phil and Hain, Thomas},
booktitle = {Proceedings of the 10th International Conference on Language Resources and Evaluation, LREC 2016},
isbn = {9782951740891},
title = {{A framework for collecting realistic recordings of dysarthric speech - The homeService corpus}},
year = {2016}
}

@article{Griffin1984,
author = {Griffin, Daniel W. and Lim, Jae S.},
doi = {10.1109/TASSP.1984.1164317},
isbn = {0096-3518 VO - 32},
issn = {00963518},
journal = {IEEE Transactions on Acoustics, Speech, and Signal Processing},
pmid = {1172092},
title = {{Signal Estimation from Modified Short-Time Fourier Transform}},
year = {1984}
}

@article{Mathieu2018,
abstract = {We develop a generalisation of disentanglement in variational auto-encoders (VAEs)---decomposition of the latent representation---characterising it as the fulfilment of two factors: a) the latent encodings of the data having an appropriate level of overlap, and b) the aggregate encoding of the data conforming to a desired structure, represented through the prior. Decomposition permits disentanglement, i.e. explicit independence between latents, as a special case, but also allows for a much richer class of properties to be imposed on the learnt representation, such as sparsity, clustering, independent subspaces, or even intricate hierarchical dependency relationships. We show that the {\$}\backslashbeta{\$}-VAE varies from the standard VAE predominantly in its control of latent overlap and that for the standard choice of an isotropic Gaussian prior, its objective is invariant to rotations of the latent representation. Viewed from the decomposition perspective, breaking this invariance with simple manipulations of the prior can yield better disentanglement with little or no detriment to reconstructions. We further demonstrate how other choices of prior can assist in producing different decompositions and introduce an alternative training objective that allows the control of both decomposition factors in a principled manner.},
archivePrefix = {arXiv},
arxivId = {1812.02833},
author = {Mathieu, Emile and Rainforth, Tom and Siddharth, N. and Teh, Yee Whye},
eprint = {1812.02833},
title = {{Disentangling Disentanglement in Variational Auto-Encoders}},
@url = {http://arxiv.org/abs/1812.02833},
year = {2018}
}


%% file: example.bib
@article{moon1996expectation,
  title={The expectation-maximization algorithm},
  author={Moon, Todd K},
  journal={IEEE Signal processing magazine},
  volume={13},
  number={6},
  pages={47--60},
  year={1996},
  doi={10.1109/79.543975},
  publisher={IEEE}
}

@article{heck2000robustness,
  title={Robustness to telephone handset distortion in speaker recognition by discriminative feature design},
  author={Heck, Larry P and Konig, Yochai and S{\"o}nmez, M Kemal and Weintraub, Mitch},
  journal={Speech Communication},
  volume={31},
  number={2-3},
  pages={181--192},
  year={2000},
  doi = {10.1016/S0167-6393(99)00077-1},
  publisher={Elsevier}
}

@article{welch2003hidden,
  title={Hidden Markov models and the Baum-Welch algorithm},
  author={Welch, Lloyd R},
  journal={IEEE Information Theory Society Newsletter},
  volume={53},
  number={4},
  pages={10--13},
  url={http://yanfenglu.net/documents/Baum-Welch_Algorithm.pdf},
  year={2003}
}

@article{field2005intelligibility,
  title={Intelligibility and the listener: The role of lexical stress},
  author={Field, John},
  journal={TESOL quarterly},
  volume={39},
  number={3},
  pages={399--423},
  year={2005},
  doi={10.2307/3588487},
  publisher={Wiley Online Library}
}

@article{trujillo2006production,
  title={The production of speech sounds},
  author={Trujillo, Fernando},
  journal={English Phonetics and Phonology},
  url={https://www.ugr.es/~ftsaez/fonetica/production_speech.pdf},
  year={2006}
}

@book{williams2006gaussian,
  title={Gaussian processes for machine learning},
  author={Williams, Christopher K and Rasmussen, Carl Edward},
  volume={2},
  number={3},
  year={2006},
  doi={10.7551/mitpress/3206.001.0001},
  publisher={MIT press Cambridge, MA}
}

@article{bishop2006pattern,
  title={Pattern recognition},
  author={Bishop, Christopher M},
  journal={Machine learning},
  volume={128},
  number={9},
  year={2006}
}

@article{abdi2007binomial,
  title={Binomial distribution: Binomial and sign tests},
  author={Abdi, Herv{\'e}},
  journal={Encyclopedia of measurement and statistics},
  volume={1},
  year={2007},
  publisher={Citeseer}
}

@article{woolson2007wilcoxon,
  title={Wilcoxon signed-rank test},
  author={Woolson, RF},
  journal={Wiley encyclopedia of clinical trials},
  pages={1--3},
  year={2007},
  publisher={Wiley Online Library}
}

@article{lee2008introduction,
  title={Introduction to statistics},
  author={Lee, Yong-Gu and Kim, Sam-Yong},
  journal={Yulgokbooks, Korea},
  pages={342--351},
  year={2008}
}

@book{darwiche2009modeling,
  title={Modeling and reasoning with Bayesian networks},
  author={Darwiche, Adnan},
  year={2009},
  publisher={Cambridge university press}
}

@book{koller2009probabilistic,
  title={Probabilistic graphical models: principles and techniques},
  author={Koller, Daphne and Friedman, Nir},
  year={2009},
  publisher={MIT press}
}

@book{denham2012linguistics,
  title={Linguistics for everyone: An introduction},
  author={Denham, Kristin and Lobeck, Anne},
  year={2012},
  publisher={Cengage Learning}
}

@book{murphy2012machine,
  title={Machine learning: a probabilistic perspective},
  author={Murphy, Kevin P},
  year={2012},
  publisher={MIT press}
}

@article{ali2012random,
  title={Random forests and decision trees},
  author={Ali, Jehad and Khan, Rehanullah and Ahmad, Nasir and Maqsood, Imran},
  journal={International Journal of Computer Science Issues (IJCSI)},
  volume={9},
  number={5},
  pages={272},
  year={2012},
  publisher={Citeseer}
}

@inproceedings{damianou2013deep,
  title={Deep gaussian processes},
  author={Damianou, Andreas and Lawrence, Neil D},
  booktitle={Artificial intelligence and statistics},
  pages={207--215},
  year={2013},
  organization={PMLR}
}

@article{minka2013expectation,
  title={Expectation propagation for approximate Bayesian inference},
  author={Minka, Thomas P},
  journal={arXiv preprint arXiv:1301.2294},
  year={2013}
}

@article{series2014method,
  title={Method for the subjective assessment of intermediate quality level of audio systems},
  author={Series, B},
  journal={International Telecommunication Union Radiocommunication Assembly},
  year={2014}
}

@book{sarkka2013bayesian,
  title={Bayesian filtering and smoothing},
  author={S{\"a}rkk{\"a}, Simo},
  number={3},
  year={2013},
  publisher={Cambridge University Press}
}

@inproceedings{lepage2014intelligibility,
  title={Intelligibility of English L2: The effects of incorrect word stress placement and incorrect vowel reduction in the speech of French and Italian learners of English},
  author={Lepage, Andr{\'e}e and Bus{\`a}, Maria Grazia},
  booktitle={Proceedings of the International Symposium on the Acquisition of Second Language Speech Concordia Working Papers in Applied Linguistics},
  volume={5},
  number={2014},
  pages={387--400},
  year={2014}
}

@article{duvenaud2014kernel,
  title={The Kernel cookbook: Advice on covariance functions, accessed on June 2022},
  author={Duvenaud, David},
  url={https://www.cs.toronto.edu/~duvenaud/cookbook/},
  year={2014}
}

@book{todhunter2014history,
  title={A history of the mathematical theory of probability},
  author={Todhunter, Isaac},
  year={2014},
  publisher={Cambridge University Press}
}

@article{lecun2015deep,
  title={Deep learning},
  author={LeCun, Yann and Bengio, Yoshua and Hinton, Geoffrey},
  journal={nature},
  volume={521},
  number={7553},
  pages={436--444},
  year={2015},
  publisher={Nature Publishing Group}
}

@article{lake2015human,
  title={Human-level concept learning through probabilistic program induction},
  author={Lake, Brenden M and Salakhutdinov, Ruslan and Tenenbaum, Joshua B},
  journal={Science},
  volume={350},
  number={6266},
  pages={1332--1338},
  year={2015},
  publisher={American Association for the Advancement of Science}
}

@article{fouz2015trends,
  title={Trends and directions in computer-assisted pronunciation training},
  author={Fouz-Gonz{\'a}lez, Jon{\'a}s},
  journal={Investigating English Pronunciation},
  pages={314--342},
  year={2015},
  publisher={Springer}
}

@inproceedings{wong2016understanding,
  title={Understanding data augmentation for classification: when to warp?},
  author={Wong, Sebastien C and Gatt, Adam and Stamatescu, Victor and McDonnell, Mark D},
  booktitle={2016 international conference on digital image computing: techniques and applications (DICTA)},
  pages={1--6},
  year={2016},
  organization={IEEE}
}

@book{goodfellow2016deep,
  title={Deep learning},
  author={Goodfellow, Ian and Bengio, Yoshua and Courville, Aaron},
  year={2016},
  publisher={MIT press}
}

@article{koyuncu2016speech,
  title={Speech and language therapy for aphasia following subacute stroke},
  author={Koyuncu, Engin and {\c{C}}am, P{\i}nar and Alt{\i}nok, Nermin and {\c{C}}all{\i}, Duygu Ekinci and Duman, Tuba Yarbay and {\"O}zgirgin, Ne{\c{s}}e},
  journal={Neural Regeneration Research},
  volume={11},
  number={10},
  pages={1591},
  year={2016},
  publisher={Wolters Kluwer--Medknow Publications}
}

@article{farrajota2012speech,
  title={Speech therapy in primary progressive aphasia: a pilot study},
  author={Farrajota, Lu{\'\i}sa and Maruta, Carolina and Maroco, Jo{\~a}o and Martins, Isabel Pav{\~a}o and Guerreiro, Manuela and De Mendonca, Alexandre},
  journal={Dementia and geriatric cognitive disorders extra},
  volume={2},
  number={1},
  pages={321--331},
  year={2012},
  publisher={Karger Publishers}
}

@article{hossin2015review,
  title={A review on evaluation metrics for data classification evaluations},
  author={Hossin, Mohammad and Sulaiman, Md Nasir},
  journal={International journal of data mining \& knowledge management process},
  volume={5},
  number={2},
  pages={1},
  year={2015},
  publisher={Academy \& Industry Research Collaboration Center (AIRCC)}
}

@article{brady2016speech,
  title={Speech and language therapy for aphasia following stroke},
  author={Brady, Marian C and Kelly, Helen and Godwin, Jon and Enderby, Pam and Campbell, Pauline},
  journal={Cochrane database of systematic reviews},
  number={6},
  year={2016},
  publisher={John Wiley \& Sons, Ltd}
}

@article{patton2016automos,
  title={AutoMOS: Learning a non-intrusive assessor of naturalness-of-speech},
  author={Patton, Brian and Agiomyrgiannakis, Yannis and Terry, Michael and Wilson, Kevin and Saurous, Rif A and Sculley, D},
  journal={arXiv preprint arXiv:1611.09207},
  year={2016}
}

@misc{unesco2016if,
  title={If you don’t understand, how can you learn?},
  author={UNESCO},
  journal={Policy Paper 24 of Global Education Monitoring Report},
  url={https://en.unesco.org/news/40-don-t-access-education-language-they-understand},
  year={2016},
  publisher={UNESCO Paris}
}

@inproceedings{radzikowski2016non,
  title={Non-native English speakers' speech correction, based on domain focused document},
  author={Radzikowski, Kacper and Wang, Le and Yoshie, Osamu},
  booktitle={Proceedings of the 18th International Conference on Information Integration and Web-based Applications and Services},
  pages={276--281},
  year={2016}
}

@article{wang2017tacotron,
  title={Tacotron: Towards end-to-end speech synthesis},
  author={Wang, Yuxuan and Skerry-Ryan, RJ and Stanton, Daisy and Wu, Yonghui and Weiss, Ron J and Jaitly, Navdeep and Yang, Zongheng and Xiao, Ying and Chen, Zhifeng and Bengio, Samy and others},
  journal={arXiv preprint arXiv:1703.10135},
  year={2017}
}

@book{ore2017cardano,
  title={Cardano: The gambling scholar},
  author={Ore, {\O}ystein},
  volume={5063},
  year={2017},
  publisher={Princeton University Press}
}

@article{czyzewski2017audio,
  title={An audio-visual corpus for multimodal automatic speech recognition},
  author={Czyzewski, Andrzej and Kostek, Bozena and Bratoszewski, Piotr and Kotus, Jozef and Szykulski, Marcin},
  journal={Journal of Intelligent Information Systems},
  volume={49},
  number={2},
  pages={167--192},
  year={2017},
  publisher={Springer}
}

@inproceedings{skerry2018towards,
  title={Towards end-to-end prosody transfer for expressive speech synthesis with tacotron},
  author={Skerry-Ryan, RJ and Battenberg, Eric and Xiao, Ying and Wang, Yuxuan and Stanton, Daisy and Shor, Joel and Weiss, Ron and Clark, Rob and Saurous, Rif A},
  booktitle={international conference on machine learning},
  pages={4693--4702},
  year={2018},
  organization={PMLR}
}

@inproceedings{rosenberg2017bias,
  title={Bias and Statistical Significance in Evaluating Speech Synthesis with Mean Opinion Scores.},
  author={Rosenberg, Andrew and Ramabhadran, Bhuvana},
  booktitle={Interspeech},
  pages={3976--3980},
  year={2017}
}

@article{cuny2017neuropsychological,
  title={Neuropsychological improvement after posterior fossa arachnoid cyst drainage},
  author={Cuny, ML and Pallone, M and Piana, H and Boddaert, N and Sainte-Rose, C and Vaivre-Douret, L and Piolino, P and Puget, S},
  journal={Child's Nervous System},
  volume={33},
  number={1},
  pages={135--141},
  year={2017},
  publisher={Springer}
}

@misc{economicforum2018,
  title={Speaking more than one language can boost economic growth},
  author={WorldEconomicForum},
  url={https://www.weforum.org/agenda/2018/02/speaking-more-languages-boost-economic-growth},
  year={2018},
  publisher={World Economic Forum}
}

@inproceedings{oord2018parallel,
  title={Parallel wavenet: Fast high-fidelity speech synthesis},
  author={Oord, Aaron and Li, Yazhe and Babuschkin, Igor and Simonyan, Karen and Vinyals, Oriol and Kavukcuoglu, Koray and Driessche, George and Lockhart, Edward and Cobo, Luis and Stimberg, Florian and others},
  booktitle={International conference on machine learning},
  pages={3918--3926},
  year={2018},
  organization={PMLR}
}

@article{botchkarev2018performance,
  title={Performance metrics (error measures) in machine learning regression, forecasting and prognostics: Properties and typology},
  author={Botchkarev, Alexei},
  journal={arXiv preprint arXiv:1809.03006},
  year={2018}
}

@article{banovic2018communication,
  title={Communication difficulties as a result of dementia},
  author={Banovic, Silva and Zunic, Lejla Junuzovic and Sinanovic, Osman},
  journal={Materia socio-medica},
  volume={30},
  number={3},
  pages={221},
  year={2018},
  publisher={The Academy of Medical Sciences of Bosnia and Herzegovina}
}

@inproceedings{merritt2018comprehensive,
  title={Comprehensive evaluation of statistical speech waveform synthesis},
  author={Merritt, Thomas and Putrycz, Bartosz and Nadolski, Adam and Ye, Tianjun and Korzekwa, Daniel and Dolecki, Wiktor and Drugman, Thomas and Klimkov, Viacheslav and Moinet, Alexis and Breen, Andrew and others},
  booktitle={2018 IEEE Spoken Language Technology Workshop (SLT)},
  pages={325--331},
  year={2018},
  organization={IEEE}
}

@article{marcus2018deep,
  title={Deep learning: A critical appraisal},
  author={Marcus, Gary},
  journal={arXiv preprint arXiv:1801.00631},
  year={2018}
}

@article{gu2018recent,
  title={Recent advances in convolutional neural networks},
  author={Gu, Jiuxiang and Wang, Zhenhua and Kuen, Jason and Ma, Lianyang and Shahroudy, Amir and Shuai, Bing and Liu, Ting and Wang, Xingxing and Wang, Gang and Cai, Jianfei and others},
  journal={Pattern Recognition},
  volume={77},
  pages={354--377},
  year={2018},
  publisher={Elsevier}
}

@article{korzekwa2019interpretable,
  title={Interpretable Deep Learning Model for the Detection and Reconstruction of Dysarthric Speech},
  author={Korzekwa, Daniel and Barra-Chicote, Roberto and Kostek, Bozena and Drugman, Thomas and Lajszczak, Mateusz},
  journal={Proc. Interspeech 2019},
  pages={3890--3894},
  year={2019},
  doi={10.21437/Interspeech.2019-1206}
}

@article{meyes2019ablation,
  title={Ablation studies in artificial neural networks},
  author={Meyes, Richard and Lu, Melanie and de Puiseau, Constantin Waubert and Meisen, Tobias},
  journal={arXiv preprint arXiv:1901.08644},
  year={2019}
}

@inproceedings{wagner2019speech,
  title={Speech synthesis evaluation—state-of-the-art assessment and suggestion for a novel research program},
  author={Wagner, Petra and Beskow, Jonas and Betz, Simon and Edlund, Jens and Gustafson, Joakim and Eje Henter, Gustav and Le Maguer, S{\'e}bastien and Malisz, Zofia and Sz{\'e}kely, {\'E}va and T{\aa}nnander, Christina and others},
  booktitle={Proceedings of the 10th Speech Synthesis Workshop (SSW10)},
  year={2019}
}

@article{ren2019fastspeech,
  title={Fastspeech: Fast, robust and controllable text to speech},
  author={Ren, Yi and Ruan, Yangjun and Tan, Xu and Qin, Tao and Zhao, Sheng and Zhao, Zhou and Liu, Tie-Yan},
  journal={arXiv preprint arXiv:1905.09263},
  year={2019}
}

@inproceedings{leung2019cnn,
  title={CNN-RNN-CTC based end-to-end mispronunciation detection and diagnosis},
  author={Leung, Wai-Kim and Liu, Xunying and Meng, Helen},
  booktitle={ICASSP 2019-2019 IEEE International Conference on Acoustics, Speech and Signal Processing (ICASSP)},
  pages={8132--8136},
  year={2019},
  organization={IEEE}
}

@article{korzekwa2019deep,
  title={Deep learning model for automated assessment of lexical stress of non-native English speakers},
  author={Korzekwa, Daniel and Kostek, Bozena},
  journal={The Journal of the Acoustical Society of America},
  volume={146},
  number={4},
  pages={2956--2957},
  year={2019},
  publisher={Acoustical Society of America},
  doi={10.1121/1.5137270}
}

@article{asha2019,
  title={American Speech-Language-Hearing Association (ASHA), accessed on June 2022},
  author={ASHA},
  url={https://www.asha.org},
  year={2022}
}

@article{sofaer2019area,
  title={The area under the precision-recall curve as a performance metric for rare binary events},
  author={Sofaer, Helen R and Hoeting, Jennifer A and Jarnevich, Catherine S},
  journal={Methods in Ecology and Evolution},
  volume={10},
  number={4},
  pages={565--577},
  year={2019},
  publisher={Wiley Online Library}
}

@article{asrifan2020effects,
  title={THE EFFECTS OF CALL (COMPUTER ASSISTED LANGUAGE LEARNING) TOWARD THE STUDENTS’ENGLISH ACHIEVEMENT AND ATTITUDE},
  author={Asrifan, Andi and Zita, Cris T and Vargheese, KJ and Syamsu, T and Amir, Muhammad},
  journal={Journal of advanced English studies},
  volume={3},
  number={2},
  pages={94--106},
  year={2020}
}

@misc{EPI2020,
  title={EF English Proficiency Index},
  author={EF-Education-First},
  url={https://www.ef.pl/assetscdn/WIBIwq6RdJvcD9bc8RMd/legacy/__/~/media/centralefcom/epi/downloads/full-reports/v10/ef-epi-2020-english.pdf},
  year={2020},
  publisher={EF Education First}
}

@inproceedings{korzekwa2021mispronunciation,
  title={Mispronunciation Detection in Non-Native (L2) English with Uncertainty Modeling},
  author={Korzekwa, Daniel and Lorenzo-Trueba, Jaime and Zaporowski, Szymon and Calamaro, Shira and Drugman, Thomas and Kostek, Bozena},
  booktitle={ICASSP 2021-2021 IEEE International Conference on Acoustics, Speech and Signal Processing (ICASSP)},
  pages={7738--7742},
  year={2021},
  organization={IEEE},
  doi={10.1109/ICASSP39728.2021.9413953}
}

@article{ Ethnologue2021,
  title={Ethnologue: Languages of the World. Twenty-fourth edition. Dallas},
  author={David M. Eberhard and Gary F. Simons and Charles D. Fennig},
  url={https://www.ethnologue.com/ethnoblog/gary-simons/welcome-24th-edition},
  year={2021},
  publisher={SIL International}
}

@inproceedings{jiao2021universal,
  title={Universal neural vocoding with parallel wavenet},
  author={Jiao, Yunlong and Gabry{\'s}, Adam and Tinchev, Georgi and Putrycz, Bartosz and Korzekwa, Daniel and Klimkov, Viacheslav},
  booktitle={ICASSP 2021-2021 IEEE International Conference on Acoustics, Speech and Signal Processing (ICASSP)},
  pages={6044--6048},
  year={2021},
  organization={IEEE},
  doi={10.1109/ICASSP39728.2021.9414444}
}

@inproceedings{gabrys2021improving,
  author={Adam Gabryś and Yunlong Jiao and Viacheslav Klimkov and Daniel Korzekwa and Roberto Barra-Chicote},
  title={{Improving the Expressiveness of Neural Vocoding with Non-Affine Normalizing Flows}},
  year=2021,
  booktitle={Proc. Interspeech 2021},
  pages={1679--1683},
  doi={10.21437/Interspeech.2021-1555}
}

@article{mu2021review,
  title={Review of end-to-end speech synthesis technology based on deep learning},
  author={Mu, Zhaoxi and Yang, Xinyu and Dong, Yizhuo},
  journal={arXiv preprint arXiv:2104.09995},
  doi = {10.48550/ARXIV.2104.09995},
  year={2021}
}

@inproceedings{ezzerg2021enhancing,
  author={Abdelhamid Ezzerg and Adam Gabrys and Bartosz Putrycz and Daniel Korzekwa and Daniel Saez-Trigueros and David McHardy and Kamil Pokora and Jakub Lachowicz and Jaime Lorenzo-Trueba and Viacheslav Klimkov},
  title={{Enhancing audio quality for expressive Neural Text-to-Speech}},
  year=2021,
  booktitle={Proc. 11th ISCA Speech Synthesis Workshop (SSW 11)},
  pages={78--83},
  doi={10.21437/SSW.2021-14}
}

@misc{wikipedia_arpabet,
  title={Arpabebet, accessed on June 2022},
  author={Wikipedia Arpabet},
  url={https://en.wikipedia.org/wiki/ARPABET},
  year={2022},
  publisher={Wikipedia}
}

@article{yan2021end,
  title={End-to-End Mispronunciation Detection and Diagnosis From Raw Waveforms},
  author={Yan, Bi-Cheng and Chen, Berlin},
  journal={arXiv preprint arXiv:2103.03023},
  year={2021}
}

@article{valizada2021development,
  title={Development and Evaluation of Speech Synthesis System Based on Deep Learning Models},
  author={Valizada, Alakbar and Jafarova, Sevil and Sultanov, Emin and Rustamov, Samir},
  journal={Symmetry},
  volume={13},
  number={5},
  pages={819},
  year={2021},
  publisher={Multidisciplinary Digital Publishing Institute}
}

@inproceedings{korzekwa21_interspeech,
  author={Daniel Korzekwa and Roberto Barra-Chicote and Szymon Zaporowski and Grzegorz Beringer and Jaime Lorenzo-Trueba and Alicja Serafinowicz and Jasha Droppo and Thomas Drugman and Bozena Kostek},
  title={{Detection of Lexical Stress Errors in Non-Native (L2) English with Data Augmentation and Attention}},
  year=2021,
  booktitle={Proc. Interspeech 2021},
  pages={3915--3919},
  doi={10.21437/Interspeech.2021-86}
}

@inproceedings{Zhang2022_interspeech,
  author={Daniel Zhang and Ashwinkumar Ganesan and Sarah Campbell and Daniel Korzekwa},
  title={L2-GEN: A Neural Phoneme Paraphrasing Approach to L2 Speech Synthesis for Mispronunciation Diagnosis},
  year=2022,
  booktitle={accepted to Interspeech 2022}
}

@inproceedings{Bilinski2022_interspeech,
  author={Piotr Bilinski and Thomas Merritt and Abdelhamid Ezzerg and Kamil Pokora and Sebastian Cygert and Kayoko Yanagisawa and Roberto Barra-Chicote and Daniel Korzekwa},
  title={Creating New Voices using Normalizing Flows},
  year=2022,
  booktitle={accepted to Interspeech 2022}
}

@article{korzekwa22_speechcomm,
  title={Computer-assisted Pronunciation Training - Speech synthesis is almost all you need},
  author={Daniel Korzekwa and Jaime Lorenzo-Trueba and Thomas Drugman and Bozena Kostek},
  journal={accepted for publication in Speech Communication Journal on June 17 ‘2022, in print},
  year=2022
}

@INPROCEEDINGS{merritt22_icassp_vc,
  author={Merritt, Thomas and Ezzerg, Abdelhamid and Biliński, Piotr and Proszewska, Magdalena and Pokora, Kamil and Barra-Chicote, Roberto and Korzekwa, Daniel},
  booktitle={ICASSP 2022 - 2022 IEEE International Conference on Acoustics, Speech and Signal Processing (ICASSP)}, 
  title={Text-Free Non-Parallel Many-To-Many Voice Conversion Using Normalising Flow}, 
  year={2022},
  volume={},
  number={},
  pages={6782-6786},
  doi={10.1109/ICASSP43922.2022.9746368}
}

@article{BeringerAMLC2020,
  title={Extending Goodness of Pronunciation to generate mispronunciationhypotheses for pronunciation assessment in L2-English},
  author={Grzegorz Beringer and Daniel Korzekwa and Ariadna Sanchez and Bofei Wang and Jaime Lorenzo-Trueba},
  journal={Amazon Machine Learning Conference, Seattle},
  year=2020
}

@article{huang2021preliminary,
  title={A Preliminary Study of a Two-Stage Paradigm for Preserving Speaker Identity in Dysarthric Voice Conversion},
  author={Huang, Wen-Chin and Kobayashi, Kazuhiro and Peng, Yu-Huai and Liu, Ching-Feng and Tsao, Yu and Wang, Hsin-Min and Toda, Tomoki},
  journal={arXiv preprint arXiv:2106.01415},
  year={2021}
}

@inproceedings{romana2021automatically,
  title={Automatically Detecting Errors and Disfluencies in Read Speech to Predict Cognitive Impairment in People with Parkinson’s Disease},
  author={Romana, Amrit and Bandon, John and Perez, Matthew and Gutierrez, Stephanie and Richter, Richard and Roberts, Angela and Provost, Emily Mower},
  booktitle={22nd Annual Conference of the International Speech Communication Association, INTERSPEECH 2021},
  pages={156--160},
  year={2021},
  organization={International Speech Communication Association}
}

@article{oneata2022improving,
  title={Improving Multimodal Speech Recognition by Data Augmentation and Speech Representations},
  author={Oneata, Dan and Cucu, Horia},
  journal={arXiv preprint arXiv:2204.13206},
  year={2022}
}

@article{fu2022improving,
  title={Improving Non-native Word-level Pronunciation Scoring with Phone-level Mixup Data Augmentation and Multi-source Information},
  author={Fu, Kaiqi and Gao, Shaojun and Wang, Kai and Li, Wei and Tian, Xiaohai and Ma, Zejun},
  journal={arXiv preprint arXiv:2203.01826, submitted to INTERSPEECH 2022},
  doi = {10.48550/ARXIV.2203.01826},
  year={2022}
}

@INPROCEEDINGS{Gong2022,
  author={Gong, Yuan and Chen, Ziyi and Chu, Iek-Heng and Chang, Peng and Glass, James},
  booktitle={ICASSP 2022 - 2022 IEEE International Conference on Acoustics, Speech and Signal Processing (ICASSP)}, 
  title={Transformer-Based Multi-Aspect Multi-Granularity Non-Native English Speaker Pronunciation Assessment}, 
  year={2022},
  volume={},
  number={},
  pages={7262-7266},
  doi={10.1109/ICASSP43922.2022.9746743}}


%% file: lexical_stress_error_detection.bib
@article{posner1990attention,
  title={The attention system of the human brain},
  author={Posner, Michael I and Petersen, Steven E},
  journal={Annual review of neuroscience},
  volume={13},
  number={1},
  pages={25--42},
  year={1990},
  publisher={Annual Reviews 4139 El Camino Way; PO Box 10139; Palo Alto; CA 94303-0139; USA}
}

@inproceedings{bergem1991acoustic,
  title={Acoustic and lexical vowel reduction},
  author={Bergem, Dick R van},
  booktitle={Phonetics and Phonology of Speaking Styles},
  year={1991}
}

@article{shattuck1994stress,
  title={Stress shift and early pitch accent placement in lexical items in American English},
  author={Shattuck-Hufnagel, Stefanie and Ostendorf, Mari and Ross, Ken},
  journal={Journal of Phonetics},
  volume={22},
  number={4},
  pages={357--388},
  year={1994},
  publisher={Elsevier}
}

@article{jordan1999introduction,
  title={An introduction to variational methods for graphical models},
  author={Jordan, Michael I and Ghahramani, Zoubin and Jaakkola, Tommi S and Saul, Lawrence K},
  journal={Machine learning},
  volume={37},
  number={2},
  pages={183--233},
  year={1999},
  publisher={Springer}
}

@inproceedings{kominek2004cmu,
  title={The CMU Arctic speech databases},
  author={Kominek, John and Black, Alan W},
  booktitle={Fifth ISCA workshop on speech synthesis},
  year={2004}
}

@inproceedings{chen2007using,
  title={Using nonlinear features in automatic English lexical stress detection},
  author={Chen, Nan and He, Qianhua},
  booktitle={2007 Intl. Conference on Computational Intelligence and Security Workshops (CISW 2007)},
  pages={328--332},
  year={2007},
  organization={IEEE}
}

@article{boersma2006praat,
  title={Praat: doing phonetics by computer},
  author={Boersma, Paul},
  journal={http://www.praat.org/},
  year={2006}
}

@inproceedings{chen2010automatic_2,
  title={Automatic lexical stress detection for Chinese learners' of English},
  author={Chen, Jin-Yu and Wang, Lan},
  booktitle={2010 7th Intl. Symposium on Chinese Spoken Language Processing},
  pages={407--411},
  year={2010},
  organization={IEEE}
}

@article{michel2011quantitative,
  title={Quantitative analysis of culture using millions of digitized books},
  author={Michel, Jean-Baptiste and Shen, Yuan Kui and Aiden, Aviva Presser and Veres, Adrian and Gray, Matthew K and Pickett, Joseph P and Hoiberg, Dale and Clancy, Dan and Norvig, Peter and Orwant, Jon and others},
  journal={science},
  volume={331},
  number={6014},
  pages={176--182},
  year={2011},
  publisher={American Association for the Advancement of Science}
}

@article{zhao2011automatic,
  title={Automatic lexical stress detection using acoustic features for computer assisted language learning},
  author={Zhao, Junhong and Yuan, Hua and Liu, Jia and Xia, S},
  journal={Proc. APSIPA ASC},
  pages={247--251},
  year={2011}
}

@inproceedings{li2013lexical,
  title={Lexical stress detection for L2 English speech using deep belief networks.},
  author={Li, Kun and Qian, Xiaojun and Kang, Shiyin and Meng, Helen},
  booktitle={Interspeech},
  pages={1811--1815},
  year={2013}
}

@article{cho2014learning,
  title={Learning phrase representations using RNN encoder-decoder for statistical machine translation},
  author={Cho, Kyunghyun and Van Merri{\"e}nboer, Bart and Gulcehre, Caglar and Bahdanau, Dzmitry and Bougares, Fethi and Schwenk, Holger and Bengio, Yoshua},
  journal={arXiv preprint arXiv:1406.1078},
  year={2014}
}

@inproceedings{panayotov2015librispeech,
  title={Librispeech: an asr corpus based on public domain audio books},
  author={Panayotov, Vassil and Chen, Guoguo and Povey, Daniel and Khudanpur, Sanjeev},
  booktitle={2015 IEEE Intl. Conference on Acoustics, Speech and Signal Processing (ICASSP)},
  pages={5206--5210},
  year={2015},
  organization={IEEE}
}

@article{ferrer2015classification,
  title={Classification of lexical stress using spectral and prosodic features for computer-assisted language learning systems},
  author={Ferrer, Luciana and Bratt, Harry and Richey, Colleen and Franco, Horacio and Abrash, Victor and Precoda, Kristin},
  journal={Speech Communication},
  volume={69},
  pages={31--45},
  year={2015},
  publisher={Elsevier}
}

@inproceedings{shahin2016automatic,
  title={Automatic Classification of Lexical Stress in English and Arabic Languages Using Deep Learning.},
  author={Shahin, Mostafa Ali and Epps, Julien and Ahmed, Beena},
  booktitle={INTERSPEECH},
  pages={175--179},
  year={2016}
}

@inproceedings{mcauliffe2017montreal,
  title={Montreal Forced Aligner: Trainable Text-Speech Alignment Using Kaldi.},
  author={McAuliffe, Michael and Socolof, Michaela and Mihuc, Sarah and Wagner, Michael and Sonderegger, Morgan},
  booktitle={Interspeech},
  volume={2017},
  pages={498--502},
  year={2017}
}

@article{porzuczek2017english,
  title={English word stress in Polish learners speech production and metacompetence},
  author={Porzuczek, Andrzej and Rojczyk, Arkadiusz},
  journal={Research in Language},
  volume={15},
  number={4},
  pages={313--323},
  year={2017},
  publisher={Sciendo}
}

@article{jung2018acoustic,
  title={Acoustic analysis of English lexical stress produced by Korean, Japanese and Taiwanese-Chinese speakers},
  author={Jung, Ye-Jee and Rhee, Seok-Chae and others},
  journal={Phonetics and Speech Sciences},
  volume={10},
  number={1},
  pages={15--22},
  year={2018},
  publisher={Korean Society of Speech Sciences}
}

@article{li2018automatic,
  title={Automatic lexical stress and pitch accent detection for L2 English speech using multi-distribution deep neural networks},
  author={Li, Kun and Mao, Shaoguang and Li, Xu and Wu, Zhiyong and Meng, Helen},
  journal={Speech Communication},
  volume={96},
  pages={28--36},
  year={2018},
  publisher={Elsevier}
}

@article{zhao2018l2,
  title={L2-ARCTIC: A non-native English speech corpus},
  author={Zhao, Guanlong and Sonsaat, Sinem and Silpachai, Alif O and Lucic, Ivana and Chukharev-Hudilainen, Evgeny and Levis, John and Gutierrez-Osuna, Ricardo},
  journal={Perception Sensing Instrumentation Lab},
  year={2018}
}

@inproceedings{latorre2019effect,
  title={Effect of data reduction on sequence-to-sequence neural tts},
  author={Latorre, Javier and Lachowicz, Jakub and Lorenzo-Trueba, Jaime and Merritt, Thomas and Drugman, Thomas and Ronanki, Srikanth and Klimkov, Viacheslav},
  booktitle={ICASSP 2019-2019 IEEE International Conference on Acoustics, Speech and Signal Processing (ICASSP)},
  pages={7075--7079},
  year={2019},
  organization={IEEE}
}

@inproceedings{paleyes2019emulation,
  title={Emulation of physical processes with emukit},
  author={Paleyes, Andrei and Pullin, Mark and Mahsereci, Maren and Lawrence, Neil and Gonzalez, Javier},
  booktitle={Second Workshop on Machine Learning and the Physical Sciences, NeurIPS},
  year={2019}
}

@inproceedings{ramanathi2019asr,
  title={ASR Inspired Syllable Stress Detection for Pronunciation Evaluation Without Using a Supervised Classifier and Syllable Level Features.},
  author={Ramanathi, Manoj Kumar and Yarra, Chiranjeevi and Ghosh, Prasanta Kumar},
  booktitle={INTERSPEECH},
  pages={924--928},
  year={2019}
}

@article{ruan2019end,
  title={An End-to-end Approach for Lexical Stress Detection based on Transformer},
  author={Ruan, Yong and Wang, Xiangdong and Liu, Hong and Ou, Zhigang and Gao, Yun and Cheng, Jianfeng and Qian, Yueliang},
  journal={arXiv preprint arXiv:1911.04862},
  year={2019}
}

@article{elias2020parallel,
  title={Parallel Tacotron: Non-Autoregressive and Controllable TTS},
  author={Elias, Isaac and Zen, Heiga and Shen, Jonathan and Zhang, Yu and Jia, Ye and Weiss, Ron and Wu, Yonghui},
  journal={arXiv preprint arXiv:2010.11439},
  year={2020}
}


%% file: pron_error_dectection_with_uncertainty_modelling.bib
@article{hieke1984linking,
  title={Linking as a marker of fluent speech},
  author={Hieke, A. E.},
  journal={Language and Speech},
  volume={27},
  number={4},
  pages={343--354},
  year={1984},
  publisher={Sage Publications Sage CA: Thousand Oaks, CA}
}

@inproceedings{graves2013speech,
  title={Speech recognition with deep recurrent neural networks},
  author={Graves, A. and Mohamed, A. and Hinton, G.},
  booktitle={2013 IEEE Intl. conference on acoustics, speech and signal processing},
  pages={6645--6649},
  year={2013},
  organization={IEEE}
}


%% file: research_methodology.bib
@article{rosenblatt1960perceptron,
  title={Perceptron simulation experiments},
  author={Rosenblatt, Frank},
  journal={Proceedings of the IRE},
  volume={48},
  number={3},
  pages={301--309},
  year={1960},
  publisher={IEEE}
}

@book{Jurafsky:2009:SLP:1214993,
 author = {Jurafsky, Daniel and Martin, James H.},
 title = {Speech and Language Processing (2Nd Edition)},
 year = {2009},
 isbn = {0131873210},
 publisher = {Prentice-Hall, Inc.},
 address = {Upper Saddle River, NJ, USA},
}

@article{hines2015visqolaudio,
  title={ViSQOLAudio: An objective audio quality metric for low bitrate codecs},
  author={Hines, Andrew and Gillen, Eoin and Kelly, Damien and Skoglund, Jan and Kokaram, Anil and Harte, Naomi},
  journal={The Journal of the Acoustical Society of America},
  volume={137},
  number={6},
  pages={EL449--EL455},
  year={2015},
  publisher={Acoustical Society of America}
}

@book{alma991002741086804901,
isbn = {1-938168-27-5},
language = {eng},
title = {University physics Volume 1},
year = {2016},
}

@article{chaudhari2021attentive,
  title={An attentive survey of attention models},
  author={Chaudhari, Sneha and Mithal, Varun and Polatkan, Gungor and Ramanath, Rohan},
  journal={ACM Transactions on Intelligent Systems and Technology (TIST)},
  volume={12},
  number={5},
  pages={1--32},
  year={2021},
  publisher={ACM New York, NY}
}

@article{piotrowska2021evaluation,
  title={Evaluation of aspiration problems in L2 English pronunciation employing machine learning},
  author={Piotrowska, Magdalena and Czy{\.z}ewski, Andrzej and Ciszewski, Tomasz and Korvel, Gra{\v{z}}ina and Kurowski, Adam and Kostek, Bo{\.z}ena},
  journal={The Journal of the Acoustical Society of America},
  volume={150},
  number={1},
  pages={120--132},
  year={2021},
  publisher={Acoustical Society of America}
}


%% file: speech_synthesis_is_almost_all_you_need.bib
@book{levy2013call,
  title={CALL dimensions: Options and issues in computer-assisted language learning},
  author={Levy, Mike and Stockwell, Glenn},
  year={2013},
  publisher={Routledge}
}

@article{golonka2014technologies,
  title={Technologies for foreign language learning: A review of technology types and their effectiveness},
  author={Golonka, Ewa M and Bowles, Anita R and Frank, Victor M and Richardson, Dorna L and Freynik, Suzanne},
  journal={Computer assisted language learning},
  volume={27},
  number={1},
  pages={70--105},
  year={2014},
  publisher={Taylor \& Francis}
}

@inproceedings{ai2015automatic,
  title={Automatic pronunciation error detection and feedback generation for call applications},
  author={Ai, Renlong},
  booktitle={International Conference on Learning and Collaboration Technologies},
  pages={175--186},
  year={2015},
  organization={Springer}
}

@phdthesis{lee2016language,
  title={Language-independent methods for computer-assisted pronunciation training},
  author={Lee, Ann and others},
  year={2016},
  school={Massachusetts Institute of Technology}
}

@inproceedings{huang2016machine,
  title={Machine translation based data augmentation for cantonese keyword spotting},
  author={Huang, Guangpu and Gorin, Arseniy and Gauvain, Jean-Luc and Lamel, Lori},
  booktitle={2016 IEEE International Conference on Acoustics, Speech and Signal Processing (ICASSP)},
  pages={6020--6024},
  year={2016},
  organization={IEEE}
}

@inproceedings{badenhorst2017limitations,
  title={The limitations of data perturbation for ASR of learner data in under-resourced languages},
  author={Badenhorst, Jaco and De Wet, Febe},
  booktitle={2017 Pattern Recognition Association of South Africa and Robotics and Mechatronics (PRASA-RobMech)},
  pages={44--49},
  year={2017},
  organization={IEEE}
}

@article{duan2019cross,
  title={Cross-lingual transfer learning of non-native acoustic modeling for pronunciation error detection and diagnosis},
  author={Duan, Richeng and Kawahara, Tatsuya and Dantsuji, Masatake and Nanjo, Hiroaki},
  journal={IEEE/ACM Transactions on Audio, Speech, and Language Processing},
  volume={28},
  pages={391--401},
  year={2019},
  publisher={IEEE}
}

@mastersthesis{eklund2019data,
  title={Data Augmentation Techniques for Robust Audio Analysis},
  author={Eklund, Ville-Veikko},
  school={Tampere University},
  year={2019}
}

@inproceedings{komatsu2019speech,
  title={Speech Error Detection depending on Linguistic Units},
  author={Komatsu, Seiya and Sasayama, Manabu},
  booktitle={Proceedings of the 2019 3rd International Conference on Natural Language Processing and Information Retrieval},
  pages={75--79},
  year={2019}
}

@article{mehri2019diagnosing,
  title={Diagnosing L2 learners’ development through online computerized dynamic assessment},
  author={Mehri Kamrood, Ali and Davoudi, Mohammad and Ghaniabadi, Saeed and Amirian, Seyyed Mohammad Reza},
  journal={Computer Assisted Language Learning},
  pages={1--30},
  year={2019},
  publisher={Taylor \& Francis}
}

@misc{statista2020,
  title={Most common languages used on the internet as of January 2020, by share of internet users},
  author={Statista},
  url={https://www.statista.com/statistics/262946/share-of-the-most-common-languages-on-the-internet/},
  year={2021},
  publisher={Statista}
}

@article{zhang2020end,
  title={End-to-end automatic pronunciation error detection based on improved hybrid ctc/attention architecture},
  author={Zhang, Long and Zhao, Ziping and Ma, Chunmei and Shan, Linlin and Sun, Huazhi and Jiang, Lifen and Deng, Shiwen and Gao, Chang},
  journal={Sensors},
  volume={20},
  number={7},
  pages={1809},
  year={2020},
  publisher={Multidisciplinary Digital Publishing Institute}
}

@inproceedings{yan20_interspeech,
  author={Bi-Cheng Yan and Meng-Che Wu and Hsiao-Tsung Hung and Berlin Chen},
  title={{An End-to-End Mispronunciation Detection System for L2 English Speech Leveraging Novel Anti-Phone Modeling}},
  year=2020,
  booktitle={Proc. Interspeech 2020},
  pages={3032--3036},
  doi={10.21437/Interspeech.2020-1616}
}

@inproceedings{korzekwa21b_interspeech,
  author={Daniel Korzekwa and Jaime Lorenzo-Trueba and Thomas Drugman and Shira Calamaro and Bozena Kostek},
  title={{Weakly-Supervised Word-Level Pronunciation Error Detection in Non-Native English Speech}},
  year=2021,
  booktitle={Proc. Interspeech 2021},
  pages={4408--4412},
  doi={10.21437/Interspeech.2021-38}
}

@article{fu2021full,
  title={A Full Text-Dependent End to End Mispronunciation Detection and Diagnosis with Easy Data Augmentation Techniques},
  author={Fu, Kaiqi and Lin, Jones and Ke, Dengfeng and Xie, Yanlu and Zhang, Jinsong and Lin, Binghuai},
  journal={arXiv preprint arXiv:2104.08428},
  year={2021}
}

@article{yan2021maximum,
  title={Maximum F1-score training for end-to-end mispronunciation detection and diagnosis of L2 English speech},
  author={Yan, Bi-Cheng and Jiang, Shao-Wei Fan and Chao, Fu-An and Chen, Berlin},
  journal={arXiv preprint arXiv:2108.13816},
  year={2021}
}

@inproceedings{fazel21_interspeech,
  author={Amin Fazel and Wei Yang and Yulan Liu and Roberto Barra-Chicote and Yixiong Meng and Roland Maas and Jasha Droppo},
  title={{SynthASR: Unlocking Synthetic Data for Speech Recognition}},
  year=2021,
  booktitle={Proc. Interspeech 2021},
  pages={896--900},
  doi={10.21437/Interspeech.2021-1882}
}

@inproceedings{shah21_ssw,
  author={Raahil Shah and Kamil Pokora and Abdelhamid Ezzerg and Viacheslav Klimkov and Goeric Huybrechts and Bartosz Putrycz and Daniel Korzekwa and Thomas Merritt},
  title={{Non-Autoregressive TTS with Explicit Duration Modelling for Low-Resource Highly Expressive Speech}},
  year=2021,
  booktitle={Proc. 11th ISCA Speech Synthesis Workshop (SSW 11)},
  pages={96--101},
  doi={10.21437/SSW.2021-17}
}

@inproceedings{huybrechts2021low,
  title={Low-resource expressive text-to-speech using data augmentation},
  author={Huybrechts, Goeric and Merritt, Thomas and Comini, Giulia and Perz, Bartek and Shah, Raahil and Lorenzo-Trueba, Jaime},
  booktitle={ICASSP 2021-2021 IEEE International Conference on Acoustics, Speech and Signal Processing (ICASSP)},
  pages={6593--6597},
  year={2021},
  organization={IEEE}
}

@inproceedings{xu21k_interspeech,
  author={Xiaoshuo Xu and Yueteng Kang and Songjun Cao and Binghuai Lin and Long Ma},
  title={{Explore wav2vec 2.0 for Mispronunciation Detection}},
  year=2021,
  booktitle={Proc. Interspeech 2021},
  pages={4428--4432},
  doi={10.21437/Interspeech.2021-777}
}

@inproceedings{peng21e_interspeech,
  author={Linkai Peng and Kaiqi Fu and Binghuai Lin and Dengfeng Ke and Jinsong Zhan},
  title={{A Study on Fine-Tuning wav2vec2.0 Model for the Task of Mispronunciation Detection and Diagnosis}},
  year=2021,
  booktitle={Proc. Interspeech 2021},
  pages={4448--4452},
  doi={10.21437/Interspeech.2021-1344}
}

@inproceedings{lin21j_interspeech,
  author={Binghuai Lin and Liyuan Wang},
  title={{Deep Feature Transfer Learning for Automatic Pronunciation Assessment}},
  year=2021,
  booktitle={Proc. Interspeech 2021},
  pages={4438--4442},
  doi={10.21437/Interspeech.2021-931}
}


%% file: weakly_supervised_pron_error_detection.bib
@article{needleman1970general,
  title={A general method applicable to the search for similarities in the amino acid sequence of two proteins},
  author={Needleman, Saul B and Wunsch, Christian D},
  journal={Journal of molecular biology},
  volume={48},
  number={3},
  pages={443--453},
  year={1970},
  publisher={Elsevier}
}

@article{garofolo1993darpa,
  title={DARPA TIMIT acoustic-phonetic continous speech corpus CD-ROM. NIST speech disc 1-1.1},
  author={Garofolo, John S and Lamel, Lori F and Fisher, William M and Fiscus, Jonathan G and Pallett, David S},
  journal={STIN},
  volume={93},
  pages={27403},
  year={1993}
}

@inproceedings{franco1997automatic,
  title={Automatic pronunciation scoring for language instruction},
  author={Franco, Horacio and Neumeyer, Leonardo and Kim, Yoon and Ronen, Orith},
  booktitle={1997 IEEE international conference on acoustics, speech, and signal processing},
  volume={2},
  pages={1471--1474},
  year={1997},
  organization={IEEE}
}

@article{witt2000phone,
  title={Phone-level pronunciation scoring and assessment for interactive language learning},
  author={Witt, Silke M and Young, Steve J},
  journal={Speech communication},
  volume={30},
  number={2-3},
  pages={95--108},
  year={2000},
  publisher={Elsevier}
}

@article{atwell2003isle,
  title={The ISLE corpus: Italian and German spoken learner's English},
  author={Atwell, ES and Howarth, PA and Souter, DC},
  journal={ICAME Journal: Intl. Computer Archive of Modern and Medieval English Journal},
  volume={27},
  pages={5--18},
  year={2003},
  publisher={The HIT Centre-Humanities Information Technologies Research Programme}
}

@inproceedings{Minematsu2004PronunciationAB,
  title={Pronunciation assessment based upon the phonological distortions observed in language learners' utterances},
  author={N. Minematsu},
  booktitle={INTERSPEECH},
  year={2004}
}

@article{neri2008effectiveness,
  title={The effectiveness of computer assisted pronunciation training for foreign language learning by children},
  author={Neri, Ambra and Mich, Ornella and Gerosa, Matteo and Giuliani, Diego},
  journal={Computer Assisted Language Learning},
  volume={21},
  number={5},
  pages={393--408},
  year={2008},
  publisher={Taylor \& Francis}
}

@inproceedings{harrison2009implementation,
  title={Implementation of an extended recognition network for mispronunciation detection and diagnosis in computer-assisted pronunciation training},
  author={Harrison, Alissa M and Lo, Wai-Kit and Qian, Xiao-jun and Meng, Helen},
  booktitle={Intl. Workshop on Speech and Language Technology in Education},
  year={2009}
}

@inproceedings{qian2010capturing,
  title={Capturing L2 segmental mispronunciations with joint-sequence models in computer-aided pronunciation training (CAPT)},
  author={Qian, Xiaojun and Meng, Helen and Soong, Frank},
  booktitle={2010 7th Intl. Symposium on Chinese Spoken Language Processing},
  pages={84--88},
  year={2010},
  organization={IEEE}
}

@inproceedings{Hongyan2011,
  author    = {Hongyan Li and
               Shen Huang and
               Shijin Wang and
               Bo Xu},
  title     = {Context-Dependent Duration Modeling with Backoff Strategy and Look-Up
               Tables for Pronunciation Assessment and Mispronunciation Detection},
  booktitle = {{INTERSPEECH} 2011, 12th Annual Conference of the International Speech
               Communication Association, Florence, Italy, August 27-31, 2011},
  pages     = {1133--1136},
  publisher = {{ISCA}},
  year      = {2011}
}

@incollection{graves2012connectionist,
  title={Connectionist temporal classification},
  author={Graves, Alex},
  booktitle={Supervised Sequence Labelling with Recurrent Neural Networks},
  pages={61--93},
  year={2012},
  publisher={Springer}
}

@inproceedings{Lee2013PronunciationAV,
  title={Pronunciation assessment via a comparison-based system},
  author={Ann Lee and James R. Glass},
  booktitle={SLaTE},
  year={2013}
}

@inproceedings{sutskever2014sequence,
  title={Sequence to sequence learning with neural networks},
  author={Sutskever, Ilya and Vinyals, Oriol and Le, Quoc V},
  booktitle={Advances in neural information processing systems},
  pages={3104--3112},
  year={2014}
}

@article{chorowski2014end,
  title={End-to-end continuous speech recognition using attention-based recurrent NN: First results},
  author={Chorowski, Jan and Bahdanau, Dzmitry and Cho, Kyunghyun and Bengio, Yoshua},
  journal={arXiv preprint arXiv:1412.1602},
  year={2014}
}

@article{chen2015mxnet,
  title={Mxnet: A flexible and efficient machine learning library for heterogeneous distributed systems},
  author={Chen, Tianqi et al.},
  journal={arXiv preprint arXiv:1512.01274},
  year={2015}
}

@inproceedings{chorowski2015attention,
  title={Attention-based models for speech recognition},
  author={Chorowski, Jan K and Bahdanau, Dzmitry and Serdyuk, Dmitriy and Cho, Kyunghyun and Bengio, Yoshua},
  booktitle={Advances in neural information processing systems},
  pages={577--585},
  year={2015}
}

@inproceedings{nicolao2015automatic,
  title={Automatic assessment of English learner pronunciation using discriminative classifiers},
  author={Nicolao, Mauro and Beeston, Amy V and Hain, Thomas},
  booktitle={2015 IEEE Intl. Conference on Acoustics, Speech and Signal Processing (ICASSP)},
  pages={5351--5355},
  year={2015},
  organization={IEEE}
}

@inproceedings{bahdanau2016end,
  title={End-to-end attention-based large vocabulary speech recognition},
  author={Bahdanau, Dzmitry and Chorowski, Jan and Serdyuk, Dmitriy and Brakel, Philemon and Bengio, Yoshua},
  booktitle={2016 IEEE international conference on acoustics, speech and signal processing (ICASSP)},
  pages={4945--4949},
  year={2016},
  organization={IEEE}
}

@article{li2016mispronunciation,
  title={Mispronunciation detection and diagnosis in l2 English speech using multidistribution deep neural networks},
  author={Li, Kun and Qian, Xiaojun and Meng, Helen},
  journal={IEEE/ACM Transactions on Audio, Speech, and Language Processing},
  volume={25},
  number={1},
  pages={193--207},
  year={2016},
  publisher={IEEE}
}

@article{van2017neural,
  title={Neural discrete representation learning},
  author={Van Den Oord, Aaron and Vinyals, Oriol and others},
  journal={Advances in Neural Information Processing Systems},
  volume={30},
  pages={6306--6315},
  year={2017}
}

@article{kroll2017benefits,
  title={The benefits of multilingualism to the personal and professional development of residents of the US},
  author={Kroll, Judith F and Dussias, Paola E},
  journal={Foreign Language Annals},
  volume={50},
  number={2},
  pages={248--259},
  year={2017},
  publisher={Wiley Online Library}
}

@inproceedings{vaswani2017attention,
  title={Attention is all you need},
  author={Vaswani, Ashish and Shazeer, Noam and Parmar, Niki and Uszkoreit, Jakob and Jones, Llion and Gomez, Aidan N and Kaiser, {\L}ukasz and Polosukhin, Illia},
  booktitle={Advances in neural information processing systems},
  pages={5998--6008},
  year={2017}
}

@article{lorenzo2018towards,
  title={Towards achieving robust universal neural vocoding},
  author={Lorenzo-Trueba, Jaime and Drugman, Thomas and Latorre, Javier and Merritt, Thomas and Putrycz, Bartosz and Barra-Chicote, Roberto and Moinet, Alexis and Aggarwal, Vatsal},
  journal={arXiv preprint arXiv:1811.06292},
  year={2018}
}

@incollection{xiao2018paired,
  title={Paired phone-posteriors approach to esl pronunciation quality assessment},
  author={Xiao, Yujia and Soong, Frank K and Hu, Wenping},
  booktitle={bdl},
  volume={1},
  number={782d},
  pages={3},
  year={2018}
}

@article{chorowski2019unsupervised,
  title={Unsupervised speech representation learning using wavenet autoencoders},
  author={Chorowski, Jan and Weiss, Ron J and Bengio, Samy and van den Oord, A{\"a}ron},
  journal={IEEE/ACM transactions on audio, speech, and language processing},
  volume={27},
  number={12},
  pages={2041--2053},
  year={2019},
  publisher={IEEE}
}

@inproceedings{wang2019child,
  title={Child Speech Disorder Detection with Siamese Recurrent Network Using Speech Attribute Features.},
  author={Wang, Jiarui and Qin, Ying and Peng, Zhiyuan and Lee, Tan},
  booktitle={INTERSPEECH},
  pages={3885--3889},
  year={2019}
}

@article{jia2019direct,
  title={Direct speech-to-speech translation with a sequence-to-sequence model},
  author={Jia, Ye and Weiss, Ron J and Biadsy, Fadi and Macherey, Wolfgang and Johnson, Melvin and Chen, Zhifeng and Wu, Yonghui},
  journal={arXiv preprint arXiv:1904.06037},
  year={2019}
}

@inproceedings{sudhakara2019improved,
  title={An Improved Goodness of Pronunciation (GoP) Measure for Pronunciation Evaluation with DNN-HMM System Considering HMM Transition Probabilities.},
  author={Sudhakara, Sweekar and Ramanathi, Manoj Kumar and Yarra, Chiranjeevi and Ghosh, Prasanta Kumar},
  booktitle={INTERSPEECH},
  pages={954--958},
  year={2019}
}

@inproceedings{Zen2019,
  author={Zen, Heiga and Dang, Viet and Clark, Rob and Zhang, Yu and Weiss, Ron J and Jia, Ye and Chen, Zhifeng and Wu, Yonghui},
  title={{LibriTTS: A Corpus Derived from LibriSpeech for Text-to-Speech}},
  year=2019,
  booktitle={Proc. Interspeech 2019},
  pages={1526--1530},
  doi={10.21437/Interspeech.2019-2441}
}

@inproceedings{plantinga2019towards,
  title={Towards Real-Time Mispronunciation Detection in Kids' Speech},
  author={Plantinga, Peter and Fosler-Lussier, Eric},
  booktitle={2019 IEEE Automatic Speech Recognition and Understanding Workshop (ASRU)},
  pages={690--696},
  year={2019},
  organization={IEEE}
}

@inproceedings{Sudhakara2019NoiseRG,
  title={Noise robust goodness of pronunciation measures using teacher's utterance},
  author={Sweekar Sudhakara and M. K. Ramanathi and Chiranjeevi Yarra and A. Das and P. Ghosh},
  booktitle={SLaTE},
  year={2019}
}

@article{kobyzev2020normalizing,
  title={Normalizing flows: An introduction and review of current methods},
  author={Kobyzev, Ivan and Prince, Simon and Brubaker, Marcus},
  journal={IEEE Transactions on Pattern Analysis and Machine Intelligence},
  year={2020},
  publisher={IEEE}
}

@article{cheng2020asr,
  title={ASR-Free Pronunciation Assessment},
  author={Cheng, Sitong and Liu, Zhixin and Li, Lantian and Tang, Zhiyuan and Wang, Dong and Zheng, Thomas Fang},
  journal={arXiv preprint arXiv:2005.11902},
  year={2020}
}

@inproceedings{Weber2020,
  title={Constructing a Dataset of Speech Recordings with Lombard Effect},
  author={D. Weber and S. Zaporowski and D. Korzekwa},
  booktitle={24th IEEE SPA},
  year={2020},
  doi={10.23919/SPA50552.2020.9241266}
}

@article{tejedor2020assessing,
  title={Assessing pronunciation improvement in students of English using a controlled computer-assisted pronunciation tool},
  author={Tejedor-Garc{\'i}a, Cristian and Escudero, David and C{\'a}mara-Arenas, Enrique and Gonz{\'a}lez-Ferreras, C{\'e}sar and Carde{\~n}oso-Payo, Valent{\'i}n},
  journal={IEEE Transactions on Learning Technologies},
  year={2020},
  publisher={IEEE}
}

@article{guo2020gluoncv,
  title={GluonCV and GluonNLP: Deep Learning in Computer Vision and Natural Language Processing.},
  author={Guo, Jian and others},
  journal={Journal of Machine Learning Research},
  volume={21},
  number={23},
  pages={1--7},
  year={2020}
}

@article{erickson2020autogluon,
  title={AutoGluon-Tabular: Robust and Accurate AutoML for Structured Data},
  author={Erickson, Nick and Mueller, Jonas and Shirkov, Alexander and Zhang, Hang and Larroy, Pedro and Li, Mu and Smola, Alexander},
  journal={arXiv preprint arXiv:2003.06505},
  year={2020}
}

@article{zhang2021text,
  title={Text-conditioned Transformer for automatic pronunciation error detection},
  author={Zhang, Zhan and Wang, Yuehai and Yang, Jianyi},
  journal={Speech Communication},
  volume={130},
  pages={55--63},
  year={2021},
  publisher={Elsevier}
}
